%% file: thesis.tex
\documentclass[12pt,a4paper,twoside]{report}
\usepackage{mystyle,paper}
\input{epsf}              
\usepackage{latexsym}     
\begin{document}
\input{title.tex}   
\pagenumbering{roman}   
\tableofcontents  
\input{acknow-alt.tex}

\input{introduction.tex}

\input{chap1.tex}
\input{chap2.tex}

\input{chap3.tex}

\input{chap4.tex}

\input{chap5.tex}

\input{chap6.tex}
\input{appendix1.tex}

\input{ref.tex}
\end{document}

%% file: title.tex
\begin{titlepage}
\vspace{7cm}
\begin{center}
{\Huge\bf Topological Aspects of \\[.4cm] Quantum Gravity}\\[4cm]

{\Large\bf Morten Weis \\
Niels Bohr Institute\\University of Copenhagen}\\[2cm]

\begin{figure}[h]
\begin{center}
\mbox{
\epsfysize6cm
\epsffile{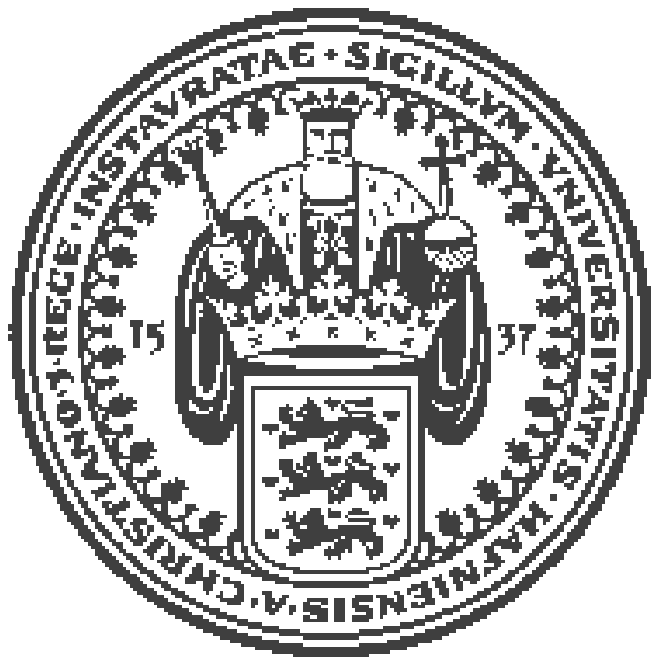}
}
\end{center}
\end{figure}
Thesis accepted for the Ph.D. degree in physics \\
at the Faculty of Science, University of Copenhagen.\\
{\bf December 1997}
\end{center}
\end{titlepage}

%% file: acknow-alt.tex
\chapter*{Acknowledgement}
\addtocontents{toc}{\protect\contentsline {chapter}{\protect\numberline {\ }Acknowledgement}{iv}}   

I would like to thank my supervisor Jan Ambj\o rn for many years
of support and guidance during my studies. Especially Jan's suggestion
that I should study the seminal paper by E. and H. Verlinde~\cite{VV} played
an important role for the last year and a half of my project. From this paper the
door opened into the fascinating world of Witten type topological field theories
and the underlying mathematical theories, which I have very much enjoyed studying.

\noindent
A warm thanks goes to
Martin G. Harris for numerous discussions on topological gravity and
for his help regarding the finer details of the English language when writing this thesis.
Both Jan and Martin are thanked for fruitful collaboration on the paper~\cite{AHW}.

\noindent
Several people have helped me with explaining and discussing topics relevant for 
this thesis. I would like to thank: J\o rgen E. Andersen (\AA rhus),
John Baez (Riverside), Mathias Blau (ENSLAP), Poul Henrik Damgaard (NBI),
William M. Goldman (Maryland), 
Nigel J. Hitchin (Cambridge), Lisa Jeffrey (McGill University),
 Kaj Roland (Nordita), Thomas Thiermann (Harvard) and Edward Witten (IAS). 

\noindent 
I thank J\o rgen E. Andersen, and the topology/geometry group at the
Department of Mathematics, \AA rhus University for inviting me to
present my project and for its kind hospitality.
I would also like to thank Abhay Ashtekar and the rest of the
Center for Geometry and
Gravitational Physics at Pennsylvania State University for  warm
hospitality during my two visits. This centre offers an extraordinarily
good working environment and I am happy at having spent four months
in this interesting and friendly place. 

\noindent
A warm thanks also goes to the staff and students here at the department
for theo\-re\-tical high energy physics at the Niels Bohr Institute. They have contributed
to creating a fun and stimulating environment, full of good discussion on all kinds
of relevant and non-relevant topics. 

\noindent
Especially I would like to thank: J\o rgen Rasmussen for
collaboration on~\cite{RW} and for many discussions of all kinds of exotic topics,
Kasper Olsen for his proof reading and discussions on topological
field theories, and Jakob Nielsen \& Juri Rolf for creating a good environment in our office. 
Finally I would like to thank Lars Jensen for nine years of friendship during our
common education and for help on the proof reading. Good luck in the
Oil industry Lars!

\noindent
This Ph.D. project was supported by grant 11-0801 from the Danish
Natural Science Research Council (SNF), to the project ``Analytical
Quantum Gravity and Knot theory''. I would like to use this opportunity
to thank the research council for its support through this project.
I also thank the administration of the Niels Bohr Institute for
supporting my project with several travel grants and 
excellent computer equipment.

\noindent
Finally I would like to thank my wife Anne and our daughter Rikke for
their love and support during my studies and for accepting that
I had to work day and night the last months instead of being home
building with Lego bricks and reading Donald Duck together with them.

%% file: introduction.tex
\setcounter{chapter}{0}
\chapter*{Introduction}
\addtocontents{toc}{\protect\contentsline {chapter}{\protect\numberline {\ }Introduction}{vi}}   

This Ph.D. thesis is based upon the project
{\it Analytical Quantum Gravity and Knot Theory}, which was intended to
pursue two goals:
\\[.5cm]
\noindent
1) To study several unanswered mathematical questions
in relation to the Ashtekar formulation of quantum gravity.
\\
\noindent
2) To investigate
the possible relation between the Ashtekar formulation and the formulation
used in two-dimensional quantum gravity, most prominently dynamical
triangulations and matrix models.

\noindent
Both of these topics have been investigated, and some progress has been
made in both areas. Work in relation to the first question has been made
by Rasmussen and the author~\cite{RW}, but since this work has no
direct connection to the main theme of this thesis, it is not discussed
here.

The second question turned out to be quite
difficult to answer in a direct manner, and it has evolved into the study
of the broader question:
{\it What is the role of gauge theoretical approaches to quantum gravity
in two and three dimensions, and how are they related to topological field
theories?}

The study of the latter question makes up most of this thesis. It is
a nontrivial question to understand why the theories of quantum gravity
in two and three dimensions, which we believe are correct, are described by
different types of topological field theories. In order to understand this,
we must study both topological field theories and quantum gravity, in the
formulations that are relevant for two and three dimensions.

The philosophy behind this thesis, has been to present a pedagogical and
detailed picture of the topics relevant for the project. Since the
major part of the research made in connection with this Ph.D. project has been 
to study and understand the involved theories, in order to investigate their
relations and differences, a large amount of space is given to the
introduction of these topics. 

Even though
there are no dynamical degrees of freedom in general relativity in three
dimensions, the quantum theory is known to be nontrivial. It is related
to knot theory through the work of Witten, formulating quantum gravity as a topological
gauge field theory of the Chern-Simons form. This theory is an example of a
wider class of topological gauge field theories, known as BF theories.
Since the Ashtekar approach to three dimensional quantum gravity is also
described by a BF theory, we reckon that this kind of topological field theory
presents the correct formulation of quantum gravity in three dimensions.

The situation is more problematic in two dimensions, when seen 
from the point of view of general relativity. The two-dimensional
Einstein-Hilbert action reduces to a topological term, namely the
Euler-Poincar\'{e} characteristic, due to the Gauss-Bonnet theorem, plus a
cosmological term involving the
cosmological constant. The only metrical dependence of the gravitational
path integral is through the cosmological term. The version of two-dimensional
quantum gravity that we consider to be correct, is the non-critical string theory
in two dimensions expressed as Liouville theory, dynamical triangulations or
as matrix models. We denote this class of theories $2D$ quantum gravity.

It is possible to formulate two different types of topological
gauge field theories in two dimensions, a Schwarz type BF theory and a
Witten type theory, where both of these define quantum theories of
gravity. Both theories are studied in detail in this thesis, and we
find that two-dimensional
BF theory fails to correspond to $2D$ quantum gravity, which instead
corresponds to the Witten type topological gravity. This is actually a special
supersymmetric version of BF theory,
based on a topological shift symmetry, which is not included in the
original BF theory.

It is discussed why it is different types of topological gauge theories which
describe quantum gravity in two and three dimensions. In addition
to this discussion, new results regarding the identification
between the Witten type theory of topological gravity and $2D$ quantum gravity are presented,
which have been a spinoff from the study of the gauge formulation of
two-dimensional quantum gravity.

\noindent
The organization of the material is as follows:
\\[.5cm]
\noindent
\underline{Chapter 1:}
covers a broad introduction to Riemann surfaces and the moduli space of Riemann
surfaces. Parts of this chapter are introductory, but there is also a more
technical discussion regarding the compactification of the moduli space of
Riemann surfaces, which is an important topic for the discussions in
chapters 4 and 5.

\noindent
\underline{Chapter 2:} introduces the mathematical background of gauge theories and
topological gauge field theories, especially Schwarz and Witten type theories.
As an example of a Witten type theory, we introduce four-dimensional
topological Yang-Mills theory and discuss the mathematical
interpretation of such Witten type theories. The mathematical theory is rather
involved, but essential
for the discussions of two-dimensional topological gravity in chapter 4.

\noindent
\underline{Chapter 3:} discusses the Jackiw-Teitelboim theory of two-dimensional
gravity and after an introduction to the first order formulation of Riemannian
geometry, we translate this theory into a BF topological gauge theory of the
Schwarz type. The quantum nature of this model is discussed in both canonical and
covariant formulations.

\noindent
\underline{Chapter 4:} introduces the Witten type theory of $2D$ topological gravity.
Various formulations of the theory are presented and the
definitions and the role of the observables and the topological invariants are discussed in great
detail. Some important relations between different observables are discussed
and a conjecture for the identification between
two different formulations of one special kind of observable in the theory is presented.

\noindent
\underline{Chapter 5:} introduces a perturbation of the theory in chapter 4.
This was studied in the paper~\cite{AHW} and the main results  
are presented together with additional comments. The
background for dynamical triangulations and matrix models is introduced in order to
discuss the identification made previously between $2D$ quantum gravity
and perturbed topological gravity. The new results regarding this
identification~\cite{AHW} are discussed in detail.

\noindent
\underline{Chapter 6:} includes discussions of the topics presented in the
first five chapters, together with additional material. We present
comparisons between the different theories and discuss why, how and if they are different.
Especially the role of Ashtekar
gravity in two dimensions, and the difference between BF theory and
super BF theory are investigated.
\\ \noindent
Final remarks and perspectives
conclude the main text of the thesis.

\noindent
\underline{Appendix 1:} 
 contains a list of conventions used in the main text.

%% file: chap1.tex
\setcounter{chapter}{0}
\chapter{Moduli Space of Riemann Surfaces}
\pagenumbering{arabic}
\section{Introduction}
The choice of space-time dimension is vital for the success of the theories of quantum gravity, which 
we study. While the lower dimensions are less interesting than four dimensions from a
physical point of view, two dimensions offers a unique mathematical situation. The usual Riemannian
space-time manifold can be replaced by a one-dimensional complex
manifold, known as a Riemann surface. This is not just any kind of Riemannian surface since 
its complex nature leads to a wide range of strong mathematical results. Much of the success of
string theory and also two dimensional topological gravity relies on the fact that
Riemann surfaces are such nice mathematical objects. In this chapter 
we give a short review of the theory of Riemann surfaces, and we present some facts regarding
the moduli space of Riemann surfaces and its compactification. This material is important
for the discussions on two dimensional topological gravity in chapter 4.

This chapter also serves to fix the notation for the main text, but since it contains no new results, 
the sections 1.2, 1.3, and 1.4 may be skipped by the experienced reader.
It is also intended to save the reader the trouble of referring to textbooks for the most common
definitions, but it is in no way a full treatment of the subject. We primarily used 
the following references~\cite{YC,Hat,GSW,LNP322,FK,SuperStringVol3}
for this chapter and will not give
references in the text unless the results are of a special nature, not
commonly given in the literature.

First we give the definitions of a Riemann surface and some of the results regarding
the uniformization theorem. Next follows a discussion on the moduli space of
Riemann surfaces, a short introduction to algebraic geometry and finally a treatment
of the Deligne-Mumford-Knudsen
compactification of moduli space.

\section{Definitions of Real and Complex Manifolds}

Complex manifolds are important for the study of
Riemann surfaces and we discuss the most important definitions and
relations with ordinary real manifolds. This
section also serves to fix the notation for the main text.
Let us first recall the usual definitions for real manifolds.

A topological manifold is a Hausdorff topological space such
that every point has a neighbourhood homeomorphic to $\R^{n}$.
A chart $(U,\phi)$ of a manifold $M$ is an open set $U\subset M$ called
the domain together with a homeomorphism $\phi: U\mapsto V$ of $U$ to a open set
$V$ in $\R^{n}$. A chart is also known as a local coordinate system, where
the local coordinates $(x_{1},\dots ,x_{n})$ are the image of
$\phi(x)\in \R^n$ for $x\in U\subset M$. An atlas of class $C^{k}$
on $M$ is a set of charts $\{ (U_{\alpha},\phi_{\alpha}) \}$ where the
set of domains $\{ U_{\alpha} \}$ must cover $M$ and where the
homeomorphisms must satisfy the following requirement: the maps
(transition functions) $\psi_{\beta\alpha} = \phi_{\beta}\circ
\phi_{\alpha}^{-1}:\phi_{\alpha}(U_{\alpha}\cap U_{\beta})
\mapsto \phi_{\beta}(U_{\alpha}\cap U_{\beta})$ must be maps of open sets
of $\R^n \mapsto \R^n$ of class $C^k$.

\noindent
To go from topology to geometry, the first step is to define a
differential manifold $M$, where the transition functions are
required to be of class $C^{\infty}$. Consider the tangent bundle $T(M)$
and assign to each tangent space $T_{x}M$ over $x\in M$, an inner product being 
a bilinear, symmetric, and positive definite functional $(\cdot,\cdot)$,
which maps two vectors to the real numbers. Such an assignment
is called a geometric structure on $M$. Denote the inner product
$(\cdot,\cdot)$ by $g$ and let $e^{i}$ be the basis vectors of $T_{x}M$.
Then $g$ will be differentiable if $ g (e^{i}(x),e^{j}(x)) $ is
a differentiable function on $\R$ in the usual sense.
The differentiable inner product $g$ is called the metric tensor
on $M$. Two atlases on a differentiable manifold are said to be compatible if
and only if their union $\{ (U_{\alpha}\cup
U_{\beta}, \phi_{\alpha}\cup \phi_{\beta}) \}$ again is a $C^{\infty}$ atlas.
This compatibility is an equivalence relation and the equivalence classes are
known as differentiable structures.

Consider a $2n$-dimensional topological manifold $M$ and construct an atlas
where the charts map open subsets of $M$ homeomorphically
to open subsets of $\C^{n}$. An analytic atlas is an atlas where all
the transition functions are holomorphic\footnote{We use the words
  holomorphic and analytic
  interchangeably.}. Two analytic atlases are
equivalent when their union remains an analytic atlas. An equivalence class of
analytic atlases is called a complex structure, and a
$2n$-dimensional real topological manifold with a complex structure is called an
$n$-dimensional complex manifold. 
A Riemann surface is a one-dimensional connected complex manifold.

Let us study a few details regarding complex manifolds in general. We require the
transition functions $\psi_{\beta\alpha} = \phi_{\beta}\circ
\phi_{\alpha}^{-1}$ to be  holomorphic maps of open sets $U \subset \C^n$
to open sets $V \subset \C^n$. Consider the coordinates in the two
open sets $(z_{i})\in U$ and $(w_{i})\in V$, then
$w_{j}=w_{j}(z_1,\dots,z_{n})$ for $j=1,2,\dots,n.$  We require that the
$w_{j}$'s are holomorphic functions of the $z_{i}$ coordinates and that
the holomorphic functional determinant
\begin{equation}
  \det \frac{\partial (w_{1},\dots,w_{n})}{\partial (z_{1}.\dots,z_{n})}
  \neq 0.
\end{equation}
Identify the map $\C^{n} \mapsto \R^{2n}$ by $z_{j}=x_{j} + i y_{j} \mapsto
(x_{j},y_{j})$. If we have a differentiable $2n$-dimensional manifold
$M$ with real coordinates organised in two pairs $(x_{j},y_{j}),(g_{j},h_{j})$
and we wanted to construct holomorphic coordinates
\begin{equation}
z_{j} = x_{j} + i y_{j} \,\,\, ; \,\,\, w_{k} = g_{k} + i h_{k},
\end{equation}
the real coordinates must satisfy the Cauchy-Riemann differential equations
\begin{equation}
  \frac{\partial g_{k}}{\partial x_{j}} = \frac{\partial h_{k}}{\partial
    y_{j}}
  \,\,\, ; \,\,\,
\frac{\partial h_{k}}{\partial x_{j}} = - \frac{\partial g_{k}}{\partial
  y_{j}},
\label{Cauchy-Riemann}
\end{equation}
where $j,k = 1,2,\dots,n$.

Orientability is an important and purely topological concept,
for understanding the relation between real and
complex manifolds. In
the case of real manifolds, we have the matrix
\begin{equation}
J_{ij} = \frac{\partial y_{i}}{\partial x^{j}}
;\,\,\, i,j = 1,2,\dots,n.
\end{equation}
If $J_{ij}>0$ for all $x\in U\cap V$, the transition functions $\psi_{ji}$
are said to be orientation preserving (holds also for $\psi_{ij}$).
A $C^{k}$ atlas is orientable if all transition functions are
orientation preserving. One can now state the following result:
\begin{description}
  \item{{\bf Theorem (1.1)}} Complex manifolds are always orientable.
  \item{{\bf Proof:}} Use the coordinates defined previously and denote
    by
\begin{equation}
  J_{hol} = \det \left( \frac{\partial w_{i}}{\partial z_{j}} \right),
\end{equation}
the holomorphic functional determinant. In the real case we have
\begin{equation}
  J = \det \left[ \begin{array}{cc} \partial w_{i}/\partial z_{j} &
      \partial w_{i}/\partial \overline{z}_{j} \\ 
\partial \overline{w}_{i}/\partial z_{j} &
      \partial \overline{w}_{i}/\partial \overline{z}_{j}
      \end{array} \right],
\end{equation}
  but
\begin{equation}
  \frac{\partial w_{i}}{\partial \overline{z}_{j}} =
  \frac{\partial \overline{w}_{i}}{\partial {z}_{j}} = 0,
\end{equation}
due to the holomorphic nature of the coordinates and we find that
\begin{eqnarray}
  J &=& \det \left( \frac{\partial w_{i}}{\partial z_{j}} \right)
  \det \left( \frac{\partial \overline{w}_{i}}{\partial
      \overline{z}_{j}} \right) 
  \det \left( \frac{\partial w_{i}}{\partial z_{j}} \right)
  \overline{\det \left( \frac{\partial w_{i}}{\partial z_{j}} \right)}
 \nonumber \\ &=& \mid J_{hol} \mid^{2} > 0.
\end{eqnarray}
\end{description}
This indicates an important result:
\begin{description}
\item{{\bf Theorem (1.2)}} A two-dimensional orientable and compact differentiable
  real manifold always admits a complex structure.
\item{{\bf Proof:}}
See~\cite{LNP322} for information on the proof.
\end{description}

\section{Riemann Surfaces: A Brief Introduction}

The theory of Riemann Surfaces is a beautiful and vast area of
mathematics, and we cover only the most important subjects in this section.
For use in the physical theories in the later chapters the
question of classification of Riemann surfaces is very important.
This is discussed in the section on moduli space, but first
an introduction to the uniformization theorem of Riemann surfaces
is needed. 
\subsection{Conformal Structures}
We can view a genus $g$ Riemann surface $\Sigma_{g}$ either as a real two-dimensional
oriented Riemannian
manifold or as a one-dimensional complex manifold. We assume that $\Sigma_{g}$ is compact and
boundary-less unless we specify otherwise. Let the real coordinates of $\Sigma_{g}$ be
$\sigma^{1},\sigma^{2}$ with a general metric of the form
\begin{equation}
  (ds)^{2} = g_{\alpha\beta}d\sigma^{\alpha}d\sigma^{\beta},\,\,\,\alpha,\beta = 1,2,
\end{equation}
where $g_{\alpha\beta}$ is the usual symmetric metric tensor. Riemann has shown that in
every local open subset $U_{i}$ in the covering $\{ U_{\alpha} \}$ of $\Sigma_{g}$, the
metric can be written in the form
\begin{equation}
  (ds)^{2} = \exp\left(\Phi(\sigma^{1},\sigma^{2})\right)
  \left( (d\sigma^{1})^{2} + (d\sigma^{2})^{2}\right),
\label{metric-isotherm-1}
\end{equation}
with $\Phi(\sigma^{1},\sigma^{2})$ being a differentiable function on $\Sigma_{g}$.
A coordinate system with a metric of this form is known as a system of isothermal coordinates.
One can change to complex coordinates
\begin{eqnarray}
  z &=& \sigma^{1} + i \sigma^{2}, \\
 \overline{z} &=& \sigma^{1} - i \sigma^{2},
\end{eqnarray}
which preserves the form of the metric in equation~(\ref{metric-isotherm-1})
\begin{equation}
 (ds)^{2} = \exp\Bigl(\Phi(z,\overline{z})\Bigr) \vert d z \vert^{2}.
\label{metric-isotherm-2}
\end{equation}
The function $\Phi(z,\overline{z})$ is real so $\Phi(z,\overline{z})=
\Phi(z,\overline{z})^{*}$, where $*$ denotes complex conjugation.

In general for an $n$-dimensional Riemannian manifold $M$, a diffeomorphism
\\ $f:M\mapsto M$ is a conformal transformation if it preserves
the metric up to a scale
\begin{equation}
 f: g_{\alpha\beta}(x) \mapsto e^{\Phi(x)}g_{\alpha\beta}(x),\,\,x\in M,
\end{equation}
where $\Phi(x)$ is a differentiable function on $M$.
The set of conformal transformations on $M$ forms a group known as the conformal group. If $g$ and
$\overline{g}$ are two different metrics on $M$, $\overline{g}$ is said to be conformally
related to $g$ if
\begin{equation}
  \overline{g}(x) = e^{\Phi(x)}g(x). \label{Weyl-1}
\end{equation}
This defines an equivalence relation on the set of metrics on $M$ under the transformation
$g(x)\mapsto \overline{g}(x)$ in equation~(\ref{Weyl-1}), known as a Weyl rescaling. An equivalence
class of metrics with regard to Weyl rescalings is known as a conformal structure and the set of
Weyl rescalings on $M$ form the
group ${\rm Weyl}(M)$. In two dimensions a special result holds. Consider a general local coordinate
transformation
$\Theta:(\sigma^{1},\sigma^{2})\mapsto (u(\sigma^{1},\sigma^{2}),v(\sigma^{1},\sigma^{2}))$,
with $u,v$ being differentiable functions on $\Sigma_{g}$, which
defines a bijective map from a region $U$ in the $(\sigma^{1},\sigma^{2})$ coordinate
plane to a region $V$ in the
$(u,v)$ plane, such that the first and second derivatives are continuous.
The coordinate transformation $\Theta$ is conformal when the complex coordinate $w=u+iv$,
is an analytic function of the complex coordinate $z=\sigma^{1}+i\sigma^{2}$ on $U$, with non-vanishing
$dw/dz$ on $U$. In this situation the
metric does not change its form and we see that a conformal structure is equivalent to a
complex structure in two dimensions.
The space of inequivalent complex structures on $\Sigma_{g}$ is known as the moduli space of
Riemann surfaces, which we discuss in detail in a later section.

It is interesting to classify the different types of Riemann surfaces. There exists a
topological classification of two-dimensional manifolds given by the genus of
the manifold. It is special for two dimensions, that one has a perfect topological
classification. In three dimensions there is no proof of a perfect classification and
in four dimensions it has been proved by Markov that such a classification can not exist. 

If two manifolds $M,N$ have the same topology, then there exists a homeomorphism
$\phi:M\mapsto N$, such that $M\simeq N$ are topological isomorphic. 
Recall that a homeomorphism is a bijection $\phi$ which is bicontinuous
(i.e. $\phi$ and $\phi^{-1}$ are continuous). A homeomorphism maps images and
inverse images of open sets to open sets. The existence of a homeomorphism between two
topological spaces is an equivalence relation. If we now equip $M,N$ with geometric
structures, where we require the topological isomorphism to respect these structures,
one can ask whether there exist a differentiable isomorphism $\phi$ between $M$ and $N$.
Let $x\in M$ and let $y=\phi(x)\in N$ be the image of $x$ in $N$, and consider
coordinate patches around $x$ and $y$
\begin{eqnarray}
  (U,\psi): \,&x\in U,& \\
  (V,\psi'): \, &y\in V.&
\end{eqnarray}
Define the function
\begin{equation}
  \phi_{UV}\equiv \psi' \circ \phi \circ \psi^{-1} \left\vert \,\, \phi_{UV}: \psi\left(
    \phi^{-1}(V)\cap U \right)\subset \R^{n} \mapsto \psi'(V)\subset \R^{n}, \right.
\end{equation}
where we say that $\phi$ is a differentiable map if $\phi_{UV}$ is differentiable for all
possible choices made, and further more that $\phi$ is a diffeomorphism if both
$\phi$ and $\phi^{-1}$ are differentiable maps.
The manifolds $M,N$ are said to be diffeomorphic if there exists a diffeomorphism
mapping the one to the other. Homeomorphic manifolds 
can be deformed continuously into each other and for diffeomorphic manifolds the
deformation is smooth. 

For two-dimensional compact and orientated manifolds there is only one differentiable structure per
genus, up to diffeomorphisms. This is not true in higher dimensions, where there exist
some exotic examples regarding higher spheres.
On the topological manifold $S^{7}$ there exist 28 inequivalent differentiable
structures and on $S^{11}$ no less than 991~\cite{LNP322,Birmingham-review}.
For the real spaces $\R^{n}$ with $n\neq 4$ there exists a unique differentiable
structure, while there exists a continuous family of non-equivalent differentiable
structures on $\R^{4}$. This has been proven by use of the Donaldson invariants,
which are discussed in chapter 2.

Two complex manifolds $M,N$ are said to be analytical isomorphic or analytical equivalent
if there exists a holomorphic homeomorphism between them.
If there exists a holomorphic homeomorphism between two Riemann
surfaces, the two surfaces are said to be conformally equivalent, since the conformal
mapping exactly is a holomorphic homeomorphism. 
The number of conformal equivalence classes depends on the genus as we will see in the
section on the moduli space of Riemann surfaces. This being the space of equivalence classes
of complex structures.
\subsection{Uniformization Theorem}
There exists an amazing result in the theory of Riemann surfaces, known
as the uniformization theorem. This states that there are only three types
of conformal inequivalent, simply connected\footnote{A topological space
  $X$ is simply connected if every loop can be continuously shrunk to a point.}
 Riemann surfaces. These are:
\begin{description}
  \item{(1)} The extended complex plane $\C \cup \{ \infty \}$, which has
  the topology of the two-sphere $S^{2}$. This correspondence is due to the
  stereographic projection, or one-point compactification of $\C$. This
  Riemann surface is also called the Riemann sphere $\Sigma_{0}$ or
  \Ci.
\item{(2)} The complex plane \C.
\item{(3)} The upper complex half plane
  $\Hu =\left\{ z\in \C \vert {\rm Im}(z)>0\right\}$.
\end{description}
The uniformization theorem also covers those Riemann surfaces which are
not simply connected, namely for any such surface $\Sigma_{g}$ the universal
covering space $\tilde{\Sigma}$ will be one of the following three Riemann surfaces $(\Ci,\C,\Hu)$.
The universal covering space is a principal fibre bundle with structure group $\pi_{1}(\Sigma_{g})$
\begin{eqnarray}
 \pi_{1}(\Sigma_{g}) \mapsto  &\tilde{\Sigma}& \nonumber \\ &\downarrow& \\
 &\Sigma_{g}&\equiv \tilde{\Sigma}/\pi_{1}(\Sigma_{g}).
\end{eqnarray}
We discuss fibre bundles in more detail in chapter 2.  The uniformization theorem states that
one and only one of the following statements can, and will be, true 
\begin{eqnarray}
  \Sigma_{g} &=& \Ci/ \pi_{1}(\Sigma_{g}), \label{genus-pi-0} \\
  \Sigma_{g} &=& \C/  \pi_{1}(\Sigma_{g}), \label{genus-pi-1}  \\
  \Sigma_{g} &=&   \Hu/ \pi_{1}(\Sigma_{g}). \label{genus-pi-2}
\end{eqnarray}
The three canonical Riemann surfaces $(\Ci,\C,\Hu)$ play a vital role and we 
discuss some of their important properties.
From the definition of the universal covering space, it is clear that
$\pi_{1}(\tilde{\Sigma})={\bf 1}$ is
trivial. This follows from $\tilde{\Sigma}$ being simply connected.
So for any surface $\Sigma_{g}$ the action of $\pi_{1}(\tilde{\Sigma})$ will be
trivial and hence free. Recall that the action of a group $G$ with
elements $g$, is free on a space $X$ if for any point $x\in X$,
$g\circ x = x$ if and only if $g=e$ is the identity.
If this was not true for a point, say $y\in X$, this
is said to be a fixed point for the group.
In order to preserve the complex structure on the covering space, one
must require the elements $\phi\in \pi_{1}({\Sigma_{g}})$,
to be analytic automorphisms on $\tilde{\Sigma}$.

It is possible to study the classification of Riemann surfaces by
classifying the representations of $\pi_{1}(\Sigma_{g})$. This is done by
investigating the fixed point free subgroups of the automorphism
group, respectively for each of the three canonical Riemann surfaces.

\begin{figure}[h]
\begin{center}
\mbox{
\epsfysize=6cm
\epsffile{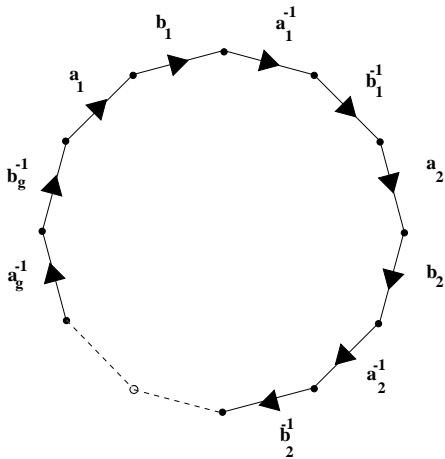}
}
\end{center}
\caption{\label{4g-polygon} A $4g$-polygon.}
\end{figure}

\begin{figure}[h]
\begin{center}
\mbox{
\epsfysize=2cm
\epsffile{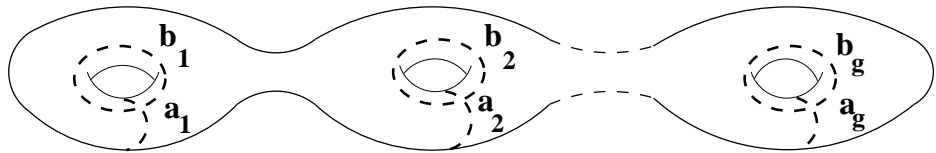}
}
\end{center}
\caption{\label{fig-genus-g}A Riemann surface of genus $g$.}
\end{figure}

We use the standard description of a genus $g$ Riemann surface $\Sigma_{g}$
as a $4g$-polygon (see figure~(\ref{4g-polygon}))
where each pair of identified sides corresponds to a homotopically 
nontrivial loop on $\Sigma_{g}$. The polygon edges are identified 
as the loops $(a_{i},b_{i}$) on figure~(\ref{fig-genus-g}), which
all for $i=1,\dots,g$ are homotopically nontrivial.

\noindent
There exists a homotopically trivial curve $\gamma$ which represents the boundary
of the $4g$-polygon
\begin{equation}
  \gamma = a_{1}b_{1}a_{1}^{-1}b_{1}^{-1}a_{2}b_{2}a_{2}^{-1}b_{2}^{-1}\dots
  a_{g}b_{g}a_{g}^{-1}b_{g}^{-1}. \label{homotopy-relation}
\end{equation}
The curve is trivial because it never encloses the nontrivial cycles, so we
can shrink $\gamma$ continuously to an interior point of the surface. The
fundamental group $\pi_{1}(\Sigma_{g})$ has $2g$
generators $\Gamma(a_{i}),\Gamma(b_{i})$ constrained by the one relation
\begin{equation}
  \prod_{i=1}^{g} \left( \Gamma(a_{i}) \Gamma(b_{i})  \Gamma^{-1}(a_{i})
    \Gamma^{-1}(b_{i}) \right) = {\bf 1}. \label{homotopy-relation-1}
\end{equation}
For the three canonical Riemann surfaces, we should represent this
group by fixed point free elements of their automorphism groups.
The fundamental group $\pi_{1}(\Sigma_{g})$ also has an interesting
classification due to the genus $g$  of the surface. It comes in three
distinct classes: For $g=0$, all curves are homotopic trivial 
$ \pi_{1}(\Sigma_{0})=0$. For $g=1$, the fundamental group
has two generators $\Gamma(a),\Gamma(b)$. Equation~(\ref{homotopy-relation-1})
reduces to the statement that the group is abelian $\Gamma(a)\Gamma(b)=
\Gamma(b)\Gamma(a)$. For higher genus $g\geq 2$,  $\pi_{1}(\Sigma_{g})$
will be non-abelian. The division into three situations is tied to the
three canonical Riemann surfaces. In the next section we show that for a non-simply
connected Riemann surface $\Sigma_{g}$, the universal covering space
is $\Ci$ for $g=0$, $\C$ for $g=1$ and $\Hu$ for $g\geq 2$.
We therefore now consider the three canonical Riemann surfaces
and find their automorphism groups.

\subsection{The Canonical Riemann Surfaces}

The Riemann sphere, or the extended complex plane as it is also called, is
the only compact surface of the three canonical Riemann surfaces. Hence
it is clear that there can not exist an analytic homeomorphism from this
to the non-compact surfaces \C, or \Hu, so it is conformally inequivalent to
these. The sphere is related to the extended plane by the usual sterographic
projection, where the equator is mapped to the unit circle in $\C$, the
south pole to the origin of \C, and the north pole to infinity. In other words
\Ci is the one point compactification of \C, where we have added the point
$\{ \infty \} $.
It can be shown that all analytic functions on \Ci are rational functions
which can be represented as the quotient of two polynomials such that
$f(z) = p(z)/q(z)$ where $q(z)$ is not identically zero.

We are interested in the automorphisms on \Ci, i.e. analytic
homeomorphisms mapping \Ci to itself. We assume that the polynomials
$p(z),q(z)$ do not have common factors, since they would cancel in any case.
So the analytic homeomorphisms will also be rational functions, but
since they must be bijective in order to be homeomorphisms, the
maximal degree of these polynomials will be one. This is due to the fact
that a complex polynomial equation of order $n$ has $n$ solutions and hence it
will not be bijective for $n>1$. So the form of a general automorphism
on \Ci is
\begin{equation}
  T(z) = \frac{az+b}{cz +d}\,;\,\,\, a,b,c,d\in \C,
\end{equation}
supplemented with the restriction $ad-cb\neq0$ ensuring that there are no
common factors in $p(z)$ and $q(z)$.
It is clear that $(a,b,c,d)\mapsto (\alpha a,\alpha b,\alpha c ,\alpha d)$, $\alpha\in \C$, $\alpha\neq 0$
defines the same transformation.
The transformations of the form $T(z)$ are called
M\"{o}bius transformations. These form a group, the automorphism group
$Aut(\Ci)$ under the composition
\begin{equation}
\left(\frac{az+b}{cz +d}\right)\cdot\left(\frac{ez+f}{gz+h}\right)=
  \frac{(ae+bg)z+(af+bh)}{(ce+dg)z+
  (cf+dh)}. \label{T-composition}
\end{equation}
The identity element is $T(z)=z$ and the inverse element
\begin{equation}
  T^{-1}(z) = \frac{dz-b}{-cz +a}. \label{T-inverse}
\end{equation}
The M\"{o}bius transformation can be written in matrix form as
\begin{equation}
  M(T) = \left( \begin{array}{cc} a & b \\ c& d\end{array} \right),
  \label{Mobius-matrix}
\end{equation}
and the composition defined in equation~(\ref{T-composition}) is now
just matrix multiplication. This also clearly explains the form of the
inverse transformation in equation~(\ref{T-inverse}).
The matrices of this form, with non-vanishing determinant (i.e. the requirement
$ad-cb\neq 0$), generate the general linear group $GL(2,\C)$. But due to the comment
above, we note that $\alpha M(T)$ defines the
same action as $M(T)$, which implies that $GL(2,\C)\neq Aut(\Ci)$.
Let $M:GL(2,\C)\mapsto Aut(\Ci)$ be a group homomorphism and denote by
$\Lambda$ those elements of $GL(2,\C)$ which are mapped to the identity
element of $Aut(\Ci)$. These elements are of the form $a=d=\lambda\neq 0$
and $b,c=0$. The automorphism group (or M\"{o}bius group) is then (almost)
\begin{equation}
  Aut(\Ci) \sim GL(2,\C)/\Lambda \equiv SL(2,\C).
\end{equation}
By setting
$\lambda^{2}=1/{\rm det} M$, the element $\lambda M(T)\in GL(2,\C)$ will
have unit determinant. But since the scaling factor is quadratic, we
also have the $\Z_{2}$ invariance under $(a,b,c,d)\mapsto(-a,-b,-c,-d)$,
which we factor out to obtain the true automorphism group
\begin{equation}
  Aut(\Ci) = SL(2,\C)/\Z_{2} \equiv PSL(2,\C). \label{psl2z-def}
\end{equation}
We conclude that the special projective group is the automorphism group
of the extended complex plane. In the literature one often ignores the
distinction between $PSL(2,\C)$ and $SL(2,\C)$.

We know that the metric can always be written in isothermal coordinates
\begin{equation}
  (ds)^{2} = \exp \Bigl( \Phi (z,\bar{z}) \Bigr) \mid dz \mid^{2}.
\end{equation}
By choosing the correct
function $\Phi$ one can obtain a metric on each of the canonical Riemann surfaces.
As a general result the scalar curvature~\cite{SuperStringVol3}
is 
\begin{equation}
  R = - e^{-\Phi(z,\bar{z})} \partial_{z} \partial_{\bar{z}} \Phi(z,\bar{z}).
\end{equation}
The correct functions for the canonical surfaces are
\begin{eqnarray}
  \Ci: & \Phi(z,\bar{z})_{+}&= \ln \left( 1 + \mid z \mid^{2}\right), \\
 \C:\,\, & \Phi(z,\bar{z})_{0}&= 0, \\
 \Hu: & \Phi(z,\bar{z})_{-}&= -2\ln \left(2 \, {\rm Im}(z)\right) .
\end{eqnarray}
We list the three corresponding metrics and their scalar curvature:
\begin{equation}
\begin{array}{||l|l|l||} \hline  & & \\
\Ci & (ds)^{2}_{+} =
\frac{\mid  dz\mid^{2}}{\left( 1 + \mid z \mid^{2}\right)^{2}} & R_{+}=2
\\[.5cm] 
\hline &&\\
\C & (ds)^{2}_{0} = \mid dz \mid^{2} & R_{0}=0 \\[.5cm]  \hline && \\
\Hu & (ds)^{2}_{-} = \frac{1}{4}
                    \frac{\mid  dz\mid^{2}}{\left({\rm Im}(z)\right)^{2}}
                    & R_{-}=-2 \\[.5cm]  \hline 
\end{array} \label{uniform-metric}
\end{equation}
Often one changes the normalization such that the curvature is $\pm1$ or $0$.
It can be shown~\cite{SuperStringVol3} that the
isometry groups for these three metrics are
$SU(2)$ for $\C_{\infty}$, $ISO(2)$ for \C, and $SU(1,1)$ for \Hu. These are
different 3-parameter subgroups of the 6-parameter group $PSL(2,\C)$. The matrices
in equation~(\ref{Mobius-matrix}) with unit determinant form the
special linear group $SL(2,\C)$ and if one then restrict the elements
to fulfill $d = \bar{a}$ and $c = - \bar{b}$, we obtain the special unitary
group $SU(2)\subset SL(2,\C)$. This is the maximal compact subgroup of
$SL(2,\C)$. The non-compact subgroup $SU(1,1)=SL(2,\R)$ is the restriction
of $SL(2,\C)$ by demanding all matrix elements to be real. 
The subgroup $ISO(2)\subset SL(2,\C)$ is spanned by the matrices of
the form
\begin{equation}
\left(
\begin{array}{cc}
e^{i\theta} & b \\ 0 & e^{-i\theta} \end{array} \right) ,
\end{equation}
where $b\in \C$ and $\theta\in \R$. For $ISO(2)$ the corresponding
M\"{o}bius transformation maps $z \mapsto ze^{i\theta} + \beta$ which is a
rotation followed by a translation. But these are not all M\"{o}bius
transformations on the complex plane, there are also the dilatations
$z\mapsto\lambda z$ where $\lambda \in \R$. So ISO(2) is not the automorphism
group of $\C$, this is instead
\begin{equation}
  Aut(\C) = ISO(2) \otimes \R.
\end{equation}
While the isometry groups for \Ci, and \C, are subsets of the automorphism
groups, it turns out that the isometry group and automorphism group for \Hu
are identical
\begin{equation}
  Aut(\Hu) = SU(1,1)=SL(2,\R).
\end{equation}

For the Riemann sphere \Ci, the automorphism group $PSL(2,\C)$ has three free complex
parameters due to the requirement $ad-bc=1$.
One can always fix these three parameters~\cite{SuperStringVol3}  in
the M\"{o}bius transformation, so there are no fixed point free subgroups for
$Aut(\Ci)$. Only the trivial homotopy group $\pi_{1}(\Sigma_{0})$ for a
genus zero Riemann surface $\Sigma_{0}$ can be represented in $Aut(\Ci)$ and
hence \Ci is the universal covering space for all genus zero Riemann surfaces.

For the complex plane \C, the automorphism group was $ISO(2)\otimes\R$. There is
a fixed point for the $ISO(2)$ rotation/dilatation part
\begin{equation}
  z\mapsto ze^{i\theta}\lambda, \,\, \mbox{ fixed point at } z=0,
\end{equation}
but not for the translation
\begin{equation}
  z\mapsto z + \beta, \,\, \mbox{ no fixed points}.
\end{equation}
So we can only represent $\pi_{1}(\C)$ in $ISO(2)\otimes\R$ by the translations, but
since the translation group is abelian it can only be $\pi_{1}(\Sigma_{1})$ for genus
one Riemann surfaces which will do the job. We thus identify \C$\,$ as the universal
covering space for all genus one Riemann surfaces.

\begin{figure}[h]
\begin{center}
\mbox{
\epsfysize=6cm
\epsffile{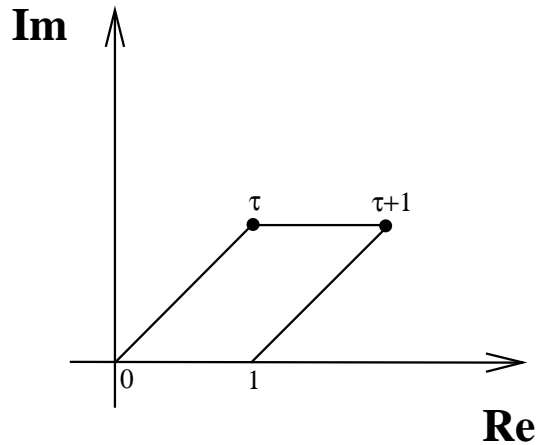}
}
\end{center}
\caption{\label{torus-1}The fundamental region of the lattice $(1,\tau)$.}
\end{figure}

\noindent
The fundamental group for a genus one surface $\Sigma_{1}$ is generated
by two elements $(n,m)$, which also generate the translation group 
\begin{equation}
  \left\{t\in\pi_{1}(\Sigma_{1}) \vert \, t: z\mapsto z + n\omega_{1} + m\omega_{2}\right\};\,\,
  n,m \in \Z,\,\,\,\,\, \omega_{1},\omega_{2}\in\C.
\end{equation}
The ratio ${\rm Im}(\omega_{1}/\omega_{2})\neq 0$ per definition.
For every choice of $(\omega_{1},\omega_{2})$, we define a genus one Riemann surface
as the equivalence class of points in \C, under the relation
\begin{equation}
  \hat{z} \sim z \,\, \mbox{ if } \,\, \hat{z} = z +  n\omega_{1} +
  m\omega_{2}\,; \, \,\,n,m \in \Z.
\end{equation}
This relation defines a lattice $\Z\times \Z$ and the corresponding torus is conformally
equivalent to the torus defined by a lattice with basis $(1,\tau)$, where the
ratio
\begin{equation}
  \tau=\omega_{2}/\omega_{1},
\end{equation}
has a positive imaginary part. Hence we identify
$\tau$ as an element of the complex upper half plane \Hu. In
figure~(\ref{torus-1}) we show the fundamental region of this lattice. 
By identifying the opposite sides of the fundamental region, we get a torus with vertices
$(0,1,\tau,\tau+1)$. The number $\tau$ is known as the moduli. But two different
moduli parameters can define the same lattice. Every lattice will have
infinitely many moduli parameters, namely one for each different choice of
basis $(\omega_{1},\omega_{2})$. Let $(\omega_{1},\omega_{2},\tau)$ and
$(q_{1},q_{2},\hat{\tau})$ be two different bases with their moduli. They define the
same lattice if they are related by
\begin{equation}
  \hat{\tau} = \frac{a\tau + b}{c\tau + d}\, , \,\, a,b,c,d\in\Z\,;\, \, ad-bc=1 \mbox{ such that }
{\rm Im}(\hat{\tau})>0.
\end{equation}
The collection of all such transformations form the modular group
$\Gamma=PSL(2,\Z)$. If $\tau$ and $\hat{\tau}$ are related by a modular transformation
the lattices $(1,\tau)$ and $(1,\hat{\tau})$ will be conformally equivalent. This shows
that the moduli space of a genus one Riemann surface must be of the form
\begin{equation}
  {\cal M}_{1} = \Hu_{\tau}/\Gamma, \label{moduli-genus-one-1}
\end{equation}
being the set of all complex structures which can be imposed on a compact genus
one Riemann surface. By $\Hu_{\tau}$ we mean the complex upper plane defined by the moduli parameter $\tau$. The modular group $\Gamma$ is a discrete subgroup of $PSL(2,\R)$
and we shall see that the elements are related to global diffeomorphisms
on $\Sigma_{1}$. The metric on $\Sigma_{1}$ can be written in isothermal coordinates,
but the function $\Phi(\sigma_{1},\sigma_{2})$, from equation~(\ref{metric-isotherm-1}),
must be periodic
\begin{equation}
  \Phi(\sigma^{1} + m + n\tau_{1}, \sigma^{2} + n\tau_{2}) =
  \Phi\left(\sigma^{1},\sigma^{2}\right), \, \tau = \tau_{1} + i \tau_{2};\, m,n\in\Z.
\end{equation}
In this representation all moduli dependence of the metric has been collected in
the conformal factor,
but this is not a universal result. Consider the following coordinate transformation
\begin{eqnarray}
  \sigma^{1} &\mapsto & \sigma^{1} + \tau_{1}\sigma^{2}, \label{diff-tau-1} \\
  \sigma^{2} &\mapsto& \tau_{2}\sigma^{2},               \label{diff-tau-2}
\end{eqnarray}
mapping the unit square of the $(\sigma^{1},\sigma^{2})$ coordinate system into
the fundamental lattice $(1,\tau)$. This maps the isothermal metric into the form
\begin{equation}
  (ds)^{2} = \exp\left(\Phi(\sigma^{1},\sigma^{2})\right) \vert d\sigma^{1} +
  \tau d\sigma^{2}\vert^{2}. \label{metric-torus-2}
\end{equation}
This is an example of a general result for Riemann surfaces.
One can always write an arbitrary metric as
\begin{equation}
  (ds)^{2} = \lambda (z,\overline{z})\mid dz + \mu \, d\overline{z} \mid^{2},
  \label{arbitrary_metric}
\end{equation}
where one requires $\mid \lambda \mid >0, \mid \mu \mid <1$ to obtain
positivity of the metric. Since $\lambda$ is always positive, we write it as the
exponential of $\Phi(z,\overline{z})$. This general metric will acquire the isothermal form
if we can find a diffeomorphism $w=f(z,\overline{z})$ that maps the
$(z,\overline{z})$-coordinates into
an isothermal coordinate system. The requirement for such a diffeomorphism is
that it should satisfy the Beltrami equation:
\begin{equation}
  \partial_{\overline{z}} f = \mu \, \partial_{z}f. \label{Beltrami}
\end{equation}
We use the standard definitions
\begin{eqnarray}
  \partial_{z} &=& \frac{1}{2}(\partial_{x} - i \partial_{y}), \\
\partial_{\overline{z}} &=& \frac{1}{2}(\partial_{x} + i \partial_{y}),
\end{eqnarray}
from which we see that if $w=f(z,\overline{z})$ satisfies equation~(\ref{Beltrami})
one can write
\begin{equation}
  \mid dw \mid^{2} = \mid \partial_{z} w \, dz +
  \partial_{\overline{z}}w \,d
  \overline{z} \mid^{2} = \mid \partial_{z}w\mid^{2}\, \mid dz +
  \mu \, d\overline{z} \mid^{2}.
\end{equation}
This shows that the diffeomorphism $f$ maps the arbitrary metric in
equation~(\ref{arbitrary_metric}) into the form of equation~(\ref{metric-isotherm-2})
\begin{equation}
  f: \lambda(z)\mid dz + \mu \, d\overline{z} \mid^{2} =
     \mid \partial_{z} w(z) \mid^{2} \mid dz + \mu \,d\overline{z} \mid^{2}
     \mapsto \mid dw \mid^{2},
\end{equation}
where $\lambda^{'}(w)=1$.
Locally one can find solutions to the Beltrami equation, provided certain
conditions are fulfilled, for example that the metric should be 
real-analytic. For details see~\cite{Nag}.

\begin{figure}[h]
\begin{center}
\mbox{
\epsfysize=5cm
\epsffile{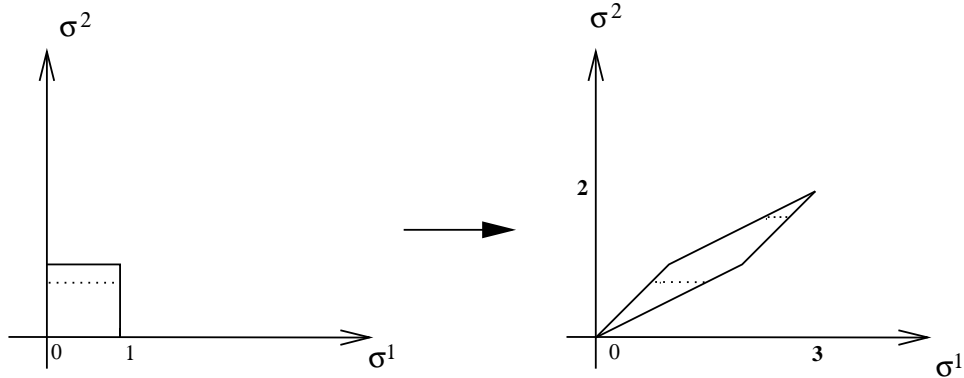}
}
\end{center}
\caption{\label{fig:torus-2}  An example of the map in equations~(\ref{sigma-map-1},\ref{sigma-map-2}).}
\end{figure}

\begin{figure}[h]
\begin{center}
\mbox{
\epsfysize=4cm
\epsffile{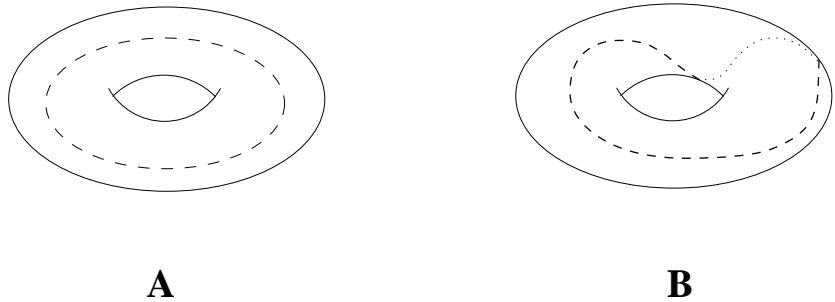}
}
\end{center}
\caption{\label{fig:torus-3}The resulting tori with the constant $\sigma^{2}$ lines.}
\end{figure}

The diffeomorphisms in equations~(\ref{diff-tau-1},\ref{diff-tau-2}) are exactly mapping the
isothermal metric into the form in equation~(\ref{arbitrary_metric})
and the Beltrami coefficient is
\begin{equation}
  \mu =\frac{1 + i \tau}{1 - i \tau}.
\end{equation}
The modular transformations are exactly those diffeomorphisms which preserve the
metric in equation~(\ref{metric-torus-2}), i.e. they should be linear in
$(\sigma^{1},\sigma^{2})$
\begin{eqnarray}
  \sigma^{1} &\mapsto & a\sigma^{1} + b\sigma^{2}, \label{sigma-map-1}\\
  \sigma^{2} &\mapsto & c\sigma^{1} + d\sigma^{2}, \label{sigma-map-2}
\end{eqnarray}
as well they should preserve the periodicity of the metric. Hence they should 
map the vertices of the unit square to the vertices of the fundamental region of the
lattice $(1,\tau)$, which forces $(a,b,c,d)$ to be integers and since the diffeomorphism
should be orientation preserving it follows that $ad-bc=1$. We see that these
diffeomorphisms define
modular transformations, and they are not homotopic to the identity.
Consider the example given in figure~(\ref{fig:torus-2}) where the unit square is mapped
into the lattice by the modular transformation given by $(a=2),(b=c=d=1)$ and
where the dotted line indicates a constant $\sigma^{2}$ value.
Identifying the opposite edges of the square and parallelogram respectively,
results in two tori,
as illustrated in figure~(\ref{fig:torus-3}), where we see from the indicated constant
$\sigma^{2}$ lines, that the diffeomorphisms mapping the torus (A) into the torus
(B) can not be homotopic to the identity.

\noindent
Hence we can identify the modular transformations with the mapping class group (defined below) on
$\Sigma_{1}$, 
\begin{equation}
  {\rm MCG}_{1} \equiv \frac{{\rm Diff}(\Sigma_{1})}{{\rm Diff}_{0}(\Sigma_{1})} 
  \simeq PSL(2,\Z),
\end{equation}
where ${\rm Diff}_{0}(\Sigma_{1})$ are the diffeomorphisms connected to the identity.
The space of inequivalent complex structures from equation~(\ref{moduli-genus-one-1}),  
can also be represented as
\begin{equation}
  {\cal M}_{1} = \Hu_{\tau}/{\rm MCG}_{1}. \label{moduli-genus-one-2}
\end{equation}
This show that $\Hu_{\tau}$, is the universal covering space for ${\cal M}_{1}$ with
covering group $PSL(2,\Z)$.

By discussions similar to that of genus one surfaces, one can prove that
\Hu is the universal covering space of all Riemann surfaces with genus $g\geq 2$, with the
covering group being the fundamental group of the surface.
We now turn to a more general discussion of the moduli spaces of Riemann surfaces.

\section{The Moduli Space of Riemann Surfaces}

The space of all conformal equivalence classes of Riemann surfaces
is known as the moduli space \m of a genus $g$ Riemann surface.
This space is very important for the study of gravity in two and three dimensions and
for string theory. One reason for this is that in those theories the path integral
involves an integration over all surfaces. The action in critical string theory is
conformally invariant and the theory reduces to an integral over moduli space. In gravity 
the moduli space is also important for reasons we will discuss in the following chapters.
We have seen an example of the moduli space of genus one Riemann surfaces and we
now extend this to a discussion of the moduli space for all genera.
\noindent
Denote by \met the set of all metrics on the Riemann surface $\Sigma_{g}$ with genus $g$.
We have seen that every metric defines a complex structure, so obviously
the set of all inequivalent complex structures \m will be a subset of \met.
But there is no reason to believe that every
metric should define a unique complex structure, so we must study how the
inequivalent complex structures can be identified.
We have learned that the Weyl transformations do not change the
complex structure and the moduli space \m must be a subset of $\met / \w$.
The group of all diffeomorphisms
mapping the surface to itself is denoted \diff. There are diffeomorphisms acting
on the metric, which preserve the complex structure and these should be factored out
to form an equivalence class of complex structures.
Thus we can conclude that the moduli space can be represented as\footnote{Warning! This is
  not quite true for genus zero and one. The explanation is given on page~\pageref{semi-direct}ff.}
\begin{equation}
  \m \sim \frac{\met}{\w \times \diff},
\label{moduli-1}
\end{equation}
characterising the metric transformations which change the complex
structure. 
The result in equation~(\ref{moduli-genus-one-2}) generalizes to all genera and the covering space to moduli space is known as Teichm\"{u}ller space \tich
\begin{equation}
  \m = \frac{\tich}{{\rm MCG}_{g}}, \label{moduli-teich-mpg}
\end{equation}
where the Teichm\"{u}ller space can be represented as
\begin{equation}
  \tich = \frac{\met}{{\rm Diff}_{0}(\Sigma_{g})\times \w}.
\end{equation}
Teichm\"{u}ller space is always smooth while moduli space
is an orbifold, since it is the quotient with a discrete group, namely the
mapping class group. Note also that moduli space is non-compact, but, as we will see,
finite dimensional.
\subsection{The Geometric Structure of Moduli Space}

It is instructive to use the identification from equation~(\ref{moduli-1}) to identify
the elements in \m and \tich.
The tangent space $T_{g}(\met)$ at the point $g_{\alpha\beta}\in\met$ is
a linear space spanned by the infinitesimal deformations of $g_{\alpha\beta}$
\begin{equation}
  g_{\alpha\beta} \mapsto g_{\alpha\beta} + \delta g_{\alpha\beta}.
\end{equation}
Since the space is linear it is endowed with a natural inner product, corresponding to a
metric on the space of metrics \met
\begin{equation}
  \langle \delta g^{(1)}, \delta g^{(2)} \rangle =
  \int_{\Sigma} d^{2}\xi \, \sqrt{g} g^{\alpha\gamma}(\xi)
  g^{\beta\delta}(\xi) \delta g_{\alpha\beta}^{(1)}(\xi)
                       \delta g_{\gamma\delta}^{(2)}(\xi).
\label{met-metric}
\end{equation}
Here $g={\rm det}(g_{\alpha\beta}(\xi))$.
The tangent space $T(\diff)$ is the Lie algebra of the diffeomorphism group, which is
spanned by the tangent vectors of $\Sigma_{g}$
\begin{equation}
  T(\diff) = T(\Sigma_{g}),
\end{equation}
since every tangent vector $t^{\alpha}(\xi)$ is a generator of an
infinitesimal coordinate transformation 
\begin{equation}
  \xi^{\alpha} \mapsto \xi^{\alpha} + t^{\alpha}(\xi).
\end{equation}
The space $T(\diff)$ is linear and thus endowed with a natural inner product
\begin{equation}
  \langle t_{(1)},t_{(2)} \rangle = \int_{\Sigma} d^{2}\xi \sqrt{g}
  g_{\alpha\beta}(\xi)t_{(1)}^{\alpha}(\xi)t_{(2)}^{\beta}(\xi).
\label{diff-metric}
\end{equation}

\noindent

A vector field on a surface $\Sigma_{g}$ is a choice of a tangent vector to every
point of the surface. Since each point has a tangent space, this amounts to a selection of
one vector from each tangent space. If this choice varies smoothly from point to point one can draw
curves on the surface such that the vector at $x\in \Sigma_{g}$ is tangent to the curve passing
through $x$.
Such curves are known as integral curves. If the vector fields never vanish at any point,
the different curves will not intersect and
there will be just one curve per point. The set of integral curves will cover the surface and is
known as the congruence of
integral curves covering $\Sigma_{g}$.
A congruence provides a natural mapping $\Sigma_{g} \mapsto \Sigma_{g}$ and if the integral curves
$(\xi^{1}(t),\xi^{2}(t))$ are infinitely differentiable with respect to $t$, the resulting congruence is
a diffeomorphism of the surface to itself.
An infinitesimal diffeomorphism can thus be parametrized by a $C^{\infty}$ vector field
$\vec{v}(\xi^{1},\xi^{2})$.

\noindent
Assume that $(\xi^{1},\xi^{2})$ are nonsingular coordinates on $\Sigma_{g}$. Let $e_{\alpha}$, for
$\alpha= 1,2$ be the basis of the tangent space at the point $(\xi^{1},\xi^{2})$, such
that $\vec{v}=v^{\alpha}e_{\alpha}$.
If $f(\xi^{1}(t),\xi^{2}(t))$ is a scalar function on $\Sigma_{g}$ then
the operation of the vector field $\vec{v}(\xi^{1},\xi^{2})$ on $f$ is  written as $\vec{v}[f]$
defined as the change in $f$ when it is displaced along the integral curves of $\vec{v}$. The change
is expressed by the directional derivative of $f$ in the direction of the
vector $\vec{v}$, tangent to the integral curve, such that
\begin{equation}
  \vec{v}[f] = v^{\alpha}e_{\alpha}[f]=v^{\alpha}\frac{\partial f}{\partial \xi^{\alpha}}.
\end{equation}
This is also known as the Lie derivative ${\cal L}_{\vec{v}}$ of $f$ along $\vec{v}$.
The similar action on a rank two covariant tensor $g_{\alpha\beta}$ on $\Sigma_{g}$ reads
\begin{equation}
  {\cal L}_{\vec{v}}(g(\vec{u},\vec{w})) =
  ({\cal L}_{\vec{v}}g)(\vec{u},\vec{w}) + g({\cal L}_{\vec{v}}\vec{u},\vec{w}) +
  g(\vec{u},{\cal L}_{\vec{v}}\vec{w}).
\end{equation}

\noindent
This enables one to describe the action of an infinitesimal diffeomorphism on the metric
as the Lie derivative along the vector field $\vec{v}$ that generates the diffeomorphism,
\begin{equation}
  \delta_{\rm diff}\,  g_{\alpha\beta} \equiv
  ({\cal L}_{\vec{v}}g)_{\alpha\beta} = v^{\gamma}(
    \partial_{\gamma} g_{\alpha\beta} ) + 
g_{\gamma\beta}( \partial_{\alpha} v^{\gamma} )
+ 
g_{\gamma\alpha}( \partial_{\beta} v^{\gamma}),
\label{lie_derivative_of_the_metric}
\end{equation}
where $\partial_{\alpha} = \partial/\partial \xi^{\alpha}$.
Using $v_{\alpha} = g_{\alpha\gamma}v^{\gamma}$ we have for example
\begin{equation}
 g_{\gamma\beta} \left( \partial_{\alpha} v^{\gamma} \right)
 =  \left( \partial_{\alpha} v_{\beta} \right)
 - v^{\gamma} \left(
    \partial_{\alpha} g_{\gamma\beta} \right) ,
\end{equation}
and equation~(\ref{lie_derivative_of_the_metric}) can be rewritten as:
\begin{equation}
 ({\cal L}_{\vec{v}}g)_{\alpha\beta} =
 \partial_{\alpha} v_{\beta} +
\partial_{\beta} v_{\alpha} +
v^{\gamma} 
    \partial_{\gamma} g_{\alpha\beta}
   - v^{\gamma} 
    \partial_{\alpha} g_{\gamma\beta}
    - v^{\gamma} 
    \partial_{\beta} g_{\gamma\alpha} .
\end{equation}
Notice that some of this form resembles that of the Christoffel symbol
\begin{equation}
  \Gamma_{\alpha\beta}^{\sigma} =
  \frac{1}{2} g^{\sigma\gamma}\left(
    \partial_{\alpha} g_{\gamma\beta} + 
    \partial_{\beta} g_{\gamma\alpha}  -
    \partial_{\gamma} g_{\alpha\beta}\right),
\label{christoffel}
\end{equation}
and by reorganising the terms we find
\begin{equation}
  ({\cal L}_{\vec{v}}g)_{\alpha\beta} = D_{\alpha}v_{\beta} +
  D_{\beta}v_{\alpha} \equiv D_{ ( \alpha}v_{\beta )}, \label{diff_compact}
\end{equation}
using that
\begin{equation}
  D_{\alpha}v_{\beta} = \partial_{\alpha} v_{\beta} -
  \Gamma_{\alpha\beta}^{\rho}v_{\rho}.
\end{equation}

\noindent
A general Weyl scaling maps the metric
\begin{equation}
  g_{\alpha\beta} \mapsto g_{\alpha\beta}' = e^{\Phi} g_{\alpha\beta},
\end{equation}
and an infinitesimal Weyl transformation by $e^{\delta\Phi}$ is then
written as
\begin{equation}
  \delta_{\rm W} g_{\alpha\beta} = g_{\alpha\beta} \delta\Phi.
\end{equation} 

\noindent
Any general transformation of the metric $\delta g_{\alpha\beta}$ can
    be decomposed into a trace part and a traceless one:
\begin{equation}
 \delta g_{\alpha\beta} = \delta \Phi g_{\alpha\beta} + \delta h_{\alpha\beta},
\label{gen_variation_decomposition_metric_1}
\end{equation}
where
\begin{eqnarray}
  \delta \Phi &=& \frac{1}{2} g^{\mu\nu}\delta g_{\mu\nu}, \\
  \delta h_{\alpha\beta}& =& \delta g_{\alpha\beta} - \frac{1}{2} g_{\alpha\beta}
  g^{\mu\nu}\delta g_{\mu\nu}.
\end{eqnarray}
Choose $\delta g_{\alpha\beta} = D_{ ( \alpha}v_{\beta )}$ for an infinitesimal
diffeomorphism and we find:
\begin{equation}
  \delta \Phi = \frac{1}{2} g^{\mu\nu}D_{ ( \mu}v_{\nu )} = D_{\mu}v^{\mu},
\end{equation}
\begin{eqnarray}
  \delta h_{\alpha\beta}& =& D_{ ( \alpha}v_{\beta )} - \frac{1}{2}
  g_{\alpha\beta}g^{\mu\nu}D_{ ( \mu}v_{\nu )} \nonumber \\ 
  &=& D_{\alpha}v_{\beta} + D_{\beta}v_{\alpha} - g_{\alpha\beta}D_{\mu}
  v^{\mu}
  \nonumber \\ &\equiv & [ P_{1}(v)]_{\alpha\beta} \label{p_1}.
\end{eqnarray}
In the last line we introduce the definition of the operator $P_{1}$ which
maps vectors into symmetric, traceless, rank two tensors. This operator
plays a central role in the following considerations.
Apply equation~(\ref{gen_variation_decomposition_metric_1}) to write the combined effect of a
Weyl transformation and a diffeomorphism on the metric as
\begin{equation}
  \delta g_{\alpha\beta} = (\delta \Phi + D_{\mu}v^{\mu})g_{\alpha\beta}
  +  [ P_{1}(v)]_{\alpha\beta}.
\label{metric_orthogonal_decomposition_1}
\end{equation}
A Weyl transformation is orthogonal to the image of $P_1$, i.e.
$\delta_{\rm Weyl} \subseteq ({\rm Im} P_1)^{\perp}$ since:
\begin{eqnarray}
 \langle g_{\alpha\beta} \delta \Phi, [P_{1}t]_{\gamma\delta}\rangle  &=&
  \int_{\Sigma} d^{2}\xi \, \sqrt{g}\, g^{\alpha\gamma}
  g^{\beta\delta} g_{\alpha\beta}\delta\Phi
      \left(D_{\gamma}t_{\delta} + D_{\delta}t_{\gamma} -
      g_{\gamma\delta} D_{\alpha}t^{\alpha}\right) \nonumber \\ &=&
  \int_{\Sigma} d^{2}\xi \, \sqrt{g}\, \delta \Phi
  \left(D_{\alpha}t^{\alpha} + D_{\alpha}t^{\alpha} - 2 D_{\rho}
  t^{\rho} \right) \nonumber \\ &=& 0.
\end{eqnarray}  
From $\langle \delta g, P_{1}t\rangle = \langle P_{1}^{\dagger}\delta g,t\rangle$, one can show
\begin{equation}
 \left[   P_{1}^{\dagger}\delta g\right]_{\alpha} =
 - 2 D^{\beta}\delta g_{\alpha\beta} + D_{\alpha}\left( (\delta g^{\gamma\delta})
   g_{\gamma\delta} \right). \label{p_1_dagger}
\end{equation}
Hence one introduces the following notation: The tangent space of $\met$ can be
decomposed as $T(\met) = T(\w) \oplus H$, where $H$ denotes the space
of all traceless ($2\times2$) matrices. The operator $P_{1}^{\dagger}$ maps
from $H$ to the space of vectors, i.e. to $T(\diff)$ and we
identify
\begin{equation}
  H = {\rm Im}\, P_{1} \oplus {\rm Ker}\, P_{1}^{\dagger},
\end{equation}
where ${\rm Ker} P_{1}^{\dagger}$ is known as the space of quadratic differentials
and ${\rm Ker} P_{1}$ as the space of conformal Killing vectors.
First we consider the latter.

\noindent
For $t^{\alpha}\in {\rm Ker} \, P_{1}$ we have
\begin{equation}
  D_{\alpha}t_{\beta} + D_{\beta}t_{\alpha} = g_{\alpha\beta}
  D_{\gamma}t^{\gamma},
\end{equation}
so any diffeomorphism induced by such a conformal Killing vector (C.K.V.) field
equals a Weyl transformation, i.e. $\delta_{{\rm Diff-C.K.V}} g_{\alpha\beta} =
\delta_{{\rm Weyl}} g_{\alpha\beta}$. Hence there is an overlap between
the groups \diff and \w if there exist conformal Killing vectors and the product
$\diff\times \w$ in equation~(\ref{moduli-1}) should really be taken as the semi-direct
product~\cite{algebra} if one is to be strictly correct. \label{semi-direct}
As we shall see this is only a problem for surfaces of genus zero and one.

The elements $t^{\alpha}\in {\rm Ker} \, P_{1}$ form a Lie sub-algebra of
$T(\diff)$, namely the Lie algebra of the group of conformal automorphisms of
the metric. The metric isomorphisms (i.e. $\delta g = 0$) are a subset of
the conformal automorphisms.

The elements $h_{\alpha\beta} \in  {\rm Ker} \, P_{1}^{\dagger}$ are the
symmetric, traceless and divergence-less (i.e. $D_{\alpha}h^{\alpha\beta}=0$)
rank 2 tensors. It is convenient to rewrite the definitions of
$P_{1},P_{1}^{\dagger}$ from
equations~(\ref{p_1},\ref{p_1_dagger}) in isothermal coordinates
\begin{equation}
  (ds)^{2} = \exp\Bigl(\Phi(z,\bar{z})\Bigr) \mid dz \mid^{2},
\end{equation}
where $g_{zz} = g_{\bar{z} \bar{z}}=g^{zz} =
g^{\bar{z} \bar{z}}=0$. Using $g^{\alpha\beta}\delta g_{\alpha\beta}
= 2 e^{-\Phi} \delta g_{z\bar{z}}$, one can write
\begin{eqnarray}
\left[ P_{1} t \right]_{zz} &=& 2 D_{z}t_{z}, \\
\left[ P_{1} t \right]_{\bar{z}\bar{z}} &=& 2 D_{\bar{z}}t_{\bar{z}}, \\
\left[ P_{1} t \right]_{z\bar{z}} &=& 0,\\
\left[   P_{1}^{\dagger}\delta g\right]_{z} &=& - 2 g^{z\bar{z}}
\left( D_{\bar{z}} \delta g_{zz} + D_{z} \delta g_{z\bar{z}}
  \right) + 2 D_{z}\left( e^{-\Phi} \delta g_{z\bar{z}} \right),
\\
\left[   P_{1}^{\dagger}\delta g\right]_{\bar{z}} &=& - 2
g^{z\overline{z}}
\left( D_{z} \delta g_{\bar{z}\bar{z}} + D_{\bar{z}}
\delta g_{z\bar{z}}
  \right) + 2 D_{\bar{z}}\left( e^{-\Phi} \delta g_{z\bar{z}} \right).
\end{eqnarray}
In the special case where $\delta g_{z\bar{z}}=0$ the elements
$\mu_{zz}\in {\rm Ker} P_{1}^{\dagger}$ must satisfy the equation
\begin{equation}
 \left[P_{1}^{\dagger}\mu\right]_{z} =g^{z\bar{z}}D_{\bar{z}}\mu_{zz} = 0.
 \label{quadraticdifferential}
\end{equation}
In the isothermal coordinate system the Christoffel symbols, defined in
equation~(\ref{christoffel}), reduce to only two non-vanishing
\label{isotherm-christoffel}
components
\begin{eqnarray}
  \Gamma^{z}_{zz}& = &\partial_{z} \Phi ,\\
  \Gamma^{\bar{z}}_{\bar{z}\bar{z}}& = &\partial_{\bar{z}} \Phi ,
\end{eqnarray}
and equation~(\ref{quadraticdifferential}) reduces to
\begin{equation}
  \partial_{\bar{z}} \mu_{zz} = 0,
\end{equation}
justifying the name quadratic differentials and we denote the space of
analytic quadratic differentials as
$H^{(2)}(\Sigma)$.

\noindent
This identifies the space of quadratic analytic differentials as the space
consisting of infinitesimal metric deformations, which are not obtained
by Weyl transformations or diffeomorphisms. Hence these are the same as
the elements of Teichm\"{u}ller space and $H^{(2)}(\Sigma) \sim \tich$.
Combining this with ${\rm Im}(P_{1}) = T(\diff)$ we have
\begin{equation}
  T(\met) = T(\w) \oplus T(\diff) \oplus H^{(2)}.
\end{equation}

\begin{figure}[h]
\begin{center}
\mbox{
\epsfysize=6cm
\epsffile{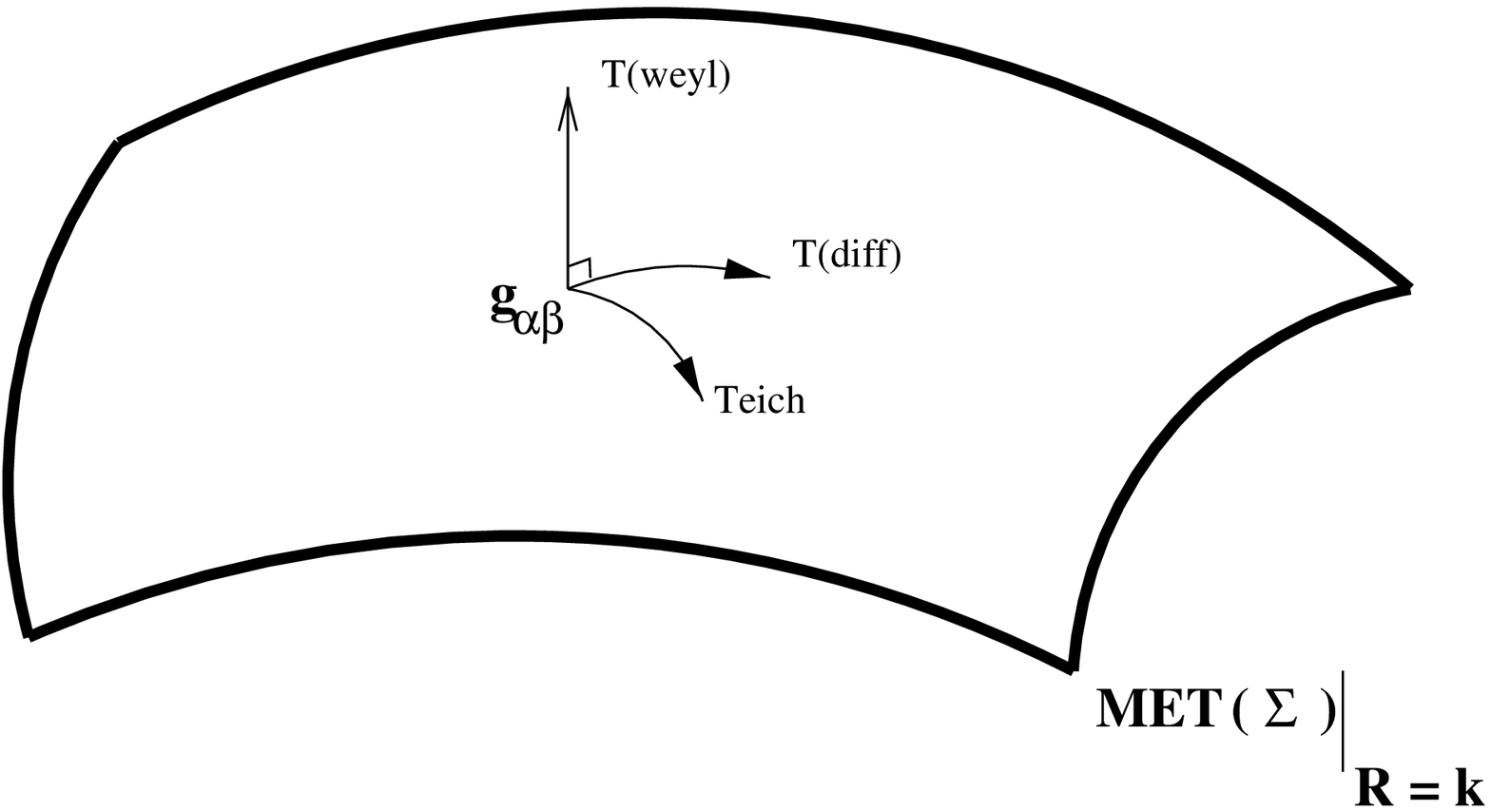}
}
\end{center}
\caption{\label{fig:const_curvature}  A constant curvature surface in \met.}
\end{figure}

We now turn to investigate how the various changes in the metric affect the
curvature of the Riemann surface. One can calculate 
the change in the scalar curvature from a general metric transformation $\delta g_{\alpha\beta}$
\begin{equation}
  \delta R = - \frac{1}{2} (\delta g_{\alpha\beta}) g^{\alpha\beta} \,R
  + D^{\gamma}D_{\gamma}\left( g^{\alpha\beta} (\delta
    g_{\alpha\beta}) \right) - D^{\alpha}D^{\beta}\left(\delta g_{\alpha\beta}
\right) \label{var_general_R} .
\end{equation}
The space of those metrics which have constant scalar curvature $R[g]=k$ is invariant under
diffeomorphisms, since a coordinate transformation can not change the value of a scalar.
If $\delta g_{\alpha\beta} \in H$, i.e. it is a so-called Teichm\"{u}ller
deformation, it must be traceless and divergence-less as e.g.
$\delta g_{\alpha\beta}=R_{\alpha\beta} = \frac{1}{2}g_{\alpha\beta}R$ and 
one also finds $\delta R = 0$. 
But since the Weyl transformations change the scale of the surface, they do not preserve
the scalar curvature and they move us away from the surface of constant curvature metrics in \met
as illustrated in figure~(\ref{fig:const_curvature}).
When dealing with functional integration over moduli space, we see that
the Weyl invariance is easily gauge fixed by restricting ourselves to a constant curvature
$R=k$, which we will do several times during the discussion of topological gravity
in two dimensions.

This concludes the introduction to metric transformations and the construction
of moduli space \m. The metric on \met in equation~(\ref{met-metric}) 
is also valid on the space of constant curvature metrics $\{\met\mid R=k\}$ and it is invariant under all
diffeomorphisms
there. In other words the diffeomorphisms are isometrics on
$\{\met\mid R=k\}$ and the metric can be used on both \m and \tich.
In the latter case it is known as the Weyl-Peterson metric~\cite{Hat}.
The dimension of the tangent space of a manifold equals the dimension of
the manifold itself and one can determine the dimension of moduli space
from its tangent space. The Riemann-Roch theorem applies here and it can be shown
~\cite{LNP322,SuperStringVol3} that the dimension of moduli space is 
\begin{equation}
{\rm dim}\m=  {\rm dim}\left({\rm Ker} P_{1}^{\dagger}\right) -
  {\rm dim}\left({\rm Ker} P_{1}\right) = -\frac{3}{2}\chi(\Sigma)
  = 3g-3,
\end{equation}
where
\begin{equation}
  \chi(\Sigma) = 2 - 2g,
\end{equation}
is the Euler-Poincar\'{e} characteristic for a boundary-less Riemann surface
$\Sigma_{g}$.  We see that there are three conformal Killing vectors
in genus zero corresponding to the six real generators of $PSL(2,\C)$. In genus one
there is one modulus (or Teichm\"{u}ller parameter) and one conformal Killing vector
corresponding to the generator of the translations.
For all higher genera are there no conformal Killing vectors and $3g-3$ or
(in real dimensions $6g-6$) moduli.

\section{A Few Lines on Algebraic Geometry}

Many of the results used in topological gravity regarding moduli space are made in the
very powerful, but also abstract, framework of algebraic geometry.
We present the
very basic ideas and explain some of the terminology used in the original work
on topological gravity, to the extent it is needed in the rest of the thesis.

Let us first introduce the complex projective space $\CP^{n}$. One can define
this space as
\begin{equation}
  \CP^{n} \equiv \left(\C^{n+1} - \{0\}\right)/\sim,
\end{equation}
where $w,z\in \C^{n+1}$ are equivalent $w\sim z$, if $z=\lambda w$ for any non-vanishing
complex number $\lambda\in \C$. One can define the map
\begin{equation}
  \pi: \left(\C^{n+1} - \{0\}\right) \mapsto \CP^{n}\, \mbox{by mapping } (z_{0},z_{1},\dots,z_{n})
  \mapsto (z_{0}:z_{1}:\dots:z_{n}).
\end{equation}
The notation $(z_{0}:z_{1}:\dots:z_{n})$ covers so-called homogeneous coordinates on
$\CP^{n}$, but due to the definition up to scalar complex multiplication these
numbers are not coordinates in the usual sense. 

One can define a topology on $\CP^{n}$ by defining $\pi$ to map into open sets of
$\CP^{n}$. By introduction of an atlas $(U_{\alpha},\psi_{\alpha})$ one can define a manifold
structure on $\CP^{n}$. Let the coordinate patches be
\begin{equation}
  U_{\alpha} \equiv \left\{  (z_{0}:z_{1}:\dots:z_{n}) \left\vert z_{\alpha}\neq 0, \alpha=0,1,\dots,n
  \right\},\right.
\end{equation}
and the homeomorphisms be given as
\begin{equation}
  \psi_{\alpha}:U_{\alpha} \mapsto \C^{n} \, \left\vert  (z_{0}:z_{1}:\dots:z_{n})
  \mapsto  (\frac{z_{0}}{z_{\alpha}},\frac{z_{1}}{z_{\alpha}},\dots,
  \frac{z_{\alpha-1}}{z_{\alpha}},  \frac{z_{\alpha+1}}{z_{\alpha}},\dots,
      \frac{z_{n}}{z_{\alpha}} ).\label{projective-1} \right.
\end{equation}
By this map the individual coordinate patches of $\CP^{n}$ are identified with $\C^{n}$, and
the space $\CP^{n}\setminus U_{\alpha}$ is an $(n-1)$-dimensional hyperplane at infinity.
The projective
space is thus a compactification of $\C^{n}$. The complex projective space can be
viewed as the space of all complex rays through the origin, excluding the intersection
point at the origin. All lines or rays in projective space intersect at one and only one
point, and even parallel lines intersect, namely at infinity. This picture makes it
possible to
map the projective spaces to the spheres. One can identify $\CP^{1}$ with the Riemann sphere
\Ci, using the covering
\begin{eqnarray}
  U_{{\rm N}} &=& \left\{ (z_{0}:z_{1}) \vert z_{0}\neq 0 \right\}, \\
  U_{{\rm S}} &=& \left\{ (z_{0}:z_{1}) \vert z_{1}\neq 0 \right\}. \\
\end{eqnarray}
Using the map in equation~(\ref{projective-1}) we can identify
\begin{eqnarray}
  U_{{\rm S}} &=& \left\{ (w:1) \vert w = \frac{z_{0}}{z_{1}} \in \C \right\}\simeq \C, \\
  \CP^{1}/U_{{\rm S}} &=& \left\{ (1:0)  \right\}\simeq \{ \infty\}, \\
\end{eqnarray}
which identifies the manifold $\CP^{1}$ with the Riemann sphere \Ci.

More generally the complex projective space is useful when one wishes to extract information
about the geometry of a Riemann surface by embedding it into a larger space. The picture in
figure~(\ref{fig-genus-g}) is the result of an embedding into $\R^{3}$. This embedding
gives a good visual picture of the surface, but it does not respect the complex
structure on the surface. The only compact complex submanifolds
of $\C^{n}$ are the points, because the coordinate functions $x^{i}$ being analytical
functions in $\C^{n}$ 
also will be analytic functions on the embedded surface $\Sigma$, but the only possible
analytic functions on a compact Riemann surface are the constants.
This implies that $\CP^{n}$ is the simplest
space we can embed our surfaces in, which respects their complex structures. One can extract
a lot of information by this embedding, but it can be problematic to separate the information
depending on the embedding, from that depending only on the surface itself. By this embedding 
Riemann surfaces will be described in terms of polynomials in projective space.

Let $f$ be a polynomial in $n+1$ variables $x_{i},\,(i=0,1,\dots,n)$, with all terms
homogeneous of order $k$. As an example take $(n=k=2)$ and let $f$ be of the form
\begin{equation}
  f(x) = x_{0}^{2} + x_{1}^{2} + x_{2}^{2}.
\end{equation}
Let $(z_{0}:z_{1}:z_{2})$ be homogeneous coordinates in $\CP^{2}$ and insert these in
the polynomial to obtain
\begin{equation}
  f(z) = z_{0}^{2} + z_{1}^{2} + z_{2}^{2}.
\end{equation}
Since the coordinates are only defined up to scaling we have:
\begin{equation}
  f(\lambda z)=\lambda^{{\rm order}(f)} f(z),\,\,\lambda\in \C.
\end{equation}
Since $\lambda\neq 0$ per definition one can always determine whether $f(z)$ is zero or not.
The set of zeros for $f(z)$
\begin{equation}
V_{f} \equiv \Biggl\{ z \in \CP^{n} \vert \, f(z) = 0 \Biggr\},
\end{equation}
defines a so-called variety. Varieties replace Riemann surfaces in algebraic geometry.
One can equip $\CP^{n}$ with Zariski topology, being the coarsest topology
in which the varieties $V_{f}$ are closed sets. Every non-empty Zariski open set is
dense\footnote{Recall: Let $X$ be a topological space:
  A point $x\in X$ is an accumulation point of $A\subseteq X$ if every neighborhood $N(x)$ of
  $x$ contains at least one point $a\in A$ different from $x:\left( N(x)- \{x\} \right) \cap A \neq
  \emptyset , \, \forall N(x)$. The closure $\overline{A}$ of $A$ in $X$ is the union of $A$
  with all its accumulation points. The set $A$ is dense in $X$ if $\overline{A}=X$.}
in $\CP^{n}$.
A subset $W\subset \CP^{n}$ is a projective variety if it is the common set of zeros of
finitely many homogeneous polynomials. These polynomials might be of different order.
A variety $W$ is said to be irreducible if every decomposition $W=V_{1}\cup V_{2}$ implies that
$V_{1}\subset V_{2}$ or $V_{2}\subset V_{1}$, where $V_{1},V_{2}$ also are varieties. A
Zariski open set of an projective variety is known as an algebraic or quasi projective variety.

Kodaire and Chow~\cite{LNP322} have shown the following:
If $\Sigma_{g}$ is a compact Riemann surface, then it admits an embedding into
some complex projective space. In addition every Riemann surface is analytically 
isomorphic to a one-dimensional non-singular variety, also known as a curve.
In algebraic geometry the moduli space of Riemann surfaces is known as the moduli space
of curves.
\noindent
A non-singular variety corresponds to the notation of a manifold in topology.
The definition is somewhat technical and we appeal to the readers' intuitions of singularities
and refer to the literature for more information~\cite{Hartshorne}. Non-singular varieties are said
to be smooth.

\noindent
Since all complex algebraic equations can be solved, algebraic geometry is a powerful theory.
Many central proofs given in topological gravity are made using
algebraic geometry and the hope is that this short list of the most central definitions will
ease the discussion of these proofs in the later chapters.

In the next section we show how the representation of Riemann surfaces as polynomials give
an extension to singular surfaces and we discuss the compactification of moduli space in terms
of these.

\section{Compactification of Moduli Space}
We begin this section by explaining how a torus can be described as a polynomial in $\CP^{2}$.
This leads us to the notation of noded Riemann surfaces, which plays a central role when we
compactify moduli space later in this section.

\begin{figure}[h]
\begin{center}
\mbox{
\epsfysize7.5cm
\epsffile{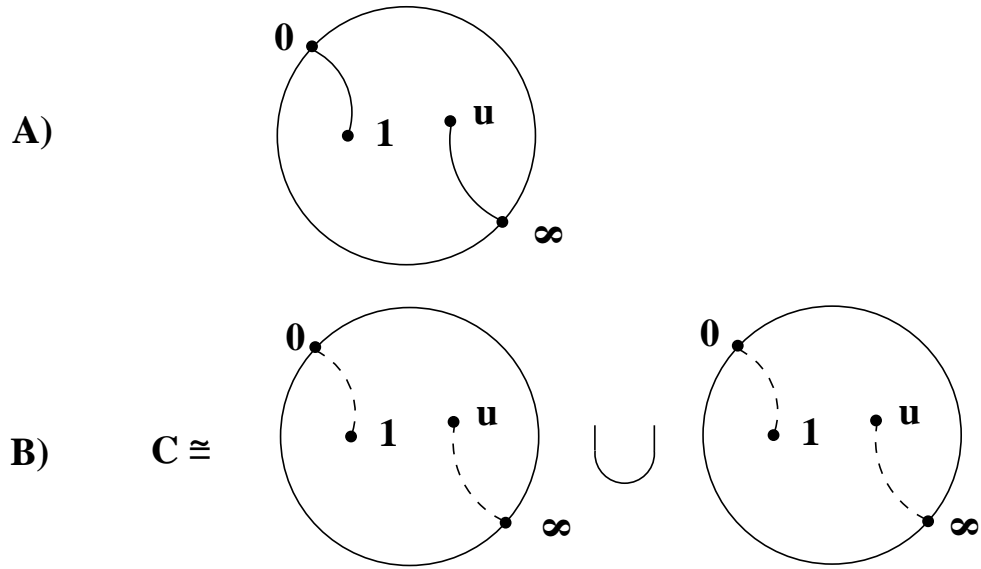}
}
\end{center}
\caption{ \label{sphere-with-cuts} (A) Two paths on the sphere.  (B) Union of two
  copies of the sphere with cuts.}
\end{figure}

\begin{figure}[h]
\begin{center}
\mbox{
\epsfysize4cm
\epsffile{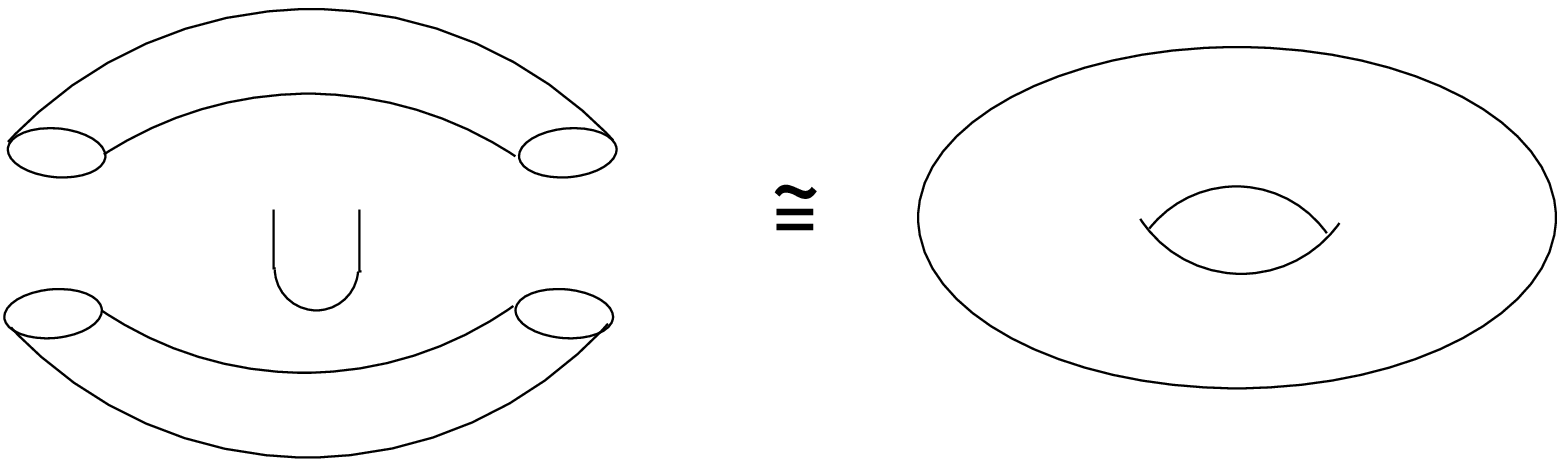}
}
\end{center}
\caption{ \label{pasting-torus} (A) Two paths on the sphere.  (B) Union of two
  copies of the sphere with cuts.}
\end{figure}

\begin{figure}[h]
\begin{center}
\mbox{
\epsfysize6cm
\epsffile{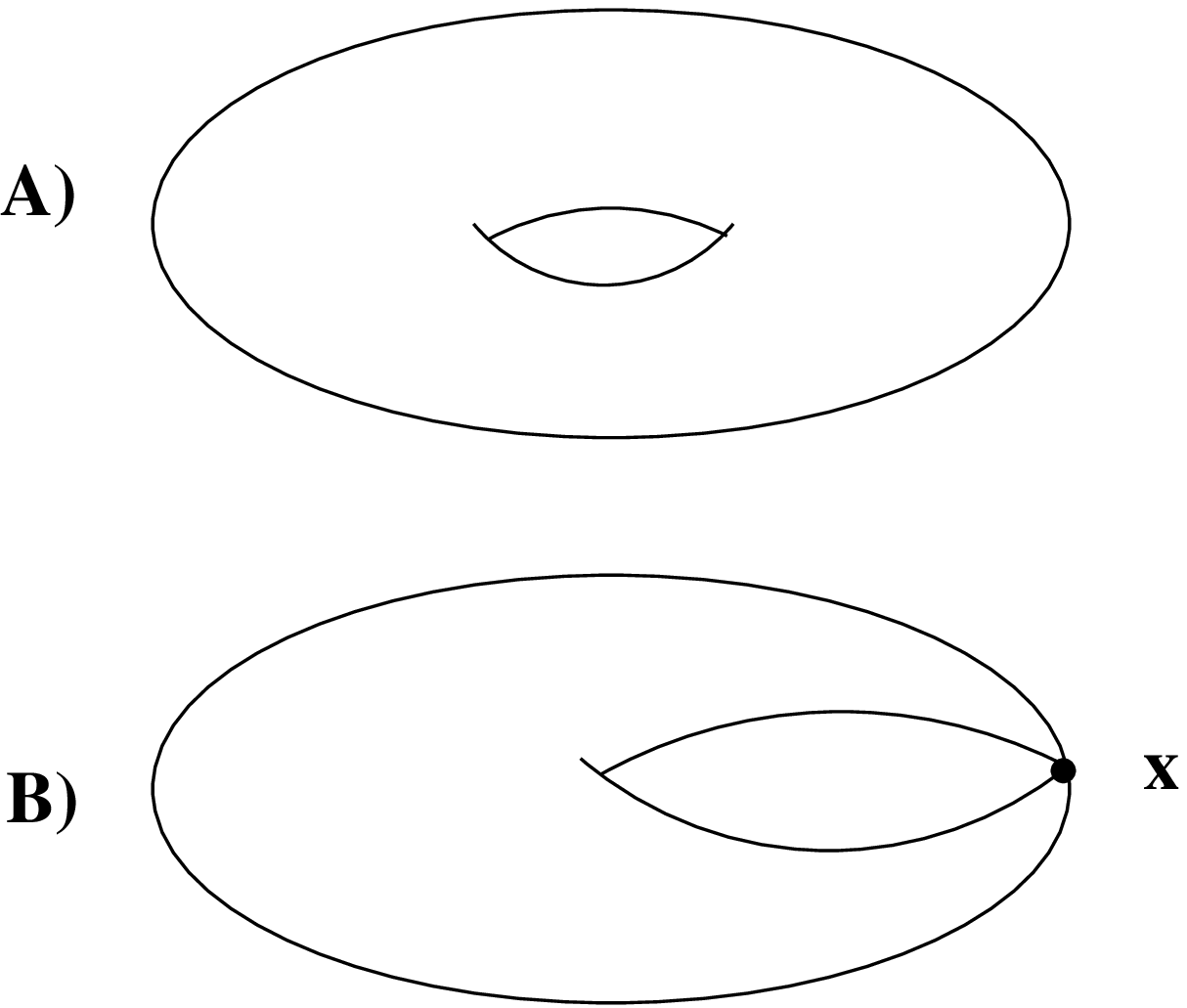}
}
\end{center}
\caption{\label{pinched-torus}(A) A normal torus (B) A pinched torus. }
\end{figure}

\begin{figure}[h]
\begin{center}
\mbox{
\epsfysize4cm
\epsffile{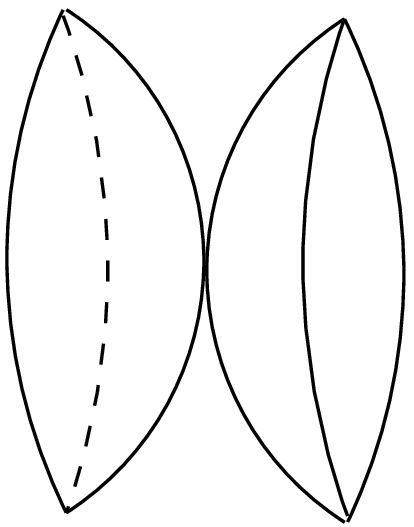}
}
\end{center}
\caption{ \label{doublepoint} The neighbourhood of a node.}
\end{figure}

\begin{figure}[h]
\begin{center}
\mbox{
\epsfysize6cm
\epsffile{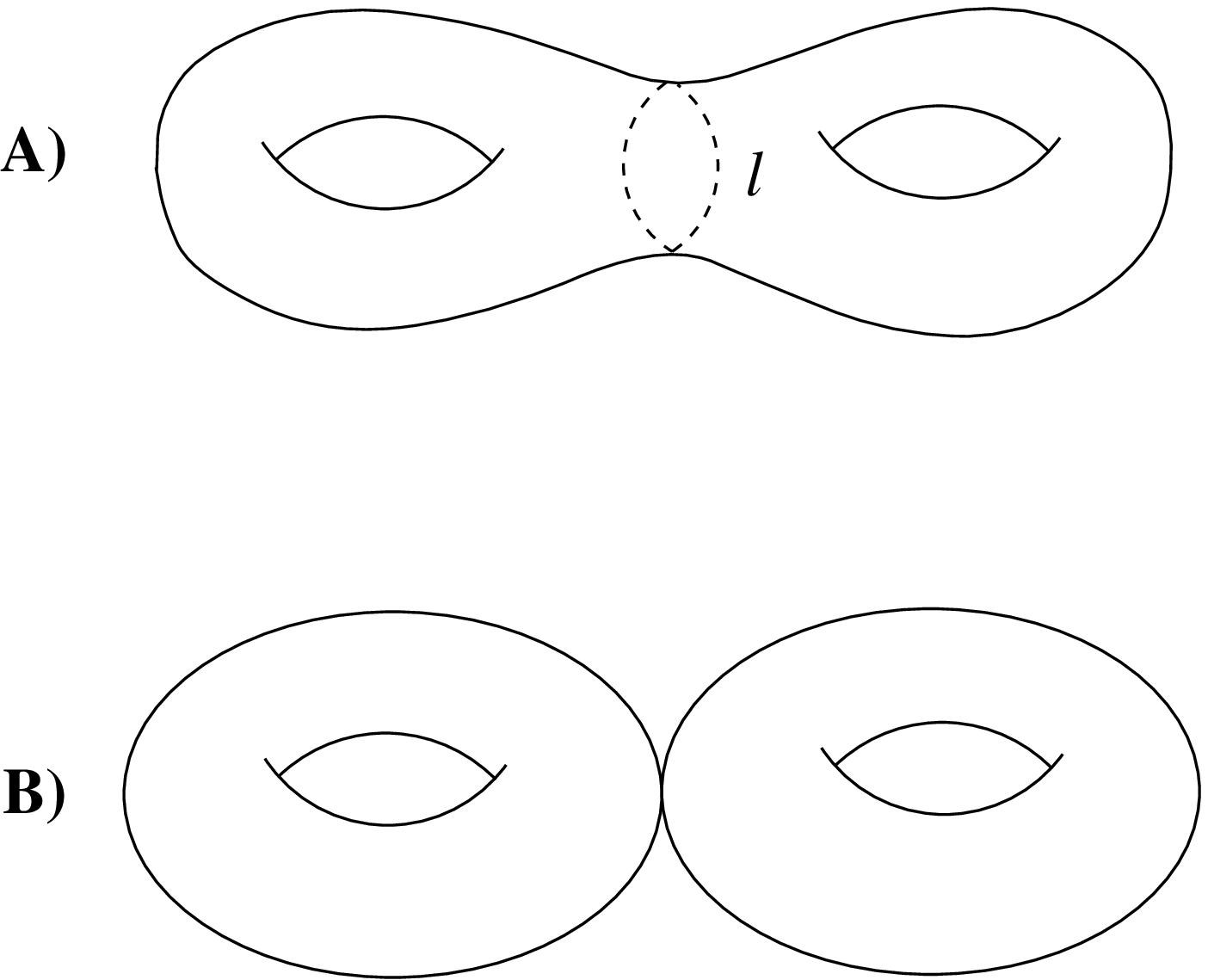}
}
\end{center}
\caption{ \label{pinched-two-torus} (A) A two-torus. (B) A pinched two-torus.}
\end{figure}

\begin{figure}[h]
\begin{center}
\mbox{
\epsfysize1.8cm
\epsffile{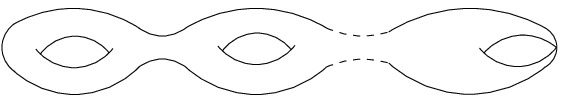}
}
\end{center}
\caption{ \label{pinched-g-torus} The general elements in $\Delta_{0}$.}
\end{figure}

A torus can be represented as a parametrized curve in $\CP^{2}$, given by the
polynomial~\cite{Reid}
\begin{equation}
  C(x,y) \equiv \Biggl( \biggl\{ y^{2}- x (x-1)(x-u)=0 \biggr\} \cup \{\infty\} \Biggr) \subset \CP^{2},
\end{equation}
where $u\in \C$ is a complex constant and $(x,y)$ are defined as
\begin{equation}
  x = \frac{z_{0}}{z_{2}} \,\, \mbox{ and }  y = \frac{z_{1}}{z_{2}},\,\,\,(z_{0}:z_{1}:z_{2})
  \in \CP^{2}.
\end{equation}
The curve is written $C(x,y) = C(z_{0}/z_{2}:z_{1}/z_{2}:1)$. We have added infinity to compactify
$\C^{2}$ to $\CP^{2}$.
To see this describes a torus, consider  the map $\pi:C\mapsto \CP^{1}$, taking
$(z_{0}:z_{1}:z_{2})$ into $(z_{0}:z_{2})$ and $\infty$ into $(1:0)$, or equivalently as
$(x,y)\mapsto x$.
This is a $2$ to $1$ map corresponding to the graph of 
\begin{equation}
  y = \pm \sqrt{ x(x-1)(x-u) }\,, \,\,\,\,x,y,u\in\C,
\end{equation}
and since the complex projective space $\CP^{1}$ corresponds to the Riemann sphere \Ci, 
we consider $y$ as a function $y=f(x)$ on \Ci. Outside the four
points $(0,1,u,\infty)$ this is a
double valued function. Mark these four points on \Ci and draw two paths between $(0,1)$ and
$(u,\infty)$, as done in picture (A) of figure~(\ref{sphere-with-cuts}). 
Now cut the sphere along the paths, which splits the double cover into two sheets as
indicated in picture (B) of figure~(\ref{sphere-with-cuts}). Then open the two cuts
to change picture (B) of figure~(\ref{sphere-with-cuts}) into the left hand side of
figure~(\ref{pasting-torus}).
By gluing the open cuts together we obtain the torus as
promised. Note that the complex number $u$ represents the moduli of the torus, but in the
algebraic geometry the limit $u\rightarrow 0$ is special. For $u=0$ $C$ is still a
perfectly regular polynomial, but the corresponding Riemann surface undertakes a radical
transformation. The torus degenerates and changes form from (A) to (B) in 
figure~(\ref{pinched-torus})
This is not a smooth Riemann surface, but it is still a polynomial of the original order in
$\CP^{2}$. We call the degenerated point, a node and the surface, a noded Riemann surface.
Almost every point on a noded Riemann surface has a neighbourhood homeomorphic to \C, but
the degenerated point has a neighbourhood homeomorphic to two discs glued together in
a double point as illustrated in figure~(\ref{doublepoint}).
The pinched torus is topologically a sphere with a single double point. For higher tori, e.g. a
two-torus, one can consider a non-contractible loop, homologous to zero, as indicated by the
dashed line, of length $l$, in picture (A) of figure~(\ref{pinched-two-torus}).
Taking the limit $l\rightarrow 0$ results in the pinched two-torus in picture (B)
of figure~(\ref{pinched-two-torus}).
The study of noded Riemann surfaces arised in string theory, where divergences occur due to
integration over \m, which is a non-compact space. One would like to compactify \m by adding
conformal equivalence classes of degenerated surfaces. A situation which would cause problems
in string theory is when the string world-sheet becomes an infinitely long cylinder. But the
infinite cylinder can be mapped to the disc with a hole, by a conformal
transformation, so the infinite cylinder is conformally equivalent to (half) the neighbourhood of a
node~\cite{Nelson}.
This was exactly two discs glued together at the hole, and this implies that the infinite
cylinder corresponds to a degenerated Riemann surface, with a node. Noded surfaces appear
to be of the nature, which corresponds to the physical degenerations in string theory.
A certain compactification scheme  known as the
Deligne-Mumford-Knudsen compactification~\cite{LNP322,Nelson,Arbarello}, based on noded
Riemann surfaces has therefore found good use in string theory. The compactified moduli
space is denoted \mb
and the added point (actually a subvariety) $\Delta$ is the complement of \m in \mb,
being of codimension one and having a decomposition
\begin{equation}
  \Delta = \bigcup_{i=0}^{g/2} \Delta_{i}.
\end{equation}
The individual $\Delta_{i}$'s are also subvarieties. A point in $\Delta$ corresponds to a
noded Riemann surface and $\Delta_{0}$ is the closed set of all irreducible curves (complex dimension one varieties) with exactly one node as singularity, as e.g. the pinched torus in
figure~(\ref{pinched-torus}) or more generally a $g$-torus with a single pinched homology
cycle as in
figure~(\ref{pinched-g-torus})
\begin{figure}[h]
\begin{center}
\mbox{
\epsfysize1.8cm
\epsffile{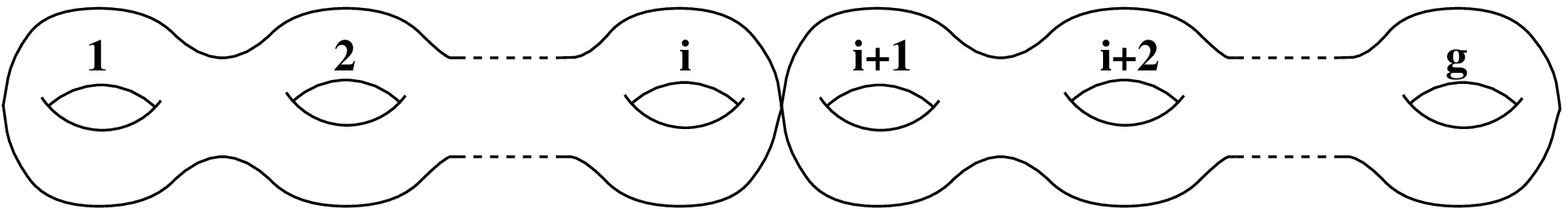}
}
\end{center}
\caption{ \label{torus-split} The general elements in $\Delta_{i}$, with a genus $i$ and
  a $g-i$ sector.}
\end{figure}
In general all $\Delta_{i}$ are irreducible components of codimension one. For $i>0$,
$\Delta_{i}$ is the closed set of stable curves, which are the union of a
nonsingular genus $i$ and a genus $(g-i)$ curve, with one common double point.
This is illustrated in figure~(\ref{torus-split}).
In this compactification  \mb is a projective variety, for which \m is an open
(in Zariski topology), irreducible subvariety, so \m is an algebraic variety. 
As in the geometric formulation, \m is singular due to nontrivial automorphisms.
That \mb is a projective variety, makes life easy since many nice theorems like the
standard index theorems and things like Poincar\'{e} duality still hold on \mb then.
This is the mathematical reason for preferring this compactification in favour of others,
and it then also happens to describe the degenerated surfaces one expects in string theory.
Therefore, it has also been used in topological gravity, as we will discuss in chapter 4.

\begin{figure}[b]
\begin{center}
\mbox{
\epsfysize5cm
\epsffile{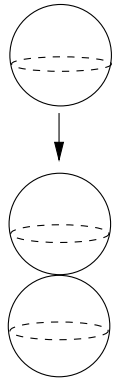}
}
\end{center}
\vspace{-.4cm}
\caption{ \label{sphere-to-doublesphere} Pinching a sphere to a double-sphere.}
\end{figure}

But not all noded curves are elements in \mb. Consider a sphere and begin to
squeeze the equator, until it becomes a node and the sphere pinches into two spheres as
illustrated in
figure~(\ref{sphere-to-doublesphere}).
The original sphere is conformally equivalent to the pinched sphere,
since all angles are preserved and this curve is not an element in \mb.
There are no moduli on a sphere, which we can kill by pinching the surface and the sphere is
said to be conformally rigid~\cite{Nelson}. These kinds of elements are not in \mb. In the
same way of reasoning one can not obtain a new element in \mb, by pinching the torus
in figure~(\ref{pinched-torus}) once more, since there are no moduli left to kill.
It is not known whether the compactified moduli space is the quotient of some
Teichm\"{u}ller-like space with some covering group $G$
\begin{equation}
  \mb \stackrel{?}{=} \overline{\tich}/G.
\end{equation}

In chapter 4 we extend this discussion to so-called punctured Riemann surfaces, with
fixed marked points on the surface, which are inert under diffeomorphisms. The nature of
the Deligne-Mumford-Knudsen compactification is vital for the map from topological gravity
to $2D$ quantum gravity.

%% file: chap2.tex
\chapter{Topological Field Theory}
\section{Introduction}
This chapter presents background material for the
relevant discussions in gravity related to topological field theories.
The chapter is organised as follows:
1) A discussion of the geometry of gauge theories. 2) A short treatment of
the most important results in Schwarz type TFT's relevant for quantum gravity.
3) A short discussion of the general features of Witten type TFT's. 4)
As an example of a Witten type TFT we discuss $4D$ topological Yang-Mills theory,
which in many ways is a model for the study of topological gravity in two
dimensions. 5)
We discuss in some detail the mathematics behind Witten type TFT's and
especially topological Yang-Mills theory.
\section{The Geometry of Gauge Theories} \label{section:gauge}
In this section we review the geometrical definitions relevant for
the discussions of gauge theories in this and following chapters. This
section defines the notation for these topics for the rest of the thesis.
A basic reference on differential geometry are the books by Kobayashi and
Nomizu~\cite{KN_1,KN_2} and for the relation to physics we mainly
use~\cite{YC,Nakahara}. The reader is assumed to be familiar with
differential geometry and I only list the most important definitions
in order to set the stage for the discussions in the rest of
the text.
\subsection{Principal Fibre Bundles}
Let $M$ be an $n$-dimensional Riemannian manifold, with either Lorentzian or
Euclidean signature. $G$ denotes a Lie group and by $P(M,G)$ we always mean
a principal $G$-bundle over $M$
\begin{eqnarray}
  G \mapsto \,\, &P& \nonumber \\  &\downarrow&\!\!\!\pi  \\ &M.& \nonumber
  \label{P->M}
\end{eqnarray}
The projection at a point $u\in P$, is $\pi(u)=x\in M$.
The inverse projection $\pi^{-1}(x)\subset P$ is the fibre in $P$
at $u$, over $ x\in M$, consisting of the points $\{ ua \vert \forall
a \in G \}$.
Every fibre in $P$ is diffeomorphic to $G$.
By $T_{u}P$ we denote the tangent space of $P$ at $u$ and $G_{u}\subset
T_{u}P$ is  the subset consisting of those tangent vectors to $P$,
which are tangent to the fibre through $u$. 

A connection $\Gamma$ in $P$ is a unique decomposition, such that
at any point $u\in P$, one can write
\begin{equation}
  T_{u}P = G_{u} \oplus H_{u}\,\, ,
\end{equation}
where $H_{u}\subset P$ is the orthogonal complement of $G_{u}$
\begin{equation}
  H_{u} = (G_{u})^{\perp}.
\end{equation}
The elements in $G_{u}$ are known as the vertical tangent vectors and those in $H_{u}$ as the
horizontal tangent vectors.
So $\Gamma$ is in a sense a metric on $P$, which defines the
orthogonality between $G_{u}$ and $H_{u}$.
Let \la be the Lie algebra of G. Recall that \la is
the space of left invariant vector fields on $G$. The diffeomorphism
$e\mapsto ge=g$ for $e$ being the identity of $G$ and $g$ any element
in $G$, induces a map $T_{e}(G)\mapsto T_{g}(G)$ being
\begin{equation}
  V_{g} = (L_{g})_{*}V_{e}.
\end{equation}
Here $V$ is a vector field on $G$ and $L_{a}g=ag$, for all $g\in G$,
is the left translation of $a\in G$. The Lie algebra consist of all
vector fields on $G$, which are invariant under all left translations. By the
definition of a fibre bundle~\cite{KN_1}, the group $G$ acts on
$P$ by right action 
\begin{equation}
  R_{u}:G \mapsto P \mbox{ by } R_{u}: a \mapsto ua \,
\mbox{ for } u \in P, a\in G,
\end{equation}
embedding a copy of $G$ at each point in $P$. This action is free, i.e.  
\begin{equation}
R_{u}: a = ua = u,  \,
  \mbox{ iff } a=e \in G .
\end{equation}
The right action induces a map $(R_{a})_{*}:T_{g}G \mapsto T_{ga}G$ for $a,g\in G$. Just as
there can be left invariant vector fields on $G$ can one define right invariant vector
fields by the requirement
$(R_{a})_{*}V = Va = V$ for all $a\in G$. 

To define the connection $\Gamma$ two requirements must hold. First,
\begin{equation}
  H_{ua} = (R_{a})_{*}H_{u},\,\,\forall u\in P\,\mbox{ and } \forall a\in G,
\end{equation}
where $(R_{a})_{*}$ is right multiplication with $a$ acting on
vectors. The existence of the connection $\Gamma$ induces a split of every vector
$X\in T_{u}P$ into vertical (tangent to fibres) and horizontal
(orthogonal to fibres) components. Secondly the space $H_{u}$ must depend differentiably on
the point $u\in P$,
such that for every differentiable vector field $X$ on $P$, the vertical and
horizontal components
will also be differentiable vector fields.

\begin{figure}[h]
\begin{center}
\mbox{
  \epsfysize=6cm
\epsffile{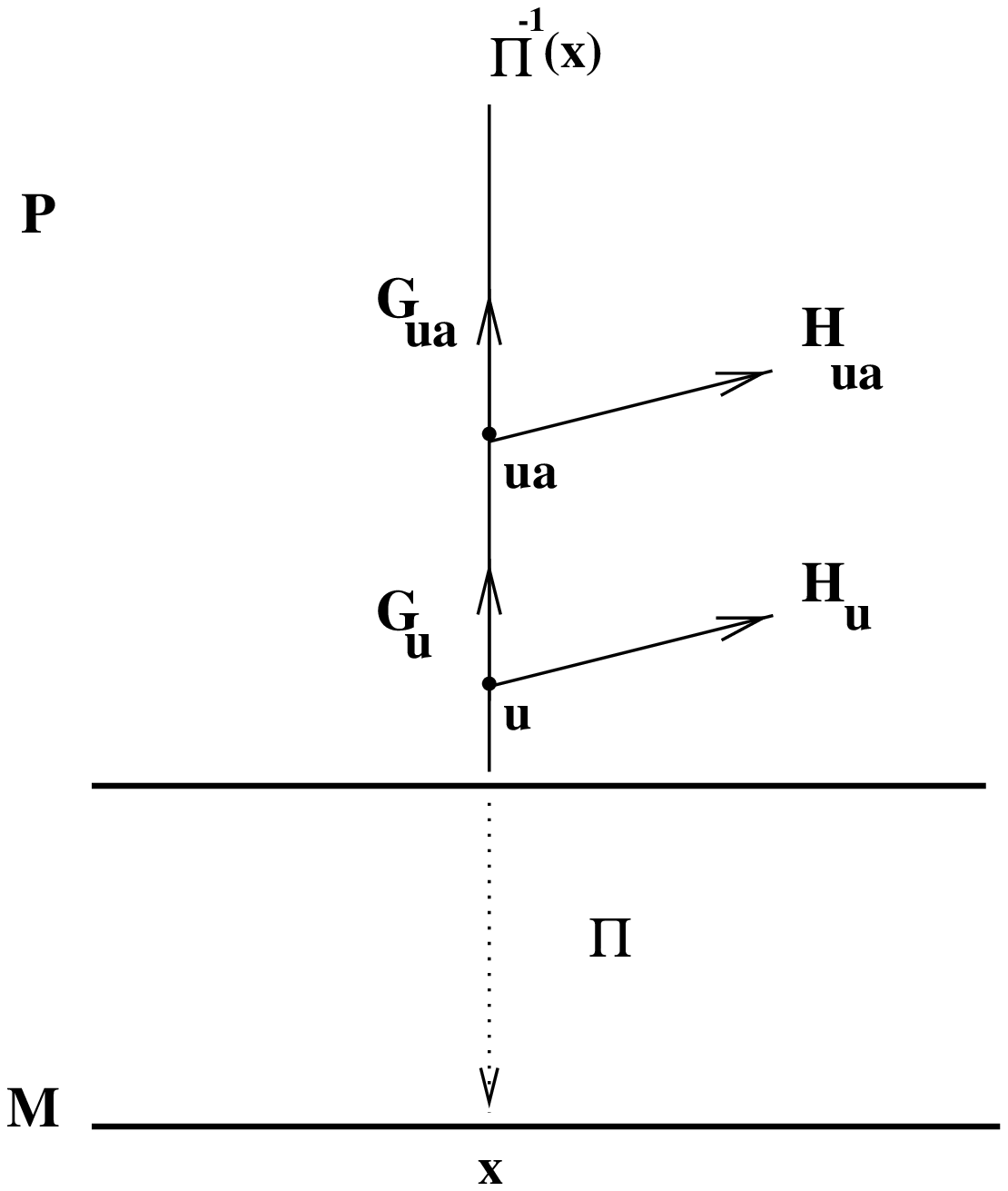}}
\end{center}
\caption{ \label{fig:fiber-bundle}
  The bundle $P(M,G)$ with a connection $\Gamma$.}
\end{figure}

Since $G$ acts on $P$ by right action, there exists a vector space isomorphism
between the Lie algebra \la and the space of vertical vectors $G_{u}$ in
every point $u\in P$. Let
\begin{equation}
  v(u) = \left. \frac{d R_{g(s)} u}{ds}\right\vert_{s=0},
\end{equation}
be an element in $G_{u}$. Such a vector field is known as a Killing vector field on $P$
relative to the action of $G$. A left invariant vector field
on $G$ satisfies the equation
\begin{equation}
  L_{g}^{'}\Bigl(v(h)\Bigr) = v(L_{g}h) = v(gh)\, , \,\, 
 \, \forall g,h \in G,
\end{equation}
which implies that $v(g)=L^{'}_{g}v(e)$. This reads in local coordinates
\begin{equation}
  v^{\alpha}(g) = \left. \frac{\partial (gh)^{\alpha}}{\partial h^{\beta}}\right\vert_{h=e} 
  v^{\beta}(e).
\end{equation}
The differential of the map $R_{u}$
\begin{equation}
  C_{u} = dR_{u}:\la \mapsto T_{u}P,
\end{equation}
defines an isomorphism $v(u)\leftrightarrow v(g)$. 
For every Lie algebra element $A$, this isomorphism defines a tangent vector to $P$ at $u$,
known as the fundamental vector fields on $P$
\begin{equation}
A^{*}_{u} = C_{u}(A), \,\, A\in \la.
\end{equation}
Since the action of $G$ preserves the fibres, $A^{*}_{u}\in\la$
is tangent to the fibres at each $u\in P$ and since $G$ acts freely on $P$,
$A^{*}$ will never vanish if $A\neq 0$. The map
$A \mapsto (A^{*})_{u}$ is a linear isomorphism from $\la$ onto $G_{u}$
for every $u\in P$. So we have the result
\begin{equation}
  {\rm Image}(C_{u}) \simeq G_{u}. \label{C-image}
\end{equation}
This global definition of a connection is often used in discussions of
the geometrical interpretation of TFT's. The more common definition
of a connection 1-form, follows from the existence of the connection
$\Gamma$.
For every element $X\in T_{u}P$ we define the
connection 1-form $\omega(X)$ to be the unique element $A\in \la$, for
which $A^{*}$ is the vertical component of $X$. 
We can collect this into the definition of the connection 1-form
$\omega$:
\begin{description}
\item{{\bf Theorem (2.1)}}   
  \begin{description}
  \item{(A)} $\omega(A^{*}) = A,\,\, \forall A\in \la$
  \item{(B)} $\omega\left((R_{a})_{*}X\right) = {\bf Ad}(a^{-1})\,\omega(X) =
    a^{-1}\omega(X)a$, 
    for all $a\in G$ and every $\,\,\,\,\,\,\,\,$ vector field $X$ on $P$. 
   \end{description}
   Conversely, given a \la- valued 1-form $\omega$ on P, satisfying (A) and
   (B), there exists a unique connection $\Gamma$ in $P$, whose connection
   1-form is $\omega$.
\item{{\bf Proof:}} See~\cite{KN_1}.
\end{description}
Note that one can define the horizontal vectors as the elements in the
kernel of $\omega$, since $\omega(X)=0$ if and only if $X$ is horizontal.
A central result is the following. The projection $\pi:P\mapsto M$ induces
a linear map $\pi: T_{u}P \mapsto T_{x}(M)$ for every $u\in P$. As always 
$\pi(u)=x\in M$. If there exists a connection $\Gamma$ in $P$, then the
projection $\pi:H_{u} \mapsto T_{x}M$ is an isomorphism. So
$H_{u}\simeq T_{x}M$, if there is a connection on P and we also just
saw that $G_{u}\simeq T_{u}(G)\simeq\la.$
If $X_{u}\in T_{u}P$ is a vector field on $P$, the connection
one-form $\omega_{u}$ at $u\in P$ can be written as
\begin{equation}
\omega_{u}(X_{u}) = C^{-1}_{u}\left(X_{u}^{{\rm vertical}}\right)
, \label{connection-C-inverse}
\end{equation}
since the horizontal vector fields constitute the kernel of $\omega$.
The general situation is
illustrated in figure~(\ref{fig:fiber-bundle}).
For differential $n$-forms $\eta$ with values in the Lie
algebra, we say that $\eta \in \Omega^{n}(P,\la)$. Hence
$\omega\in\Omega^{1}(P,\la)$. Next we discuss the curvature
$2$-form of $\omega$, but first we need the definition of a (pseudo)tensorial $r$-form.
\begin{description}
  \item{{\bf Definition (2.1)}} \label{definition_2.1}
    Let $P(M,G)$ be a principal fibre bundle and let
    $\rho$ be a representation of $G$ on a finite dimensional vector space
    $V$. That $\rho(a)$ is a representation, means that it is a linear
    transformation of $V$ for every $a\in G$ and there is a composition
    of these transformations $\rho(ab)=\rho(a)\rho(b)$, for $a,b\in G$.
    A pseudo-tensorial $r$-form on $P$ of type $(\rho,V)$, is a $V$-valued, $
    r$-form $\eta$ on $P$, such that\footnote{The $*$ supscript indicates that
      we perform the right translation on a form and not a vector field, where the $*$ is in subscript.}
    $R_{a}^{*}\eta = \rho(a^{-1})\eta$, for all $a\in G$.
    We say that $\eta$ is
    horizontal if $\eta\left(X_{1},\dots,X_{r}\right)=0$, whenever
    one or more of the tangent vectors $X_{i}\in T_{u}P$, $i=(1,\dots,r)$
    are vertical $(\mbox{i.e. } X_{i}\in G_{u}P)$.
    If $\eta$ is a pseudo-tensorial
    $r$-form and it is horizontal, then it is said to be a tensorial $r$-form
    on $P$.
\item{{\bf Theorem (2.2)}} 
  If $\eta$ is a pseudo-tensorial $r$-form on $P$ of type
  $(\rho,V)$, and we by $h$ denote the projection $h:T_{u}P\mapsto H_{u}$,
  then
\begin{description}
  \item{(A)} The form $\eta h$ defined by
    \begin{equation}
      (\eta h)\left(X_{1},\dots,X_{r}\right) =
      \eta\left(hX_{1},\dots,hX_{r}\right),
     \end{equation}
  $\,\,$  where $X_{i}\in T_{u}P,\, i=(1,\dots,r)$, is a tensorial $r$-form of
    type $(\rho,V)$.
  \item{(B)} $d\eta$ is a pseudo-tensorial $(r+1)$-form of type $(\rho,V)$,
    where $d$ is the $\,\,\,\,\,\,$ exterior derivative on $P$.
  \item{(C)} The $(r+1)$-form $D\eta \equiv (d\eta)h$ is the exterior
    covariant derivative of $\eta$, $\,\,\,\,\,\,\,\,\,\,\,\,\,$ and it is a tensorial form of type
    $(\rho,V)$.
\end{description}
\item{{\bf Proof:}} See~\cite{KN_1}
\end{description}
The relevant representation of $G$ for gauge theories, is when $\rho$ is
the adjoint representation of $G$ in \la, and $V$ is \la.
Then we say that the (pseudo-) tensorial forms are of type \AdG.
The curvature $\Omega$ of the connection one-form $\omega$ is
defined to be
\begin{equation}
  \Omega \equiv D\omega,\,\,\,\, \Omega\in \Omega^{2}(P,\la),
\end{equation}
which is a tensorial form of type \AdG, following 
theorem~(2.2).
It satisfies the famous structure equation
\begin{equation}
  \Omega(X,Y) = d\omega + \frac{1}{2}[ \omega(X),\omega(Y)],
\end{equation}
where $X,Y\in T_{u}P$. The proof of this well-known equation can also
be found in~\cite{KN_1}.

The reason for going through these seemingly dull definitions of
well-known objects in physics, is the following: First we wish to stress
the difference between expressing geometry on $P$ and on $M$.
Second, many of these technical details regarding e.g. whether a form is
horizontal or vertical, are important for the geometric interpretation of
topological field theories.

To go from the mathematical language to the notation commonly used in physics,
we need to express the connection and curvature on $M$ instead of $P$.
This requires some additional knowledge of $M$ which we list below
\begin{description}
  \item{(A)} We must specify an open covering $\{ U_{\alpha} \},\alpha\in I $
    of $M$, for $I$ being some index set, and
    $U_{\alpha}$ is an open subset of $M$.
  \item{(B)} We must choose a family of isomorphisms
    $\psi_{\alpha}:\pi^{-1}(U_{\alpha}) \mapsto U_{\alpha}\times G$, and
transition functions $\psi_{\alpha\beta}:U_{\alpha}\cap U_{\beta}\mapsto G$,
for $\alpha,\beta\in I$.
\item{(C)} Finally we need a set of local sections $\sigma_{\alpha}:U_{\alpha}\mapsto
 P.$
\end{description}
For every $\alpha \in I$ one defines the connection one-form
$\omega_{\alpha}$ as the pull back of $\omega$ on $P$ along the local section
\begin{equation}
  \omega_{\alpha} \equiv \sigma_{\alpha}^{*}\omega.
\end{equation}
On nonempty intersections $U_{\alpha}\cap U_{\beta} \neq \emptyset$ the 
pull back of the connection to different local sections is related by 
\begin{equation}
  \omega_{\beta} = \psi^{-1}_{\alpha\beta} \omega_{\alpha} \psi_{\alpha\beta}
  + \psi_{\alpha\beta}^{-1} d\psi_{\alpha\beta},
\label{omega-beta-alpha}
\end{equation}
This is of course just the statement that $\omega_{\beta}$ is related to
$\omega_{\alpha}$ by a local gauge transformation. The shift to physics
notation, is to patch together a one-form $A=A_{\mu}(x)dx^{\mu}$ from the
different $\omega_{\alpha}$'s, and if the bundle $P(M,G)$ is trivial
(i.e. $P\simeq M\times G$), the form $A$ will be a global one-form. Otherwise
we will have several local forms, e.g. for $S^{n}$ $(n>1)$ we would need at least two versions
of $A$, since the sphere can not be covered by a single coordinate patch. 
The physics version of equation~(\ref{omega-beta-alpha}) is the well-known
non-Abelian gauge transformation formula
\begin{equation}
  \tilde{A}_{\mu}(x) = g^{-1}A_{\mu}(x)g + g^{-1}dg,
\label{nonabelian-gt}
\end{equation}
where $g:M\mapsto G$ is a local gauge transformation, i.e. a function
assigning to every point in $M$ an element of $G$. We also must specify
the relation to the Lie algebra by introducing a basis $\{T_{i}\}$ of
\la with commutation relations
\begin{equation}
  [T_{i},T_{j}] = f_{ij}^{\,\,\,k}T_{k},
\end{equation}
where $f_{ij}^{\,\,\,k}$ are the structure constants of \la and
$i=(1,\dots,{\rm dim}(G))$. Now we can
write the one-form $A$ as a Lie algebra valued form
\begin{equation}
  A = A_{\mu}^{i}(x)T_{i}dx^{\mu}.
\end{equation}
The curvature two form read
\begin{equation}
  F[A] = \frac{1}{2} F_{\mu\nu}^{i}(x)T_{i}dx^{\mu}\wedge dx^{\nu},
  \label{curvature-F}
\end{equation}
where $F_{\mu\nu}^{i}$ is the field-strength tensor
\begin{equation}
F_{\mu\nu}^{i}(x) = \partial_{\mu}A_{\nu}^{i}(x) - \partial_{\nu}A_{\mu}^{i}(x)
+ f_{jk}^{\,\,\,i}A_{\mu}^{j}(x)A_{\nu}^{k}(x). \label{field-strength}
\end{equation}

\noindent
A relevant question is how the forms on $P$ and $M$ are related? One can
approach this from various angles, but the one presented here, is
the one most useful for later discussions. We state the following two
lemmas used by Kobayashi and Nomizu, related to their proof of the
important Weil theorem, which is discussed later in this section.
\begin{description}
  \item{{\bf Lemma (2.1)}}
    An $r$-form $\eta$ on $P$ projects by $\pi:P\mapsto M$, to
    a unique $r$-form $\tilde{\eta}$ on $M$ if
   \begin{description}
     \item{(A)}
       $\eta\left(X_{1},\dots,X_{r}\right)=0$, if one or more 
       $X_{i}\in G_{u}$, for $X_{i}\in T_{u}P$, $i=(1,\dots,r)$ and
       $u\in P$.
     \item{(B)}
       $\eta\left(R_{a}X_{1},\dots,R_{a}X_{r}\right)=
       \eta\left(X_{1},\dots,X_{r}\right)$, for every $a\in G$.
   \end{description}
 \item{{\bf Lemma (2.2)}}
   If an $r$-form $\eta$ on $P$ projects to an $r$-form $\tilde{\eta}$
   on $M$, that is if $\eta = \pi^{*}(\tilde{\eta})$, then
   \begin{equation}
     d \eta = D \eta.
   \end{equation}
 \item{{\bf Proofs}} For proofs see~\cite{KN_2}.
\end{description}
Several important papers mention a result, without proof, of very
similar nature. We define the interior product and the Lie derivative in order to present this result as
the next lemma. The interior product is a map
  $\imath : \Omega^{r} \mapsto \Omega^{r-1}$ written as
  \begin{equation}
    \imath_{X}\omega(X_{1},\dots,X_{r-1}) \equiv
    \omega(X,X_{1},\dots,X_{r-1}).
   \end{equation}
Write $\omega = (1/r!)\omega_{\mu_{1}\dots\mu_{r}}dx^{\mu_{1}}\wedge
\dots\wedge dx^{\mu_{r}}$ for the $r$-form and $X=X^{\mu}\partial_{\mu}$
for the vector field. The interior product of $\omega$ with $X$ is 
\begin{eqnarray}
 \imath_{X}\omega &=& \frac{1}{(r-1)!}X^{\nu}\omega_{\nu\mu_{2}\dots\mu_{r}}
dx^{\mu_{2}}\wedge\dots\wedge dx^{\mu_{r}}\nonumber \\
&=& \frac{1}{r!}\sum_{s=1}^{r}X^{\mu_{s}}
\omega_{\mu_{1}\dots\mu_{s}\dots \mu_{r}}(-1)^{s-1}dx^{\mu_{1}}\wedge
\dots d\tilde{x}^{\mu_{s}}\wedge \dots \wedge dx^{\mu_{r}},
\end{eqnarray}
where we omit $d\tilde{x}^{\mu_{s}}$ in each term in the sum. 
The Lie derivative which we introduced in chapter 1, acts on forms ${\cal L} : \Omega^{r} \mapsto
\Omega^{r}$ as a degree preserving map
\begin{equation}
  {\cal L}_{X}\omega = (d \imath_{X} + \imath_{X}d )\omega,
\end{equation}
\begin{description}
  \item{{\bf Lemma (2.3)}}
    Let $\omega$ be an $r$-form on $P$. A form on $P$ is said to be horizontal if
\begin{description}
\item{(A)} $\imath_{X} \omega = 0$ for all $X\in \la\simeq G_{u}P$.
\end{description}
and is said to be invariant if:
\begin{description}
\item{(B)} ${\cal L}_{X} \omega = 0$ for all $X\in \la\simeq G_{u}P$.
\end{description}
If a form $\omega$ on $P$ is both horizontal and invariant, it is said to
be a basic form on $P$. Every basic form $\omega$
on $P$ projects to an unique form $\tilde{\omega}$ on $M$, such that
$\omega= \pi^{*}(\tilde{\omega})$.
\item{{\bf Proof: }} Since we have no knowledge of any 
  proof of this lemma in the literature, we show that it is equivalent to the
  previous lemma, for which proof is given in~\cite{KN_2}. First we show
  that  lemma (2.1:A) implies lemma (2.3:A).
  
Lemma (2.1:A), states that  $\omega
\left(X_{1},\dots,X_{r}\right)=0$ if one or more 
$X_{i}\in G_{u}$, for $X_{i}\in T_{u}P$, $i=(1,\dots,r)$ and $u\in P$. Is it
clear that this is equivalent to  $\imath_{X}\omega = 0$ for all $X\in \la$,
since the elements in the Lie algebra are the vertical vectors on $P$.
The opposite way,  
we have that $\imath_{X}\omega=0$ is a vanishing $(r-1)$-form on $P$,
$\omega(X,X_{1},\dots,X_{r-1})=0$. At least one of the arguments
of this $(r-1)$-form, namely $X$ 
is vertical. So lemma (2.1:A) implies lemma (2.3:A) and vice versa.

To show that lemma (2.3:B) implies lemma (2.1:B) we 
assume that $G$ is a connected Lie group. When this is true, we can always
represent a group element by an algebra element using the exponential
map. Let $g\in G$ and $X\in \la$ and for $t\in \R$ write
\begin{equation}
  g_{t} = \exp(tX).
\end{equation}
Let $\phi_{t}:P\mapsto P$ be a map defined by the right translation by
$g_{t}$
\begin{equation}
  \phi_{t} p = p g_{t},\, \mbox{ for } p \in P.
\end{equation}
The statement in lemma (2.1:B) is that $\phi^{*}_{t}(\omega) = \omega$
for all $t$. We calculate
\begin{equation}
\frac{d}{dt} \phi_t^*(w) |_{t=t_0} = 
lim_{s\rightarrow 0} \frac{ \phi_{t_0+s}^*w - \phi_{t_0}^*w}{s} =
\phi_{t_0}^*lim_{s\rightarrow 0} \frac{ \phi_{s}^*w - w}{s} =
\phi_{t_0}^*({\cal L}_{X}w) = 0.
\end{equation}
Here we use that $\phi$ is a linear transformation and the definition
of the Lie derivative. This show that $\phi_{t}^{*}(\omega)$
is independent of $t$ so $\phi_{1}^{*}(\omega)=\omega$, which is
lemma (2.1:B). The other way is along the same lines. Assume
\begin{eqnarray}
\phi_{t}^{*}(\omega) &=& \omega \nonumber \\
&\Downarrow & \nonumber \\
\omega ( (\phi_{t})_{*}X_{1},\dots,(\phi_{t})_{*}X_{r} )
&=& \omega (X_{1},\dots,X_{r}) \nonumber \\
& \Downarrow & \nonumber \\
\omega ( X_{1}e^{tX},\dots, X_{r}e^{tX} )
&=& \omega (X_{1},\dots,X_{r}) \nonumber \\
& \Downarrow & \nonumber \\
\frac{d}{dt} \omega ( X_{1}e^{tX},\dots, X_{r}e^{tX} ) |_{t=t_0} 
&=& \frac{d}{dt}\omega (X_{1},\dots,X_{r}) |_{t=t_0} \nonumber \\
& \Downarrow & \nonumber \\
\phi^{*}_{t_{0}}({\cal L}_{X}\omega) &=& 0,
\end{eqnarray}
Since this holds for all $t_{0}$, ${\cal L}_{X}\omega = 0$ for
all $X \in \la$. Hence we have shown that lemma (2.3) is equivalent to 
lemma (2.1), for which we know a proof in the literature. Lemma (2.3)
is important in relation to the discussion of equivariant cohomology in
section~(\ref{section:math-behind-phys}).\footnote{I thank J.E. Andersen for
  help on the proof.}
\end{description}

\begin{figure}[h]
\begin{center}
\mbox{
\epsfysize=6cm
\epsffile{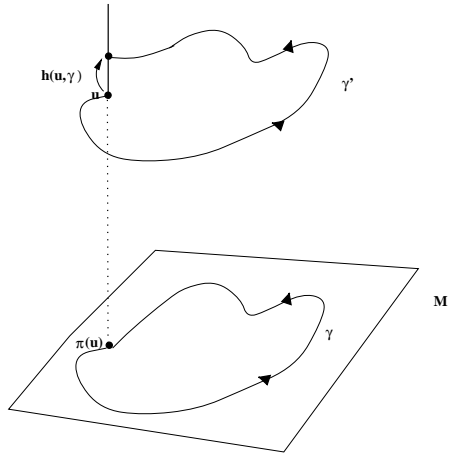}
}
\end{center}
\caption{\label{fig:holonomy}  The horizontal lift of $\gamma$ to $P$.}
\end{figure}

Consider a closed loop $\gamma(t), \,t\in [0,1]$ in $M$, such that
$\gamma(0)=\gamma(1)$. The horizontal lift of $\gamma$ to
$P$ is the horizontal curve $\gamma'(t)$ given by $\pi(\gamma'(t)) =
\gamma(t)$ for $t\in [0,1]$. The horizontal nature stems from the fact that
the tangent vectors to $\gamma$ are lifted to horizontal vectors in $P$,
since $H_{u}\simeq T_{x}(M)$ where $\pi(u)=x \in M$. In
figure~(\ref{fig:holonomy}) we have indicated that the lifted curve does
not need to be closed.
The lack of closure is the existence of the holonomy $h(u,\gamma)$.
If $P\mapsto M$ is a principal bundle then   
\begin{equation}
  \gamma'(1) = \gamma'(0)\cdot h(u,\gamma) = u \cdot h(u,\gamma),
\end{equation}
for some group element $h(u,\gamma)\in G$. The holonomy elements are
the parallel displacement of $\gamma'(0)\in \pi^{-1}(x)$ along the
curve $\gamma$, into $\gamma'(1)\in\pi^{-1}(x) $, and
represent an isomorphism of the fibre $\pi^{-1}(x)$ to itself.
That it is an isomorphism follows from the result that parallel displacement
commutes with right multiplication on $G$~\cite{KN_1}. It follows from
the fact that the 
horizontal subspaces $H_{u}\subset T_{u}P$ are invariant under right
multiplication, that the holonomy will transform under adjoint action of $G$
under a change of base point
\begin{equation}
  h(ug,\gamma) = g^{-1}h(u,\gamma)g.
\end{equation}
An obvious consequence is that the trace of the holonomy
$tr(h(u,\gamma))$ will be
gauge invariant. This is exactly what is known as the Wilson loop
in physics, which can be represented on $M$ as
\begin{equation}
  W(\gamma,A) = tr \left( {\cal P}\exp(\oint_{\gamma}
    A_{\mu}^{i}T_{i} dx^{\mu})\right),
\end{equation}
where ${\cal P}$ stands for path ordering. The gap between the start and
end points of $\gamma'$ is due to the curvature of the connection
on $P$. The Ambrose-Singer theorem~\cite{KN_1} states that the knowledge of
the holonomy group, which is the collection of all isomorphisms
$h(u,\gamma):\pi^{-1}(x)
\mapsto \pi^{-1}(x)$ defined above, is enough to reconstruct the curvature
$\Omega$ on $P$.
\subsection{The Moduli Space of Connections}
We introduce the following concepts. By \A we denote the space of all
connections  $A\in \Omega^{1}(P,\la)$. This space is infinite dimensional
and affine\footnote{It has no fixed origin.} so the difference between two
elements $A,A'\in \A$ reads~\cite{Mitter_Viallet,Babelon_Viallet}
\begin{equation}
A-A' = \tau, \,\,\,\, \tau\in \Omega^{1}(M,\la), \label{A-affine}
\end{equation}
where $\tau$ transforms in the adjoint representation.
Since $\tau$ is a one-form, the equation above just states
that $\delta A=A'-A=D\chi$ for some zero-form $\chi\in\Omega^{0}(M,\la)$
and $D$ being the covariant exterior derivative on $M$.
But connections (or families of horizontal subspaces $H_{u}$) which
differ by a gauge transformation, defined as the map $\phi$ below, should
be considered equivalent. Let $\phi:P\mapsto P$ be a diffeomorphism,
mapping $u \mapsto \phi(u)$ for $u\in P$, such that 
\begin{equation}
   \phi(ug) = \phi(u)g, \, \forall g\in G,   
\end{equation}
and that the map preserves base points of fibres
\begin{equation}
  \pi(\phi(u)) = \pi(u), \, u\in P. \label{basepoint}
\end{equation}
In order to be compatible with the right action of $G$ on $P$ we demand 
\begin{equation}
u \mapsto \phi(u)= u \tilde{\phi},  \, \label{commute}
\end{equation}
where $\tilde{\phi}:P\mapsto G$ such that
\begin{equation}
\tilde{\phi}(ua) = a^{-1}\tilde{\phi}(u)a, \, \forall a\in G.
\end{equation}
The set of all such $\phi$'s is known as the vertical automorphism group
on $P$, and they form the group of gauge transformations \G.
Before we discussed gauge transformations as local maps $g:U\mapsto G$,
where $U\subset M$.
Recall that the local section
$\sigma_{\alpha}:U_{\alpha}\mapsto P$ is ``inverse'' to $\pi:P\mapsto M$
such that $\sigma_{\alpha} \cdot \pi = {\bf 1} \vert_{U_{\alpha}}$ and that
the local sections are related via
$\sigma_{\beta}(x)=\sigma_{\alpha}(x)\psi_{\alpha\beta}(x), \forall 
x\in U_{\alpha}\cap U_{\beta}$.
with smooth transition functions
$\psi_{\alpha\beta}:U_{\alpha}\cap U_{\beta} \mapsto G$.  Now
apply equation~(\ref{commute}) to show that under a gauge transformation
\begin{equation}
  \phi( \sigma_{\alpha}(x)) = \sigma_{\alpha}(x)\tilde{\phi}_{\alpha}(x),
  \,\mbox{where } \tilde{\phi}_{\alpha}(x)\equiv \tilde{\phi}
(\sigma_{\alpha}(x)),
\end{equation}
we recover equation~(\ref{omega-beta-alpha})
\begin{equation}
  A_{(\alpha)} \mapsto  A_{(\alpha)}\tilde{\phi}_{\alpha}(x) =
  Ad_{\tilde{\phi}^{-1}}
  A_{\alpha} + \tilde{\phi}_{\alpha}^{-1}d\tilde{\phi}_{\alpha},
\end{equation}
for the connection one-form $A_{(\alpha)}=\sigma_{\alpha}^{*}A$, defined
on $U_{\alpha}\subset M$.
\noindent
Under a change of local sections we have $\tilde{\phi}_{\beta}=
Ad_{\psi_{\alpha\beta}^{-1}} \tilde{\phi}_{\alpha}$
pointwise for all $x\in U_{\alpha}\cap U_{\beta}$. We 
identify \G with the set of families $\{ \tilde{\phi}_{\alpha} \}$,
consisting of maps $\tilde{\phi}_{\alpha}:U_{\alpha}\mapsto G$, or more
precisely $\pi^{-1}(U_{\alpha})\mapsto
U_{\alpha}\times G$, where $\alpha\in I$.

Consider the associated bundle ${\bf Ad}P = P \times_{{\bf Ad}}
G=(P\times G)/G$,
where $G$ acts on $G$ by adjoint action. Every $\tilde{\phi}$ defines a
section $\hat{\phi}$ of this adjoint bundle
\begin{equation}
  \hat{\phi}(x) = [(u,\tilde{\phi}(u))], \, \pi(u)=x\in M,\, u\in P,
\end{equation}
where $[\,\cdot\,]$ denotes an equivalence class in $P\times G$ under the
projection $P\times G \mapsto {\bf Ad}P$. Since $\tilde{\phi}$
transforms in the adjoint representation, $[(u,\tilde{\phi}(u))]$ is 
independent of the choice of $u$.
In this framework \G is the space of sections on the
adjoint bundle ${\bf Ad} P$ and the Lie algebra of \G is the space of
sections on the associated bundle ${\bf ad} P = P\times_{{\bf ad}} \la$~\cite{Mitter_Viallet}.
In the rest of the thesis, when denoting a differential
$r$-form on $M$ as having values in the Lie algebra \la, i.e. being an
element in $\Omega^{r}(M,\la)$, it is understood to be an
$r$-form on the adjoint bundle ${\bf ad} P$. These forms transform in the adjoint
representation of \la under gauge transformations.

The action of \G on \A reads
\begin{equation}
  A \mapsto \phi^{*}A = \phi^{-1} A \phi + \phi^{-1}d\phi , \,  A\in \A ,\,
\phi\in \G.
\end{equation}
The elements $\phi\in \G$ for which $A=\phi^{*}A$, form a subgroup $I_{A}$ of
\G, known as the isotropy group of \G. The centre $Z(G)$ of the gauge
group $G$ (i.e. those elements which
commute with all other elements in $G$)
is a subgroup of $I_{A}$. Let $z\in Z(G)$ and consider
the gauge transformation
\begin{equation}
  \phi_{z}: u \mapsto \phi_{z}(u) = uz, \,\, u\in P,
\end{equation}
which represent a global gauge transformation. For this $\phi_{z}$ we
have $A = \phi^{*}_{z}A$ and hence $Z(G)\subseteq I_{A}$. We only consider
the subset of \A, consisting of those connections for which
$Z(G)=I_{A}$. We denote this space \A as well,
and the elements are known as
irreducible connections. Also
for irreducible connections, the construction~\cite{Mitter_Viallet}
\begin{eqnarray}
  \G \mapsto &\A & \nonumber \\ &\downarrow& \, \pi \\ &\AMG&,\nonumber
\label{AMG}
\end{eqnarray}
where we now by \G understand $\G/Z(G)$, will be a principal fibre bundle.
In this situation \AMG can be given a smooth manifold structure, but
the presence of reducible connections would make \AMG into an orbifold with
conical singularities~\cite{Birmingham-review}. 

Now we want to study the geometry of the principal bundle $\A\mapsto \AMG$.
This is central for understanding section~\ref{section:math-behind-phys}.
The space of vertical vector fields in equation~(\ref{C-image}) is given
as the image of the map $C_{u}:\la \mapsto T_{u}P$. Since we have a
connection we also have a metric, which defines the orthogonal complement
to $G_{u}$ as the space of horizontal vectors. The metric allows us to
define the
adjoint map
\begin{equation}
  C^{\dagger}_{u}: T_{u}P \mapsto \la
, \label{C-dagger}
\end{equation}
as a Lie algebra valued
one-form. It is then possible to define the following projection operator
\begin{equation}
  \Pi^{{\rm vertical}} \equiv C\frac{1}{C^{\dagger}C}C^{\dagger}:T_{u}P
  \mapsto G_{u}\subset T_{u}P.
\end{equation}
If it is a projection operator, it must be idempotent such that $\Pi^{2}=\Pi$
\begin{equation}
C\frac{1}{C^{\dagger}C}C^{\dagger} C\frac{1}{C^{\dagger}C}C^{\dagger} = 
C\frac{C^{\dagger}C}{C^{\dagger}C}\frac{1}{C^{\dagger}C}C^{\dagger} =
C\frac{1}{C^{\dagger}C}C^{\dagger}.
\end{equation}
Recall equation~(\ref{connection-C-inverse}), which we now can write as
\begin{equation}
  \omega(X_{u}) = C^{-1}\Pi^{{\rm vertical}}(X_{u}) =
  \frac{1}{C^{\dagger}C}C^{\dagger}, \, X_{u}\in T_{u}P.
\label{omega-som-C-invers-Pi}
\end{equation}

We now apply this general formalism on $\A \mapsto\AMG$.
The tangent space $T_{A}\A$ at a connection $A\in \A$  can be
identified with $\Omega^{1}(M,\la)$ since equation~(\ref{A-affine})
showed that the difference between two connections is a Lie algebra valued
one-form. The gauge group \G acts on \A as gauge transformations
on the points 
\begin{equation}
  A \mapsto A + D\chi, \chi \in \Omega^{0}(M,\la),
\end{equation}
where $D \chi = d \chi + [A,\chi]$.
Using the projection above we can split the tangent space of
\A into a vertical and a horizontal part
\begin{equation}
  T_{A}\A = V_{A} \oplus H_{A},
\end{equation}
where $V_{A}= {\rm Image}(D)$ is the vertical part being tangent to the
orbits of \G through $A$. The horizontal part is as always the
orthogonal complement with respect to the inner product between forms on $M$
\begin{equation}
  \langle X,Y\rangle = \int_{M} tr( X\wedge * Y), \,\,\,\,\, X,Y \in \Omega^{*}(M,
  \la). \label{metric-M-forms}
\end{equation}
The trace is taken as the Killing metric on the Lie algebra.
A generic one-form $X\in \Omega^{1}(M,\la)$ will split into vertical and
horizontal parts as
\begin{eqnarray}
  X &=& \Pi^{{\rm vertical}}X + (X -  \Pi^{{\rm vertical}}X) \nonumber \\
    &=& DG_{A}D^{\dagger} X + (X - DG_{A}D^{\dagger} X) \nonumber \\
    &\equiv & v_{A}X + h_{A}X, \label{split}
\end{eqnarray}
where $G_{A}$ is the Greens function of the covariant scalar Laplacian
\begin{equation}
  G_{A} \equiv \frac{1}{\Delta_{0}^{'}}.
\end{equation}
The general covariant Laplacian $\Delta_{i}^{'}$, which acts on $i$-forms,
is the operator
\begin{equation}
  \Delta_{i}^{'} \equiv D^{\dagger}D + DD^{\dagger}.
\end{equation}
When acting on zero-forms $D^{\dagger}$ projects to a minus one-form and the
last term is excluded in the definition of the scalar Laplacian
\begin{equation}
  \Delta_{0}^{'} \equiv D^{\dagger}D.
\end{equation}
Using that $H_{u}\simeq T_{x}M$, we will identify
\begin{equation}
  T_{[A]}\AMG \simeq H_{A}, \, \mbox{ for } A\in [A], \mbox{ where} [A]
  \in \AMG. \label{tangent-horizontal-AMG}
\end{equation}
Since the tangent space is identified with the Lie algebra valued
one-forms on $M$, the metric in equation~(\ref{metric-M-forms})
also defines a metric $g_{{\cal A}}$ on \A. Using the identification in
equation~(\ref{tangent-horizontal-AMG}) this metric induces a
metric on \AMG
\begin{equation}
  g_{{\cal A}/{\cal G}}( [X], [Y] ) = g_{{\cal A}}( h_{A}X, h_{A}Y).
\end{equation}
Here  $X,Y\in \Omega^{1}(M,\la)$ projects to  $[X],[Y]\in T_{A}\AMG$, using 
the split in equation~(\ref{split}). Application of 
equation~(\ref{omega-som-C-invers-Pi}) in this situation, where the
connection $\omega_{\!\A}\in \Omega^{1}(\A)$ on the bundle
$\A\mapsto\AMG$, can be viewed as mapping
\begin{equation}
  \omega_{\!\A}: T_{A}\!\A \mapsto \Omega^{0}(M,\la) \mbox{ such that }
  X \mapsto \omega_{\!\A}(X) = \frac{1}{D^{\dagger}D}D^{\dagger}X.
\end{equation}
The connection is vertical and it assigns to the fundamental vector field
$D\tau$ the Lie algebra element
\begin{equation}
  \omega_{\!\A}(D\tau) = \frac{1}{ D^{\dagger}D} D^{\dagger} (D\tau) = \tau,
\end{equation}
in agreement with the definition of the connection one-form in
theorem (2.1). The curvature $\Omega_{\!\A}$ is a horizontal two-form
\begin{equation}
  \Omega_{\!\A} = d_{\!\A}\omega_{\!\A} + \frac{1}{2}[ \omega_{\!\A},
                  \omega_{\!\A}], \label{curvature-on-A}
\end{equation}
with $d_{\!\A}$ denoting the exterior derivative on \A.
If we evaluate the curvature on horizontal vectors, in order to get a
curvature which projects to a curvature on \AMG, only the first
term in equation~(\ref{curvature-on-A}) will contribute, since the
horizontal vectors are in the kernel of the connection. One can
view\footnote{I thank Matthias Blau for explaining this fact to me.}
the connection $\omega_{\!\A}$ as the following one-form on
\A
\begin{equation}
  \omega_{\!\A} = \frac{1}{D^{\dagger}D}D^{\dagger}d_{\!\A} A \in
  \Omega^{1}(\A),\,\, A\in \A. \label{A-connection}
\end{equation}
In this notation the curvature becomes 
\begin{equation}
  \Omega_{\!\A} = d_{\!\A} \left(
    \frac{1}{D^{\dagger}D}D^{\dagger}d_{\!\A} A \right) =
   \frac{1}{D^{\dagger}D}  d_{\!\A}\left(D^{\dagger}d_{\!\A} A \right),
\label{A-connection-2}
\end{equation}
since the horizontal vectors are in the kernel of $D^{\dagger}$. Using that
$ D^{\dagger} = *D* $ and the definition of the covariant exterior derivative,
one finds 
\begin{equation}
  \Omega_{\!\A} =\frac{1}{D^{\dagger}D} \left( *[d_{\!\A} A, *d_{\!\A} A]
    \right). 
\end{equation}
When evaluated on tangent vectors and using $T_{A}(\A)
\simeq \Omega^{1}(M,\la)$ we can write
\begin{equation}
  \Omega_{\!\A}(X,Y) = \frac{1}{D^{\dagger}D} \left( *[\tilde{X},*\tilde{Y}]\right),
  \label{curvature-commutator}
\end{equation}
where $\tilde{X},\tilde{Y} \in \Omega^{1}(M,\la)$. These forms corresponds to the
vector fields $X,Y$ on \A, and by evaluating $\Omega_{\!\A}$ on these fields and
translating the vector fields to forms on $M$, we obtain a two-tensor. The forms
$\tilde{X},\tilde{Y}$ are subject to the horizontal condition
$\tilde{X},\tilde{Y} \in {\rm ker}( D^{\dagger} )$, written in components
$D_{\mu}\tilde{X}^{\mu}=D_{\mu}\tilde{Y}^{\mu}=0$. This formula is very important for
interpreting the geometry of Witten type TFT's.

\subsection{Definition of Characteristic Classes}
Finally we list a few definitions~\cite{KN_2} regarding
characteristic classes which will be good to remember in the
following sections.
\begin{description}
  \item{{\bf Definition (2.2) }} Let $f$ be a multi-linear map
    \begin{equation}f:
      \overbrace{\la \times \cdots \times \la}^{ \mbox{k-times}}
      \mapsto \R,
     \end{equation}
such that
   \begin{equation}
     f\left( {\bf Ad}(a)t_{1},\dots, {\bf Ad}(a)t_{k} \right) =
     f\left( t_{1},\dots,t_{k}\right), \mbox{ for } a\in G \mbox{ and }
     t_{1},\dots,t_{k}\in \la.
    \end{equation}
Then $f$ is said to be a $G$- invariant map.
   \item{{\bf Definition (2.3)}} Let $I^{k}(G)$ be the set of all
     symmetric multi-linear, $G$- invariant maps $f$, defined above.
     $I^{k}(G)$ is a vector space over \R, and one defines
     $I(G)=\sum_{k=0}^{\infty}I^{k}(G).$
\item{{\bf Definition (2.4)}} Let $f\in I^{k}$ and $g\in I^{g}$, and define
  the product $fg\in I^{k+g}$ as
  \begin{equation}
    fg\left( t_{1},\dots,t_{k+g}\right)
 = \frac{1}{(k+g)!} \sum_{\sigma} f
 \left( t_{\sigma(1)},\dots,t_{\sigma(k)}\right)
 g\left( t_{\sigma(k+1)},\dots,t_{\sigma(k+g)}\right).
 \end{equation}
Where the sum is taken over all permutations $\sigma$ of $(1,\dots, k+g)$.
Endowed with this product structure, $I(G)$ is a commutative algebra.
\end{description}
The following theorem is due to Weil.
\begin{description}
  \item{{\bf Theorem (2.3)}}
    Let $\pi:P\mapsto M$ be a principal $G$ bundle with a connection one-form
    $\omega$ and its curvature two-form $\Omega$
    For each $f\in I^{k}(G)$, construct the following $2k$-form
        \begin{equation}
          f(\Omega)(X_{1},\dots,X_{2k}) = \frac{1}{(2k)!}
          \sum_{\sigma}\, {\rm sign}\sigma
          f\left((\Omega(X_{\sigma(1)},X_{\sigma(2)}),\dots,
          \Omega(X_{\sigma(2k-1)},X_{\sigma(2k)})\right),
        \end{equation}
        where $X_{1},\dots,X_{2k}\in T_{u}P$ and the sum is taken over
all permutations $\sigma$, such that:
\begin{description}
      \item{(A)}      
   \noindent The $2k$-form $f(\Omega)$ on $P$ projects
        to a unique closed $2k$-form $\overline{f}(\Omega)$ on $M$, $\,\,$i.e.
        there exists a unique $\overline{f}(\Omega)$ such that
        \begin{equation}
        f(\Omega)=\pi^{*}(\overline{f}(\Omega)).
        \end{equation}  
\item{(B)}
        Denote by $w(f)$ the element of the de Rham cohomology group
        $H^{2k}(M,\R)$ defined by the closed $2k$-form
        $\overline{f}(\Omega)$.
        The class $w(f)$ is independent of the choice of connection $\omega$
        in $P$ and
        \begin{equation}
          w: I(G)\mapsto H^{2k}(M,\R),
        \end{equation}
        is an algebra homomorphism. (The Weil homomorphism).
     \end{description}
     \item{{\bf Proof:}} For a proof see~\cite{KN_2}.
\end{description}
The two lemmas (2.1) and (2.2) are used in this proof. One uses that
$\Omega$ is a tensorial two-form of type \AdG, and since $f$ by definition
is $G$-invariant $f(\Omega)$ satisfies lemma (2.1). Lemma (2.2) states
that $df(\Omega)=Df(\Omega)$ and since the Bianchi identity
$D\Omega =0$ implies that $D(f(\Omega))=0$, $f(\Omega)$ will be a
closed form on $M$. The de Rham cohomology group is build on the de Rham
complex, which consists of the complex $\Omega^{*}(M)=\{ C^{\infty} \mbox{
  functions on } M \}\otimes \Omega^{*}(M)$ together with the exterior
differential $d$ on $M$. The kernel of $d$ are the closed forms and the
image of $d$, the exact forms. The cohomology group $H^{q}(M,d)$ is then
the vector space of closed $q$-forms modulo the exact $q$-forms.

Next we define the invariant polynomials on the Lie algebra \la.
\begin{description}
  \item{{\bf Definition (2.5)}} Let $V$ be a vector space over \R.
    Define $S^{k}(V)$ to be the space of all
  symmetric multi-linear mappings
  \begin{equation}
    f: \overbrace{V\times\cdots\times V}^{\mbox{k-times}} \mapsto \R.
  \end{equation}
This can be endowed with the same product structure as $I(G)$ which defines a
commutative algebra over \R
\begin{equation}
  S(V) = \sum_{k=0}^{\infty} S^{k}(V).
\end{equation}
 \item{{\bf Definition (2.6)}} Denote by $\xi^{1},\dots,\xi^{k}$ a basis for the
   dual of $V$. The map $p:V\mapsto \R$ is a known as a polynomial function
   if it can be expressed as a polynomial of the basis
   vectors $\xi_{i}\in V^{*}$ for $i=(1,\dots,k)$. By $P^{k}(V)$ we denote the space of all
   homogenous polynomials of degree $k$ on $V$. By the natural product
   structure 
\begin{equation}
  P(V) = \sum_{k=0}^{\infty}P^{k}(V),
\end{equation}
is the algebra of polynomial functions on $V$.
\end{description}
With these definitions the following results can be found
\begin{description}
  \item{{\bf Theorem (2.4)}}
    The map $\phi:S(V)\mapsto P(V)$ defined by
    \begin{equation}
      (\phi f)(t) = f\left(t_{1},\dots,t_{k}\right), \mbox{ for }
      f\in S^{k}(V) \mbox{ and } t_{1},\dots,t_{k}\in V,
    \end{equation}
is an isomorphism of $S(V)$ onto $P(V)$.
\item{{\bf Theorem (2.5)}} Let $G$ be a group of linear
  transformations of $V$,
  and let $S_{G}(V)$ and $P_{G}(V)$ be the sub-algebras of $S(V)$ and $P(V)$
  respectively, consisting of those elements which are $G$- invariant.
  Then the isomorphism $\phi$ defined above induces an isomorphism
  $\phi:S_{G}(V)\mapsto P_{G}(V)$.
\item{{\bf Corollary (2.1)}} Let $G$ be a Lie group and \la its Lie
  algebra. The algebra $I(G)$ of \AdG-invariant symmetric mappings
  of $\la$ into \R $\,$ then can be identified with the algebra of $({\bf Ad}G)$-invariant
  polynomial functions $P_{G}(\la)$.
\item{{\bf Proofs:}} See~\cite{KN_2}.  
\end{description}
By the Weil theorem and the definitions and results quoted above, we
can define characteristic classes as the cohomology classes on $M$ defined
by the ${\bf Ad}G$-invariant polynomials $P_{G}(\la)$ of the curvature
$\Omega\in \Omega^{2}(P,\la)$.
The most important examples are the Chern classes defined via
complex vector bundles $E\mapsto M$ with fibre $\C^{k}$ associated
to $P$, where $G\subseteq GL(k,\C)$, and the Euler classes, defined via
a real orientable vector bundle $E\mapsto M$ associated to $P$.
Both will define elements in $H^{k}(M,\R),\,(k=2n)$. We discuss the
definition of the Euler class in more detail in section~\ref{loc} and 
for further information we refer to the literature~\cite{YC,KN_2,Nakahara}.
All characteristic classes are topological invariants since they do not
depend on the connection from which the curvature is defined
(by the Weil theorem),  and they are by definition gauge invariant.
These cohomology classes ``measure'' the non-triviality of the bundle,
from which they are constructed~\cite{YC}.

\section{Schwarz Type TFT's.}

In this section we discuss the oldest type of topological field theories,
now named after Schwarz, who found a number of significant results in the
late seventies. We discuss some of these results in the context of
BF theories, which are generalisations of the well-known Chern-Simons
theory. At the end we discuss Chern-Simons theory as a special case of
BF theory and also the relation between BF theory and Yang-Mills theory.

The defining property of Schwarz type TFT's, is that the classical action
is independent of the space-time metric $g_{\mu\nu}$ on $M$. The standard
example is the Chern-Simons action
\begin{equation}
  S_{{\rm class}} = \frac{k}{4\pi}\int_{M} tr( A\wedge dA + \frac{2}{3}
  A\wedge A\wedge A), 
\end{equation}
where $M$ is a three dimensional manifold and $k$ is a constant.
We return to the details of Chern-Simons theory later,
but the important thing is, that one can write down the action without
reference to the metric. The non-trivial part of Schwarz type
theories is to show that the resulting quantum theory is topological. This is
the reason for this type of theories to be known as topological
quantum field theories or just as TQFT's.
\subsection{BRST Quantization}
Let us for a moment recall how one quantizes theories with gauge symmetries.
We use the theory of BRST quantization~\cite{Henneaux-book},
which the reader is assumed to be familiar with. Let us fix the
notation used in this thesis regarding BRST transformations.
Assume the existence of a classical local gauge symmetry,
under which a field $\Phi^{i}$ in the classical action changes, 
\begin{equation}
  \delta_{{\rm gauge}} \Phi^{i}(x) = f[\Phi^{i}(x),\lambda^{i}(x)],
  \label{general-gaugetransformation}
\end{equation}
where $f[\Phi^{i}(x),\lambda^{i}(x)]$ is some functional of the fields
and gauge parameters $\lambda^{i}$. When writing like this, we understand
that $\delta_{{\rm gauge}}$ is
an infinitesimal transformation, where $ \delta_{{\rm gauge}} \Phi^{i}(x)
\equiv \tilde{\Phi}^{i}(x) - \Phi^{i}(x)$ and $\tilde{\Phi}^{i}(x)$ is the
result of an infinitesimal gauge transformation with parameter
$\lambda^{i}$. The BRST transformation of $\Phi^{i}$ is obtained by
replacing $\lambda^{i}$ in equation~(\ref{general-gaugetransformation})
by a grassmann field $c^{i}(x)$ 
\begin{equation}
  \delta_{{\rm BRST}} \Phi^{i}(x) \equiv f[\Phi^{i}(x),c^{i}(x)].
\end{equation}
To illustrate this, consider as an example the result of an infinitesimal
gauge transformation on the Yang-Mills field $A_{\mu}^{i}$
\begin{equation}
   \delta_{{\rm gauge}} A_{\mu}^{i}(x) = D_{\mu}\lambda^{i}(x),
\end{equation}
where $\lambda^{i}(x)\in \Omega^{0}(M,\la)$. The covariant derivative has the
usual form:
\begin{equation}
  D_{\mu}\lambda^{i} =
\partial_{\mu}\lambda^{i} + [A_{\mu},\lambda]^{i} = 
\partial_{\mu}\lambda^{i} + A_{\mu}^{j}\lambda^{k}f_{jk}^{\,\,\, i}.
\end{equation}
The BRST transformation
of $A_{\mu}^{i}(x)$ is then
\begin{equation}
  \delta_{{\rm BRST}} A_{\mu}^{i}(x) = D_{\mu}c^{i}, \label{BRST-A:gen}
\end{equation}
where $c^{i}\in \Omega^{0}(M,\la)$ with odd grassmann parity and
ghost number one. This transformation
looks like a local gauge transformation but if we are careful
the actual form should be written as 
\begin{equation}
  \delta_{{\rm BRST}}^{\epsilon} A_{\mu}^{i}(x) = D_{\mu}c^{i}\delta\epsilon,
\end{equation}
where $\delta\epsilon$ is a global infinitesimal grassmann parameter.
Since $\epsilon$ is grassmannian the transformed $A_{\mu}^{i}$ is
still bosonic. The BRST transformation is a global transformation,
but since it has the form of a local gauge transformation, a
gauge invariant function will also be BRST invariant. It is common to write 
the BRST transformation like equation~(\ref{BRST-A:gen}), and
then think of the transformation as a kind of supersymmetry which
maps bosons into fermions and vice versa. One also has the
notation of the BRST operator $Q$, which is a global grassmann odd operator
, which is the ( BRST ) charge associated with the symmetry transformation
$\delta_{{\rm BRST}}$. We write
\begin{equation}
   \delta_{{\rm BRST}} \Phi^{i}(x) \equiv \{ Q, \Phi^{i}(x) \},\label{Q-def}
\end{equation}
where we define the meaning of the ( BRST ) bracket with $Q$ on a field
as the action of $\delta_{{\rm BRST}}$ on this field. In some texts
the infinitesimal parameter is included as $\delta_{{\rm BRST}}(\cdot )
= \epsilon \{ Q, \cdot \} $ and often also an imaginary unit
$\delta_{{\rm BRST}}(\cdot ) = i\epsilon \{ Q, \cdot \} $. We use the
form of equation~(\ref{Q-def}) in the rest of the thesis and the
infinitesimal parameter is never written. Both the BRST transformation
$\delta_{{\rm BRST}}$ and $Q$ are
nilpotent.

From the general theory of BRST quantization, the physical states of
a theory are known to be the cohomology classes of the BRST operator $Q$.
This leads back to the work of Kugo and Ojima~\cite{Kogu-Ojima} from 1979.
So if one can determine the BRST cohomology, one has found the
physical interesting objects. We write in symbolic notation 
the physical states as $\mid {\rm phys} \rangle$ and they are defined by
being $Q$-closed
\begin{equation}
  Q \mid {\rm phys} \rangle = 0,
\end{equation}
where physical states differing by an $Q$-exact term are equivalent
\begin{equation}
  \mid {\rm phys}' \rangle \sim \mid {\rm phys} \rangle + Q\mid \chi
  \rangle,
\end{equation}
with $\mid\chi\rangle$ being an arbitrary state. By making
the important physical choice of demanding the vacuum to be BRST
invariant $Q\mid 0 \rangle=0$,
all vacuum expectation values of BRST exact terms vanish
\begin{equation}
  \langle 0 \mid \, \{Q, X\} \, \mid 0 \rangle \equiv \langle  \, \{Q, X\} \
  \rangle = 0, \label{vev-0}
\end{equation}
where $X$ symbolises any functional of fields and metric.
The next step is to apply the Fradkin-Vilkovisky theorem (for proof and
further explanation see~\cite{Batalin-Vilkovisky,Henneaux-phys.rep}),
where the partition function of the theory is constructed using the notation
of a gauge fermion $\Psi$.
The general situation is, that the gauge fixed action of a theory
has the form
\begin{equation}
  S_{\rm q} = S_{\rm class} + \int_{M} \{Q, \Psi\} , \label{S_q:general}
\end{equation}
where $\Psi$
can be any combination of fields, ghosts and Lagrangian multipliers. The
main result of the theorem is, that the partition function is independent of
the choice of $\Psi$. It is common to construct $\Psi$ in a way such that
$\{ Q, \Psi\}$ produces only the relevant gauge fixing terms.
The BRST exact term $\{Q, \Psi \}$
is obviously also BRST invariant, and hence the whole effective action is BRST
invariant. This is true since the classical action $S_{{\rm class}}$ is
gauge invariant and thus BRST invariant.
Note also that since $Q$ raises the the ghost number by one,
the net ghost number of $\Psi$ should be minus one,
for $S_{\rm q}$ to have ghost number zero. 

\subsection{Abelian BF Theory}
We introduce the abelian BF theory as in~\cite{BT,
  Blau-Thompson:ATF}.
We follow the standard set up of gauge theories given in
section~\ref{section:gauge}, where $\pi:P\mapsto M$ is a principal bundle,
with abelian gauge group $G$.
Let $n$ be the dimension of $M$. The classical action of this theory is
\begin{equation}
  S_{{\rm class}} = \int_{M} B_{p}\wedge dA_{n-p-1},
\end{equation}
where $B_{p}\in \Omega^{p}(M,\la)$ and $A_{n-p-1}\in  \Omega^{n-p-1}(M,\la)$.
The equations of motion  for this action read
\begin{eqnarray}
  d B_{p} &=& 0, \label{BF:B-eqm-abelian} \\
  d A_{n-p-1} &=& 0. \label{BF:A-eqm-ablian}
\end{eqnarray}
The classical
action is invariant under the gauge transformation
\begin{eqnarray}
  \delta A_{n-p-1} &=& d \Lambda'_{n-p-2}, \label{BF-ablian-A-sym} \\
  \delta B_{p} &=& d \Lambda_{p-1} \label{BF-ablian-B-sym},
\end{eqnarray}
where the gauge parameters $ \Lambda_{p-1} \in \Omega^{p-1}(M,\la)$
 and $ \Lambda^{'}_{n-p-2} \in \Omega^{n-p-2}(M,\la)$.
The space ${\cal N}$ of classical solutions can be written as
\begin{equation}
  {\cal N} = H^{p}(M,\R)\oplus H^{n-p-1}(M,\R). \label{cal-N-ablian}
\end{equation}

The quantization of systems like this, was studied by Schwarz in a
series of papers~\cite{Schwarz,Schwarz-1,Schwarz-Tyupkin}. He considered
a more general setting where the action was viewed as a quadratic
functional on a pre-Hilbert space $\Gamma_{0}$. Recall~\cite{Lang-FA}
that a pre-Hilbert space $\Gamma_{0}$ is a vector space with a positive
definite hermitian scalar product
$\langle \cdot,\cdot \rangle$. $\Gamma_{0}$ can be made a Hilbert space if
one takes the completion in the $L^{2}$ norm. The action is of the form
\begin{equation}
  S[f] = \langle Kf, f \rangle
\end{equation}
where $K$ is a self-adjoint operator acting on the elements
$f\in\Gamma_{0}$. We say that 
$S[f]$ is non-degenerate if $f$ only is the the kernel of $K$ when it is
identical to zero. In
that case one can define the partition function, which will be a
gaussian integral and thus expressible in terms of determinants.
These must be given meaning for infinite dimensional operators, which
is done by zeta function regularization. The zeta function
of an operator $A$ is
\begin{equation}
  \zeta_{s}(A) = \sum_{\lambda >0} \lambda_{i}^{-s},
\end{equation}
where one sums over the positive
eigenvalues $\lambda_{i}$. Since the zeta function
is only analytic for ${\bf Re}(s)>1$, one may need to perform an
analytic continuation for ${\bf Re}(s)\leq 1$.
The regularized determinant formula for a so-called regular operator $A$, is
\begin{equation}
  \log {\rm det}(A) = \left. - \frac{ d\, \zeta_{s}(A)}{d\, s}\right\vert_{s=0}, 
\end{equation}
and in the following ${\rm det(\cdot )}$ will be understood in that
sense. The operators which are regular can be taken as
those for which this formula
is valid~\cite{Schwarz-1}.
If $A:\Gamma_{i} \mapsto \Gamma_{j}$ maps one (pre) Hilbert space
$\Gamma_{i}$ into another $\Gamma_{j}$, its determinant can 
be written~\cite{Schwarz-1} as 
\begin{equation}
  {\rm det}(A) = \exp \left( -\frac{1}{2} \frac{d}{ds}(
    \zeta_{s}(A^{\dagger} A))\vert_{s=0} \right) = {\rm det}(A^{\dagger} A)^{\frac{1}{2}},
\label{det-dagger}
\end{equation}
and if $A$ is self-adjoint and regular, the two
definitions will be the same.
When $K^{2}$ is regular, the partition
function can formally be written as
\begin{equation}
  Z = \int {\cal D}f e^{-S[f]} = {\rm det}(K)^{-\frac{1}{2}}.
\end{equation}
In addition to $S[f]$ being a quadratic functional on $\Gamma_{0}$,
let there exist a sequence of pre-Hilbert spaces $\Gamma_{i},\,i=1,2\dots,N $
and linear operators $T_{i}:\Gamma_{i}\mapsto\Gamma_{i-1}$. These
satisfy the following relations
\begin{eqnarray}
  {\rm Ker}(K) &=& {\rm Im} T_{1}, \label{KT_1} \\
  {\rm Ker}(T_{i-1}) &=& {\rm Im} T_{i}.
\end{eqnarray}
The first requirement states that the action is invariant under
the transformation
\begin{equation}
  S[f + Th] = S[f], \,\,  \forall h\in \Gamma_{1}.
\end{equation}
This can only be the satisfied if $KT_{1}=0$. This transformation is
trivial if and only if $h=T_{2}g$, for $g\in \Gamma_{2}$ and so forth.
If $T_{1}$ is non-zero, the action $S[f]$ will be a degenerate functional,
i.e. it has a gauge symmetry, and the partition function may be defined as
\begin{equation}
   Z \equiv {\rm det}(K)^{-\frac{1}{2}} \prod_{i=1}^{N}
       {\rm det}(T_{i})^{(-1)^{i-1}}.
\end{equation}
Schwarz introduced the operators
\begin{eqnarray}
  \Box_{0} &=& K^{2} + T_{1}T_{1}^{\dagger} \\
  \Box_{i} &=&  T_{i}^{\dagger}T_{i} + T_{i+1}T_{i+1}^{\dagger}
\end{eqnarray}
and under the assumption that $K^{2}, T_{i}^{\dagger}T_{i}$ are regular
the partition function can be written as
\begin{equation}
  Z = \prod_{i=0}^{N} {\rm det}(\Box_{i})^{\nu_{i}}, \,\,\,
  \nu_{i}=(-1)^{i+1}\frac{(2i+1)}{4}. \label{Z-schwarz}
\end{equation}
The determinants of the box operators are expressed as
\begin{equation}
  {\rm det}(\Box_{0}) =  {\rm det}(K^{2}){\rm det}(T_{1}T_{1}^{\dagger}) =
  {\rm det}(K^{2}){\rm det}(T_{1})^{2},
\end{equation}
using equation~(\ref{det-dagger}), and similarly
\begin{equation}
  {\rm det}(\Box_{i}) = {\rm det}(T_{i})^{2}{\rm det}(T_{i+1})^{2}.
\end{equation}

We now apply this abstract formalism to three-dimensional
abelian BF theory, where both $B$ and $A$ are 1-forms
\begin{equation}
  S_{{\rm class}} = \int_{M} B_{1}\wedge dA_{1},
\end{equation} 
so $\Gamma_{0}=\Omega^{1}(M)\oplus\Omega^{1}(M)$. The gauge symmetries
of this action are
\begin{eqnarray}
  \delta B_{1} &=& d \Lambda_{0},\\
  \delta A_{1} &=& d \Lambda'_{0},
\end{eqnarray}
which means that one identifies $\Gamma_{1}=\Omega^{0}(M)\oplus\Omega^{0}(M)$
and $T_{1}= d \oplus d$ ( so $T^{\dagger}_{1}=
d^{\dagger}\oplus d^{\dagger}$) and thus Schwarz's sequence reads
\begin{equation}
  0 \mapsto \Omega^{0}(M)\oplus\Omega^{0}(M)
  \stackrel{T_{1}}{\mapsto} \Omega^{1}(M)\oplus\Omega^{1}(M)
  \stackrel{*d}{\mapsto}  \Omega^{1}(M)\oplus\Omega^{1}(M)
  \mapsto 0.
\end{equation}
The last step in the sequence is the operator $K$, which by the
existence of the inner product can be viewed as a map
$K^{*}:\Gamma_{0} \mapsto \Gamma_{0}^{*}$ from the pre-Hilbert
space on which $K$ acts to its dual space.
Under the assumption that there are no zero modes in $A,B$ we apply
equation~(\ref{Z-schwarz}) and write the partition function as
\begin{eqnarray}
  Z &=& \prod_{i=0}^{1} {\rm det}(\Box_{i})^{\nu_{i}} \nonumber , \\
    &=& {\rm det}(\Box_{0})^{-\frac{1}{4}}
        {\rm det}(\Box_{1})^{\frac{3}{4}} .
\end{eqnarray}
Since the laplacian acting on $i$-forms is
$\Delta_{i} = d_{i}^{\dagger}d_{i} +d_{i}d_{i}^{\dagger}$ we can 
rewrite the partition function as
\begin{equation}
  Z = {\rm det}(\Delta_{0}\oplus\Delta_{0})^{-\frac{1}{4}}
    {\rm det}(\Delta_{1}\oplus\Delta_{1})^{\frac{3}{4}} =
    {\rm det}(\Delta_{0})^{-\frac{1}{2}}{\rm det}(\Delta_{1})^{\frac{3}{2}},
\label{schwarz-Z}
  \end{equation}
using equation~(\ref{det-dagger}).  We now show how to obtain the same 
partition function by BRST gauge fixing of the classical action.
The BRST algebra reads
\begin{eqnarray}
  \delta_{{\rm BRST}} A_{1} &=& d c, \label{delta-BRST-A:BF_abelian} \\ 
  \delta_{{\rm BRST}} B_{1} &=& d \omega, \label{delta-BRST-B:BF_abelian} \\
  \delta_{{\rm BRST}}\,\, c &=& 0 ,\label{delta-BRST-c:BF_abelian}\\
  \delta_{{\rm BRST}}\,\,  \omega &=& 0. \label{delta-BRST-omega:BF_abelian}
\end{eqnarray} 
The ghost fields are $c,\omega$, which are scalar grassmann fields on $M$.
We choose a gauge fermion
\begin{equation}
  \Psi = *\overline{\omega} \, d^{\dagger}B + *\overline{c}\, d^{\dagger}A,
\end{equation}
where we have introduced two anti-ghost multiplets
$(\overline{\omega},G),(\overline{c},E)$ which trivially extend the
BRST algebra
\begin{eqnarray}
  \delta_{{\rm BRST}}\,\, \overline{\omega} &=& G,
  \label{delta-BRST-overlineomega:BF_abelian} \\ 
  \delta_{{\rm BRST}} G &=& 0 , \label{delta-BRST-G:BF_abelian} \\
  \delta_{{\rm BRST}}\,\, \overline{c} &=& E ,
  \label{delta-BRST-overlinec:BF_abelian}\\
  \delta_{{\rm BRST}} E &=& 0. \label{delta-BRST-E:BF_abelian}
\end{eqnarray} 
where $E,G$ are Lagrange multipliers.
This leads to the following quantum action
\begin{equation}
  S_{q} = \int_{M} \left( B_{1}\wedge d A_{1} + * G\wedge d^{\dagger}B_{1}
+ * E\wedge d^{\dagger}A + \* \overline{\omega}d^{\dagger}d \omega
+ * \overline{c}d^{\dagger}d c\right).
\end{equation}
Here we used that $Q$ commutes with the Hodge star and anti-commutes 
with $d^{\dagger}$. The term $d^{\dagger}d$ is just $\Delta_{0}$. The
gaussian ghost terms contribute to the partition
function with ${\rm det}(\Delta_{0})^{2}$. The $(A_{1},B_{1},E,G)$ system
is not gaussian, but it can be transformed into a gaussian form, by a
so-called Nicolai map which we discuss in the case of non-abelian
BF theory. One can also square the kinetic operator, diagonalize
it and then take the square-root of its determinant,
and the result is~\cite{Blau-Thompson:ATF}
\begin{equation}
  Z = {\rm det}( \Delta_{1})^{-\frac{1}{2}}
  {\rm det}( \Delta_{0})^{-\frac{1}{2}}
    {\rm det}( \Delta_{0})^{2}
  = {\rm det}( \Delta_{1})^{-\frac{1}{2} }
    {\rm det}( \Delta_{0})^{\frac{3}{2}}.
\end{equation}
This is the same as obtained by Schwarz formula in
equation~(\ref{schwarz-Z}). What is interesting, is the fact that
Schwarz has proved that $Z =({\cal T}_{3})^{-1}$, where 
\begin{equation}
  {\cal T}_{3} =
  {\rm det}( \Delta_{1})^{ \frac{1}{2} }
  {\rm det}( \Delta_{0})^{ - \frac{3}{2} },
\end{equation}
is the so-called Ray-Singer torsion of $M$, which he showed to be a
topological invariant of $M$. Hence the partition function of
abelian BF theory is a topological invariant, if there are no zero modes.
The zero modes can also be taken into account and the topological
result is also extended to that situation. In that situation the
gauge symmetry will be reducible, and following the
BFV formalism~\cite{Henneaux-book,Henneaux-phys.rep}, 
higher generation ghost fields must be introduced. These higher ghosts,
or ghost for ghosts, as they are also known have alternating
grassmann parity for each generation and the ghost number rises
with +1 per generation. One
enlarges the BRST algebra with higher ghosts until all symmetries are
gauge fixed. Zero modes or not, the quantum action is always of the
same form as in equation~(\ref{S_q:general}), and from this we can
conclude that $Z$ is topological, since the only metric dependence is 
in $\Psi$. Under a metric transformation $g_{\mu\nu}\mapsto
g_{\mu\nu} + \delta g_{\mu\nu}$ the partition function will 
change as
\begin{eqnarray}
  \delta_{g}Z &=&
  - \int {\cal D}[X]e^{-S_{{\rm q}}}\left( \{ Q, \delta_{g}\Psi\} \right)
  \nonumber \\ &=& \left\langle  \{ Q, \delta_{g}\Psi\} \right\rangle
  \nonumber \\ &=& 0,
\end{eqnarray}
following equation~(\ref{vev-0}). Since $Z$ only depends on the
metric in a BRST exact manner, it is topologically invariant.
We return to this kind of arguments in the section on Witten type TFT's.

Until now we have assumed that the action was a quadratic functional, but
if that is not the case, the results of Schwarz can be applied in the
stationary phase approximation, i.e. in 1-loop calculation. For more
information see~\cite{Schwarz-1}.

\subsection{Non-Abelian BF Theory}

In this section we discuss the most relevant version of BF theory, in
relation to the development of topological gravity. Namely the
two-dimensional non-abelian BF theory. The classical action reads
\begin{equation}
  S_{{\rm class}} = \int_{M} tr( B \wedge F[A] ),
  \label{nonabelian:BF-action}
\end{equation}
where $B\in \Omega^{0}(M,\la)$ and $A\in\Omega^{1}(M,\la)$. Under a
gauge transformation the fields transform as
\begin{eqnarray}
  \delta A_{\mu}^{i}(x)& =& D_{\mu}\lambda^{i}, \\
 \delta B^{i}(x)& =& [\lambda, B]^{i}.
\end{eqnarray}
In dimensions $n>2$, $B$ also possesses an individual gauge symmetry
$\delta^{'}B = D \eta$, where $\eta$ has form degree one less than $B$.
$A$ remains unaffected by this transformation and the two gauge symmetries
can be gauge fixed independently of each other, i.e. there are no
mixed terms for the $A$ and $B$ ghosts in the action when $n>2$. This
situation requires use of the BFV formalism of ghost for ghosts, and
it is discussed in detail in~\cite{Wallet}. Since we only study
two-dimensional BF theory, we do not need this.

The equations of motion for the classical action in
equation~(\ref{nonabelian:BF-action}) read
\begin{eqnarray}
  F[A] &=& 0, \\
  D B &=& 0.
\end{eqnarray}
The first equation states that the solutions to the $B$ equations of motion,
remembering the gauge symmetry, are the elements in the moduli space
of flat connections
\begin{equation}
  {\cal M}_{F} = \frac{ \{ A \in {\cal A} \vert F[A] = 0\} }{\G }.
\end{equation}
The $A$ equation states that $B$ lies in ${\cal B}$, the
space of gauge-equivalence classes of covariant constants.
\begin{equation}
  {\cal B} = \frac{ \{ B \in \Omega^{0}(M,\la)\vert D B = 0\} }{\G}.
\end{equation}
The reduced phase space ${\cal N}$ is a fibre bundle over ${\cal M}_{F}$
\begin{eqnarray}
  &{\cal N}& \nonumber \\ &\downarrow& \, \pi \\ &   {\cal M}_{F}&,\nonumber
  \label{reduced-phase-space-1}
\end{eqnarray}
with fibre $\pi^{-1}(A) = {\cal B}, \, A\in {\cal M}_{F}$.
So locally has ${\cal N}$ the decomposition
$ {\cal M}_{F}\times {\cal B}$ and the tangent space at a
point $(A,B)\in  {\cal N}$ is
\begin{equation}
  T_{(A,B)}{\cal N} = H^{0}_{A}(M, \la) \oplus H_{A}^{1}(M,\la)
    \label{tangent-cal-N}
\end{equation}
where the de Rham cohomology groups $H^{*}_{A}$ are labelled by $A$ since they
consists of the $D=d+[A,\cdot]$-closed forms modulo the $D$-exact
forms. Equation~(\ref{tangent-cal-N}) is the nonlinear generalization
of equation~(\ref{cal-N-ablian}).

The quantum action is of the same form as in the abelian case, but
since $n=2$ we only gauge fix the symmetry of $A$ 
\begin{equation}
  S_{{\rm q}} = \int_{M} tr\Biggl( B\wedge F[A] + * E\wedge D^{\dagger}A
  + \overline{\omega}* \Delta^{'}_{0}\omega \Biggr) ,
\end{equation}
where $\Delta^{'}_{0} = D^{\dagger}D$ is the covariant laplacian.
One can apply a so-called Nicolai
map~\cite{Nicolai-1,Nicolai-2} which transforms the $(B,A,E)$ terms of the
action into a gaussian form. The nontrivial information is in the
Jacobian of the transformation. The Nicolai map is a coordinate
transformation in the space of fields
\begin{equation}
  A\mapsto (\xi,\eta): \left\{
    \begin{array}{lcl}
                       A_{1} &=\eta(A) &= *D^{\dagger} A, \\
                       A_{2} &= \xi(A) &= F[A].
    \end{array} \right.
\end{equation}
The Jacobian is
\begin{equation}
 \vert J\vert = \left\vert {\rm det }\left( \frac{\delta A_{\mu}}{\delta (\eta,\xi)}
    \right) \right\vert =
    \left\vert \left[{\rm det }\left( \frac{\delta \eta}{\delta A_{1}}
       \frac{\delta \xi}{\delta A_{2}}
    \right)\right]^{-1} \right\vert
    =  \left\vert {\rm det }\left( * D^{\dagger}D \right)^{-1}\right\vert.
\end{equation}
We see that this exactly cancels the contribution $*\Delta^{'}_{0}$
coming from the ghost term in the action. Hence the partition
function is reduced to the free action
\begin{equation}
  S_{{\rm q}} = \int_{M} tr(B\xi + E\eta) = \int_{M} \frac{1}{2}tr\biggl(
  B\xi + \xi B + E\eta + \eta E\biggr),
\end{equation}
where the integrand can be written as 
\begin{equation}
  (B\,\,E\,\,\xi\,\,\eta) \left( \begin{array}{cccc}
      0&0& \frac{1}{2} &0 \\
      0&0&  0&  \frac{1}{2} \\
 \frac{1}{2} &0 & 0& 0\\
 0&  \frac{1}{2} & 0&0 \end{array} \right) \left( \begin{array}{c}
 B \\ E\\ \xi\\ \eta \end{array} \right),
\end{equation}
that is in a gaussian form. It should be stressed that this
transformation only reduces the integration over the space of fields,
to those elements which are in the kernel of the Nicolai maps. These are
exactly the elements in the moduli space of flat connections, since
$F=0$ localizes to the flat connections and the gauge fixing condition
$D^{\dagger}A=0$ restricts the integration over \A to \AMG.

It is a common feature of topological field theories that the
reduced phase space is finite dimensional, in contrast to ordinary
QFT's as e.g. Yang-Mills theory, where \AMG is infinite dimensional.
This is reflected by the fact that the partition function reduces to an
integral over moduli space, which is finite dimensional. Hence 
a topological field theory only has a finite number of degrees of freedom.
This is also the case for the Witten type TFT's which we introduce in a moment.

The fact that the ratio of determinants above is one, reflects 
the fact that the Ray-Singer torsion is trivial in even
dimensions~\cite{Schwarz-1} and the partition function is then just a
sum of ``1'''s if ${\rm dim}({\cal M}_{F})=0$ or an integral
\begin{equation}
  Z = \int_{{\cal M}_{F}} d\mu,
\end{equation}
if ${\rm dim}({\cal M}_{F}) > 0$. In general we do not have a generic
measure $d\mu$ on the moduli space of flat connections. We study the
general features of moduli space in some detail. Let us just at this
stage mention, that the topological nature of the non-abelian BF theories
has only been proven for $n\leq 4$~\cite{Blau-Thompson:ATF}.

\subsection{Moduli Space of Flat Connections}
In this section we give a brief introduction to the structure of the moduli
space of flat connections.
In chapter 3 we will discuss this in more detail for the gauge
groups relevant for gauge theories of gravity.

Recall the notation of the holonomy in
section~(\ref{section:gauge}). If $P\mapsto M$
is a flat $G$ bundle, so we have a flat connection $A$, then the
holonomy will only depend on the homotopy class~\cite{KN_1}.
Thus the holonomy defines a homomorphism from $\pi_{1}(M)$ into
$G$, and since the holonomy is invariant under the \AdG-action, one can 
identify
\begin{equation}
{\cal M}_{F} =  {\rm Hom}\biggl( \pi_{1}(M), G\biggr) / G,\label{moduli-flat-A-1}
\end{equation}
where the quotient of $G$ is taken as the adjoint action of $G$.
Another way to see this is the following: Given a homomorphism
$h \in {\rm Hom}( \pi_{1}(M),G)$, we can construct a fibre bundle over $M$
\begin{eqnarray}
  &\tilde{M}& \nonumber \\  &\downarrow& \\ &M&,\nonumber
\end{eqnarray}
with fibre $\pi_{1}(M)$. This makes $\tilde{M}$ the universal covering
space on $M$. Viewing this as an associated bundle to $P$, 
the homomorphism $h$ defines a flat
connection in $P$ by setting $h=h(u,\gamma)$~\cite{Birmingham-review}.
There is a lot of information on the
spaces ${\rm Hom}(\pi_{1}(M),G))$ and
${\rm Hom}(\pi_{1}(M),G))/G$ in the literature, and we return to this
in chapter 3 in relation to gauge theories of gravity. If
$M$ is a compact, orientated two dimensional manifold of genus $g$, written
as $\Sigma_{g}$, we can find the dimension of ${\cal M}_{F}$. The elements
of ${\cal M}_{F}$ are conjugacy classes of $\pi_{1}(\Sigma_{g})$
into $G$. We know from
chapter 1, that the fundamental group has $2g$ generators
$\Gamma(a_{i}),\Gamma(b_{i})$, where $(a_{i},b_{i}),i=1,2,\dots,g$ are
the homology cycles on $\Sigma_{g}$. These generators are subject to
the relation given in equation~(\ref{homotopy-relation-1})
and we have that 
\begin{equation}
  \prod_{i=1}^{g} \left( \Gamma(a_{i})\Gamma(b_{i})
                         \Gamma^{-1}(a_{i})\Gamma^{-1}(b_{i})
                   \right) = {\bf 1},
\end{equation}
The elements in  ${\rm Hom}(\pi_{1}(M),G)$ are given by the set
\begin{equation}
  \left\{ h_{i},k_{i}\in G\Biggl| \,\, \prod_{i=1}^{g}
    h_{i}k_{i}h_{i}^{-1}k_{i}^{-1} = 1 \right\}, 
\end{equation}
and ${\rm Hom}(\pi_{1}(M),G)/G$ is given by~\cite{Jeffrey-Weitsman}
\begin{equation}
  \left\{ h_{i},k_{i}\in G\vert \prod_{i=1}^{g}
   Ad_{q}( h_{i})Ad_{q}(k_{i})Ad_{q}(h_{i}^{-1})Ad_{q}(k_{i}^{-1}) =
 1 \right\}\,\,\forall q\in G. 
\end{equation}
Hence the dimension of ${\cal M}_{F}$ will be
\begin{equation}
{\rm dim}\left({\cal M}_{F}(\Sigma_{g},G)\right) = (2g-1){\rm dim}(G),
\label{dim-M_F}
\end{equation}
since the relation fixes one of the $2g$ generators of $\pi_{1}(\Sigma_{g})$.
Under most conditions the moduli space of flat connection will not be
a manifold and some times not even an orbifold.

\subsection{BF Theory and Yang-Mills Theory}\label{section:YM-BF}
As a short side track, let us present a connection between the
topological BF theory and the non-topological Yang-Mills theory.
The first proposal of transforming
BF theory into a Yang-Mills theory was given by Blau and Thompson~\cite{BT}.
But the first application was made by Witten~\cite{Witten-2dGT}
where he used that the path integrals of Yang-Mills theory and BF theory
are related
\begin{equation}
\int {\cal D}[A] e^{-\int_{M} \sqrt{g}d^{n}x\,
  tr(F\wedge * F)} = \int {\cal D}[A,B] e^{-\left(\int_{M} tr(B\wedge F) +
  \epsilon \int_{M} \sqrt{g}d^{n}x\, tr(B^{2})\right)}. 
\end{equation}
to prove several important results. This can be shown by gaussian integration
over $B$ or by applying the $B$ equation of motions.
Hence BF theory can be seen as the $\epsilon\rightarrow 0$
limit of Yang-Mills theory, and this has played an important role
in Witten's studies of two dimensional Yang-Mills theory~\cite{Witten-2dGT,
  Witten-2dGTR}. Notice that the term
$tr(B^{2})$ requires the metric to make the contraction of indices
$B_{\mu}B^{\mu}$ or equivalently in order to write $B\wedge * B$. This
reflect the well-known fact that the Yang-Mills action is not topological.

\subsection{Chern-Simons Theory}
Let us as one of the last issues on general Schwarz theories, present the
Chern-Simons action and discuss the proof of
the topological invariance of the partition function. The reason
for considering Chern-Simons theory alone is the relation to knot theory and
quantum gravity.

The Chern-Simons action is special in the sense that the action is
only defined in three dimensions
\begin{eqnarray}
  S_{{\rm CS}}[A] &=&
  \frac{k}{4\pi} \int_{M} tr\left(A\wedge dA + \frac{2}{3}A\wedge A
    \wedge A\right) \nonumber \\ &=&
    \frac{k}{8\pi}\int_{M} d^{3}x \epsilon^{\mu\nu\gamma}tr \biggl( A_{\mu}(x)
    \partial_{\nu}A_{\gamma}(x) - A_{\mu}(x)\partial_{\gamma}A_{\nu}(x)
  \nonumber \\
    && \qquad +\frac{2}{3} A_{\mu}(x)[A_{\nu}(x),A_{\gamma}(x)]\biggr),
\end{eqnarray}
where the gauge indices are suppressed. This action defines a classical
Schwarz type TFT and it is gauge invariant for gauge transformations
connected to the identity, and changes with a constant factor
for so-called large gauge transformations 
\begin{equation}
  S_{{\rm CS}} \mapsto S_{{\rm CS}} + 2\pi k S(g),
\end{equation}
where $S(g)\in \Z$. This is related to the fact that for compact simple Lie
groups $G$, the third homotopy class $\pi_{3}(G)=\Z$, and by viewing
the gauge transformations as maps $g:M\mapsto G$ the winding number of
this map is 
\begin{equation}
  S(g) = \frac{1}{24 \pi^{2}}\int_{M} tr(g^{-1}dg)^{3}.
\end{equation}
The exponential of the action $S_{{\rm CS}}$ need to be a single valued
function in the path integral, and this requirement
forces $k$ to be an integer. The action can be viewed as the special
case $A=B$ of BF theory, and BF theory is a kind of generalization of
Chern-Simons theory to arbitrary dimensions. 

It is interesting to study the weak coupling limit of the Chern-Simons
action in the path integral. This
semi-classical phase of the theory will be dominated by
a sum of contributions from the points of stationary phase, i.e.
the classical solutions of the theory. Obviously the classical
solutions are the flat connections, and the reduced phase space is
${\cal M}_{F}$. Witten proved in an important paper~\cite{Witten:Jones}
that Chern-Simons theory generates the Jones polynomials~\cite{Jones}
from knot theory. This important link between Schwarz type TFT's and
knot invariants, which are topological invariants of three manifolds,
started intense research into this topic, which is still on-going. The
reviews~\cite{Birmingham-review,Kauffman} can be used as an entry to the
literature on this topic.

Here I present the semi-classical treatment by Witten~\cite{Witten:Jones}
which show that the partition function is topological and that it
fits nicely with the method due to Schwarz discussed above. Witten
did his calculation in Lorentzian signature and in order
to follow the literature we write the partition function as
\begin{equation}
  Z= \int {\cal D}[A]  \exp\left( i S_{{\rm CS}}[A] \right),
\end{equation}
and in the stationary phase approximation, we take the large $k(=1/\hbar)$
limit, which is of the form
\begin{equation}
  Z = \sum_{\alpha} \mu(A^{(\alpha)}),
\end{equation}
if ${\rm dim}({\cal M}_{F})=0$.
Here $A^{(\alpha)}$ is a complete set of gauge equivalence classes of
flat connections and $\mu(A^{(\alpha)})$ is the contribution from the
partition function when expanding around $A^{(\alpha)}$. Now introduce a
change in variables, so that the gauge field is written as a
quantum correction $B$ around a classical $A^{(\alpha)}$ solution
\begin{equation}
  A_{\mu}(x) = A^{(\alpha)}_{\mu}(x) + B_{\mu}(x).
\end{equation}
The gauge fixing condition $D^{\dagger}B= D^{\mu}B_{\mu}=0$ requires
a metric on $M$ in order to make the contraction of indices. This
breaks the metric independence of the classical action. The quantum
action is then
\begin{equation}
  S_{{\rm q}} =  S_{{\rm CS}}[ A^{(\alpha)}] +
                   \frac{k}{4\pi}\int_{M} tr( B\wedge dB + \frac{2}{3}
                   B\wedge B\wedge B) +
                   \int_{M} tr( * E\wedge D^{\dagger}B + \overline{c}
                   \wedge * \Delta_{0} c),
\end{equation}         
where the Chern-Simons action for the classical solution
$S_{{\rm CS}}[ A^{(\alpha)}]$ ,  is a topological invariant on $M$.  The
quadratic part (in the quantum fields) of the action, reads
in components
\begin{equation}
  S_{{\rm q}}^{(2)} = \int_{M} d^{3}x \, tr(\epsilon^{\mu\nu\gamma}
  B_{\mu}D_{\nu}B_{\gamma} + 2e D^{\mu}B_{\mu} +\overline{c}D_{\mu}D^{\mu}c ),
\end{equation}
where $e=E/2$. The $(B,e)$ pars can be written in matrix form
\begin{equation}
  S_{{\rm q}}^{(B,e)} = \int_{M} d^{3}x\, tr\left(
    (B_{\mu}^{i}\,\,e^{i}) \left\{ \begin{array}{cc} -\epsilon^{\mu\nu\gamma}
        D_{\nu} & -D^{\alpha} \\ D^{\gamma} & 0 \end{array} \right\}^{ij}
      \left( \begin{array}{c} B_{\nu}^{j}\\ e^{j} \end{array}
        \right)\,\,\right) .
\end{equation}
The first order operator in the bracket will be denoted $H$, and the
partition function in the stationary phase approximation is
\begin{equation}
  Z = \sum_{(\alpha)} \exp \left( i\frac{k}{4\pi} S[ A^{( \alpha )} ] \right)
\frac{{\rm det} ( -D^{2} ) }{ \sqrt{ {\rm det}(H)} }.
\end{equation}
There is a slight complication in relating this ratio of determinants to
the Ray-Singer torsion. This is due to the existence of the so-called
(Atiyah) Eta phase~\cite{Witten-TFT}
\begin{equation}
  (\sqrt{{\rm det}(H)})^{-\frac{1}{2}} =
  ({\rm det}(H^{2}))^{-\frac{1}{4} } \exp\left(i\frac{\pi}{4} \eta(0)\right) ,
\end{equation}
leading to the partition function
\begin{equation}
  Z = \sum_{ (\alpha)} \exp \left(
      i \frac{k}{4\pi} S[A^{ ( \alpha ) } ] \right)
 \frac{ {\rm det} (-D^{2} ) }{ ({\rm det}(H^{2}))^{ \frac{1}{4} }}
\exp\left( i \frac{\pi}{4} \eta (0) \right).
\end{equation}
Witten proved that this is a topological invariant only if one specifies a
trivialization of the tangent bundle over $M$. A manifold with such a
structure, is known as a framed manifold and the Chern-Simons partition
function is a topological invariant on such a manifold. Notice though that
this is only a semi-classical calculation. Witten went further and proved
the relationship to Jones polynomials non-perturbatively using conformal
field theory~\cite{Witten:Jones}. The ratio of determinants above is now the
inverse Ray-Singer torsion of $M$.

This short discussion shows that one must be careful when stating that
a Schwarz type TFT is topological, since, as we just saw, one can be forced to
specify some additional data on $M$ for this to be true. Framing plays
an important role in knot theory and framed knots (known as ribbons)
are also topological invariants on $M$. For further information
see~\cite{Kauffman}. Witten showed that the expectation values of
products of Wilson loops along knots embedded in $M$ are related
to the Jones polynomials~\cite{Witten:Jones}.
We now consider the similar discussion in two dimensional abelian BF theory.

\subsection{Observables in $2D$ Abelian BF Theory}
\label{section:BF-observables}

We briefly discuss the possibility of defining generalized Wilson
loops, whose expectation values are related to link invariants, like
in the case of Chern-Simons theory.

The discussion is valid for BF theory without zero modes in
the action. Assume that $\Sigma\subset M$, is a $p+1$ dimensional
sub-manifold of $M$ and that $\Sigma^{'}\subset M$ is a sub-manifold of
dimension $n-p$. The boundaries $\partial\Sigma,\partial \Sigma^{'}$ are
of dimension $n-1$ and $n-p-1$ respectively and are taken to be compact
(sub)manifolds. We assume that $\partial\Sigma
\cap \partial \Sigma^{'} = \emptyset$.

The concept of Wilson loops can be generalized to so-called Wilson
surfaces~\cite{Blau-Thompson:ATF}, which are the gauge invariant functionals
\begin{eqnarray}
 W_{B}(\Sigma) & = & \exp \left( \int_{\partial \Sigma} B \right),
 \label{wilson-surface-1} \\
 W_{A}(\Sigma^{'}) & = & \exp \left( \int_{\partial \Sigma^{'}} A \right).
 \label{wilson-surface-2}
\end{eqnarray}
Since we only consider abelian BF theory, we do not have the usual trace
in the definition. The most interesting object is the ``two point'' function
\begin{equation}
  W(\Sigma,\Sigma^{'} ) = \lan W_{B}(\Sigma) W_{A}(\Sigma^{'}) \ra,
\end{equation}
since it can be related to the linking number between $(\partial\Sigma,\partial \Sigma^{'})$. We briefly review the definitions of intersection- and
linking numbers.

\begin{figure}[h]
\begin{center}
\mbox{
\epsfysize=4cm
\epsffile{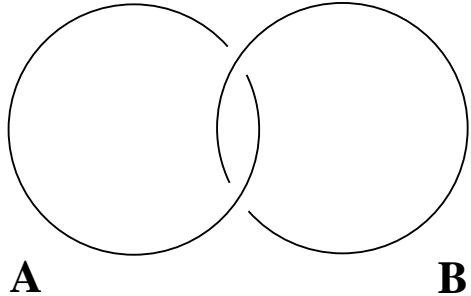}
}
\end{center}
\caption{\label{fig:link}The linking between $A$ and $B$}
\end{figure}

\begin{figure}[h]
\begin{center}
\mbox{
\epsfysize=4cm
\epsffile{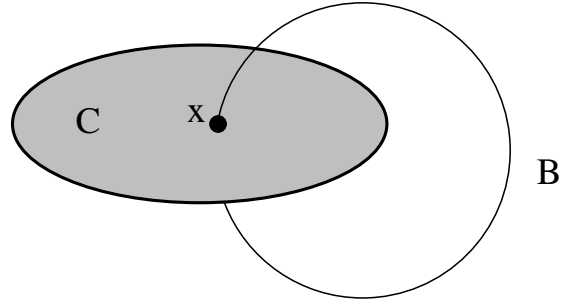}
}
\end{center}
\caption{\label{fig:intersect}The intersection between $C$ and $B$}
\end{figure}

Let $C,D$ be two sub-manifolds of $M$ of dimension $p+1$ and $n-p$
respectively and let $A,B$ be their non-intersecting boundaries. The linking
number of $A$ and $B$ is a topological
invariant (actually the simplest knot invariant~\cite{Kauffman})
and the geometric picture is given in figure~(\ref{fig:link}).
This invariant
is also related to the transversal intersection number between one of the
sub-manifolds $A$ or $B$, and the manifold of which the other is a boundary.
This is shown for the case $C,B$, where $\partial C=A$
in figure~(\ref{fig:intersect}).
The transversal intersection number (from now on just called
the intersection number) is defined below:
\begin{description}
  \item{{\bf Definition (2.7)}} Let $X,Y$ be sub-manifolds of $M$ such that
    \begin{equation}
      {\rm dim}(X) + {\rm dim}(Y) = {\rm dim}(M).
    \end{equation}
   Consider for all $x\in X\cap Y$ the tangent spaces $T_{x}X$, $T_{x}Y$
   and $T_{x}M$. The intersection number of $X,Y$ is
   \begin{equation}
     {\rm Int}(X,Y) = \sum_{x\in X\cap Y}\pm 1,
   \end{equation}
   where we get $+1$ if
   \begin{equation}
     {\rm Orientation}(T_{x}X\oplus T_{x}Y) = {\rm Orientation}(T_{x}M),
   \end{equation}
   and $-1$ if the orientation is opposite, for all the points $x\in X\cap Y$.
\end{description}
The linking number can be defined~\cite{Bott-Tu} if: (1)
${\rm dim}(A)+{\rm dim}(B)={\rm dim}(M)-1$, (2) $A\cap B=\emptyset$ and
(3) the fundamental homology classes $[A],[B]$ are homologous to
zero in $H_{*}(M)$. The last requirement is fulfilled since one can
``lift'' the boundary loop into the sub-manifold of which it is a boundary,
where it will be a trivial loop. 

The surfaces $\Sigma,\Sigma^{'}$ and their boundaries $\partial\Sigma,
\partial\Sigma^{'}$, fit into these definitions and we have that
\begin{equation}
  \#_{\rm intersection} \left(\Sigma,\partial\Sigma^{'}\right) = 
   \#_{\rm linking} \left(\partial\Sigma,\partial\Sigma^{'}\right).
\end{equation}
As mentioned before, products of Wilson loops in
Chern-Simons theory are related to knot invariants. The linking number is just
the simplest knot invariant. Actually a knot can link to itself or
other knots.
The two point function, i.e. the product of two Wilson loops over distinct
knots, can be given a perturbative expansion in the $k$- parameter in the
Chern-Simons action and the first terms are related to the 
Gauss self linking number and the linking number~\cite{GMM:Wilson}.
A similar result can be derived for BF theories, but expressed in
the Wilson surfaces from equations~(\ref{wilson-surface-1},
\ref{wilson-surface-2})
\begin{equation}
  \log \left( W(\Sigma,\Sigma^{'}) \right) =
  \#_{\rm linking} \left(\partial\Sigma,\partial\Sigma^{'}\right).
\end{equation}
For the situation of interest where $n=2$ we can take $B\in\Omega^{0}(M)$
and $A\in \Omega^{1}(M)$. The geometric situation is thus tied to the
question of whether a point $x$ (a zero cycle, against which, the zero-form $B$ is
integrated) will intersect the surface $\Sigma^{'}$ (of which the loop
$\partial\Sigma^{'}$ is boundary) or not. The answer is given by
\begin{equation}
  W(x,\Sigma^{'}) = \exp\left( B(x) + \oint_{\partial \Sigma^{'}} A \right)
  = \left\{ \begin{array}{ll} 0 & \mbox{ if }  \#_{\rm intersection}
      (x,\partial\Sigma^{'}) = 0 \\  \pm 1 & \mbox{ if }
      \#_{\rm intersection} (x,\partial\Sigma^{'}) = \pm 1 \end{array}
  \right.
\end{equation}
A nice proof of this using the variational method can be
found in~\cite{Oda}, but since it is not very relevant for the work
done in topological gravity, we refer the reader to the
literature for more information on this proof.

What is important for the later comparison between the different
types of topological gravity of Schwarz vs. Witten type, is the
fact that the observables of BF theory can be related to
intersection numbers of
sub-manifolds of $M$. In contrast to this, as we shall see, 
the Witten type observables are connected to intersection numbers of
sub-manifolds of moduli space.

\section{General Features of Witten Type TFT}
As we have seen it is a non-trivial task to ensure that a Schwarz type TFT
will be topological when quantized. For the Witten type
TFT's this is not the case. In this section we give a short proof of the
metric independence of both the partition function and certain observables in all
Witten type TFT's, following simple arguments related to the BRST invariance
of the vacuum.

A Witten type topological field theory is a
quantum field theory described by a BRST exact quantum action
\begin{equation}
  S_{q}[\Phi^{i}, g_{\mu\nu}] = \{ Q, \Psi(\Phi^{i},g_{\mu\nu}) \},
  \label{S-exact}
\end{equation}
where $\Phi^{i}$ symbolises the fields in the theory and $g_{\mu\nu}$
is the metric of the underlying manifold $M$. $Q$ denotes the BRST operator
and $\Psi$ denotes the gauge fermion, i.e. an arbitrary
function of the fields and the metric.
The relevant BRST transformations are determined from the
symmetries one wishes to study. A Witten type theory is a kind
of field theoretical model which enables one to study 
topological invariants of a moduli space, relevant for
the theory one wishes to study. The moduli space is defined as elements in
the space of fields, which are solutions
to certain equations, modulo the symmetries one chooses. In symbolic
notation we have
\begin{equation}
  {\cal M} = \frac{ \{ \Phi^{i} \in {\cal S}\vert \,
    D\Phi^{i}=0\}}{{\rm symmetries}}, \label{Witten-modulo-1}
\end{equation}
where ${\cal S}$ denotes the space of fields, and $D$ is some 
operator acting on the fields. The action in equation~(\ref{S-exact}) is
constructed such that the path integral ``localizes'' to ${\cal M}$. Generally
${\cal M}$ will be finite dimensional and the theory only has a
finite number of degrees of freedom.
This ``localization'' nature of the action is vital for all Witten type
TFT's and we discuss it in detail in section~\ref{loc}.

The first Witten type TFT was the seminal work by
Witten~\cite{Witten-TFT} on $4D$ topological Yang-Mills theory,
while the more general scheme described above, was advocated by Witten
in~\cite{Witten-intro}. 
As we shall see in the case of topological Yang-Mills, one can add a
classical topological ( i.e. metric independent ) action to the BRST exact
action, as long as this
topological action is BRST invariant, but for the moment let us
stay with the definition in equation~(\ref{S-exact}).

The Energy-momentum tensor $T_{\mu\nu}$ is defined as the change of the
action under an infinitesimal metric variation $g_{\mu\nu} \mapsto
g_{\mu\nu} + \delta g_{\mu\nu}$
\begin{equation}
  \delta_{g} S \equiv \frac{1}{2}
  \int d^{n}x \sqrt{g} \delta g^{\mu\nu} T_{\mu\nu}.
\end{equation}
As a general result the Energy-momentum
tensor is BRST exact in Witten type theories
\begin{equation}
  T_{\mu\nu} = \{ Q, V_{\mu\nu}(\Phi^{i},g_{\mu\nu}) \}, \label{T_exact}
\end{equation}
where $V_{\mu\nu}(\Phi^{i},g_{\mu\nu})$ is some function of the fields and
the metric. Combining equations~(\ref{vev-0},\ref{T_exact}) makes it
possible to show, that the partition function is independent under the
metric transformation above. By direct calculation we find
\begin{eqnarray}
  \delta_{g} Z &=& \int {\cal D}[\Phi^{i}] e^{-S_{q}}\left(\delta_{g}S_{q}
\right) \nonumber \\ &=& \int {\cal D}[\Phi^{i}] e^{-S_{q}} 
\left( -\frac{1}{2} \int d^{n}x \sqrt{g} \delta g^{\mu\nu} T_{\mu\nu}\right)
\nonumber \\ &=&  \int {\cal D}[\Phi^{i}] e^{-S_{q}} \{Q,X\} \nonumber
\\ &=& \langle  \{Q,X\} \rangle = 0,
\end{eqnarray}
where $X$ is
\begin{equation}
  X = -\frac{1}{2} \int d^{4}x\, \sqrt{g} \delta_{g}^{\mu\nu}V_{\mu\nu}.
\end{equation}
This proof excludes the
possible dependence of the measure ${\cal D}[\Psi^{}]$ on the
metric~\cite{Witten-TFT} and is thus in some sense only a classical
calculation. Up to this question of anomalies, we have shown that the
partition function of a Witten
type TFT is independent of metric transformations and is thus said
to be topological. Actually metric-independence is a stronger requirement than
that of being topological, since it requires a differentiable structure on $M$.
Since the action always is BRST exact this is enough to ensure the
metric independence, in contrast to e.g. non-abelian BF theory and
Chern-Simons theory.

By a very similar calculation we now show that the partition function
is also independent of the coupling constant. Consider a rescaling of
the action $S_{q} \mapsto t S_{q}$ where $t\in \R$ and is 
considered to be the same as the inverse coupling constant $t=1/\hbar$. The change
of the partition function under this transformation reads
\begin{eqnarray}
  \delta_{t} Z &=& -  \int {\cal D}[\Phi^{i}] e^{-S_{q}} S_{q} \delta t
  \nonumber \\ &=& - \int {\cal D}[\Phi^{i}] e^{-S_{q}} \{Q,\Psi\}\delta t
  \nonumber \\ &=& 0,
\end{eqnarray}
where we have applied equations~(\ref{vev-0}, \ref{S-exact}). 
This calculation shows that the partition function is independent of the
change $\delta t$ because it changes in a BRST exact manner. Since $Z$
is independent of $t$ one can just as well take the semi-classical
limit $t\rightarrow \infty$ (i.e. $\hbar\rightarrow 0$). It is an important
result, that for a Witten type TFT, the semi-classical
approximation is exact. In the mathematical language one says, that the
stationary phase approximation is exact. For the Witten type 
$2D$ topological Yang-Mills theory, this statement is equivalent to the famous
Duistermaat-Heckman theorem~\cite{Jeffrey} and the argument above gives
a physics ``proof'' of this theorem.

The next step is to define the notion of observables in a Witten
type TFT. These are defined as functions $O(\Phi^{i})$ of the fields, which
are elements in the BRST cohomology $O\in H^{*}(M,Q)$, i.e. $\{Q,O\}=0$.
If one assumes that the metric dependence of $O$ is BRST exact
manner, i.e.
\begin{equation}
  \delta_{g} O = \{ Q, R(\Phi^{i},g_{\mu\nu}) \},
\end{equation}
then can one prove that the expectation value of O is
independent of the metric. $R(\Phi^{i},g_{\mu\nu})$ symbolises an
arbitrary function of the fields and the metric.
One has
\begin{eqnarray}
  \delta_{g} \langle O \rangle &=& \int {\cal D}[\Phi^{i}] e^{-S_{q}}
  \Biggl( (\delta_{g} O) - (\delta_g S_{q}) O \Biggr) \nonumber \\ &=&
   \langle \{ Q, R + X\cdot O \} \rangle = 0,
\end{eqnarray}
since $\delta_{g} S_{q} \cdot O = \{Q, X \} O = \{Q, X \cdot O\} $ because $O$ is
$Q$-closed. Hence expectation values of observables and products of
observables will be topological invariants in a Witten type TFT.

We see that the general feature is, that $S_{q}$ can depend on the metric
$g_{\mu\nu}$ on $M$, but $Z$ and $\langle O\rangle $ change only under
$\delta_{g}$ inside the cohomology class of $Q$, and hence the expectation
value of the change vanishes.
\section{A Physicist's approach to Witten's $4D$ Top.YM} \label{section:TYM}
In this section I present the general steps in the construction of the theory
of 4-dimensional topological Yang-Mills theory. I have chosen not to follow
the historical development of this subject, but
instead tried to focus on the most important issues, which help us
understand the later development of topological gravity. For a
detailed discussion of the historical development of this subject, I refer
to the literature (e.g.~\cite{Baal}), and will just outline the main steps
here.  In 1987 Atiyah~\cite{Atiyah-Weyl} suggested that it should be
possible to find a Lagrangian
description, which would correspond to the Hamiltonian formalism used
by Floer, to describe the new invariants of four-manifolds. These
invariants were found in 1985 by Donaldson, and they represent a
revolution in the classification of four-manifolds, or more precisely of
the differentiable structures of four-manifolds. As mentioned in
chapter 1, $\R^{4}$ stands alone in $\R^{n}$, with $n=1,2,\dots$
as the only case where there is more than one differentiable structure.
The breakthrough due to Donaldson, let us partially classify
the different differentiable structures, and the most impressive
thing for a physicist is, that the construction of these invariants are
closely related to the classical solutions of Yang-Mills theory.
In the seminal work by Witten~\cite{Witten-TFT},  a quantum field theory
approach to Yang-Mills theory gave a Lagrangian 
description of the Donaldson polynomials.
It was not the old fashioned version of Yang-Mills theory, but a
(twisted) supersymmetric version of Yang-Mills theory.
It was clear for Witten that his supersymmetry was a kind of BRST symmetry,
and the topological invariance of the partition function and the
correlators of
observables depended firmly on this fact. But is was not possible for Witten
to construct his Lagrangian as a BRST gauge fixing of an gauge invariant
action. This was first done by Baulieu and
Singer~\cite{Baulieu-Singer-1} and I follow their approach, since it offers
a clearer way to the results found by Witten. When the Lagrangian is
firmly in place, we discuss topological invariance and construction
of the observables, based on a number of different references.

\subsection{The Lagrangian of Topological Yang-Mills Theory}
Following~\cite{Baulieu-Singer-1}, consider the usual Yang-Mills connection
$A=A_{\mu}^{i}(x)T_{i}dx^{\mu}\in \Omega^{1}(M,\la)$.
We will absorb the Lie algebra index $i,j,\dots$ from time to time for
simplicity.
The usual change of $A$ under an infinitesimal gauge
transformation is
\begin{equation}
  \delta A_{\mu}(x) =  D_{\mu} \epsilon(x), \,\,\, \epsilon\in\Omega^{0}(M,\la).
\end{equation}
Now extend the transformation such that the change in $A$ reads
\begin{equation}
\delta A_{\mu}(x) =  D_{\mu} \epsilon(x) + \epsilon_{\mu}(x), \label{A_shift}
\label{A-top-trans}
\end{equation}
where $\epsilon_{\mu}(x)dx^{\mu}\in\Omega^{1}(M,\la)$ is an
infinitesimal one-form. This new
transformation is much larger that the usual local gauge transformations.
The curvature $F$ of $A$ changes accordingly as
\begin{equation}
  \delta F_{\mu\nu}(x) = D_{[\mu}\epsilon_{\nu]} - [\epsilon,F_{\mu\nu}].
\end{equation}
This new large gauge symmetry
will not leave the Yang-Mills action invariant
\begin{equation}
  \delta \int_{M} tr\left( F\wedge * F \right) \neq 0,
\end{equation}
but instead the topological action
\begin{equation}
  S_{{\rm top}} = \int_{M} tr\left( F\wedge F\right), \label{FwedgeF}
\end{equation}
is invariant under the transformation in equation~(\ref{A-top-trans}).
The integrand is the first Pontryagin class and the fields $\epsilon_{\mu},
\epsilon$ and $A_{\mu}$ must be in the same topological sector for
consistency.

One could actually turn the argument around and start from the
topological action above and search for the gauge invariance of this action.
It will clearly be much larger that the Yang-Mills action since one does
not need the space-time metric to write down $F\wedge F$, in contrast to
$F\wedge * F$ which involves $g_{\mu\nu}$ through the Hodge star.
The larger invariance is found to be the gauge transformations introduced
here. Since the new infinitesimal one-form $\epsilon_{\mu}(x)$ is
proportional to the (change) of the connection $A_{\mu}$ it will also
be subject to gauge transformations
\begin{equation}
  \epsilon_{\mu}(x) \mapsto \epsilon_{\mu}(x)^{'} = \epsilon_{\mu}(x) +
  D_{\mu}\lambda(x), \label{additional-symmetry}
\end{equation}
where $\lambda(x)$ is an infinitesimal scalar gauge parameter, i.e.
$\lambda\in\Omega^{0}(M,\la)$.
The enlarged gauge symmetry alters the original BRST algebra for Yang-Mills
theory. Recall that the usual prescription for obtaining the action of the
BRST operator is to exchange the gauge parameters with ghosts in the
gauge transformations. The original Yang-Mills BRST algebra, with
the Faddeev-Popov ghost $c^{i}(x)$ reads
\begin{eqnarray}
  \delta_{BRST} A^{i}_{\mu} &=&  D_{\mu}c^{i}, \\
  \delta_{BRST} c^{i} &=& - \frac{1}{2}[c,c]^{i} = -\frac{1}{2}
  f_{jk}^{\,\,\,i}c^{j}c^{k},\\
  \delta_{BRST} F^{i}_{\mu\nu} &=& = [F_{\mu\nu},c]^{i}.
\end{eqnarray}
Note that the ghost $c$ and the curvature transform in the adjoint
representation of \la. We should now introduce a one-form ghost $\psi=
\psi_{\mu}^{i}T_{i}dx^{\mu}\in\Omega(M,\la)$
associated with $\epsilon_{\mu}$, but the
additional gauge invariance expressed in equation~(\ref{additional-symmetry})
makes this ghost possess a gauge invariance of its own. This redundancy of the
BRST algebra requires the introduction of a ghost for
ghost. The second generation ghost
is denoted $\phi^{i}(x)\in \Omega^{0}(M,\la)$, and it is a
bosonic scalar field with ghost number two. It is a general feature that for each
generation in the ghost tower, the ghost number and grassmann parity is raised
by one unit. This new ghost fixes the gauge symmetry in $\psi_{\mu}$ and
Baulieu and Singer claim~\cite{Baulieu-Singer-1} that the only possible BRST algebra with
nilpotent BRST operator $\br$ is
\begin{eqnarray}
  \br A_{\mu}^{i} &=& D_{\mu}c^{i} + \psi_{\mu}^{i}, \label{br_A:TYM} \\
  \br c^{i} &=& - \frac{1}{2}[c,c]^{i} + \phi^{i}, \label{br_c:TYM} \\
  \br \psi_{\mu}^{i} &=& - D\phi^{i} - [c,\psi_{\mu}]^{i},\label{br_psi:TYM} \\
  \br \phi^{i} &=& - [c,\phi]^{i}, \label{br_phi:TYM}\\
  \br F_{\mu\nu}^{i} &=& D_{\mu}\psi_{\nu}^{i} - [c,F_{\mu\nu}]^{i}. \label{br_F:TYM}
\end{eqnarray}
In the next section we show that this is an example of what is known as a
Weil algebra, and that the differential in such algebras, which in this case
is \br, is always nilpotent.
The BRST algebra is now free of additional symmetries and requires
no third generation ghost. It is often useful to express the BRST algebra
in term of diffeential forms, but one should be careful regarding the signs
of the individual terms in this translation. The reason for this is the
fact that $\br$ anticommutes with the exteriour algebra~\cite{Baulieu-Singer-1,Kanno}
because $d$ can be viwed as a fermionic derivative, mapping the commuting
coordinates $x^{\mu}$ into anticommuting differentials $dx^{\mu}$. This implies
\begin{equation}
  \br ( A^{i}  ) \equiv \br ( dx^{\mu}A_{\mu}^{i}  ) =
  - dx^{\mu} \br A_{\mu}^{i} = - dx^{\mu} D_{\mu} c^{i} -dx^{\mu}\psi_{\mu}^{i} = - D c^{i} -
  \psi^{i},
\end{equation}
but by change of coordinates $\psi_{\mu}^{i} \mapsto \tilde{\psi}_{\mu}^{i} =-\psi_{\mu}^{i}$
we find
\begin{equation}
  \br (A^{i} ) = - Dc^{i} + \tilde{\psi}^{i},
\end{equation}
in addition to
\begin{eqnarray}
  \delta (\tilde{\psi}^{i}) &=& \delta( dx^{\mu}\tilde{\psi}_{\mu}^{i}) \nonumber \\ &=&
                                dx^{\mu}\delta \psi_{\mu}^{i} \nonumber \\
                             &=& dx^{\mu}\left( -D_{\mu}\phi^{i} -
                               c^{j}\psi^{k}f_{jk}^{\,\,\,i} \right) \nonumber \\
                             &=& - D\phi^{i} + c^{j}dx^{\mu}\psi_{\mu}^{k}f_{jk}^{\,\,\,i}
                             \nonumber \\ &=& - D\phi^{i} - [c, \tilde{\psi}]^{i}.
\end{eqnarray}
By relabeling $\tilde{\psi} = \psi$ we obtain the differential form version of the
BRST algebra for topological Yang-Mills theory~\cite{Birmingham-review,Baulieu-Singer-1,Kanno}
\begin{eqnarray}
  \br A^{i} &=& -Dc^{i} + \psi^{i}, \label{br_A:TYM-form} \\
  \br c^{i} &=& - \frac{1}{2}[c,c]^{i} + \phi^{i}, \label{br_c:TYM-form} \\
  \br \psi^{i} &=& - D\phi^{i} - [c,\psi]^{i},\label{br_psi:TYM-form} \\
  \br \phi^{i} &=& - [c,\phi]^{i}, \label{br_phi:TYM-form}\\
  \br F^{i} &=& D\psi^{i}- [c,F]^{i}. \label{br_F:TYM-form}
\end{eqnarray}
This algebra has a deep geometric meaning
explained independently by Baulieu and Singer~\cite{Baulieu-Singer-1} and
Kanno~\cite{Kanno}. Inspired
by Baulieu and Bellon's work on a geometric interpretation of the original
Yang-Mills BRST algebra~\cite{Baulieu-Bellon-1}, Baulieu and Singer could
write the BRST algebra above as the defining equations for a connection and
its curvature, in a generalized setting. Extend the exterior derivative
$d \mapsto \tilde{d} = d + \br$ and construct a generalized connection
$A \mapsto \tilde{A} = A+tc$. The $t$ is just an real integer keeping track
of the ghost numbers, in the expressions we derive.
By direct inspection the generalized curvature is 
\begin{equation}
  \tilde{F} = \tilde{d}\tilde{A} + \frac{1}{2}[\tilde{A},\tilde{A}]
  = F + t(\psi) + t^{2}(\phi), \label{gen_F-def:TYM}
\end{equation}
with the Bianchi identity
\begin{equation}
  \tilde{D}\tilde{F} = \tilde{d}\tilde{F} + [\tilde{A},\tilde{F}]
  = 0 \label{gen_Bianchi:TYM}.
\end{equation}
At first glance these equations look rather strange.
We add zero-, one- and two-forms, bosons
and fermions as though they were of the same nature. But the proper way to
view these equations is to expand them on both sides of the equality
in form-degree and ghost number and then identify the lhs and rhs of
the various combinations of these numbers. For example 
the ghost number 1 and form-degree 1 relation from
equation~(\ref{gen_F-def:TYM}) is the same as equation~(\ref{br_A:TYM-form}), 
and second the form-degree 2 and ghost number 1 relation from
equation~(\ref{gen_Bianchi:TYM}) is the same as equation~(\ref{br_F:TYM-form}).
The rest of the BRST algebra follows in the same way. These relations show
that there must exist some principal fibre bundle with connection $\tilde{A}$
and curvature $\tilde{F}$, which is closely related to the BRST algebra. We explain this
abstract construction in the next section.

Consider for a moment the action in equation~(\ref{FwedgeF}) and recall that
the classical solutions to the Yang-Mills equations
\begin{equation}
  D* F=0,
\end{equation}
are the self-dual field configurations $A_{\rm classical}$ for which
$F = \pm * F$. If these are inserted in the Yang-Mills action one
obtains our topological action
\begin{equation}
  S_{\rm classical} = -\frac{1}{2}\int_{M} tr(F\wedge * F)\left\vert_{A_{\rm classical}}\right.
  = \mp
  \frac{1}{2}\int_{M} tr(F\wedge F).
\end{equation}
One could ask whether the arbitrary deformation of
$A_{\mu}$ in equation~(\ref{A_shift}) would transform a self-dual
connection into a non-self-dual one, but it has been shown~\cite{Daniel} that
the self-dual solutions of the Yang-Mills equations are stable under these
transformations. This shows that the
transformations in equation~(\ref{A_shift})
are gauge symmetries of the action in equation~(\ref{FwedgeF}).

In order to fix the gauge invariance, three gauge fixing conditions
are used in~\cite{Baulieu-Singer-1}, two covariant conditions fixing the
transformations of $A$ and $\psi$
\begin{equation}
  \partial_{\mu} A^{\mu} = D_{\mu}\psi^{\mu} = 0, \label{covariantgauge:TYM}
\end{equation}
together with a condition fixing the connections to be self-dual
\begin{equation}
  F^{\mu\nu} \pm \epsilon^{\mu\nu\gamma\kappa}F_{\gamma\kappa} = 0.
  \label{F=*F}
\end{equation}
We have three gauge conditions and should hence introduce three
anti-ghost multiplets, each consisting of an anti-ghost and its
Lautrup-Nakanishi field companion. We choose two scalar anti-ghosts
$(\overline{c},\overline{\phi})$ and an antisymmetric tensor ghost
$\overline{\chi}_{\mu\nu}$. We list the ghost multiplets and the
relevant ghost numbers:
\begin{equation}
  \begin{array}{||r|c|c|c||}\hline &&& \\
    \mbox{ghost:} & \overline{c} &\overline{\phi}&
    \overline{\chi}_{\mu\nu}\\[.2cm] \hline &&& \\[.01cm]
    \mbox{L-N field:} & b & \eta& b_{\mu\nu}\\[.2cm]
    \hline
  \end{array}
\end{equation}
\begin{equation}
  \begin{array}{||r|c|c|c|c|c|c||}\hline & &&&&& \\
    \mbox{fields}:& \overline{c} &b & \overline{\phi} & \eta &
    \overline{\chi}_{\mu\nu} & b_{\mu\nu} \\[.2cm] \hline &&&&& &
    \\[.01cm]
    \#_{\rm ghost}: & -1 & 0 &-2 & -1 & -1 & 0\\[.2cm]
    \hline
  \end{array}
\end{equation}  
The BRST algebra is extended to these fields
(where $\overline{c},\eta,\overline{\phi},b\in\Omega^{0}(M,\la)$
and $b_{+-}\equiv b_{\mu\nu}dx^{\mu}\wedge dx^{\nu},
\overline{\chi}\equiv \overline{\chi}_{\mu\nu}dx^{\mu}\wedge dx^{\nu}\in \Omega^{2}(M,\la)$.
This extension is trivial in the sense that it is a closed sub-algebra
of the full BRST algebra and clearly \br is  nilpotent on these fields.
The quantum action for topological Yang-Mills reads generally
\begin{equation}
  S_{{\rm q}} = \mp
  \frac{1}{2}\int \sqrt{g} d^{4}x\, tr(F\wedge F) + \int \sqrt{g}
  d^{4}x\, \{Q, \Psi \}.
  \label{S_q:TYM}
\end{equation}
The central part in the work of Baulieu and Singer was to choose $\Psi$
so cleverly that the action reproduces the one given by
Witten in~\cite{Witten-TFT}. It reads
\begin{eqnarray}
  \Psi &=& tr\biggl( \overline{\chi}_{\mu\nu}(F^{\mu\nu}\pm
    \epsilon^{\mu\nu\gamma\kappa}F_{\gamma\kappa}) \pm \frac{1}{2} \rho
    \overline{\chi}_{\mu\nu}
    b^{\mu\nu} + \overline{\phi}D_{\mu}\psi^{\mu} + \overline{c}\partial_{\mu}
    A^{\mu} + \nonumber \\ && \qquad \frac{1}{2}\overline{c}b + 
    c[\overline{\chi}_{\mu\nu},\overline{\chi}^{\mu\nu}] + \overline{c}[\phi,\overline{\phi}]
    \biggr), \label{TYM:gaugefermion}
\end{eqnarray}
where $\rho$ is a real gauge parameter. By direct inspection we find
\begin{eqnarray}
\{ Q, \Psi \} & = & tr \biggl( b_{\mu\nu}(F^{\mu\nu} \pm
  \epsilon^{\mu\nu\gamma\kappa} F_{\gamma\kappa}) - \overline{\chi}_{\mu\nu}
  (D^{[\mu}\psi^{\nu]} \pm \epsilon^{\mu\nu\gamma\kappa}D_{[\gamma}
  \psi_{\kappa]})-  \nonumber \\  && \qquad    \overline{\chi}_{\mu\nu}
  ([c,F^{\mu\nu}\pm \epsilon^{\mu\nu\gamma\kappa}F_{\gamma\kappa}])
  \pm \frac{1}{2} \rho b_{\mu\nu} b^{\mu\nu} + \frac{1}{2}b^{2} +
  b \partial_{\mu}A^{\mu}  \nonumber \\ && 
  \qquad \eta D_{\mu}\psi^{\mu} - \overline{c}\partial_{\alpha}D^{\alpha} c -
  \overline{c}\partial_{\mu}\psi^{\mu} + \overline{\phi}[\psi_{\mu},
  \psi^{\mu}] + \overline{\phi}D_{\mu}D^{\mu}\phi + \nonumber \\ && \qquad
  \phi[\overline{\chi}_{\mu\nu},\overline{\chi}^{\mu\nu}] -\frac{1}{2}[c,c]
  [\overline{\chi}_{\mu\nu},\overline{\chi}^{\mu\nu}] 
  + 2b_{\mu\nu}[c,\overline{\chi}^{\mu\nu}]     +\nonumber \\ && \qquad
  b[\phi,\overline{\phi}] - \overline{c}[[\phi,c],\overline{\phi}] + \overline{c}[\phi,\eta]
  \biggr),
\label{S-gaugefix:TYM}
\end{eqnarray}
where one applies the Jacobi identity, remembering that all fields have
hidden gauge indices. There are two special values of $\rho$ which are
interesting. First the case $\rho=0$ where the equations of motion for
the $b_{\mu\nu}$ fields are $F^{\mu\nu}= \mp \epsilon^{\mu\nu\gamma\kappa}
F_{\gamma\kappa}$ or equivalently $F = \mp * F$. So in this case we
single out the (anti)self-dual configurations, which are the classical
solutions to the Yang-Mills equations, just as we wished to do by our
choice of gauge fixing function in equation~(\ref{F=*F}). Secondly
the choice $\rho=1$ is interesting, since by applying the equations of motion for
$b_{\mu\nu}$, we can eliminate this multiplier field from the action just
as in the discussion between BF theory and Yang-Mills theory in section~\ref{section:YM-BF}. We find
\begin{eqnarray}
tr\left( b_{\mu\nu}(F^{\mu\nu} \pm \epsilon^{\mu\nu\gamma\kappa} F_{\gamma\kappa}) \pm \frac{1}{2} b_{\mu\nu}b^{\mu\nu} \right) &\rightarrow& \nonumber \\
tr\left( F^{\mu\nu}F_{\mu\nu} + \epsilon^{\mu\nu\gamma\kappa}F_{\gamma\kappa}
  F_{\mu\nu} \right),
\end{eqnarray}
which in short hand notation reads
\begin{equation}
  b\wedge (F \pm * F) + b\wedge * b \rightarrow
  F \wedge * F + F \wedge F.
\end{equation}
The special reason for discussing $\rho=1$, is that the last $F\wedge F$ term
above cancels the classical action, and the quantum action is now identical to
Witten's action~\cite{Witten-TFT}, which is a Yang-Mills action plus
fermionic terms (ghosts and multipliers). The main importance of these
manipulations is that they provide a geometrical interpretation of the
theory described with Witten's action. Especially explaining the 
supersymmetry $\delta A = \psi$ introduced by Witten as a topological ghost term
for the additional gauge invariance of the action $F\wedge F$. Moreover,
the geometrical interpretation which will follow in the next section,
makes it possible to identify the
individual terms in the quantum action $S_{{\rm q}}$ as
components of the ``curvature'' of a universal fibre bundle.
There exists a firm geometrical interpretation of the individual terms in
equation~(\ref{S-gaugefix:TYM}). 

\subsection{Observables in Topological Yang-Mills}
We now turn to the discussion of observables in topological Yang-Mills.
Recall that in the previous section we defined observables to be elements
in $H^{*}(M,Q)$ with a BRST exact dependence on the space-time metric.
At first it does not look like we have any obvious candidates for
observables, since no elements in the BRST algebra in
equation~(\ref{br_A:TYM-form},..,\ref{br_phi:TYM-form}) are $Q$-closed, but the
solution to this is to consider the two symmetries involved in the BRST
algebra as individual symmetries. We have a local gauge invariance which
is fixed with the ghost $c^{i}(x)$ and then a topological symmetry which
is fixed via the ghosts $\psi_{\mu}^{i}(x)$ and $\phi^{i}$. A lot of discussions
have appeared in the literature regarding the so-called problem of
``triviality of observables'' in Witten type TFT's, but here I try to
cut through this discussion and present the result. In
the next section we find the mathematical reason for this
problem, which actually is no problem at all. The correct thing to do when
writing down observables is to consider closed forms under the topological
part of $Q$, i.e. to reduce the situation to the case where the
Faddeev-Popov ghost $c^{i}(x)$ is set to zero. We then get the reduced
BRST algebra introduced by Witten in~\cite{Witten-TFT}
\begin{eqnarray}
\dw A^{i} &=& \psi^{i}, \label{cartan:TYM-A}\\
\dw \psi^{i} &=& - D \phi^{i}, \label{cartan:TYM-PSI}\\
\dw \phi^{i} &=& 0 \label{cartan:TYM-PHI},  \\
\dw F^{i} &=& D\psi^{i}, \label{cartan:TYM-F}
\end{eqnarray}
where $\dw = \br\vert_{c=0}$. Note though that \dw is only nilpotent
up to gauge transformations. We return to this issue in the next section.
After this reduction we see that $\phi^{i}$ is now $Q_{W}$-closed (where
$\dw = \{Q_{W},\cdot \}).$  To obtain an observable, we must 
make a gauge invariant expression in $\phi$, and consider
the trace of $\phi^{2}$, which is gauge invariant since $\phi$ transforms
in the adjoint representation of \la. Hence we define the first observable
of topological Yang-Mills as
\begin{equation}
  W_{(0)}(x) \equiv tr\phi^{2}(x)\in \Omega^{0}(M,\la),
\end{equation}
which is just the quadratic Casimir of the gauge group $G$. 
The expectation value of products of observables is then
of the form
\begin{equation}
  \lan \prod_{i=1}^{k} W_{(0)}(x_{k}) \ra =
  \int {\cal D}[X] e^{-S_{q}} \prod_{i=1}^{k} W_{(0)}(x_{k}),
\end{equation}
which by the general arguments in the previous section is known to be
a topological invariant. Here $X$ denotes the collection of all relevant
fields in the path integral. 
We can test that the correlators do not depend on the positions
$x_{1},\dots, x_{k}$ simply by differentiating
$W_{(0)}(x_{\mu})$ with respect to its position $x_{\mu}\in M$
\begin{equation}
  \partial_{\mu} W_{(0)}(x_{\mu}) = \partial_{\mu}\left(tr\phi^{2}(x_{\mu})
  \right)= 2tr(\phi D_{\mu}\phi ) = -2\{Q_{W}, tr (\phi\psi_{\mu} )\},
\end{equation}
since $tr[A_{\mu},\phi]=0$ due to $tr(T_{i})=0$. Hence while $W_{(0)}$ per definition not is
BRST exact,
$\partial_{\alpha}W_{(0)}$ is BRST exact. By this one can construct
the difference in $W_{(0)}$ when defined at two distinct
points $x_{\mu},x_{\mu}^{'}$ and find
\begin{equation}
  W_{(0)}(x_{\mu}) - W_{(0)}(x_{\mu}^{'}) = \int_{x_{\mu}}^{x_{\mu}^{'}}
  \partial_{\mu}W_{(0)} dx^{\mu} =  \left\{ Q_{W},
    \int_{x_{\mu}}^{x_{\mu}^{'}}W_{(1)}\right\},
\end{equation}
where we have defined
\begin{equation}
  W_{(1)}(x_{\mu}) \equiv - tr(\phi\psi_{\mu})dx^{\mu} \in\Omega^{1}(M,\la).
\end{equation}
The index $(1)$ signals that it is a one-form. Consider now the correlators
\begin{equation}
  \left\langle \left(W_{(0)}(x_{\mu}) - W_{(0)}(x_{\mu}^{'})\right)
  \prod_{i=1}^{k} W_{(0)}(x_{k}) \right\rangle =
  \left\langle \left\{ Q_{W},  \int_{x_{\mu}}^{x_{\mu}^{'}}
  \partial_{\mu}W_{(0)} dx^{\mu} \cdot
  \prod_{i=1}^{k} W_{(0)}(x_{k})\right\}
  \right\rangle = 0,
\end{equation}
by the general arguments from the previous section. Since the difference
$W_{0}(x_{\mu})-W_{0}(x^{'}_{\mu})$ is
BRST exact, the $W_{0}$'s stay in the same cohomology class under
coordinate transformations and the correlator is a topological invariant.
We now have two important results
\begin{equation}
  \{ Q_{W}, W_{(0)} \} = 0 \,\,\mbox{ and }\,\, dW_{(0)} =
  \{ Q_{W}, W_{(1)} \} .
\end{equation}
By recursion Witten wrote down the famous descent equations
\begin{eqnarray}
 d W_{(0)} &=& \{ Q_{W}, W_{(1)} \}, \label{TYM-desent1} \\
 d W_{(1)} &=&  \{ Q_{W}, W_{(2)} \}, \\
 d W_{(2)} &=&  \{ Q_{W}, W_{(3)} \}, \\
 d W_{(3)} &=&  \{ Q_{W}, W_{(4)} \}, \\
 d  W_{(4)} &=& 0, \label{TYM-desent2}
\end{eqnarray}
where we get zero in the last equation, since otherwise we would have
to integrate a five form on $M$, which can not be done. In more detail:
Let $\gamma_{k}$ be an $k$'th-
homology cycle in $M$, against which we can
integrate a $k$-form to obtain the function 
\begin{equation}
  I(\gamma_{k}) = \int_{\gamma_{k}\subset M} W_{(k)}.
\label{I-gamma}
\end{equation}
The right hand side is BRST invariant
\begin{equation}
  \{ Q, I(\gamma_{k})\} =  \int_{\gamma_{k}\subset M}
  \{ Q, W_{(k)} \} =  \int_{\gamma_{k}\subset M}
  d W_{(k-1)} = 0,
\end{equation}
since $Q_{W}^{2}=0$ (up to gauge invariance).  Let $\gamma_{k}=\partial
\beta_{k}$ be a boundary, then 
\begin{equation}
  I(\gamma_{k}) = \int_{\gamma_{k}\subset M} W_{(k)} =
  \int_{\beta_{k}}dW_{(k)} =  \int_{\beta_{k}} \{Q_{W}, W_{(k+1)} \}
  =  \left\{ Q_{W}, \int_{\beta_{k}} W_{(k+1)} \right\} ,
\end{equation}   
by Stokes theorem. We see that $I(\gamma_{k})=I([\gamma_{k}])$ for
$[\gamma_{k}] \in H_{k}(M,\R)$, up to a BRST commutator. 
Witten derived the exact form of the forms $W_{(k)},\, k=1,2,3,4$
\begin{eqnarray}
  W_{(1)} &=& -tr(\phi\wedge \psi), \label{donaldson-1}\\
  W_{(2)} &=& tr( \frac{1}{2}\psi \wedge \psi -  \phi \wedge F), \\
  W_{(3)} &=& tr (\psi\wedge F), \\
  W_{(4)} &=& \frac{1}{2}tr (F\wedge F). \label{donaldson-4}
\end{eqnarray}
These satisfy the descent equations~(\ref{TYM-desent1},\dots,\ref{TYM-desent2})
which can be seen directly by inspection. These relations can also be derived
in the full BRST algebra. Recall that we could write the generalized
curvature $\tilde{F}=F+t(\psi)+t^{2}(\phi)$ in
equation~(\ref{gen_F-def:TYM}), as
the curvature of $\tilde{A}=A+tc$ with generalized exterior derivative
$\tilde{d}= \br + d.$ As will become clear in the next section, the
generalized fields $\tilde{A},\tilde{F}$ form a Weil algebra
with a nilpotent derivative, which in this case is $\tilde{d}$.
We can construct gauge invariant
polynomials in the generalized fields, and we define~\cite{Kanno} the
$N$'th Chern class
\begin{equation}
  {\cal W}_{N} \equiv c_{N}\,tr ( \overbrace{
    \tilde{F}\wedge\cdots\wedge
  \tilde{F}}^{\mbox{$N$-times}} ),\,\, N=1,2,\dots, \label{calW-w:expansion}
\end{equation}
with a normalization constant $c_{N}$. Since $\tilde{F}$ transforms in the
adjoint representation this is clearly gauge invariant. The geometrical
situation is discussed in detail in the next section.
The Weil theorem ensures that ${\cal W}_{N}$ is closed since
it projects down from  a $G$-invariant form on $P$
\begin{equation}
  \tilde{d}{\cal W}_{N}=0. \label{closed-W:TYM}
\end{equation}
Only it is not a closed form on $M$, but on the base space of the
universal bundle, which we introduce in the next section. We now make an
expansion of ${\cal W}_{N}$ in ghost number,
\begin{equation}
  {\cal W}_{N} = \sum_{k=0}^{2N}t^{2N-k}w_{k,N}(F,\psi,\phi). \label{W-expansion}
\end{equation}
Here we use that $\tilde{F}$ is a polynomial of order $t^{2}$ in
the fields $(F,\psi,\phi)$ and $w_{k,N}$ is then a $k$-form with
ghost number $2N-k$, representing the individual terms
which comes from taking the $N$'th power of $\tilde{F}$.
Insert this in equation~(\ref{closed-W:TYM})
\begin{equation}
  (d+\br)\left\{\sum_{k=0}^{2N}t^{2N-k}w_{k,N}(F,\psi,\phi)\right\}=0.
\end{equation}
This gives a sum of $4N$ terms, where $d$ and \br act on the various
$w_{k,N}$'s and the total sum is zero. When $d$ acts on $w_{k,N}$, we get
an element $w_{k+1,N}$ and when \br acts on an element of order $t^{i}$
we get an element of order $t^{i+1}$, since \br raises the ghost number by
$+1$. By reading off elements in the sum, which have the same ghost number
and form degree, we can obtain the descent
equations~(\ref{TYM-desent1},\dots,\ref{TYM-desent2})
\begin{eqnarray}
  \br w_{0,N} &=& 0 \nonumber ,\\
  \br w_{1,N} + dw_{0,1} &=& 0, \nonumber \\
  \vdots && \\
  \br w_{2N,N} + d w_{2N-1,N} &=&0, \nonumber \\
  dw_{2N,N} &=&0,\nonumber
\end{eqnarray}
but expressed by \br instead of \dw.
As an example consider the second from last equation. The term 
$w_{2N,N}$ has ghost number zero and form degree $2N$, so $\br w_{2N,N}$
has ghost number one and form degree $2N$. This can thus be added
with the term $d w_{2N-1,N}$ which has form degree $2N$ and ghost number
one, since $w_{2N-1,N}$ has form degree $2N-1$ and ghost number 1,
according to equation~(\ref{W-expansion}).

How do the descent equations for \br relate to those of \dw? According to
equations~(\ref{br_A:TYM},\dots,\ref{br_F:TYM}) and~(\ref{cartan:TYM-A},
\dots,\ref{cartan:TYM-F}) 
the difference between \br and \dw 
for $X=(F,\psi,\phi)$ is of the form
\begin{equation}
  \br X = \dw X + [X,c], 
\end{equation}
since these fields transform in the adjoint representation under
gauge transformations. Hence they have the same transformation
properties with regard to the Faddeev-Popov ghost $c$, so if we
restrict ourselves to gauge invariant polynomials in the fields
$F,\psi,\phi$ like $w_{k,N}$ we can take
\begin{equation}
\br w_{k,N} = \dw w_{k,N}.
\end{equation}
This means that the descent equations for \br and for \dw are equivalent.

In the seminal paper by Witten~\cite{Witten-TFT}, he considered the
second Chern class (i.e. $N=2$) for reasons explained in the next section and
the constant $c_{2}=-1/2$. This gives by equation~(\ref{calW-w:expansion})
\begin{eqnarray}
  {\cal W}_{2} &=& - \frac{1}{2} \left(F + t\psi + t^{2}\phi\right)^{2}
  \nonumber \\ &=&
  W_{(0)} + t W_{(1)} + t^{2}W_{(2)} - t^{3}W_{(3)} - t^{4}W_{(4)}.
\end{eqnarray}
Where $W_{(i)},\, i=(0,1,2,3,4)$ are given in equations~(\ref{donaldson-1},
\dots,\ref{donaldson-4}). So in our notation $w_{k,2}=W_{(k)}$.
These are exact copies of the Donaldson
polynomials which are related to topological invariants of
four dimensional manifolds. We return briefly to the discussion
of these invariants in the next section.

The last thing we shall discuss is how one can see that the expectation
values of the observables we have been considering, e.g. $W_{(0)}$ and
products of these, are related to closed differential forms on moduli space.
By integrating these forms over the moduli space we produce numbers, known
as the intersection numbers, which are topological
invariants. We proceed as follows: We show how the path integral reduces
to an integral over moduli space. In the next section we then discuss how
this is tied to intersection numbers.

Recall, that the quantum action in equation~(\ref{S_q:TYM}) is independent of
the form of the gauge fermion in equation~(\ref{TYM:gaugefermion}). We
use this fact to introduce a background gauge fixing~\cite{BBT}, here
written in the more transparent differential form notation. The background
gauge fermion we use is
\begin{equation}
  \Psi = tr\left(\overline{\chi}(F^{+} - \frac{\alpha}{2}b_{+-}) +
  \overline{\phi}D_{0}^{\dagger} \psi + \overline{c}(D_{(0)}^{\dagger} A -
  \frac{\alpha^{'}}{2}b)\right), \label{background-fermion}
\end{equation}
where $\alpha,\alpha^{'}$ are gauge parameters and $D_{(0)}$ denotes 
the covariant derivative around a fixed
classical solution $A_{{\rm clas}}$. So this solution corresponds to a
fixed element in 
\begin{equation}
{\cal M}_{{\rm ins}} = \frac{ \left\{ A\in {\cal A}\vert (F[A]\pm
    * F[A])= 0 \right\}}{\G}.
\end{equation}
The classical solutions are known as instantons and ${\cal M}_{{\rm ins}}$ is
the moduli space of instantons. By $F^{+}$ we denote
the self-dual part of the curvature for which $F = * F$.
This gauge fermion leads us to a quantum action
\begin{eqnarray}
S_{q} &=& \int_{M}
tr\biggl( b_{+-}(F^{+}-\frac{\alpha}{2}b_{+-} )-\overline{\chi}
  \frac{1}{2}(D\Psi)^{+} - \overline{\chi}[c, F^{+}] - \nonumber
 \\   &&  \qquad
   \frac{\alpha^{'}}{2}b^{2} + bD_{(0)}^{\dagger} A + \eta(D_{(0)}^{\dagger} \psi) -
  \overline{c} D_{(0)}^{\dagger} D c + \overline{c}D_{(0)}^{\dagger}\psi + 
\nonumber \\ && \qquad  \overline{\phi}D_{(0)}^{\dagger}[c,\psi] +
  \overline{\phi}(D_{(0)}^{\dagger} D)\phi \biggr).
\label{S_q:TYM-form-version}
\end{eqnarray}
For the choice of delta function gauges, which are convenient for our
purpose, we set $\alpha=\alpha^{'}=0$.
We expand the path integral in the semi-classical
approximation, around the classical solution. This expansion
is exact for Witten type TFT's, following the general arguments in the
last section. Hence the gauge field is of the form
\begin{equation}
  A = A_{{\rm class}} + A_{{\rm q}},
\end{equation}
where we have a classical and a quantum contribution. All other fields are
assumed to be of pure quantum nature. In the semi-classical approximation we
go to first order, i.e. only quadratic terms in the quantum fields contribute
and this contribution to the quantum action is
\begin{eqnarray}
  S_{{\rm q}-(2)} &=& \int_{M} tr\left(
    b_{+-}D_{(0)}A_{q} + b D_{(0)}^{\dagger} A_{q} -
    \overline{c}D_{(0)}^{\dagger}
    Dc + \nonumber \right. \\ &&\left. \qquad 
    \overline{\phi}D_{(0)}^{\dagger} D \phi + (\eta+\overline{c})D_{(0)}^{\dagger}\psi
    + \overline{\chi}D_{(0)}\psi \right).
\end{eqnarray}
Redefining $\eta^{'}=\eta + \overline{c}$, which has trivial Jacobian
leads to a cancellation between the Faddeev-Popov determinant in
the $\overline{c} c$ term against the $\overline{ \phi} \phi$ terms,
since they are
of opposite grassmann parity.  
When we consider the partition function
\begin{equation}
  Z = \int {\cal D}[A,\psi,\phi,b,b_{+-},c,\overline{c},
  \eta,\overline{\chi},\overline{\phi}] e^{-S_{{\rm q}}},
\end{equation}
we see, that if there are any zero modes in the fermionic field $\psi$,
i.e. nontrivial solutions to the equations
\begin{eqnarray}
  D_{(0)}^{\dagger} \psi(x) &=&  0, \label{psi-zeromode-0}
  \label{psi-zeromode} \\
  D_{(0)} \psi(x) &=& 0,  \label{psi-zeromode-1}
\end{eqnarray}
then the rules of grassmann integration
\begin{equation}
  \int \prod_{x\in M} d\psi_{\mu}(x) f(X) = 0,
\end{equation}
where $f(X)$ is any function of all fields excluding $\psi$, will
kill the partition function totally. This is due to the fact that the $\psi$
has dropped out of $S_{q}$, since it is a solution to
equation~(\ref{psi-zeromode}) or equation~(\ref{psi-zeromode-1}).

Assuming that there do not exist any symmetries in the
moduli space of instantons,
Witten~\cite{Witten-TFT} concludes that only $A$, and hence by supersymmetry
$\psi$, can have zero modes. Witten's notation of $\psi$ being the
super partner to $A$, is identical to our notation of $\psi$ being
the topological ghost for $A$. So one should worry about the zero modes
in $\psi$ since it is the only fermionic field with this possibility.

First it might be instructive to consider the semi-classical
partition function in the case with no zero modes in $\psi$.
The integration over $b$ and $b_{+-}$
imposes two delta functions
\begin{equation}
  \delta \left(D_{(0)}A\right)  \delta\left( D_{(0)}* A \right),
\end{equation}
and the effect of integrating these with respect to $A$ is to
restrict the integration
over \A to ${\cal M}_{{\rm ins}}$. The integral
can be calculated along the lines of similar expressions in Schwarz
theories and it has been shown to give~\cite{Ouvry-Thompson, BBT}
\begin{equation}
  \int_{{\cal M}_{{\rm ins}}} \left\vert \frac{1}{{\rm det}T_{0}} \right\vert,
\end{equation}
where $T_{0}$ is the appropriate differential operator mapping connections
into zero forms and self-dual 2-forms. We do not need the specific form of
this operator and refer to~\cite{Schwarz-1,Ouvry-Thompson,BRT}
for further arguments.
The integration over $\overline{\chi},\eta^{'}$ gives a contribution
\begin{equation}
  \delta\left( D_{(0)}\psi \right)
  \delta\left(D_{(0)}^{\dagger} \psi \right),
\end{equation}
which can be represented as another determinant
\begin{equation}
 \int_{T_{A_{{\rm c}}}{\cal M}_{{\rm ins}}}
{\cal D}[\psi]  {\rm det}T,\,\,\, A_{{\rm c}}=A_{{\rm class}}.
\label{psi-som-tangent-til-A}
\end{equation}
The operator $T$ is related to $T_{0}$ and can be studied
in~\cite{Ouvry-Thompson}. Thus the partition function will be an integral
over ${\cal M}_{{\rm ins}}$ of the ratio of these determinants. We have
indicated that $\psi$ should be viewed as an element in the tangent space of
the moduli space. The reason for this is explained in the next section.

While the are no zero modes, moduli space is
just a collection of points, but it becomes a manifold (actually an
orbifold, due to singularities) with dimension $n$ when there are zero modes.
When ${\cal M}_{{\rm ins}}$ is a manifold, we need to have forms of the
top dimension of  ${\cal M}_{{\rm ins}}$, in order to be able to
perform the integral. The integration over the zero modes is missing
due to the rules of grassmann integration and we must therefore insert
functionals \OO of the various fields in the action to get a non-vanishing
result. The functionals must have a ghost number which sums up to $n$ such that they
can ``soak up'' the fermionic zero modes~\cite{Blau-Thompson:ATF}. We show how the observables are
expressed in terms of the zero modes in the action.

The strategy is to integrate out all non-zero modes in the path integral, 
such that the measure of all fields reduces to
\begin{equation}
  {\cal D}[X] \mapsto d\mu \equiv \prod_{i=1}^{n}d\tilde{A}_{i}\, d\tilde{\psi}_{i},
\end{equation}
where $\tilde{A}_{i},\tilde{\psi}_{i}$ are the zero modes. Since $\tilde{A}_{i}$ and
$\tilde{\psi}_{i}$ have opposite grassmann parity, $d\mu$ will be invariant under
change of basis in the space of $(A_{\mu},\psi_{\mu})$ zero modes.
The integration over non-zero modes in the semi-classical
approximation gives a cancellation of the determinants mentioned above, and
these integrations give $\pm 1$~\cite{Witten-TFT}. Under the integration, all
non-zero modes in \OO drop out, and thus \OO is effectively only a functional
\Op, of the zero modes 
\begin{equation}
  \Op = \Phi_{m_{1},\dots,m_{n}}(\tilde{A}_{i})\tilde{\psi}^{m_{1}}
  \cdots \tilde{\psi}^{m_{n}}.
\end{equation}
By $\Phi_{m_{1},\dots,m_{n}}(\tilde{A}_{i})$ we denote an
antisymmetric tensor on moduli space, i.e. a form of top dimension
on ${\cal M}_{{\rm ins}}$. The trick is now to compute the expectation
value of \OO and replace \OO by \Op, due to the integration of all
non-zero modes to obtain~\cite{Witten-TFT}
\begin{eqnarray}
  \langle \OO \rangle &=& \int {\cal D}X e^{-S_{{\rm q}-(2)}} \OO 
\nonumber \\ &=& \int d\mu \Op = \int \prod_{i=1}^{n}d\tilde{A}_{i}\,
d\tilde{\psi}_{i} \Phi_{m_{1},\dots,m_{n}}(\tilde{A}_{i})\tilde{\psi}^{m_{1}}\cdots
\psi^{m_{n}} \nonumber \\ &=& \int_{{\cal M}_{{\rm ins}}} \Phi(\tilde{A}).
\label{vev-OO-1}
\end{eqnarray}
By $\Phi(\tilde{A})$ we denote the top form on ${\cal M}_{{\rm ins}}$ and
the last equality is due to the rules of grassmann integration.

Witten states~\cite{Witten-TFT} that it is possible to view \OO as a
product of, say $k$ different terms $\OO_{i},\, i=1,\dots,k$ and that
this sub-division holds after integrating out non-zero modes. So we
consider
\begin{equation}
  \Op = \prod_{i=1}^{k} \Op_{i},
\end{equation}
where $\Op_{i}$ has a ghost number $n_{i}$ and the total ghost number
still adds up to $n=\sum_{i=1}^{k} n_{i}$. The form of an individual term is
\begin{equation}
  \Op_{i} = \Phi^{(i)}_{j_{1}\dots j_{m_{i}}} \tilde{\psi}^{j_{1}}\dots
  \tilde{\psi}^{j_{m_{i}}},
\end{equation}
which we view as an $m_{i}$-form on the moduli space. Hence the form of top
dimension on moduli space is
\begin{equation}
  \Phi = \Phi^{(1)}\wedge \dots\wedge \Phi^{(k)}.
\end{equation}
This changes equation~(\ref{vev-OO-1}) to
\begin{eqnarray}
  \langle \OO \rangle &=& \langle \OO_{i_{1}}\cdots \OO_{i_{k}}
  \rangle \nonumber \\
  &=& \int {\cal D}X e^{-S_{{\rm q}-(2)}}\, \OO_{i_{1}}\cdots
    \OO_{i_{k}}
  \nonumber \\ &=& \int_{{{\cal M}_{{\rm ins}}}}
    \Phi^{(i_{1})}\wedge \dots\wedge \Phi^{(i_{k})}
\end{eqnarray}
where the dimension of moduli space is $n=\sum_{j=1}^{k}i_{j}$. 

\noindent
We introduce the candidates for $\OO_{i_{k}}$ based on the discussions
in the previous section. Recall from equation~(\ref{I-gamma}) that
\begin{equation}
  I(\gamma_{k}) = \int_{\gamma_{k}\subset M}W_{(k)},
\end{equation}
is BRST invariant and since $W_{(k)}$ is a $k$-form with
ghost number $4-k$, is it obvious to try to relate $I(\gamma_{k})$ to a $
(4-k)$-form
$\Phi^{(\gamma_{k})}\in {\cal M}_{{\rm ins}}$. The expectation value of
\begin{eqnarray}
\left\langle I(\gamma_{1})\cdots I(\gamma_{s}) \right\rangle &=&
\left\langle \int_{\gamma_{1}} W_{(k_{\gamma_{1}})} \cdots
\int_{\gamma_{s}} W_{(k_{\gamma_{s}})} \right\rangle\nonumber \\
&=& \int_{{\cal M}_{{\rm ins}}} \Phi^{(\gamma_{1})}\wedge \dots \wedge
\Phi^{(\gamma_{s})}.\label{vev-I-gamma}
\end{eqnarray}
According to the discussion above, we should integrate out
non-zero modes of the fields present in the $W_{(k_{\gamma_{i}})}$'s
on the left hand side of the last equality sign in order to obtain
the explicit form of the
$\Phi^{({\gamma_{i}})}$'s. In the semi-classical limit we can
replace the $F[A]$ in the $W_{(k_{\gamma_{i}})}$'s with $F[A_{{\rm class}}]$
and $\psi_{\mu}$ with
\begin{equation}
  \psi_{\mu}^{a} = \sum_{i=1}^{n} \tilde{\psi}^{i}u_{(i),\mu}^{a},
\end{equation}
where $u_{(i),\mu}^{a}$ is an element in the basis $u_{(1),\mu}^{a}\cdots
u_{(n),\mu}^{a}$ in the space of
$(A_{\mu},\psi_{\mu})$ zero modes. Since the $A_{\mu}$ zero modes and the
$\psi_{\mu}$ zero modes are in the same basis, the discussion is independent
of the specific choice of $u_{(i)}$, since $d\mu$ is invariant under shift
of basis. Regarding $\phi$, which is the third field present in the 
$W_{(k_{\gamma_{i}})}$'s, consider the form of the quantum action in
equation~(\ref{S-gaugefix:TYM}) with $\phi,\overline{\phi}$ terms
\begin{equation}
  S_{{\rm q}} = \int \sqrt{g} \,d^{4}x \,tr
  \left( \dots \, + \overline{\phi}[\psi_{\mu},
  \psi^{\mu}] + \overline{\phi}D_{\mu}D^{\mu}\phi + \dots \right).
\end{equation}
When performing the integration over $\overline{\phi}$ we get a delta
function
\begin{equation}
  \delta \left( \int \sqrt{g} \,d^{4}x \,tr
    ( D_{\mu}D^{\mu}\phi+ [\psi_{\mu},
  \psi^{\mu}] )\right).
\end{equation}
The remaining terms in the action are independent of $\phi$ so the
expectation value of $\phi$ is
\begin{eqnarray}
  \langle \phi^{i} \rangle &=&
  \int {\cal D}[\phi,\dots] e^{-\tilde{S}_{{\rm q}}} \phi^{i}
  \delta \left( \int \sqrt{g} d^{4}x tr ( D_{\mu}D^{\mu}\phi+ [\psi_{\mu},
  \psi^{\mu}] )\right)  \nonumber \\ &=&
-\int \sqrt{g}\,d^{4}x\, (D_{\mu}D^{\mu})^{-1}[\psi_{\mu},\psi^{\mu}]^{i}.
\label{vev-phi:TYM}
\end{eqnarray}
Here $\tilde{S}_{{\rm q}}$ signals that we have performed the
$\overline{\phi}$ integration such that $\phi$ is no longer present
in the action. So the expectation value of $\phi$ is the integral
over the Greens function of the laplacian $D_{\mu}D^{\mu}$ acting on
the $[\psi_{\mu},\psi^{\mu}]$. This has a direct geometric interpretation
in the context of the universal bundle which we discuss in the next section.

When expressing $\langle \phi^{i} \rangle$ in terms of the zero modes
$ \psi^{a}$ we can write down Witten's expressions for
$\Phi^{(\gamma_{k})}$ by inserting $(F[A_{{\rm class}}],\psi^{a},
\langle \phi^{i} \rangle)$ in the expressions for $W_{(k_{\gamma_{i}})}$
in equation~(\ref{donaldson-1} ,\dots, \ref{donaldson-4}). We have 
\begin{eqnarray}
  \Phi^{(\gamma_{0})} &= & \frac{1}{2} tr( \langle \phi(x) \rangle )^{2} ,\,\,
  \gamma_{0} \in H_{0}(M,\R), \, x\in M,\label{witten-inv-1} \\
  \Phi^{(\gamma_{1})} &= & \int_{\gamma_{1}} tr(-\langle \phi \rangle \wedge
  \psi^{a}
  ),\,\, \gamma_{1} \in H_{1}(M,\R), \\
  \Phi^{(\gamma_{2})} &= & \int_{\gamma_{2}} tr(\frac{1}{2}
  \psi^{a}\wedge \psi^{a} - \langle \phi \rangle \wedge
  F[A_{{\rm class}}] ),\,\, \gamma_{2} \in H_{2}(M,\R), \\
  \Phi^{(\gamma_{3})} &= & \int_{\gamma_{3}} tr( \psi^{a}\wedge
  F[A_{{\rm class}}] ),\,\, \gamma_{3} \in H_{3}(M,\R), \\
 \Phi^{(\gamma_{4})} &= & \int_{\gamma_{4}} \frac{1}{2}
 tr( F[A_{{\rm class}}] \wedge  F[A_{{\rm class}}]),\,\,
  \gamma_{4} \in H_{4}(M,\R). \label{witten-inv-5}
\end{eqnarray}
As mentioned before $ \Phi^{(\gamma_{k})}$ is a $(4-k)$-form on
${\cal M}_{{\rm ins}}$ and hence it represents a map, first
introduced by Donaldson, from homology on $M$ to cohomology on
${\cal M}_{{\rm ins}}$
\begin{equation}
  \Phi^{(\gamma_{k})}: H_{k}(M,\R) \mapsto H^{4-k}({\cal M}_{{\rm ins}},{\bf Q})
  .
\end{equation}
The cohomology on moduli space is rational since moduli space is an
orbifold and not truly a manifold. The original Donaldson invariants
were exactly defined as the wedge product (giving a form of top dimension)
of such maps, integrated over the moduli space of Yang-Mills instantons.
In that sense the expressions in equation~(\ref{vev-I-gamma})
represent the QFT representation of the Donaldson
invariants~\cite{Witten-TFT}. The rules of BRST theory singled out the
BRST invariant functions as observables and we now see that the elements
of the BRST cohomology are related to the cohomology of the moduli space,
i.e. the space of classical solutions to the Yang-Mills equations.
When there are no fermionic zero modes the partition function is a
topological invariant and by introducing observables with the proper
ghost number one can construct non-vanishing invariants in the presence
of fermionic zero modes.

In the next section we discuss several issues from this section again and
put them into the geometrical picture given by the universal bundle
construction. For further information about four dimensional
topological Yang-Mills theory we refer to the vast literature on this
topic. The above presentation is in no way complete, but serves to
help the understanding of similar constructions in topological gravity.

\section{The Mathematics behind the Physics} \label{section:math-behind-phys}
In this section, we review the most important mathematical constructions
and results, which are related to the material presented in the previous
section. After reading this, it will be clear what amazing symbiosis between
mathematics and physics, the Witten type TFT's represent. I have as far as
possible tried to use the original material as the basis for this section, but
also gained a lot of understanding from the long list of reviews written by
physicists working in this area. The most prominent reviews I know, are
those by
Birmingham et.al.~\cite{Birmingham-review}, Cordes et.al.~\cite{CMR},
Blau~\cite{Blau-MQ-TFT} and Blau \& Thompson~\cite{Blau-Thompson-Trieste},
and
all of these have contributed to my understanding of this field. When I use
a result specifically based on a review text, I will cite it, but otherwise
I hereby acknowledge the importance of these reviews.

\subsection{The Universal Bundle and Equivariant Cohomology}
One of the most prominent features in the theories studied in this thesis, is
the study of manifolds on which there is a group action. This is either
the action of gauge transformations or diffeomorphism. Here we present the
important theory of equivariant cohomology, which will enable us to study
cohomology in relation to this group action.

Let $M$ be a manifold on which the group $G$ acts. Then $M$ is called a
$G$ space and one can form the principal $G$-bundle
\begin{eqnarray}
  &M& \nonumber \\  &\downarrow&  \\ &M/G&. \nonumber \label{M->M/G}
\end{eqnarray}
If there are fixed points for the action of $G$ on $M$, the space $M/G$ is
complicated and it is hard to calculate its cohomology because
$M/G$ might not even be a manifold.

The notation of the ``Universal $G$- Bundle'', due to Atiyah and
Singer~\cite{AS}, gives us a tool to overcome this problem.
The universal bundle
is a kind of ``Mother of all $G$-bundles'' as we shall see. Let $EG$ be a
manifold on which $G$ acts and construct the universal (principal)
bundle over $BG$
\begin{eqnarray}
  &EG& \nonumber \\  &\downarrow&  \\ &BG&= EG/G, \nonumber \label{EG->BG}
\end{eqnarray}
with two defining properties. First that $EG$ is a contractible space
such that $\pi_{1}(EG)=\{ {\bf id} \}$, and second that $G$ acts
freely on $EG$.
The universal $G$ bundle is a principal $G$ bundle over the base space $BG$,
which in general is a nontrivial bundle. There exist
characteristic classes on $BG$ which measure the ``twist'' of the bundle,
i.e. the obstruction to triviality and these are elements in $H^{*}(BG)$.
At the moment this may seem unconnected with the $G$ bundle $
M\rightarrow M/G$, but any $G$-bundle can be obtained from this
universal $G$-bundle as the pull back of a certain map known as the
classifying map. For every Lie group (actually every topological group)
$G$, the loop space of $BG$ is homotopic to $G$. The loop space $L(BG)$ is
the space of all maps $\phi:S^{1}\mapsto BG$ and
$L(BG)\simeq_{\rm homotopic} G$. Consider two different $G$-bundles
\begin{equation}
  \begin{array}{ccccc}G \mapsto & E &\hspace{1cm}  &G\mapsto & F \\ &
    \downarrow & & \hspace{1cm} &\downarrow \\ & M & & \hspace{1cm} & M
  \end{array}
\end{equation}
and consider two maps $f_{E}:M \mapsto BG$ and $f_{F}:M\mapsto BG$ from
each of these bundles. These maps are known as classifying maps and $BG$
is also known as the classifying space of $G$. This is due to the following
result. Let $c_{1},\dots,c_{n}\in H^{*}(BG,\Z)$ be characteristic classes,
for example the Chern Classes, which are cohomology classes on $BG$. Make
the pull back of these cohomology classes via the classifying maps
\begin{equation}
  f_{E}^{*}c_{i}\in H^{*}(M,\Z) \,\,\mbox{ and } \,\, f_{F}^{*}c_{i} \in
  H^{*}(M,\Z).
\end{equation}
If the pull back of the cohomology classes agree for the two classifying
maps $f_{E}$ and $f_{F}$, then the maps are homotopic and the bundles
$E$ and $F$ will be isomorphic. So isomorphism classes of $G$-bundles are in
one to one correspondence with homotopy classes of maps $M\mapsto BG$.
Every principal $G$-bundle up to isomorphism,
can be obtained by a pull back from the universal bundle.

As we stated, the universal bundle is a tool which is used to study the
cohomology on $M/G$, whereas a direct approach does not work.
Due to the fact that $EG$ is contractible, its de Rham cohomology is
trivial~\cite{Nakahara,Bott-Tu}. This follows from Poincar\'{e}'s lemma,
which states that if a manifold $M$ is contractible to a point in a coordinate
neighbourhood $U\subset M$, then every closed form on $U$ is also exact.
When the whole of $EG$ is contractible, any globally closed form will also
be exact and hence $EG$ has trivial de Rham cohomology
\begin{equation}
  H^{*}(EG) = \delta_{*0}\R.
\end{equation}
But there exists the framework of equivariant cohomology~\cite{Atiyah-Bott}
written as $H_{G}^{*}(M)$ which is ordinary de Rham cohomology of a
larger space $H^{*}(M_{G})$, where
\begin{equation}
  M_{G} = EG \times_{G} M \equiv \left(EG \times M\right)/G.
\end{equation}
The elements in $M_{G}$ are equivalence classes under the equivalence relation
$(pg,q)\sim (p,gq)$ where $p\in EG,q\in M, g\in G.$ A central feature in this
construction is the following result: $M_{G}$ is a fibre bundle with fibre $M$
over $BG$, associated with the universal bundle 
\begin{eqnarray}
  &M_{G}& \,\, = (EG \times M)/G \nonumber \\ \pi &\downarrow& \,\,\mbox{with fibre }
  M  \\ &BG& \,\, =EG/G. \nonumber \label{M_G->BG}
\end{eqnarray}
In addition to the projection $\pi$, there exists a map $\sigma:M_{G}
\rightarrow M/G$, which in contrast to $\pi$ is not always 
a fibration. The so-called Cartan-Borel mixing diagram
illustrates this construction
\begin{equation}\begin{array}{ccccc}
  EG & \leftarrow & EG\times M & \rightarrow & M \\[.6cm]
  \downarrow & & \downarrow & &\downarrow \\[.6cm]
  BG & \leftarrow & EG \times_{G} M & \rightarrow & M/G. \\[.05cm]
  & \pi & & \sigma &
\end{array}
\end{equation}
Using $\sigma$ one can define a map in cohomology
\begin{equation}
  \sigma^{*}: H^{*}(M/G) \mapsto H^{*}_{G}(M),
\end{equation}
which is an isomorphism~\cite{Atiyah-Bott} if $G$ acts freely on $M$.
This links the cohomology of $M/G$ to the equivariant cohomology on
$M$.

The equivariant cohomology of a point $q\in M$ is $H^{*}_{G}(q) =
H^{*}(BG)$, which follows from ``integration over the fibre''
$\pi_{*}: H^{*}(EG\times_{G}M)\mapsto H^{*}(BG)$.

It is possible to give an algebraic description of this
construction which enables us to determine $H^{*}(BG)$.
This leads back to the work by Weil, Cartan and others,
but we follow the treatment in~\cite{Atiyah-Bott,MQ}.

The Weil algebra \W is based on the Lie algebra \la of $G$.
It is the tensor product between the exterior algebra $\Lambda(\dla)$ and
the symmetric algebra $S(\dla)$ on the dual Lie algebra \dla
\begin{equation}
  \W = \Lambda(\dla) \otimes S(\dla). \label{Weil-algebra}
\end{equation}
The exterior algebra~\cite{Karoubi}
\begin{equation}
\Lambda(\dla) = \bigoplus_{n=0}^{\infty} \Lambda^{n}(\dla ),
\end{equation}
is a graded algebra of differential $n$-forms and 
\begin{equation}
  S(\dla) = \bigoplus_{n=0}^{\infty} S^{n}(\dla),
\end{equation}
is the graded algebra of symmetric polynomials on \la.
Let $\{X_{a}\}$ be a basis for \la dual to $\{\theta^{a} \}$, the
basis for \dla.
We assign the degree 1 to elements
$\theta^{a} \in \Lambda(\dla)$ and degree 2 for $S(\dla)$, which we
denote $\phi^{a}$.
Then \W will be a commutative graded algebra with product rule
\begin{equation}
  w^{p}w^{q} = (-1)^{pq}\,w^{q}w^{p}, \,\,\,p,q=(1,2), \label{theta-phi}
\end{equation}
for elements $w^{p}\in \W$. The range of the indices is 
$a,b,\dots = 1,\dots,{\rm dim}(G).$ By introducing a differential $d_{W}$ on
\W, defined by the action on the generators
\begin{eqnarray}
  d_{W}\theta^{a}&=& \phi^{a} -\frac{1}{2} f^{\,\,\,a}_{bc}
  \theta^{b}\theta^{c},
  \label{d_W-theta}\\
  d_{W}\phi^{a} &=&- f^{\,\,\,a}_{bc}\theta^{b}\phi^{c}, \label{d_W-phi}
\end{eqnarray}
\W is promoted to a differential graded algebra. As usual $f^{\,\,\,a}_{bc}$
are the structure constants of \la via
$[X_{b},X_{c}]=f_{bc}^{\,\,\,a}X_{a}$.
We now show that $d_{W}^{2}=0$ by acting
with $d_{W}$ twice on both generators.
\begin{eqnarray}
  d_{W}(d_{W}\theta^{a}) &=& d_{W}\phi^{a} - \frac{1}{2}f_{bc}^{\,\,\,a}
  d_{W}(\theta^{b}\theta^{c}) \nonumber \\ &=&
  f_{bc}^{\,\,\,a}\phi^{b}\theta^{c} - \frac{1}{2}f_{bc}^{\,\,\,a}
  \left( d_{W}\theta^{b}\cdot \theta^{c} - \theta^{b}d_{W}\theta^{c}\right)
  \nonumber \\ &=& f_{bc}^{\,\,\,a}\phi^{b}\theta^{c} -
  \frac{1}{2}f_{bc}^{\,\,\,a}\left( \phi^{b}\theta^{c}-\frac{1}{2}\phi^{b}
    \theta^{c} + \frac{1}{2}\theta^{b}\phi^{c} \right)
  - \frac{1}{4}f_{bc}^{\,\,\,a}\left( f_{ij}^{\,\,\,b}\theta^{i}\theta^{j}
    \theta^{c} + f^{\,\,\,c}_{de}\theta^{b}\theta^{d}\theta^{e}\right)
  \nonumber \\ &=& 0.
\end{eqnarray}
Where we have used equation~(\ref{theta-phi}) and that $d_{W}$ is an
anti derivation so it obeys the anti-Leibniz rule
\begin{equation}
  d_{W} (w^{p}\wedge w^{q}) = d_{W}w^{p}\wedge w^{q} + (-1)^{p}w^{p}
  \wedge d_{W}w^{q}.
\end{equation}
In the same way we find
\begin{eqnarray}
  d_{W}(d_{W}\phi^{a}) &=& d_{W}(f_{bc}^{\,\,\,a}\phi^{b}\theta^{c}) \nonumber \\ &=& f_{bc}^{\,\,\,a} (d_{W}\phi^{b}) \theta^{c} + f_{bc}^{\,\,\,a}
  \phi^{b} d_{W}\theta^{c} \nonumber \\ &=& 
f_{bc}^{\,\,\,a}f_{ij}^{\,\,\,b}\phi^{i}\theta^{j}\theta^{c} +
f_{bc}^{\,\,\,a}\phi^{b}\phi^{c} - \frac{1}{2}f_{bc}^{\,\,\,a}
f_{ij}^{\,\,\,c}\phi^{b}\theta^{i}\theta^{j} \nonumber \\ &=&
[[X_{a},X_{b}],X_{c}]\phi^{a}\theta^{b}\theta^{c} + \frac{1}{2}
[[X_{b},X_{c}],X_{a}]\phi^{a}\theta^{b}\theta^{c} \nonumber \\ &=&
[[X_{a},X_{b}],X_{c}]\phi^{a}\theta^{b}\theta^{c} - \frac{1}{2}
\left( [[X_{c},X_{a}],X_{b}] + [[X_{a},X_{b}],X_{c}]\right)
\phi^{a}\theta^{b}\theta^{c}  \nonumber \\ &=&
\frac{1}{2}\left( [[X_{a},X_{b}],X_{c}]- [[X_{c},X_{a}],X_{b}] \right)
\phi^{a}\theta^{b}\theta^{c}  \nonumber \\ &=&
0.
\end{eqnarray}
In the fourth equality we switch to commutator notation and relabel the
indices to improve the clarity, and the term
$f_{bc}^{\,\,\,a}\phi^{b}\phi^{c}=0$ due to contraction between the antisymmetric structure constant ( in $b,c$ ) and the symmetric expression in $\phi$.
In the fifth equality we use the Jacobi identity
and in the last step the expression inside the brackets is symmetric in
$b,c$ so it vanishes when contracted with the antisymmetric term
$\theta^{b}\theta^{c}$. We conclude that the so-called Weil differential
$d_{W}$ is nilpotent.
It is now possible to show that the cohomology of \W is trivial
by changing to a new set of generators $\theta^{a},(\kappa^{a}= d_{W}
\theta^{a})$,
and it then follows that $\kappa=d_{W}\theta^{a}$ is
exact and $d_{W} \kappa = 0$ closed, since $d_{W}^{2}=0$.
This hints that \W is the algebraic model of $EG$. To find a nontrivial
cohomology, we must find the algebraic model of $BG$. This is done by studying
the action of $G$ on elements in \W and on differential forms.

Recall our general construction: $P$ is a smooth principal $G$ bundle
$\pi: P \rightarrow M$ and the elements of the Lie algebra $X\in \la$ are
the vertical vector fields. The elements $X$ act on the
differential forms $\Omega^{*}(P)$ in two ways. First by the interior product
$\imath(X):\Omega^{k}(P) \mapsto \Omega^{k-1}(P)$ and next by the Lie
derivative ${\cal L}_{X}:\Omega^{k}(P) \mapsto \Omega^{k}(P)$, which are
related by the infinitesimal homotopy relation
\begin{equation}
  {\cal L}_{X} = \imath(X) \, d + d \, \imath(X), \label{inft-homotopy}
\end{equation}
where $d$ is the exterior derivative. Recall that lemma (2.3) said that the
basic forms on $P$, were the forms in the kernel of $\imath_{X}$ and
${\cal L}_{X}$ for all $X\in \la$.
In addition it stated that the basic forms on $P$ can be mapped uniquely
to the forms on $M$ by the bundle
projection
\begin{equation}
  \Omega^{*}(M) = \pi \left( \Omega^{*}(P)\vert_{{\rm basic}} \right).
  \label{omega(M)-omega(P)}
\end{equation}
So as de Rham complexes we identify 
$(\Omega^{*}(M),d) = (\Omega^{*}(P)\vert_{{\rm basic}},d)$.
Since $P\times M$ is a principal $G$ bundle over $P\times_{G}M$ we also
have
\begin{equation}
  \Omega^{*}(P\times_{G} M) = \Omega^{*}(P\times M)\vert_{{\rm basic}}.
\end{equation}

On the principal $G$-bundle $\pi:P\rightarrow M$ we assume the existence
of a connection $A$ with curvature $F$. They are 
one and two-forms on $P$ with values in the Lie algebra, i.e.
forms in the adjoint bundle
\begin{equation}
  A \in ( \Omega^{1}(P) \times_{{\bf ad}} \la) \,\,;\,\, F\in
  (\Omega^{2}(P) \times_{{\bf ad}} \la).
\end{equation}
They satisfy the following standard relations
\begin{eqnarray}
  \imath_{X} A &=& X, \,\, X\in \la;\,\,\,\,\mbox{the connection is vertical}
  \label{A-vertical}\\
  \imath_{X} F &=& 0,  \,\, X\in \la;\,\,\,\,\,\mbox{the curvature is horizontal}
  \label{F-horizontal} \\
F &=& d A + \frac{1}{2}[A,A] \label{def-F-weil} \\
d F &=& [F,A];\,\,\,\, \mbox{the Bianchi identity} \label{bianchi-F-weil}.
\end{eqnarray}
The Weil algebra is actually a model of the relations defining a vertical
connection and a horizontal curvature, in terms of the the generators of \W 
\begin{eqnarray}
  I_{X_{a}}\theta^{b} &=& \delta_{a}^{\,\,b},   \label{I-theta} \\
  I_{X_{a}}\phi^{a}   &=& 0,      \label{I-phi} \\
  L_{X_{a}} \theta^{a} &=& 0, \label{L-theta}\\
  L_{X_{a}} \phi^{a} &=& 0, \label{L-phi}
\end{eqnarray}
where $I_{X_{a}} d_{W} + d_{W}I_{X_{a}} = L_{X_{a}}$. The sub-algebra of the
elements in \W satisfying these relations is denoted $B\!\la$, and the elements
are called the basic elements in \W.
By defining the elements $A^{a} \in \W^{1}$ and $F^{a}\in\W^{2}$ as the images
of $\{ \theta^{a} \}$ in the $\Lambda(\dla)$ and $S(\dla)$ respectively,
the universal connection and curvature over the Weil algebra is defined~\cite{MQ} as
\begin{eqnarray}
  A &=& A^{a}X_{a} \in \W^{1} \otimes \la , \\
  F &=& F^{a}X_{a} \in \W^{2} \otimes \la.
\end{eqnarray}
These are $G$-invariant and thus in the kernel of $L_{X_{a}}$ for $a=1,\dots,
{\rm dim}(G).$
Combining equations~(\ref{d_W-theta},\ref{d_W-phi},\ref{I-theta},\ref{I-phi}) we see that these relations expressed in $A^{a},F^{a}$ reproduce the
identities in equations~(\ref{A-vertical},\dots,\ref{bianchi-F-weil})
\begin{eqnarray}
  I_{X_{a}} A^{b} &=& \delta_{a}^{b}, \\
  I_{X_{a}} F^{a} &=& 0, \\
  d_{W} A^{a} & = & F^{a} - \frac{1}{2}f_{bc}^{\,\,\,a}A^{b}A^{c}, \\
  d_{W} F^{a} &=& - f_{bc}^{\,\,\,a}A^{b}F^{c},
\end{eqnarray}

So the connection $A$ in $P$ determines a map $\dla \mapsto \Omega^{1}(P)$
and the same for $F$ in $P$ with $\dla \mapsto \Omega^{2}(P)$, which
induces a unique homomorphism of graded algebras
\begin{equation}
  w: \W \mapsto \Omega(P).
\end{equation}
This is (almost) the Weil homomorphism from theorem (2.3), which
maps the universal connection and curvature
$A^{a},F^{a}$ into $A,F$. The importance of this construction is
seen when we consider the tensor product $\W \otimes \Omega^{*}(M)$, which
again is a differential graded algebra. The sub-algebra
\begin{equation}
  \Omega_{G}(M) = \left\{ \W \otimes \Omega(M) \right\} \vert_{{\rm basic}} \subset \W \otimes \Omega(M),
\end{equation}
is the algebra of equivariant differential forms. Here the basic elements
are those $\eta\in  \W \otimes \Omega(M)$ which satisfy
\begin{eqnarray}
  \eta &\in & {\rm ker} \bigcap_{a=1}^{{\rm dim} G} \left(I_{X_{a}}\otimes
    {\bf 1} + {\bf 1}\otimes \imath_{X_{a}}\right), \\
  \eta &\in & {\rm ker} \bigcap_{a=1}^{{\rm dim} G} \left(L_{X_{a}}\otimes
    {\bf 1} + {\bf 1}\otimes {\cal L}_{X_{a}} \right).
\end{eqnarray}  
  
The homomorphism $w$ and equation~(\ref{omega(M)-omega(P)})
determine a new homomorphism $W$ (which is the Weil homomorphism from
theorem (2.3)), and also follows $\overline{W}$ on the
basic sub-algebras
\begin{equation}\begin{array}{ccc}
    & \mbox{{\tiny $W$}} & \\[.01cm]
  \W \otimes \Omega(M)& \rightarrow & \Omega(P\times M) \\[.6cm]
  \bigcup & &  \bigcup \\[.1 cm] & \mbox{{\tiny $\overline{W}$}}& \\[.01cm]
 
  \Omega_{G}(M) &\rightarrow & \Omega(P \times_{G} M ).
\end{array}
\end{equation}
The homomorphism $\overline{W}$ between basic sub-algebras is known as the
Chern-Weil homomorphism and it induces a homomorphism in cohomology
\begin{equation}
  H^{*}(\Omega_{G}(M)) \mapsto H^{*}(P\times_{G} M), \label{cohom-hom}
\end{equation}
which is independent of the choice of $A\in P$, from which $\overline{W}$
is determined~\cite{MQ}.
 
By $H^{*}(\Omega_{G}(M))$ we mean $H^{*}(\Omega_{G}(M), d_{T})$ where
\begin{equation}
  d_{T} \equiv d_{W} \otimes {\bf 1} + {\bf 1} \otimes d.
\end{equation}
So what have we obtained? The step $M\mapsto EG \times M$ is mirrored by
$\Omega(M) \mapsto \W \times \Omega(M)$. The equivariant cohomology of $M$
was identified with the cohomology of $BG$ and in the same way by setting
$P=EG$, we now have from equation~(\ref{cohom-hom})
\begin{equation}
  H^{*}( B\!\la \otimes \Omega(M),d_{T}) \rightarrow
  H^{*}\left((EG\times M)/G,d\right) \mapsto_{\mbox{{\tiny integration over
        fibre}}} H^{*}\left( BG, d \right).
\label{alg=form}
\end{equation}
The first step is actually an isomorphism if $G$ is compact\cite{Atiyah-Bott},
and hence we have found an algebraic description of the
$H^{*}(BG,d)$.
We now just need to identify the basic elements in \W, in order to have
characterised the equivariant cohomology of $M$. Notice that a horizontal
element in \W must be in the kernel of $I_{X_{a}}$ but
equations~(\ref{I-theta},\ref{I-phi}) show that only the elements in the
symmetric algebra satisfy this requirement. Invariance under $G$, i.e.
that the elements must be in the kernel of $L_{X_{a}}$ identifies
\begin{equation}
  B\!\la \simeq S(\dla)^{G},
\end{equation}
with the invariant polynomials on the Lie algebra. The Weil derivative
vanishes on the basic sub-complex, due to equation~(\ref{inft-homotopy}).
Hence one identifies $H^{*}(B\!\la,d_{W})$ with $B\!\la$.

There exist other representations of equivariant cohomology.
The so-called Cartan representation of equivariant
cohomology~\cite{Jeffrey,CMR} which is based on the basic sub-complex of \W.
So as a vector space the Cartan model is based on $S(\dla)\otimes\Omega(M)$
with the differential $d_{C}$, which is defined by its action on the generators
\begin{eqnarray}
  d_{C} \phi^{a} &=&  0 \\
  d_{C} \eta &=& \left( {\bf 1}\otimes d - \phi^{a}\otimes \imath_{X_{a}}
\right) \eta = \left( d - \imath_{\phi^{a}} \right) \eta, \label{cartan}
\end{eqnarray}
where $\eta\in\Omega(M)$. In contrast to $d_{W}$, $d_{C}$ does not always
square to zero. One finds
\begin{eqnarray}
 \left( {\bf 1}\otimes d - \phi^{a}\otimes \imath_{X_{a}}
\right) \left( {\bf 1}\otimes d - \phi^{a}\otimes \imath_{X_{a}}
\right)&=& - \phi^{a}\otimes d \imath_{X_{a}} - \phi^{a}\otimes
\imath_{X_{a}} d \nonumber \\ &=&
- \phi^{a}\otimes{\cal L}_{X_{a}},
\end{eqnarray}
since $\imath_{X_{a}}^{2} = 0$ and $d^{2}=0$. By restricting to the
invariant sub-complex
\begin{equation}
  \Omega_{G}(M) = \left( S(\dla)\otimes\Omega(M)\right)^{G},
\end{equation}
$d_{C}^{2}$ becomes
nilpotent. In cohomology we have the isomorphism~\cite{CMR}
\begin{equation}
  H^{*}\left( \left\{ W(\dla)\otimes \Omega(M) \right\}\vert_{{\rm basic}},
    d_{T} \right) \simeq H^{*}\left( (S(\dla)\otimes\Omega(M))^{G},
    d_{C}\right).
\end{equation}

Finally let us mention the BRST representation of equivariant
cohomology~\cite{CMR}, which is built upon the same complex as the
Weil algebra, but with a new differential 
\begin{equation}
  d_{B} = d_{W}\otimes {\bf 1} + {\bf 1}\otimes d + \theta^{a}\otimes
  {\cal L}_{X_{a}} - \phi^{a}\otimes\imath_{X_{a}},
  \label{BRST-model-general}
\end{equation}
where $d_{B}^{2}=0$, so the cohomology is trivial as in the case of \W.
The basic sub-complex is the same as the invariant sub-complex of the Cartan
algebra, and on this $d_{C}=d_{B}$ and $d_{C}^{2}=0$.
There also exists an algebra isomorphism mapping $d_{T}$ into $d_{B}$ by
conjugation with $\exp(\theta^{a}\imath_{X_{a}})$ and the cohomology
groups $H^{*}(W(\underline{{\bf g}}),d_{B}) \simeq
H^{*}(W(\underline{{\bf g}}),d_{T})$ are isomorphic~\cite{Kalkman}

\subsection{Translation into Physics - Part 1}

The grading of the Weil algebra signals that it may be described by
the framework of supersymmetry. One can construct~\cite{CMR} a
superspace (for a general introduction see e.g.~\cite{Manin}) 
$\widehat{ \underline{{\bf g}}}$ built on the tangent bundle over \la
\begin{eqnarray}
  &T\!\la & \nonumber \\ &\downarrow & \pi \\ &\la&, \nonumber
\end{eqnarray}
where the coordinates of the fibres
\begin{equation}
  \xi^{i} \in \pi^{-1}(X), i=1,\dots, {\rm dim}(G); \,\, X\in \la,
\end{equation}
now are taken to be grassmann odd. The superspace
\begin{equation}
  \widehat{\underline{{\bf g}}} = \Pi T\la,
\end{equation}
is identified with this bundle, where $\Pi$ indicates that the fibres are
taken to be odd. One can identify the Weil algebra with
$\widehat{\underline{{\bf g}}}$, and a generic function on
$\widehat{\underline{ {\bf g} }}$ is represented by a superfield
$\Phi(c^{i},\gamma^{i})$, where $c^{i}$ are the odd generators of functions
on the fibre and $\gamma^{i}$ is the generator of symmetric polynomials on
\la. By introducing the dual to these generators, namely $b_{i}$ to $c^{i}$
and $\beta_{i}$ to $\gamma^{i}$, with commutation relations
\begin{eqnarray}
  \bigl[ \beta_{i},\gamma^{j}\bigr] &=& \delta_{i}^{\,\,j} ,\\
  \bigl[ b_{i}, c^{j} \bigr] &=&  \delta_{i}^{\,\,j}.
\end{eqnarray}
We get the indication that the Weil algebra is rather closely related
to the description of the $(b,c,\beta,\gamma)$ ghost systems known from
superstring theory, which we discuss in chapter 4. For
more details on this connection see~\cite{CMR}, where it is shown how
equivariant cohomology on \la is the same as supersymmetrized Lie algebra
cohomology.

In more general terms we can write a superspace $\widehat{M}$ as the
tangent bundle with odd fibres over $M$
\begin{equation}
  \widehat{M} = \Pi TM,
\end{equation}
where the coordinates $(x^{i},\psi^{i})\in \widehat{M}$ are the coordinates
$x^{i}\in M$ and $\psi^{i}$ are the odd fibre coordinates, where
$i=1,2,\dots,n$. This setup gives the possibility of identifying
functions on $\widehat{M}$ with differential forms on $M$, by the
identification
\begin{equation}
  \psi^{i} \leftrightarrow dx^{i},
\end{equation}
since the differential forms on $M$ are anti-commuting objects. Let
$\eta \in \Omega^{*}(M)$ be a form on $M$ and $\hat{\eta}$ a function
on $\widehat{M}$. In general we do not have a generic measure on a
manifold, and are left to integrate differential forms, but by
construction one has a generic measure $d\mu$ on $\widehat{M}$
\begin{equation}
d\mu = dx^{1}\wedge\dots\wedge dx^{n} \wedge d\psi^{1}\wedge \dots\wedge d\psi^{n}.
\end{equation}
This measure is invariant under coordinate transformations on $\widehat{M}$ because 
the coordinates $(x^{i},\psi^{i})$ transform in an opposite way to each other.
So one can interchange integration over forms on $M$ with integration
of functions on $\widehat{M}$
\begin{equation}
  \int_{\widehat{M}}d\mu\, \hat{\eta} = \int_{M}\eta.
\end{equation}

We now show how this formalism is used in topological Yang-Mills theory,
discussed in section~(\ref{section:TYM}). The configuration space is
the space of connections \A and using the identification above, one can 
represent forms $\Omega^{*}(\A)$ as functions on superspace generated by 
\begin{equation}
  A \in \Omega^{0}(\A) \,\, \mbox{ and } \psi = d A \in \Omega^{1}(\A).
\end{equation}
These forms can be pulled back to the underlying space-time manifold $M$,
using a local section and they are described by local fields
\begin{eqnarray}
  A &=& A_{\mu}(x)^{i}T_{i} dx^{\mu} \in \Omega^{1}(M,\la) \\
  \psi &=& \psi_{\mu}^{i}(x)T_{i} dx^{\mu} \in \Omega^{1}(M,\la)
\end{eqnarray}
with ghost number zero and one respectively. In this sense $A$ represents
a point in \A, and $\psi$ an element in $T_{A}\A$. This explains the notation
in equation~(\ref{psi-som-tangent-til-A}).

The relevant group action on \A is of course the group of gauge
transformations \G, and the Weil algebra is thus modelled on the
Lie algebra $\underline{\cal G}$. The dual of this Lie algebra,
on which we should form the exterior and symmetric algebras, can be
viewed as the space $\Omega^{n}(M,\la)$ where $(n=4)$ is the dimension
of $M$. This follows from Poincar\'{e} duality and the fact that gauge
transformations can be viewed as functions from $M$ into $G$.
The Faddeev-Popov ghost $c^{i}$ and the ghost for ghost $\phi^{i}$
are zero forms on $M$ with values in the Lie algebra \la, and they may
be considered as analogues to $\theta^{a},\phi^{a}$ in the Weil algebra
since they respectively have ghost number one and two, and hence odd and
even grassmann parity.

The BRST model from equation~(\ref{BRST-model-general}) is now built on
the complex $W(\underline{{\cal G}})\otimes\Omega^{*}(\A)$ and
Kalkman~\cite{Kalkman} has proved that the BRST
differential $d_{B}$, is exactly the BRST differential given in
equations~(\ref{br_A:TYM},\dots, \ref{br_phi:TYM}). His proof is quite
complicated and independent of the models one chooses to consider.
The reader is referred to the literature for more details on this proof. 
In addition it also follows~\cite{Jeffrey} that the Cartan model is
equivalent to the reduced BRST algebra introduced
by Witten in equations~(\ref{cartan:TYM-A},\dots,\ref{cartan:TYM-PHI}).
The important thing to note is that there exists a framework, namely
the algebraic representations of equivariant cohomology due to Weil and
Cartan, which corresponds to the form of the BRST algebra in the Witten
type TFT's. In this way equivariant cohomology hints at what
kind of geometric structure these theories
describe and we now turn our interest to this.

We have in the principal $G$- bundle $\pi:P\mapsto M$ where $G$ is the
gauge group and $M$ is the space-time manifold. The (Atiyah-Singer)
universal bundle constructed over $P$, is the principal \G bundle
\begin{eqnarray}
  & P \times \!\A &\nonumber \\
  & \downarrow & \\
  & (P\times\! \A)/\G&. \label{universalbundle:TYM-1} \nonumber
\end{eqnarray}
The gauge group $G$ has an action on the base space of this bundle and we
can construct the principal $G$ bundle
\begin{eqnarray}
  & (P \times \!\A)/\G &\nonumber \\
  & \downarrow & \\
  & M \times \!\AMG &. \label{universalbundle:TYM-2} \nonumber
\end{eqnarray}
The $G$ action corresponds to global gauge transformations after the
local gauge transformations have been gauge fixed.
The action of both groups is taken to be free. One often writes
$L= (P\times\!\A)/G$. If there exists a $G$-invariant metric on $L$,
we can then isolate the horizontal vector fields on $L$ as the orthogonal
complement to the vertical vector fields. The vertical vector fields
are isolated by studying the Lie algebra \la of $G$.
This is of course identical to having a connection $\Gamma$ on the
principal bundle $L\mapsto M\times \!\AMG$.

Assume the existence of a metric $\hat{g}$ on $M$ and let $tr$ denote the
Killing metric on $G$. To have a metric on $L$ consider the
tangent vector
\begin{equation}
  (X+\lambda) \in T_{(u,A)}(P\times\! \A),
\end{equation}
where $X$ is tangent along $P$ and $\lambda\in\Omega^{1}(M,\la)$ since
\A is affine. The metric $g$ on $L$ acts at tangent vectors
\begin{equation}
  \left[ g \left( X+\lambda,Y+\mu \right) \right]_{(u,A)} \equiv
    (X,Y)_{(u,A)} + (\lambda,\mu),\label{metric-L}
\end{equation}
where
\begin{equation}
  (X,Y)_{(u,A)} \equiv \left[ \hat{g}\left(\pi_{*}(X),\pi_{*}(Y)\right)
    \right]_{\pi(u)} + tr(\omega_{A}(X)\omega_{A}(Y))
\end{equation}
is the metric on the tangents $X,Y$ along $P$, defined via the projection
$\pi:P\mapsto M$ and $\hat{g}$, and where $\omega_{A}$ is the
connection one form on $P$ defined via the point $A\in\A$. The remaining
metric in equation~(\ref{metric-L}) is the usual inner product between
forms on $M$
\begin{equation}
  (\lambda,\mu) = \int_{M} tr( \lambda \wedge * \mu ),
  \label{innerproduct-forms}
\end{equation}
where $*$ is defined via the metric $\hat{g}$ on $M$. The metric
$g(\cdot,\cdot)$ in equation~(\ref{metric-L}) is $G\times\G$ invariant,
so it induces a $G$- invariant metric on $L$, since $P\times\A$ is
a \G bundle over $L$. This induced metric fixes the connection $\Gamma$ on
$L$ and the curvature of this connection $F_{\Gamma}$ can be viewed as
an element in $\Omega^{2}(M\times \AMG,\la)$. Since $F_{\Gamma}$ is
a two form it takes arguments on two vector fields on $M\times \AMG$.
There are three components of the curvature at $(x,[A])\in M\times \AMG$
\begin{equation}
  F_{\Gamma} = F_{(2,0)} + F_{(1,1)} + F_{(0,2)},
\end{equation}
where the indices $(a,b)$ label the number of vector fields living on
$M$ and $\AMG$ respectively. These components were calculated by
Atiyah and Singer~\cite{AS} and the first component is the ordinary curvature of $A$
on $M$
\begin{equation}
   F_{(2,0)} = F[A]_{(X,Y)}(u); \,\, X,Y\in T_{x}(M), \,\, \pi(u)=x.
\end{equation}
The $(1,1)$ component is
\begin{equation}
   F_{(1,1)} = F_{\Gamma}(X,\tau) = \tau(X); \,\, X\in T_{x}M, \,\,
   \tau \in T_{[A]}\AMG.
\end{equation}
We know that the elements in $T_{A}\AMG$ correspond to horizontal elements in $T_{A}\A$
or equivalently to one-forms $\psi$ on $M$ with values in the Lie algebra which are subject
to the horizontal condition $D^{\dagger}\psi=0$. This is just the topological ghost
$\psi$ satisfying the zero mode requirement in equation~(\ref{psi-zeromode-0}).
The last $(0,2)$ component is the curvature on \AMG, which from
equation~(\ref{curvature-commutator}) reads
\begin{equation}
  F_{(0,2)} = G_{A}\left(*[\tau_{1},*\tau_{2}]\right),\,\,
  \tau_{1},\tau_{2} \in  T_{[A]}\AMG,
\end{equation}
where the tangent vectors of $ T_{[A]}\AMG$ are identified with
one-forms $\tau_{1},\tau_{2}\in\Omega^{1}(M,\la)$ subject to the
horizontal condition
\begin{equation}
  \tau_{1},\tau_{2} \in {\rm Ker}(D^{\dagger}).
\end{equation}
This is the same as $\langle \phi^{i} \rangle$ expressed in terms of the
zero mode contributions of $\psi$ in equation~(\ref{vev-phi:TYM}).

\noindent
Consider the BRST cohomology as
\begin{equation}
  H^{*}(M,Q) = H^{*}([ W(\underline{\G})\otimes \Omega(P)]\vert_{{\rm basic}},d_{T}),
\end{equation}
equation~(\ref{alg=form}) then shows that
\begin{equation}
  H^{*}(M,Q) \simeq H^{*}\left( (\A\times P)/\G \right) \simeq
    H^{*}(\AMG \times M) \mapsto  H^{*}(\AMG),
\end{equation}
where the last step is obtained by integration over fibre $M$. 
This identification follows from the fact that \G acts on $P$ as
a $G$-bundle automorphism, and hence induces the action of \G on \A.
In the universal bundle notation $E\G = \A$ and $B\G = \AMG$.
\A is only contractible when we 
consider it as the space of irreducible connections~\cite{CMR}.

We should identify the
BRST invariant observables with the closed forms on \AMG.
In equation~(\ref{gen_F-def:TYM}) the BRST algebra was identified with the defining
equations for the generalized curvature
\begin{equation}
  \tilde{F} = F[A] + \psi + \phi.
\end{equation}
We can now see, that this is exactly the same form as the curvature
two-form on $M\times \! \AMG$. The conclusion of the Witten approach to
Donaldson theory was that the classical solutions to the equations
of motion and the ghost fields $\psi,\phi$ were the building blocks for
the closed forms on moduli space, corresponding to Donaldson's invariants.
Recall, that the requirement for the zero modes for the topological ghost
in equations~(\ref{psi-zeromode},\ref{psi-zeromode}) can be written as
\begin{eqnarray}
   P_{-}D \psi &=& 0\label{psi-zeromode-2} \\
  D^{\dagger}\psi &=& 0 \label{psi-zeromode-3} 
\end{eqnarray}
where $P_{-}$ is the projection to self-dual forms, i.e.
$P_{-}:\Omega^{2}(\AMG) \mapsto \Omega^{2}({\cal M}_{{\rm ins}})$.
We saw earlier that $\psi\in T_{[A]}\AMG$, and the requirement in
equation~(\ref{psi-zeromode-2}) states that $\psi$ projects to a
tangent vector to
${\cal M}_{{\rm ins}}$, while equation~(\ref{psi-zeromode-3}) states that
$\psi$ is a horizontal tangent vector. We note that the zero mode requirements
for $\psi$ are precisely identifying the curvature component $F_{(1,1)}$
on $M\times \AMG$ with the topological ghost. The form of the expectation
value of $\phi$ in equation~(\ref{vev-phi:TYM}) is exactly of the same form
as the curvature component $F_{(0,2)}$
\begin{equation}
  \langle \phi(x) \rangle = -\int_{M}\sqrt{g}
  G_{A}(x,y)[\psi_{\mu},\psi^{\mu}](y).
\end{equation}
One should note that this expression is non-local in space time due to the
$G_{A}$ term.
Recall also the BRST transformation of $A$ from equation~(\ref{br_A:TYM-form})
\begin{equation}
  \br A = \psi - Dc.
\end{equation}
When $\psi$ is forced to be horizontal due to
equation~(\ref{psi-zeromode-3}), the BRST transformation of
$A$ split into a horizontal and a vertical
part (recall that $V_{A}\in {\rm Image}(D)$). So $\psi$ and $Dc$ are
orthogonal with respect to the metric on \A. In this way 
$\br$ is interpreted as the exterior derivative $d_{\!\A}$ on \A. Projecting
down to \AMG in the bundle $\A\mapsto\AMG$ we would like to determine
the exterior derivative on \AMG and hence on ${\cal M}_{{\rm ins}}$. The
split into horizontal and vertical vectors splits the BRST derivative 
\begin{equation}
  \br = \br^{{\rm horizontal}} + \br ^{{\rm vertical}} =
  \dw + \delta_{{\rm BRST}}, \label{BRST-split}
\end{equation}
where $\delta_{{\rm BRST}}$ is the usual Faddeev-Popov BRST derivative.
Recall from lemma (2.2) that for a horizontal form $\eta$,
the covariant exterior derivative is equal to the exterior derivative. So
$D\eta = dh_{A}(\eta) = d\eta$, where $h_{A}$ is the projection
to horizontal forms.
So while $\delta_{{\rm BRST}}$ squares to zero, 
\dw squares to the curvature on \AMG. The derivative $\delta_{{\rm BRST}}$
is the exterior derivative in the \G direction, and \dw is thus the
covariant exterior derivative on \AMG, and for self-dual forms also on
the instanton moduli space. When $\dw^{2} = {\cal L}_{\phi}$, according
to the identification between Witten's \dw  and the
Cartan derivative $d_{C}$, it is clear, that $\phi$ is the curvature for
$\dw$ on \AMG and ${\cal M}_{{\rm ins}}$.
We now understand how Witten could derive the Donaldson invariants from
the second Chern Class of the curvature $\tilde{F}$. This is simply the
second Chern class $tr(F_{\Gamma}\wedge F_{\Gamma})$ on $M\times \AMG$.
There are five different components, namely $(4,0)$,$(3,1)$,$(2,2)$,$(1,3)$
,$(0,4)$, representing
the form degree on $M$ and \AMG respectively. When $M$ has homological
nontrivial $i$-cycles $\gamma_{i}\subset M$, we can integrate over fibres
of the bundle $M\times \AMG\mapsto\AMG$
\begin{equation}
\int_{\gamma_{i}\subset M} tr(F_{\Gamma}\wedge F_{\Gamma})_{i}^{4-i}, \,\, i=0,1,2,3,4
\end{equation}
resulting in a closed $(4-i)$-form on \AMG. These are exactly the forms in
equations~(\ref{witten-inv-1},\dots,\ref{witten-inv-5}), namely the
four-form
\begin{equation}
  tr(F_{(0,2)}\wedge F_{(0,2)}) \sim tr( \langle\phi\rangle^{2}) 
\end{equation}
evaluated at any point $x\in M$, since we integrate over a zero-cycle
$\gamma_{0}$ or a point.
Integration over a one-cycle $\gamma_{1}$ gives a three-form
\begin{equation}
tr (F_{(0,2)})\wedge F_{1,1}) \sim tr( \langle \phi\rangle \wedge \psi).
\end{equation}
Next, follow the two-forms:
\begin{equation}
  tr(F_{(1,1)}\wedge F_{1,1}) +
      tr(F_{(2,0)}\wedge F_{0,2}) \sim tr(\psi\wedge\psi +  F[A] \wedge
     \langle\phi\rangle ),
\end{equation}
by integration over $\gamma_{2}$.
Integration over $\gamma_{3}$ results in a one-form
\begin{equation}
  tr( F_{(2,0)}\wedge F_{(1,1)}) \sim  tr(F[A] \wedge \psi),
\end{equation}
and finally integration over the (fundamental)
four-cycle $\gamma_{4}=M$ gives a zero-form
\begin{equation}
\int_{M} tr( F_{(2,0)}\wedge F_{(2,0)}) \sim \int_{M} tr (F[A]\wedge F[A]).
\end{equation}
By restricting $(A,\psi)$ to their zero modes, these forms
give the Donaldson invariants on ${\cal M}_{{\rm ins}}$.
Actually the moduli space of instantons is non-compact and the
finiteness of these invariants is a tricky question in mathematics
which requires great care. One of the reasons for the great
interest in the Seiberg-Witten invariants in mathematics, is
due to the fact, that there the moduli space is automatically 
compact~\cite{Szabo}.

We have shown that the framework of equivariant cohomology and of the
universal bundle helps to explain how Witten's Lagrangian describes
Donaldson invariants. We now give the explanation on how the form of
the BRST exact Lagrangian for topological Yang-Mills can be explained
in this geometrical formalism. All these results generalize to all Witten
type TFT's and will be important for the theories of topological gravity.

\subsection{Localization and Projection} \label{loc}
In the following pages we introduce the general tools for constructing
the so-called localization and projection terms in the action of
Witten type TFT's. The main goal of this section is to be able to
write the cohomology classes on the configuration space as an integral over 
closed forms on moduli space.

So far we have studied the de Rham cohomology $H^{*}(M,d)$ as the
closed forms on $M$, modulo the exact forms. The equivariant cohomology
was also the de Rham cohomology, just on $EG\times_{G}M$. In the general
theory of cohomology~\cite{Bott-Tu}, there is another type of
de Rham cohomology on $M$, namely that of the forms which have compact
support on a subset of $M$. The complex for compact cohomology is
\begin{equation}
  \Omega^{*}_{{\rm C}}(M) = \left\{ C^{\infty} \mbox{ functions on } M
    \mbox{ with compact support} \right\} \otimes \Omega^{*}(M),
\end{equation}
and we write $H^{*}_{{\rm C}}(M,d)$ for the cohomology groups.
We now consider the cohomology on a vector bundle $E\mapsto M$,
where we recall that a vector bundle of rank $n$
is a fibre bundle with fibre $\R^{n}$ and $G=GL(n,\R)$. We assume
for simplicity that the bundle is trivial, i.e. $E=M\times \R^{n}$ and that
$n={\rm dim}(M)$.
The Poincar\'{e} lemma for vector bundles states
\begin{eqnarray}
  H^{*}(M\times \R^{n}) &=& H^{*}(M), \\
  H^{*}_{{\rm C}}(M\times \R^{n}) &=&H^{*-n}_{{\rm C}}(M), \label{Poincar-compact}
\end{eqnarray}
where we absorb the $d$ in the notation of the cohomology groups.
The pull back of a form on $M$ to a form on $M\times \R^{n}$, does not have compact
support and hence  $\pi^{*}: \Omega^{*}_{{\rm C}}(M)
\not\rightarrow \Omega^{*}_{{\rm C}}(M\times \R^{n})$.
However there exists an important push-forward map~\cite{Bott-Tu}
\begin{equation}
  \pi_{*}: \Omega^{*}_{{\rm C}}(M\times \R^{n}) \mapsto
           \Omega^{*-n}_{C}(M),
\end{equation}
known as ``integration over fibre''. Let $(t_{1},\dots,t_{n})$ be
coordinates on $\R^{n}$. A form in $ \Omega^{*}_{{\rm C}}(E)$ is
a linear combination of two types of forms
\begin{equation}
\begin{array}{rl} \mbox{(A)} & \pi^{*}\phi\cdot f(x,t_{1},\dots,t_{n}), \\
                  \mbox{ (B)} & \pi^{*}\phi\cdot f(x,t_{1},\dots,t_{n})
                  dt_{1}\wedge\dots\wedge dt_{n},
\end{array}
\end{equation}
where $\phi\in\Omega^{*}(M)$ and $f(x,t)$ is a function with compact
support on $\R^{n}$, for every $x\in M$.
The (B) type forms include a $n$-form
$dt_{1}\dots dt_{n}$ along the fibre. The integration over fibre maps
is given by 
\begin{equation}
  \pi_{*}:\left\{
\begin{array}{rl} \mbox{(A)} & (\pi^{*}\phi)f(x,t_{i_{1}},\dots,t_{i_{r}})
  \mapsto 0, \,\, \mbox{ if } r < n \\
                    \mbox{ (B)} & (\pi^{*}\phi)\cdot f(x,t_{1},\dots,t_{n})
                     dt_{1}\wedge\dots\wedge dt_{n}\\ & \mapsto
                     \phi \int_{\R^{n}} f(x,t_{1},\dots,t_{n}) dt_{1}
                     \wedge\dots \wedge dt_{n}. \end{array}\right.
\end{equation}
So only the forms of top dimension on the fibre are non-vanishing under
this map. This can of course be generalized to non-trivial bundles. The
integration over fibre lowers the form degree by the dimension of the fibre
\begin{equation}
  \pi_{*}: \Omega^{*}_{{\rm C}}(M\times \R^{n}) \mapsto \Omega^{*-n}_{C}(M),
\end{equation}
and this map is an isomorphism in cohomology
\begin{equation}
  \pi_{*}: H_{C}^{*}(M\times \R^{n})\simeq H^{*-n}_{C}(M). \label{thom-0}
\end{equation}
This explains the Poincar\'{e} lemma for compact cohomology in
equation~(\ref{Poincar-compact}).
We introduce some terminology.
If the covering of $M$ by a family $\{U_{\alpha}\}_{\alpha\in I}$
of open subsets $U_{i}\subset M$ is of the form where all nonempty
finite intersections obey 
\begin{equation}
  U_{\alpha_{0}}\cap  U_{\alpha_{1}} \cap \dots \cap  U_{\alpha_{p}}
  \simeq_{{\rm diffeomorphic}} \R^{n},
\end{equation}
the covering is said to be good, and for $M$ being compact one can
always take a finite covering. The manifold $M$ is then said to be of
finite type. This implies that the compact cohomology groups of $M$ are finite
dimensional. We assume from here on, that $M$ is orientable.

The notation of Poincar\'{e} duality for an orientable manifold
states~\cite{Bott-Tu} that the integration over the wedge product
of two forms $\eta,\xi\in \Omega^{q}(M)$ induces a cohomology pairing
\begin{equation}
  \int_{M} \eta \wedge \xi: H^{q}(M) \otimes H^{n-q}_{{\rm C}}(M)
  \mapsto \R,
\end{equation}
following the usual inner product between forms in
equation~(\ref{innerproduct-forms}). This cohomology pairing is
non-degenerated if $M$ has a finite good covering and one writes
\begin{equation}
   H^{q}(M) \simeq \left(H^{n-q}_{{\rm C}}(M)\right)^{*}.
\end{equation}
We now extend this to the vector bundle $E \mapsto M$. The zero section $s_{0}$
of this bundle is a map
\begin{equation}
  s_{0}: M \mapsto E \mbox{ such that } x\mapsto (x,0)\in M\times \R^{n},
\end{equation}
which embeds $M$ diffeomorphically into $E$. The subspace
$M\times\{0\}\subset E$, is said to be a deformation retract of $E$. This
means that there exists a continuous map $H:E\times [0,1]\mapsto E$ for which
\begin{eqnarray}
  H(e,0) &=& e, \,\, \mbox{ and } H(e,1)\in  M\times\{0\}, \mbox{ for any }
  e\in E,\\
  H(e,t) &=& e, \mbox{ for any } t\in [0,1] \mbox{ and any } e \in
  M\times\{0\}.
\end{eqnarray}
$H$ is a homotopic map between ${\bf Id}(E)$ and the retraction $f:E\mapsto
M \times \{0\}$, which leaves all points in $M \times \{0\}$ fixed during the
deformation.
This implies that the cohomology on $E$ and $M$ agree
\begin{equation}
  H^{*}(E) \simeq H^{*}(M),
\end{equation}
since homotopic maps give rise to the same cohomology. The same type of
result holds for compact cohomology if $E$ and $M$ are
orientable~\cite{Bott-Tu}
\begin{equation}
   H^{*}_{{\rm C}}(E) \simeq H^{*-n}_{{\rm C}}(M).
\end{equation}

\begin{figure}[h]
\begin{center}
\mbox{
\epsfysize=6cm
\epsffile{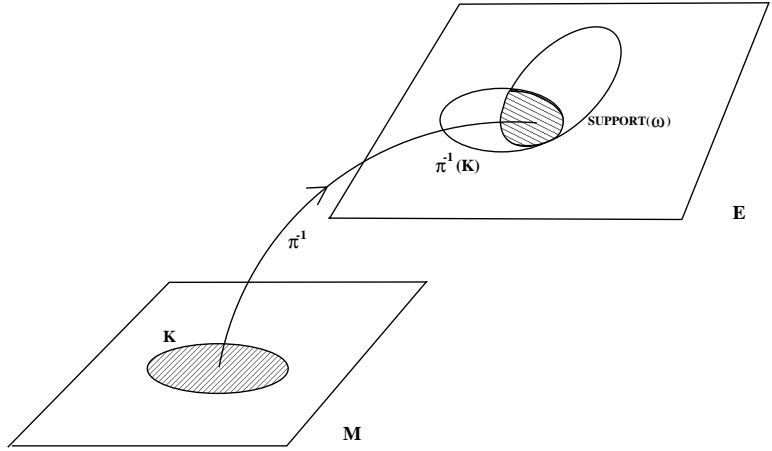}
}
\end{center}
\caption{\label{fig:CVC-1}  The definition of $\Omega^{*}_{{\rm CV}}(E)$}
\end{figure}

A third and important type of cohomology exists on vector bundles, known
as the compact vertical cohomology. A
smooth $n$-form $\omega$ on $E$ is an element in
$\Omega^{n}_{{\rm CV}}(E)$ if and only if for every compact set
$K \subset M$, the set $\pi^{-1}(K)\cap {\rm support}(\omega)\subset E$
is also compact. The situation is illustrated in figure~(\ref{fig:CVC-1}).
If a form $\omega\in \Omega^{*}_{{\rm CV}}(E)$ then the set
\begin{equation}
  {\rm support}(\omega)\vert_{\pi^{-1}(x)} \subset \pi^{-1}(x)\cap
   {\rm support}(\omega), \,\, x\in M
\end{equation}
will be compact since it is a closed subset of a compact set. So
even though a form in $\omega\in \Omega^{*}_{{\rm CV}}(E)$ need not have
compact support in $E$, its restriction to every fibre in $E$
will have compact support. The cohomology of the forms of this complex is
denoted $H^{*}_{{\rm CV}}(E)$.
The integration over fibre map also applies for compact vertical cohomology~\cite{Bott-Tu}
\begin{equation}
  \pi_{*}: \Omega^{*}_{{\rm CV}}(M\times \R^{n}) \mapsto
           \Omega^{*-n}(M).
\end{equation}
Let $(t_{1},\dots,t_{n})$ be
coordinates on $\R^{n}$. A form in $ \Omega^{*}_{{\rm CV}}(E)$ is
a linear combination of two types of forms
\begin{equation}
\begin{array}{rl} \mbox{(A)} & \pi^{*}\phi\cdot f(x,t_{1},\dots,t_{n}), \\
                  \mbox{ (B)} & \pi^{*}\phi\cdot f(x,t_{1},\dots,t_{n})
                  dt_{1}\wedge\dots\wedge dt_{n},
\end{array}
\end{equation}
where $\phi\in\Omega^{*}(M)$ and $f(x,t)$ is a function with compact
support on $\R^{n}$ for every $x\in M$.
The (B) type forms include a $n$-form
$dt_{1}\dots dt_{n}$ along the fibre. The integration over fibre maps
is given by 
\begin{equation}
  \pi_{*}:\left\{
\begin{array}{rl} \mbox{(A)} & (\pi^{*}\phi)f(x,t_{i_{1}},\dots,t_{i_{r}})
  \mapsto 0, \,\, \mbox{ if } r < n \\
                    \mbox{ (B)} & (\pi^{*}\phi)\cdot f(x,t_{1},\dots,t_{n})
                     dt_{1}\wedge\dots\wedge dt_{n}\\ & \mapsto
                     \phi \int_{\R^{n}} f(x,t_{1},\dots,t_{n}) dt_{1}
                     \wedge\dots \wedge dt_{n}. \end{array}\right.
\end{equation}
Again only the forms of top dimension on the fibre are non-vanishing under
this map. This can of course be generalized to non-trivial bundles. The
integration over fibre lowers the form degree by the dimension of the fibre
\begin{equation}
  \pi_{*}: \Omega^{*}_{{\rm CV}}(M\times \R^{n}) \mapsto \Omega^{*-n}(M),
\end{equation}
and this map is an isomorphism in cohomology
\begin{equation}
  \pi_{*}: H_{CV}^{*}(M\times \R^{n})\simeq H^{*-n}(M). \label{thom-1}
\end{equation}
This is exactly the map used in the universal bundle discussion.

Let $M$ be oriented and of dimension $m$. Consider then
the forms $\omega\in\Omega_{CV}^{q}(E)$ and $\tau\in\Omega^{m+n-q}(M)$
where $n$ is the fibre dimension in $E$. Then the following projection
formula holds
\begin{equation}
\int_{E} (\pi^{*}\tau)\wedge \omega = \int_{M} \tau\wedge (\pi_{*}\omega),
\label{project-1}
\end{equation}
which is of great importance for the projection from configuration space to
moduli space in Witten type TFT's.
The isomorphism in equation~(\ref{thom-1}) holds in general
\begin{equation}
 H_{CV}^{*}(E)\simeq H^{*-n}(M),
\end{equation}
for a rank $n$ vector bundle over $M$, with finite good cover. This
isomorphism is known as the Thom isomorphism
\begin{equation}
  {\cal T}:  H^{*}(M)\simeq H^{*+n}_{{\rm CV}}(E ),
\label{thom-2}
\end{equation}
and the image  of $1$ in $H^{0}(M)$ determines a cohomology
class $\Phi\in H^{n}_{{\rm CV}}(E)$, known as the Thom class of $E$.
Using that $\pi_{*}\pi^{*}={\bf 1}$, one can use the projection formula in
equation~(\ref{project-1}) to show that $\pi_{*}\Phi=1$
\begin{equation}
  \int_{E} \pi_{*}(\pi^{*}\tau\wedge \Phi) = \int_{M} \tau\wedge \pi_{*}\Phi
  = \int_{M} \tau, \,\,\tau\in\Omega^{*}(M),
\end{equation}
since $\Phi$ is the image of $1$. So the Thom isomorphism is
inverse to integration over fibres, and it acts on forms
$\tau\in\Omega^{*}(M)$ by
\begin{equation}
  {\cal T}(\tau) = \pi^{*}(\tau)\wedge \Phi.
\end{equation}
Note that $\pi_{*}\Phi=1$ can be written in the nice form
\begin{equation}
  \int_{{\rm fibre}} \Phi = 1.
\end{equation}
The Thom class is uniquely determined on $E$ as the cohomology class in
$H_{{\rm CV}}^{*}(E)$ which restricts to the generator of
$H_{{\rm C}}^{n}(F)$ on each fibre $F$ in $E$.
If $\pi_{1}:E\mapsto M$, $\pi_{2}:F\mapsto M$ are two oriented vector
bundles over $M$ of rank $(n,m)$ respectively 
\begin{equation}
  \begin{array}{ccccc} && E\oplus F && \\
                       \pi_{1} & \swarrow && \searrow & \pi_{2} \\
                       E &&&& F ,
   \end{array}
\end{equation}
the Thom class will satisfy~\cite{Bott-Tu}
\begin{equation}
  \Phi(E\oplus F) = \pi^{*}_{1}\Phi(E)\wedge  \pi^{*}_{2}\Phi(F)
  \in H_{{\rm CV}}^{n+m}(E\otimes F).\label{Thom-addition}
\end{equation}
Consider $S$ to be a closed sub-manifold of dimension $k$ in $M$, where
$M$ is orientable and of dimension $n$. To any inclusion $i:S\hookrightarrow
M$, one can associate a unique cohomology class $[\eta_{S}]\in H^{n-k}(M)$
, denoted the Poincar\'{e} dual to $S$. Using Poincar\'{e} duality
$(H_{{\rm C}}^{k}(M))^{*}\simeq H^{n-k}(M)$, integration over $S$
corresponds to $[\eta_{S}]\in H^{n-k}(M)$
\begin{equation}
  \int_{S} i^{*}\omega = \int_{M} \omega \wedge \eta_{S},
  \,\, \forall \omega \in H_{{\rm C}}^{k}(M).
\end{equation}
How does this fit with the above results which were formulated using
vector bundles? 

To see this we need the notation of the normal bundle $N\mapsto M$
, of $M$ embedded in $\R^{m+k}$, which is of the
form
\begin{equation}
  N = \bigcup_{x\in M} N_{x}M,
\end{equation}
where $N_{x}$ is normal to $T_{x}M$ at every point
$x\in M$. The spaces $N_{x}M$ are isomorphic to $\R^{k}$.
The normal bundle is a rank $k$ vector bundle with typical fibre $\R^{k}$.

The trick is to consider a tubular neighbourhood $T$ of $S$, which
is an open neighbourhood of $S\subset M$, diffeomorphic to a vector bundle
of rank $n-k$ over $S$, such that $S$ is diffeomorphic to the
zero section. This vector bundle is the normal bundle
\begin{equation}
  N_{S} = T_{M}\vert_{S} / T_{S},
\end{equation}  
of $S$ embedded in $M$, where $T_{M}$ is the tangent bundle of $M$ and
$T_{S}$ is the tangent bundle over $S$. Since $M$ and $S$
are orientable, the normal bundle will also be oriented as
\begin{equation}
  N_{S} \oplus T_{S} = T_{M}\vert_{S}.
\end{equation}
Let $j:T \hookrightarrow M$ be the inclusion of $T$ in $M$. We apply the
Thom isomorphism on the normal bundle $T=N_{S}$ and have a sequence
of maps
\begin{equation}
  H^{*}(S) \simeq H_{{\rm CV}}^{*+n-k}(T) \stackrel{j_{*}}{\mapsto}
  H^{*+n-k}(M).
\end{equation}
The first step is made by wedging with $\Phi$, the Thom class of the bundle
$T$. By $j_{*}$ we extend the compact vertical cohomology with $\{0\}$
to the whole of $M$. One considers only forms which vanish near the
boundary of the tubular neighbourhood, since it only plays a technical role
in this construction. The Poincar\'{e} dual of $S$ is the Thom class of
the normal bundle of $S$~\cite{Bott-Tu}
\begin{equation}
  \eta_{S} = j_{*}\left( \Phi\wedge 1 \right) = j_{*}\Phi\in H^{n-k}(M).
\end{equation}

The main conclusion is that the Poincar\'{e} dual of a closed
sub-manifold $S$ in an oriented manifold $M$, and the Thom class of the
normal bundle of $S$ can be represented by the same forms.
Since the normal bundle $N_{S}$ is diffeomorphic to any tubular
neighbourhood $T$, the support of the Poincar\'{e} dual $\eta_{S}$ of a
sub-manifold $S$ can be shrunk to any tubular neighbourhood of $S$
arbitrarily close to $S$.

If we have two closed sub-manifolds $R$ and $S$, embedded in $M$ and these
intersect transversally as defined in section~(\ref{section:BF-observables}),
we have
\begin{equation}
  {\rm codim}(R\cap S) = {\rm codim}(R) + {\rm codim}(S),
\end{equation}
and the normal bundle has the composition
\begin{equation}
  N_{R\cap S} = N_{R}\oplus N_{S}.
\end{equation}
Apply the result to the Thom class on a oriented vector bundle from
equation~(\ref{Thom-addition})
\begin{equation}
  \Phi(  N_{R\cap S} ) = \Phi( N_{R}\oplus N_{S}) = \Phi(N_{R}) \wedge
  \Phi( N_{S}), \label{Thom-intersection-1}
\end{equation}
which implies that
\begin{equation}
\eta_{R\cap S} = \eta_{R}\wedge \eta_{S}.
\end{equation}
Under Poincar\'{e} duality, the transversal intersection of
closed oriented sub-manifolds corresponds to the wedge product of forms.

This result is used again and again in Witten type TFT's and is vital
for the map between the physical and the mathematical approach to the
invariants one studies. It generalizes to the intersection
of $N$-sub-manifolds
\begin{equation}
  S_{1}\cap\dots\cap S_{N}\subset M \mapsto \eta_{S_{1}}\wedge \dots
  \wedge \eta_{S_{N}},
\end{equation}
under Poincar\'{e} duality.

The Thom class is also related to the Euler class of the (even dimensional) oriented vector
bundle $E\mapsto M$. The Euler class $e(E)$ is a characteristic
class of this bundle and an element in $H^{2m}(M)$, where $2m$ is the
rank of $E$ (i.e. the fibre dimension) and $n={\rm dim}(M)$.

There are three ways to define $e(E)$~\cite{Bott-Tu,MQ}
\begin{description}
  \item{(A)} Considering sections $s:M\mapsto E$. In general 
    there will not be any nowhere vanishing sections $s$ and the Euler
    class is
    the homology class of the set of zeros of $s$. This is also known as
    the zero locus ${\cal Z}(s)\subset M$ of $s$. The Poincar\'{e} dual
    is the cohomology class $e(E)\in H^{2m}(M)$.
  \item{(B)} If there is a connection $\Gamma$ on $E$ with curvature
    two-form $\Omega_{\Gamma}$ written as a matrix of two-forms, 
   $e(E)$ can be represented as
    \begin{equation}
      e_{\Gamma}(E) = \frac{1}{(2\pi)^{m}}{\rm Pf}(\Omega_{\Gamma}),
    \end{equation}
    where the Pfaffian of a generic antisymmetric two tensor $A_{ij}$ is
    \begin{equation}
    {\rm Pf}(A) = \frac{(-1)^{m}}{2^{m}m!}\sum_{\sigma(a_{1},\dots,a_{m})}
      \epsilon_{a_{1}\dots a_{2m}}A_{a_{1}a_{2}}\cdots A_{a_{2m-1}a_{2m}},
    \end{equation}
    where one sums over all permutations of the indices. The Pfaffian
    is related to the determinant of $A$ as 
    \begin{equation}
     \Biggl( {\rm Pf}(A) \Biggr)^{2} = {\rm det}(A).
     \end{equation}
    The cohomology class of $e_{\Gamma}(E)$ is independent of $\Gamma$.
    By introducing $(2m)$ real grassmann variables
    $\xi^{a}$, the Euler class
    can be written as a grassmann integral
    \begin{equation}
      e_{\Gamma}(E) = (2\pi)^{-m}\int \prod_{a=1}^{2m}
      d\xi_{a} \exp\left( \xi_{a}\Omega_{\Gamma}^{ab}\xi_{b}\right).
    \end{equation}
     
\item{(C)} The Euler class $e(E)$ is the pull back of the Thom class
   $\Phi(E)$ via the zero section $s_{0}:M\mapsto E$
  \begin{equation}
    e(E) = s_{0}^{*}\Phi(E).
   \end{equation}
\end{description}
If the rank of $E$ coincides with the dimension of $M$ (i.e. $2m=n$)
one can also introduce the Euler number $\chi(E)$.
According to definition (A) this is defined as the Euler class evaluated
on the fundamental homology cycle $[M]$
\begin{equation}
  \chi(E) = e(E)[M].
\end{equation}
The Euler number can be calculated as the number of zeros of a
so-called generic section $s$ (transverse to $s_{0}$) counted with 
signs according to whether it preserves orientation when mapping
from $M$ to $E$~\cite{Hirsch}
\begin{equation}
  \chi(E) = \sum_{\{x_{k}\in M \vert s(x_{k})=0\}}(\pm 1), \label{xi-1}
\end{equation}
or by the more familiar relation as the integral over $M$
\begin{equation}
  \chi(E) = \int_{M} e_{\Gamma}(E). \label{xi-2}
\end{equation}
If the vector bundle is the tangent bundle $E=TM$, equation~(\ref{xi-1})
is just the Poincar\'{e}-Hopf theorem
\begin{equation}
  \chi(TM) = \chi(M) \equiv \sum_{k}(-1)^{k}b_{k}(M),
\end{equation}
where $b_{k}={\rm dim}(H^{k}(M))$ are the Betti numbers.
Equation~(\ref{xi-2}) is just the Gauss-Bonnet Theorem
\begin{equation}
  \chi(M) = \frac{1}{4\pi}\int_{M}d^{2}x\,\sqrt{g} R[g].
\end{equation}

The projection formula in equation~(\ref{project-1}) can be translated
into the notation of the Euler class, and for any form ${\cal O}\in
\Omega^{*}(M)$ the integral of this form over $M$ can be localized as
an integral over the set of zeros ${\cal Z}(s)$ for the generic
section $s$ defined earlier~\cite{CMR}
\begin{equation}
  \int_{M} s^{*}(\Phi(E))\wedge {\cal O} = \int_{{\cal Z}(s)}i^{*}{\cal O}.
\label{Thom-localization-1}
\end{equation}
The $i$ denotes the inclusion $i:{\cal Z}(s) \hookrightarrow M$.
We have that
\begin{equation}
  s^{*}(\Phi(E)) = \eta[{\cal Z}(s) \mapsto M] \label{Thom-localization-2}
\end{equation}
by Poincar\'{e} duality. When the
section is not generic one must make a small correction to this equation.
We do not enter this technical discussion and refer the reader
to~\cite{CMR,Witten-matrix-N}. 

We see that the localization of integrals on $M$ to a subset defined by the
zero locus to the section, is deeply related to the Thom class and
hence to the Euler class. In physics this is exactly the kind of
situation we encounter. Recall the discussion around
equation~(\ref{Witten-modulo-1}). We
want to go from the space of all fields ${\cal S}$, to the subset
defined by some equation $D\Phi^{i}=0$ where $\Phi^{i}$ symbolises
an element in the space of fields. We will view this equation as
a section of the bundle ${\cal S}\mapsto {\cal M}$ modulo
the action of the gauge transformations. This can
be done in a two step manoeuvre known as localization and
projection. 

It happens that one can give an explicit
representation of the Thom Class, which turns out to be (almost) the
whole action of the Witten type TFT. This discussion is sadly quite long
and complicated. An excellent discussion is given in~\cite{CMR} but
other reviews also discuss these techniques (see
e.g.~\cite{Blau-MQ-TFT,Blau-Thompson-Trieste}). Due to the complicated
nature of this subject, several of the discussions in the literature
do not quite agree. I follow the discussion in~\cite{CMR} and will present
only the most important results.

Combining
equations~(\ref{Thom-intersection-1},\ref{Thom-localization-1},\ref{Thom-localization-2}) we can write the wedge product of forms
${\cal O}_{i}\in H^{*}(M)$ as the intersection number of their Poincar\'{e}
duals $\eta_{{\cal O}_{i}}$
\begin{eqnarray}
  \#_{{\rm intersection}}\left( \eta_{{\cal O}_{1}} \cap \dots \cap
    \eta_{{\cal O}_{k}}\right) &=& \int_{{\cal Z}(s)} i^{*}{\cal O}_{1}\wedge
    \dots \wedge i^{*}{\cal O}_{k} \nonumber \\ &=&
    \int_{M} s^{*}\biggl(\Phi_{\Gamma}(E)\biggr) \wedge {\cal O}_{1}\wedge \dots
    \wedge {\cal O}_{k}.
\label{intersection-correlator-1}
\end{eqnarray}
Here $i:{\cal Z}(s)\hookrightarrow M$ is the inclusion, and $s:M\mapsto E$ the section. The main idea is, that one can represent the Thom class above by a fermionic
integral of an exponential, which turns out to be a part of the action.
In the end one will find
\begin{equation}
  \int_{M} s^{*}(\Phi_{\Gamma}(E)) \wedge {\cal O}_{1}\wedge \dots
    \wedge {\cal O}_{k} = \int_{{\rm superspace}} e^{-S_{{\rm q}}}
  \tilde{{\cal O}}_{1}\dots  \tilde{{\cal O}}_{k},
\end{equation}
where $ \tilde{{\cal O}}_{i}$ are superspace functions representing
the forms on $M$. This is the link between the mathematical structure
and the QFT formulation. We give some details on this construction below.

First of all it is useful to not only work with $E$, but instead
construct a tower of bundles over the principal $G$-bundle $\pi_{3}:P\mapsto M$
\begin{equation}
  \begin{array}{ccc} \tilde{P}\times V & \mapsto & \tilde{P} \\
       \downarrow \tilde{\pi}_{2} & & \downarrow \tilde{\pi}_{3} \\
       {\cal V} = \tilde{P}\times_{SO(V)} V & \stackrel{\mapsto}{\pi_{1}} &
       P = \tilde{P}/SO(V) \\  \downarrow \pi_{2} & & \downarrow \pi_{3} \\
      E = {\cal V}/G & \stackrel{\mapsto}{\pi_{4}} & M = P/G =
      \tilde{P}/(SO(V)\times G) . \end{array}
\label{bundlemap}
\end{equation}
We have constructed an (almost always) trivial vector
bundle $\pi_{1}:{\cal V}\mapsto P$
with typical fibre $V$. $V$ is a $2m$ dimensional vector space, and by the
metric on $V$, one can define orthonormal coordinates. The
group $SO(V)=SO(2m)$ acts on $V$ as linear transformations
mapping $V\mapsto V$.
The bundle $\tilde{\pi}_{3}:\tilde{P}\mapsto P$ is a principal $SO(V)$ bundle
and ${\cal V}$ is an associated vector bundle to it. 
We will denote the sections between the relevant bundles with the
same indices as the projections.

Mathai and Quillen~\cite{MQ} have given name to the so-called Mathai-Quillen
form, which is an explicit representative $U$ for the Thom class on a vector
bundle. The Thom class is always taken to be the universal Thom class,
which means that the Thom class is an element in the equivariant cohomology
of the bundle in question. In order for it to be non-trivial, it must be
a basic form, and it must of course be closed.

As a technical detail the forms $\omega\in \Omega_{{\rm CV}}^{*}$ of
compact vertical support on this
bundle can just as well be represented as
forms  $\omega\in \Omega_{{\rm RVD}}^{*}$ of rapid
vertical decrease (${\rm RVD}$) along the fibres in the bundle.

The Mathai-Quillen form on ${\cal V}$ can be considered in either
the Weil model or the Cartan model. There are technical differences
related to this choice. Consider first the Weil model built on
$\la_{s}=\mbox {Lie algebra of } SO(2m)$, such that
\begin{equation}
  U\in W(\la_{s})\otimes \Omega(V)_{{\rm RVD}},
\end{equation}
satisfying three requirements: (1) $U$ must be a basic form, (2) $QU=0$
where $Q = d_{W} + d$ and (3) The integral of $U$ over the fibre $V$
must be one.
Choose a metric $(\cdot,\cdot)_{V}$ on the fibre and a connection $\Gamma$
on the bundle. The metric is taken to
be compatible to this connection. Let $(t^{1},\dots,t^{2m})$ be orthonormal
coordinates on $V$ and let the action of the covariant derivative with
respect to $\Gamma$ act as
\begin{equation}
  (D_{\Gamma}t)^{a} = dt^{a} + \Gamma^{i}(T_{i})^{a}_{\,\,b}t^{b},
\end{equation}
where we take the Lie algebra in the adjoint representation. Let
$(\rho_{1},\dots,\rho_{2m})$ be the orthonormal coordinates on $V^{*}$ defined
by the dual-paring $\langle t^{a},\rho_{b}\rangle_{V} = \delta^{a}_{\,\, b}$.
These dual coordinates are always taken to be anti-commuting and they generate 
$\Pi V^{*}$, where we use the superspace convention to stress that the
fibre has odd coordinates. The coordinates
$\rho_{a}$ are zero-forms on $\Pi V$, but they are assigned ghost number
or degree minus one. In TFT they correspond to the anti-ghosts. 
The Weil algebra is generated by $\Gamma^{i}\in \Lambda(\la_{s}^{*})$ and
its curvature $\Omega_{\Gamma}^{i}$ is represented by $\phi^{i}\in
S(\la_{s}^{*})$ and the Mathai-Quillen form is
the integral
\begin{equation}
  U_{\Gamma} = \pi^{-m}e^{-(t,t)_{V}} \int_{\Pi V^{*}}
  \prod_{a=1}^{2m} d\rho_{a} \exp \left\{
    \frac{1}{4}(\rho, \phi\rho)_{V^{*}} + i \langle Dt,\rho\rangle_{V}
  \right\}. \label{MQ-1}
\end{equation}
The exponential lies in
$W(\la_{s})\otimes \Omega^{*}(V)\otimes \Omega^{*}(\Pi V^{*})$.
It is clear that $U$ is a form of rapid decrease in the fibre direction
due to the gaussian term $-(t,t)_{V}$. 
Integration of the part of $U$ which is a $2m$-form on $V$ gives
\begin{eqnarray}
  \int_{V} U_{\Gamma} &=& \pi^{-m}\int_{V}\int_{\Pi V^{*}} e^{-(t,t)_{V}}
  \frac{(i)^{2m}}{2m!}(dt^{a}\rho_{a})^{2m}  \prod_{a=1}^{2m} d\rho_{a}
  \nonumber \\ &=&
  \pi^{-m}\int_{V}\prod_{a=1}^{2m}dt_{a}e^{-(t,t)_{V}} \int_{\Pi V^{*}}
   \frac{(i)^{2m}}{2m!} \prod_{a=1}^{2m} (d\rho_{a} \rho_{a}) = 1.
\end{eqnarray}
One must also show that $U_{\Gamma}$ is closed, and it is convenient
to extend the derivative $Q$ to be 
\begin{equation}
  Q_{W} = d_{W}\otimes  1\otimes  1 + 1\otimes d \otimes 1 +
    1 \otimes 1 \otimes \delta, \,\mbox{on}
 \,   W(\la_{s})\otimes \Omega^{*}(V)\otimes \Omega^{*}(\Pi V^{*}).
\end{equation}
The derivative $\delta$ is the exterior derivative on $\Pi V^{*}$ and
by introducing orthonormal commuting coordinates 
$(\pi_{1},\dots,\pi_{2m}$) on $V^{*}$, this is given by
\begin{eqnarray}
  \delta \rho_{a} &= &\pi_{a}, \\
  \delta \pi_{a} &= & 0.
\end{eqnarray}
This pair form an anti-ghost multiplet and the ghost number of $\pi_{a}$ is
zero. One can form a function playing the role of the gauge fermion
\begin{equation}
  \Psi = -i \langle \rho, t\rangle_{V} + \frac{1}{4}(\rho,\Gamma\rho)_{V^{*}}
  - \frac{1}{4}(\rho,\pi)_{V^{*}} \in  W(\la_{s})\otimes
  \Omega^{*}(V)\otimes \Omega^{*}(\Pi V^{*}).
\end{equation}
The Mathai-Quillen form can be represented~\cite{CMR} as
\begin{equation}
  U_{\Gamma} =
  \int_{V^{*}\times \Pi V^{*}} \prod_{a=1}^{2m} \frac{d\pi_{a}}{\sqrt{2}}
    \frac{d\rho_{a}}{\sqrt{2}} e^{ Q_{W}  \Psi},
\end{equation}
which is shown by expanding the term $Q_{W}\Psi$, performing the
gaussian $\pi$ integral and using the duality map
between $t^{a}$ and $\rho_{a}$.
Since the integrand is $Q_{W}$ exact, it is also $Q$ closed.
One should also prove, that $U_{\Gamma}$ is horizontal but we refer to
the literature for this~\cite{MQ}.

The Mathai-Quillen form $U_{\Gamma}$ is a basic form on
$W(\la_{s})\otimes \Omega^{*}(V)$ and the
Chern-Weil homomorphism $\overline{w}$ maps it into a basic
form on ${\cal V}=P\times V$. Since $E$ is the base space for the
principal $G$ bundle ${\cal V}\mapsto ({\cal V}/G)=E$, the
forms on $E$ correspond to basic forms on ${\cal V}$, due to
the existence of the connection $\Gamma$ on ${\cal V}$. The form
\begin{equation}
  \overline{w}(U_{\Gamma}) = \pi_{2}^{*}\Phi_{\Gamma}(E),
\label{Thom-MQ}
\end{equation}
descents from the Thom class $\Phi_{\Gamma}(E)\in H_{RVD}^{2m}(E)$ on $E$.

The Mathai-Quillen form can also be discussed in the Cartan model, which
actually was the original approach by Mathai and Quillen~\cite{MQ}
\begin{equation}
  U_{\Gamma}^{{\rm Cartan}} = \left(S(\la_{s}^{*})\otimes
    \Omega(V)\right)^{G}.
\end{equation}
The differential $Q_{{\rm C}}$ on $S(\la_{s}^{*})\otimes
\Omega^{*}(V)\otimes\Omega^{*}(\Pi V)$ acts as usual
\begin{eqnarray}
  Q_{{\rm C}} t &=& (d - \imath_{\phi})t, \\
  Q_{{\rm C}} \phi & = & 0, \\
  Q_{{\rm C}} \rho_{a} &=& \pi_{a}, \label{cartan-rho}\\
  Q_{{\rm   C}} \pi_{a} &=& - {\cal L}_{\phi}\rho_{a} \label{cartan-pi}.
\end{eqnarray}
The Mathai-Quillen form can be written as
\begin{eqnarray}
  U_{\Gamma}^{{\rm Cartan}}\!\! &=&(2\pi)^{-2m}\int_{V^{*}\times\Pi V^{*}}
  \prod_{a=1}^{2m}d\pi_{a}d\rho_{a} e^{\left(Q_{{\rm C}}\{\rho_{a}(-t^{a} -
  \kappa \pi_{a}) \} \right) } \nonumber \\  \!\!&=&
 (4\pi\kappa)^{-2m}\int_{\Pi V^{*}}\! \prod_{a=1}^{2m} d\rho_{a}
 e^{\left( -\frac{1}{4\kappa}(t,t)_{V} + i\langle \rho, dt\rangle +
   \kappa (\rho, \phi \rho)_{V^{*}}\right)}.\label{MQ-2}
\end{eqnarray}
Here $\kappa$ is a real constant.
The main difference between the representations of the Mathai-Quillen
forms in equations~(\ref{MQ-1},\ref{MQ-2}) is that we have $Dt$ in the
Weil representation and $dt$ in the Cartan representation. When we use
the Chern-Weil homomorphism to go from the algebraic to the form representation,
we send $W(\la_{s})\otimes\Omega^{*}(V)\vert_{{\rm basic}}$
into $\Omega(P\times V)\vert_{{\rm basic}}$. But
in the Cartan representation we should take Chern-Weil homomorphism
together with horizontal projection so we get $(S(\la_{s}^{*})\otimes
\Omega^{*}(V))^{G}\mapsto \Omega(P\times V)\vert_{{\rm basic}}$.
Recall that $\phi^{i}$ is the algebraic version of the
curvature $\Omega_{\Gamma}^{i}$, but $Q_{{\rm C}}\phi^{i}=0$
while it is mapped into $F$, for which $dF=-[A,F]$. This is why we
should make a horizontal projection to obtain a basic form. The
gauge fermion is simpler in the Cartan representation and we often use
this representation in TFT's, but the projection is problematic
in gauge theories since the connection on $P\times V$, required to define the
horizontal direction, is non-local in space time. For example 
if $P=\A$, equation~(\ref{A-connection}) shows that the connection is
nonlocal due to the $1/D^{\dagger}D$ part. This projection procedure
can also be built into the ``action'' to make a total Mathai-Quillen form,
which we study below.

The projection step is a general problem. If $\pi:P\mapsto M$ is some
principal fibre bundle and $\tau\in H^{*}(M)$, how can we then write
the integral of $\tau$ over $M$ as an integral over $P$? This can be
solved by having a cohomology class $\Phi(P\mapsto M)$ on $P$ such that
\begin{equation}
  \int_{M} \tau = \int_{P} \pi^{*}\tau\wedge \Phi(P\mapsto M).
  \label{projection-1}
\end{equation}
We want to integrate over $P$ instead of $M$, because in the physical
interesting cases, we have a much better control over the space $P$. An
example is in gauge theories where $P$ could be \A and $M$ would then be
\AMG. The observables for which we seek the expectation values are
gauge invariant, but the path integral is done by gauge fixing
on \A instead of writing the path integral directly on \AMG.
In general we do not have a generic measure on \AMG. An illustration
is the well-known picture from Yang-Mills theory
\begin{equation}
  \langle {\cal O} \rangle = \int_{\AMG}d\mu {\cal O} =
  \frac{1}{{\rm Vol}\G}\int_{\A}dA e^{-S_{{\rm q}}}{\cal O}.
\end{equation}
This is the picture to have in mind when we discuss projection, even though
it is not quite the same as the ordinary Faddeev-Popov procedure we discuss.
But when we write down a local QFT with an action to describe the observables
on moduli space, this is done by writing the integral on the configuration
space. 

When we construct the form $\Phi(P\mapsto M)$, it is important to recall
that the projection $\pi$\footnote{If there exists a connection on $P$.} is an isomorphism
\begin{equation}
  \pi: H_{u} \simeq T_{x}M, \,\,\, \pi(u)=x,
\end{equation}
and we know that the forms on $M$ are identified with the basic forms on $P$
which can be written as the pull back by $\pi$. So all vertical directions
on $P$ must be in $\Phi(P\mapsto M)$.
The projection form must be a top form in the vertical direction of $P$
\begin{equation}
  \Phi(P\mapsto M) \in \Omega^{{\rm top}}(TP)^{{\rm vertical}},
\end{equation}
in order for it to ``soak up'' all vertical direction when integrated over
$P$. It should also be an equivariant closed form,
\begin{equation}
  d_{C} \Phi(P\mapsto M)=0.
\end{equation}
Finally is there a technical requirement~\cite{CMR} regarding its
normalization which we skip for simplicity. The projection form is
constructed using the Lie algebra valued one-form $C^{\dagger}$ defined
in equation~(\ref{C-dagger}). Recall that
\begin{equation}
  {\rm ker}(C_{u}^{\dagger}) \simeq H_{u} = (T_{u}P)^{{\rm horizontal}},
\end{equation}
so a top form on $(TP)^{{\rm vertical}}$ should be a wedge product
\begin{equation}
\bigwedge_{i}^{{\rm dim}(TP)^{{\rm vertical}}} C^{\dagger}.
\end{equation}
Again one can form a fermionic integral representation of this construction,
and the complex we consider is of the form
\begin{equation}
  S(\dla)\otimes\Omega(P)\otimes \left(S(\la)\otimes \Lambda(\la)\right).
\end{equation}
The first two terms are the usual Cartan representation for the form
$\Phi(P\mapsto M)$ and the last two terms are generated by the elements
in the needed anti-ghost multiplet. These are $\lambda_{a}$ with ghost number
minus two, generating $S(\la)$ and $\eta_{a}$ with ghost number minus one,
which generates $\Lambda(\la)$.
These are as usual Lie algebra valued zero-forms.
The derivative on this complex acts as
\begin{equation}
  Q_{C} = d_{C}\otimes 1 + 1\otimes \delta,
\end{equation}
where
\begin{eqnarray}
  \delta \lambda_{a} &=& \eta_{a}, \\
  \delta \eta_{a} &=& - {\cal L}_{\phi}\lambda_{a} \equiv [ \lambda, \phi]_{L}\in \dla.
\end{eqnarray}
We can use the metric $(\cdot ,\cdot)_{\la}$,  to identify $\phi$ and
  $[\lambda,\phi]_{L}$ as elements in the Lie algebra\footnote{$[\cdot,\cdot]_{L}$ is
    the Lie bracket~\cite{YC}} \la.
Only by restricting it to the $G$ invariant sub-complex, will the extended
Cartan derivative be nilpotent.

The projection form can be defined from a gauge fermion
\begin{equation}
\Psi_{{\rm projection}} = i(\lambda, C^{\dagger})_{\la} =
\lambda_{a}(C^{\dagger})^{a}_{i}T^{i}dp^{\alpha} \in \Omega^{1}(P),
\end{equation}
where $T_{i}$ is a basis element in \la and $dp^{\alpha}$ is a basis element of
$\Omega^{1}(P)$. The contraction between $\lambda_{a}$ and $C^{\dagger}$ is
obtained using the metric on \la. The projection form is then defined
by the path integral
\begin{eqnarray}
  \Phi(P\mapsto M) &\equiv & \left(\frac{1}{2\pi i}\right)^{{\rm dim G}}
  \int_{\la\times\Pi \la} \prod_{a=1}^{{\rm dim}G}
d\lambda_{a} d\eta_{a} e^{Q_{C}\Psi_{{\rm projection}}} \nonumber \\
  &=& \left( \frac{1}{2\pi i}\right)^{{\rm dim G}}
  \int_{\la\times\Pi \la} \prod_{a=1}^{{\rm dim}G}
d\lambda_{a} d\eta_{a} \exp \left\{ \eta_{a}(C^{\dagger})^{a}_{i}T^{i}dp^{i}
  + \lambda (dC^{\dagger})^{a}_{i}T^{i}dp^{i} \right. \nonumber \\
  &&  \left. \mbox{\hspace{6cm}} +\lambda_{a}(C^{\dagger})^{a}_{i}
    T^{i}\imath_{\phi}dp^{i} \right\} \nonumber \\
&=& \left(\frac{1}{2\pi i}\right)^{{\rm dim G}}\int_{\la\times\Pi \la} \prod_{a=1}^{{\rm dim}G}
d\lambda_{a} d\eta_{a} \exp \left\{ (\eta,C^{\dagger})_{\la} +
  (\lambda, dC^{\dagger})_{\la} + (\lambda, C^{\dagger}C\phi)_{\la} \right\}
\nonumber \\ &=& \left(\frac{1}{2\pi}\right)^{{\rm dim G}}
\int_{\Pi \la} \prod_{a=1}^{{\rm dim}G}d\eta_{a} \delta\left(C^{\dagger}C\phi
+ dC^{\dagger}\right)\exp\left\{ i(\eta,C^{\dagger})_{\la}\right\}.
\end{eqnarray}
It can be proved that this form is $d_{C}$- closed and obeys its normalization
requirement~\cite{CMR}. Note that $\phi$ is localized on
\begin{equation}
  - \frac{1}{(D^{\dagger}D)}dC^{\dagger},
\end{equation}
resembling the result for the curvature on \A in
equation~(\ref{A-connection-2}).

The intersection numbers in equation~(\ref{intersection-correlator-1})
can now be given as an integral over $P$
\begin{eqnarray}
 \#_{{\rm intersection}}\left( \eta_{{\cal O}_{1}} \cap \dots \cap
    \eta_{{\cal O}_{k}}\right) &=&
    \int_{M} s^{*}(\Phi_{\Gamma}(E)) \wedge {\cal O}_{1}\wedge \dots
    \wedge {\cal O}_{k} \nonumber \\ &=&
    \int_{P}\left\{ \pi_{3}^{*}s^{*}(\Phi_{\Gamma}(E)) \wedge \Phi(P\mapsto M)
    \nonumber \right.  \\ && \left. \mbox{\hspace{1cm}}
    \wedge \pi_{3}^{*}({\cal O}_{1})\wedge \dots \wedge
    \pi_{3}^{*}({\cal O}_{k}) \right\}.
\label{intersection-correlator-2}
\end{eqnarray}
The pull back of the cohomology classes ${\cal O}_{i}$ on $M$ to $P$,
will be represented by local operators on $P$ in the topological
field theory. 

As one could see from the bundle map in equation~(\ref{bundlemap}),
there will be two connections on $E$. First the gauge connection
$\Gamma_{{\rm gauge}}$ tied to the $G$ connection on $P\mapsto M$.
And secondly the $SO(V)$ connection $\Gamma_{SO(V)}$ descending from
the vector bundle ${\cal V}\mapsto P$.
This implies that the total Mathai-Quillen form on $E$ is made with
the direct sum of these connections $\Gamma_{{\rm total}}
=\Gamma_{{\rm gauge}}\oplus
\Gamma_{SO(V)}$ and lies in  $G_{{\rm total}} = G + SO(V)$ equivariant
cohomology. It is studied in a mixed representation, where the Weil
model is used for the $SO(V)$ part and the Cartan model for the
$G_{{\rm gauge}}$ part
\begin{equation}
  U_{{\rm total}}\in W(\la_{so(V)})\otimes\Omega^{*}(V)\otimes
                                    S(\la_{{\rm gauge}} ).
\end{equation}
There are technical reasons~\cite{CMR} for this choice that we will not
discuss. Apply the Chern-Weil homomorphism on the $SO(V)$ part, realizing
that $\Omega^{*}(V)$ plays a double role in the complex above, and we
obtain
\begin{equation}
  \overline{w}U_{{\rm total}} = U_{{\rm gauge}} \in S(\la_{{\rm gauge}})
  \otimes \Omega^{*}({\cal V}).
\end{equation}
Using the result in equation~(\ref{Thom-MQ}), we have that
\begin{equation}
  \overline{w}U_{{\rm gauge}} = \pi^{*}_{2}\Phi_{\Gamma_{{\rm total}}}(E).
\end{equation}

When we wish to use $\Phi_{\Gamma_{{\rm total}}}(E)$ in
equation~(\ref{intersection-correlator-2}) we should apply both the
Chern-Weil homomorphism and a horizontal projection on $U_{{\rm gauge}}$, but
because we wedge the form together with the projection form, which is
fully vertical, this is not needed in practice. 

At last are we at the end of this rather abstract discussion. The
intersection numbers in equation~(\ref{intersection-correlator-1})  
can be represented by a QFT expression
\begin{equation}
  \langle \hat{{\cal O}}_{1}\cdots  \hat{{\cal O}}_{k}\rangle =
  \int_{{\cal Z}(s)} i^{*}{\cal O}_{1}\wedge\dots \wedge  i^{*}{\cal O}_{k},
\end{equation}
where the left hand side is a quantum field theory correlator of
superfield representatives $\hat{{\cal O}}_{i}$ of the forms
${\cal O}_{i}\in H^{*}(M)$. The correlator is defined via
equation~(\ref{intersection-correlator-2})
\begin{equation}
  \langle \hat{{\cal O}} \rangle \equiv
  \frac{1}{{\rm vol}(\G)}\int_{\la\times\hat{{\cal S}}}
  d\phi \hat{\mu} e^{Q\Psi_{{\rm total}}} \hat{{\cal O}}.
\end{equation}
In this notation $\hat{{\cal S}}$ is a superspace of form ${\cal V}^{*}\times
\hat{\la}$ generated by $(A,\psi)$ for the base space $P$ of ${\cal V}$,
together with $(\rho,\pi)$ for the fibres of ${\cal V}^{*}$ and finally
by $(\lambda,\eta)$ for $\hat{\la}$. The measure is symbolically written as
\begin{equation}
  \hat{\mu} = dA d\Psi d\pi d\rho d\lambda d\eta,
\end{equation}
and the derivative is
\begin{equation}
  Q = d_{C}\otimes 1\otimes 1 + 1\otimes Q_{C} \otimes 1 +
  1\otimes 1\otimes \delta,
\end{equation}
using the notation from the above discussion. This derivative acts on
\begin{equation}
  S(\la_{{\rm gauge}^{*}}) \otimes \left(\hat{{\cal F}}(P)
    \otimes \hat{{\cal F}}(P)\right)\otimes\hat{{\cal F}}
  (\la_{{\rm gauge}}),
\end{equation}
where $\hat{{\cal F}}(\cdot)$ symbolises functions on superspace and where
we have assumed that ${\cal V}=P\times_{SO(V)} V$ is trivial.
The total gauge fermion
\begin{equation}
  \Psi_{{\rm total}} = \Psi_{{\rm localization}} + \Psi_{{\rm projection}},
\end{equation}
where
\begin{eqnarray}
  \Psi_{{\rm localization}} &=& i \langle \rho, t \rangle -
  (\rho, \Gamma_{{SO(V)}} \rho )_{V^{*}} + (\rho,\pi)_{V^{*}} ,\\
   \Psi_{{\rm projection}} &=& i(\lambda, C^{\dagger})_{\la_{{\rm gauge}}}.
\end{eqnarray}
\subsection{Translation into Physics - Part 2}
The above discussion presents the mathematical view on Witten type TFT's.
Since these topological field theories are local quantum field theories,
which model some moduli space in order to calculate intersection numbers,
it is not so strange that such a general description exists. But in
practice one does not need to think in terms of Mathai-Quillen forms and
Thom classes when doing topological field theories.
The straight forward approach presented in
section~(\ref{section:TYM}) is just as effective as the mathematical approach.
Actually the physics methods are more powerful that the mathematical ones,
since the tricks of QFT's enable one to actually calculate intersection
numbers, which can be quite hard to compute using the abstract methods.
A warning is also needed at this stage.
Several of the mathematics results are
only formally true, when applied to interesting physics
examples. If the gauge group is non-compact, several of the results
in equivariant cohomology will be of formal nature. Also the application
of the universal bundle only acts as a formal guide, when the gauge group
does not act freely. Both things happen in topological Yang-Mills theory if
one reintroduces reducible connections, and are always true in topological
gravity.

Even though one does not need to use the methods explained in the
last section, and though they might only be of a formal nature, they
do give some helpful hints on how to proceed when writing down actions for
new TFT's.

As an example we translate the general discussion to the special case
of topological Yang-Mills theory. The general set up is the same as in
section~(\ref{section:TYM}), but now we construct in addition
the vector bundle
\begin{equation}
  \left({\cal V}\mapsto P\right) \,\,\sim \,\,  \left({\cal V}_{+} =
  \A \times \Omega^{2,+}(M,\la)\right) \mapsto P,
\end{equation}
where the fibres are the self-dual two-forms on $M$, with values in the
Lie algebra. This bundle has a section $s_{1}= F+*F$. The group of
gauge transformations \G acts on the fibres of ${\cal V}_{+}$ and
we form the vector bundle
\begin{equation} 
  E_{+} = \A \times_{\G}  \Omega^{2,+}(M,\la),
\end{equation}
associated to the principal bundle $\pi_{3}:\A\mapsto \AMG$. The section
$s_{1}$ transforms equivariantly under \G, and it descends to a section
$s_{4}:\AMG\mapsto E_{+}$.

The fibre coordinates $t^{a}$ on ${\cal V}$ will be replaced by the
section $s_{1}$ so the gaussian term gives  
\begin{equation}
  (t,t)^{2} \mapsto (s_{1},s_{1})^{2} = \mid\mid s(A)\mid\mid =
  \int tr( F + * F)^{2} = \int tr(F^{+})^{2},
\end{equation}
in the action. The localization term is
\begin{equation}
  \Psi_{{\rm localization}} = \int_{M} tr\Bigl[\overline{\chi}_{\mu\nu}\biggl(
  F^{\mu\nu}\pm\epsilon^{\mu\nu\gamma\kappa}F_{\gamma\kappa} +
  \frac{1}{2}b^{\mu\nu}\biggr)\Bigr],
\end{equation}
corresponding to the first two terms of equation~(\ref{TYM:gaugefermion}).
This can be seen from equation~(\ref{MQ-2}) where $t\mapsto s_{1} = F^{+}$
and $(\rho_{a},\pi_{a})\mapsto (\overline{\chi}_{\mu\nu},b^{\mu\nu})$.
The projection gauge fermion is
\begin{equation}
  \Psi_{{\rm projection}} = (\lambda, C^{\dagger}) = \int_{M} tr(\lambda
  D^{\dagger}\psi) = \int_{M} tr(\lambda D_{\mu}\psi^{\mu}),
\end{equation}
where $\psi\leftrightarrow dA$ is used. Letting $(\lambda, \eta)\mapsto
(\overline{\phi},\eta)$ this is the third term of
equation~(\ref{TYM:gaugefermion}). The last terms of
equation~(\ref{TYM:gaugefermion}) are related to the gauge fixing of the
path integral, which is always needed if we want to do any computations.
These two terms can be put into a gauge fixing gauge fermion
\begin{equation}
  \Psi_{{\rm gauge fixing}} = \int_{M} tr(\overline{c}\partial_{\mu}A^{\mu}
  + \frac{1}{2}\overline{c}b +c[\overline{\chi}_{\mu\nu},\overline{\chi}^{\mu\nu}] + \overline{c}[\phi,\overline{\phi}]).
\end{equation}
The full action of topological Yang-Mills is thus given by acting with
the BRST derivative on
\begin{equation}
  \Psi = \Psi_{{\rm gauge fixing}} + \Psi_{{\rm projection}}+
  \Psi_{{\rm localization}}.
\end{equation}
Note that the projection term ensures that $\psi$ is in the kernel of
$D^{\dagger}$ such that it is a horizontal tangent vector to \AMG, and that
the curvature on \AMG is localized around
\begin{equation}
  \phi = - \frac{1}{D^{\dagger}D} \left( *[\psi,*\psi]\right),
\end{equation}
by integration over $\lambda$. In this sense the action of a Witten type
TFT is just a fancy way of writing several delta functions, which localize
the theory to the relevant moduli space. Let us just mention that Atiyah and
Jeffrey~\cite{AJ} have used this to show, that the partition function
of topological Yang-Mills theory equals the (regularized) Euler number,
using the relation between the Thom class and the Euler number.

In chapter 4 we discuss the Witten type approach to
topological gravity in two dimensions. We will not use the full machinery
of this section, but just rely on the discussion given in this chapter, now
that we know there exists a general framework for Witten type TFT's.

%% file: chap3.tex
\chapter{$2D$ Topological BF Gravity} 
\section{Introduction}
The most successful approach to the quantization of gravity has been
to lower the space-time dimension, such that gravity simplifies. One
can learn many lessons from lower dimensional theories, but it is not
clear which results are stable under the change of dimension. In both
four and three dimensions, gauge theories of general relativity have
been formulated by Ashtekar and Witten.
In four dimensions
Ashtekar~\cite{Ashtekar-1,Ashtekar-2} introduced a complex
self-dual $SL(2,\C)$ connection, in terms of which the
Wheeler-DeWitt equation simplified considerably.
In three dimensions, Witten showed~\cite{Witten-2+1:1,Witten-2+1:2}
that an $ISO(2,1)$ Chern-Simons theory could describe general relativity.
Both approaches have had a great impact on the study of quantum gravity.
Especially Witten's work proved that since
general relativity can be describes as a $ISO(2,1)$ Chern-Simons theory,
it must be renormalizable in three dimensions, contrary to the general belief.
It appears that the correct way to describe gravity in three dimensions is through a
Schwarz type topological field theory. 

In two dimensions it is more difficult to know what approach is correct,
since Einstein gives no guidelines on what to do. The Einstein tensor
is identically zero in two dimensions and the Einstein-Hilbert action
\begin{equation}
  S_{{\rm EH}} = \frac{1}{4\pi}\int_{M} d^{2}x\, \sqrt{g} R[g] = \chi(M),
\end{equation}
is just a constant, namely the Euler characteristic of the
space-time manifold $M$. Therefore, as Jackiw states~\cite{Jackiw-9511048},
``When it comes to gravity in (1+1) dimensions, it is necessary to invent
a model''. In this chapter we present
the BF-theory of two-dimensional gravity, which is a generalization of
Witten's work in three dimensions on Chern-Simons theory. The model is
also connected to the Jackiw-Teitelboim model, which we introduce first.
In the later chapters a Witten type theory of two-dimensional gravity is 
presented and we discuss how the different theories are  related.

\section{The Jackiw-Teitelboim Model}

Due to the lack of guidelines from general relativity in two-dimensions,
one must seek inspiration from other subjects in order to write down an
action for gravity. One theory from which we seek inspiration is
Liouville theory, which is a completely
integrable field theory expressed in a scalar field $\Phi(x)$. In
Minkowski space, the Lagrangian density of Liouville theory reads
\begin{equation}
  {\cal L} = \frac{1}{2} \partial_{\alpha} \Phi \partial^{\alpha}\Phi -
  \frac{m^{2}}{\beta^{2}} e^{\beta\Phi},\,\,\,\,\,\,\,\,\, \beta,m >0,
\end{equation}
and the equation of motion is
\begin{equation}
  \Box \Phi + \frac{m^{2}}{\beta}e^{\beta\Phi}=0. \label{Liouville-1}
\end{equation}
Recall that $\Box = \partial^{x}\partial_{x} - \partial^{t}\partial_{t}$
for space-time coordinates $(x,t)$ on $M$ where the speed of light is set to
one $(c=1)$.
Let  $\hat{g}_{\alpha\beta}$ be a flat background metric on $M$ and
consider the metric $g_{\alpha\beta}$ in the conformal gauge
\begin{equation}
  g_{\alpha\beta} = e^{\beta\Phi} \hat{g}_{\alpha\beta}.
\end{equation}
The scalar curvature $R[g]$ is then of the form~\cite{Jackiw-1}
\begin{equation}
  R = \beta e^{-\beta\Phi}g^{\alpha\beta}\partial_{\alpha}\partial_{\beta}
  \Phi.
\end{equation}
We see that the curvature is a constant if $\Phi$ is a solution to the
Liouville equation~(\ref{Liouville-1}). Classical Liouville theory is
invariant under conformal transformations, which led Jackiw to propose
the Liouville equation as a replacement for the missing Einstein equations.
If one introduces the cosmological constant $\Lambda$ in the Einstein equations
\begin{equation}
  R_{\alpha\beta} - \frac{1}{2}g_{\alpha\beta}R + \Lambda g_{\alpha\beta}=0
  \label{Einstein}
\end{equation}
the nature of two-dimensions $(R_{\alpha\beta}=\frac{1}{2}g_{\alpha\beta}R)$,
signals that either the metric is vanishing for $\Lambda \neq 0$ or totally
undetermined when $\Lambda=0$.
The Liouville equation
\begin{equation}
  R + \Lambda = 0, \label{Liouville-2}
\end{equation}
could be a nice alternative to equation~(\ref{Einstein}). The form in
equation~(\ref{Liouville-2}) is the same as in equation~(\ref{Liouville-1}) if the metric is
conformally flat (we can always choose isothermal coordinates) and
$\Lambda = m^{2}$. 
We must stress that this is a choice made by hand, not something
proposed by nature. Taking this as a replacement for the Einstein equations
in two-dimensions, Jackiw suggested an action to replace the
Einstein-Hilbert action, from which the Liouville equation~(\ref{Liouville-2})
could be derived.
This action was also proposed independently by Teitelboim~\cite{Teitelboim-1}
and it read
\begin{equation}
  S_{JT} = \int d^{2}x\, \sqrt{g} N(x)(R[g] + \Lambda), \label{JT-action}
\end{equation}
with the Liouville equation as the equation of motion for the
scalar Lagrangian multiplier $N(x)$.
The equation of motion for $g_{\alpha\beta}$ results in the equation 
\begin{equation}
  \left( D_{\alpha}D_{\beta} - g_{\alpha\beta} \Box  \right) N(x) = \Lambda
  \,g_{\alpha\beta} N(x). \label{N-equation}
\end{equation}
The price for obtaining an action from which one can derive the Liouville
equation as an alternative for Einstein's equations in two dimensions, is
this very complicated and non-transparent constraint on  the Lagrange
multiplier $N$. This equation for $N$ is hard to interpret in a geometrical
way, but we note that it does not put any new constraints on the metric.
The action in equation~(\ref{JT-action}) is known as the
Jackiw-Teitelboim action and it is generally covariant.

In the next section we introduce the first order formalism of Riemannian
geometry, which makes it possible to transform the Jackiw-Teitelboim
action into a BF theory.

\section{First Order Framework} \label{first-order}
In this section we review the definitions of the first order formalism of
Riemannian geometry~\cite{Nakahara,Eguchi}.
Let $M$ be an $m$-dimensional differentiable manifold with metric $g$.
The tangent space $T_{x}M$ at the point $x \in M$ is spanned by the
derivatives $ \{ \partial_{\alpha} \}$ and the cotangent
space $T_{x}^{*}M$ by the differentials $\{dx^{\alpha}\}$.

\noindent
The metric $g_{\alpha\beta}$ can be decomposed by introducing $m$-beins
$e^{a}_{\,\, \alpha}(x)$
\begin{equation}
  g_{\alpha\beta} = e^{a}_{\,\,\alpha}e^{b}_{\,\,\beta} \eta_{ab},\label{metric-ee}
\end{equation}
where we use the notation
\begin{equation}
  \eta_{ab} = \left\{  \begin{array}{ll} \delta_{ab} & \mbox{if $(M,g)$ is Euclidean,} \\
       \eta^{\rm m}_{ab} & \mbox{if $(M,g)$ is Lorentzian.} \end{array} \right. 
\end{equation}
The Minkowski metric $\eta^{\rm m}_{ab}$ is the diagonal $m\times m$ matrix with entries
$(-1,1,\dots, 1)$.
We can also write equation~(\ref{metric-ee}) as
\begin{equation}
  \eta^{ab} = g^{\alpha\beta} e^{a}_{\,\,\alpha}e^{b}_{\,\,\beta}.
\end{equation}
The inverse $m$-bein is defined as
\begin{equation}
  E_{a}^{\,\,\alpha} = \eta_{ab}g^{\alpha\beta} e^{b}_{\,\,\beta},
\end{equation}
where the internal indices $(a,b,\dots)$ are lowered and raised by $\eta_{ab}$ and
its inverse, and the space-time indices $(\alpha,\beta,\dots)$ by $g_{\alpha\beta}$
and its inverse. The $m$-bein and its inverse are orthogonal in both Latin and Greek
indices
\begin{equation}
  E_{a}^{\,\,\alpha} e^{b}_{\,\,\alpha} = \delta_{a}^{\, b} \,\, ; \,\,
  E_{a}^{\,\,\alpha} e^{a}_{\,\,\beta} = \delta_{\beta}^{\, \alpha}.
\end{equation}
Hence one can identify the $m$-bein $e^{a}_{\,\,\alpha}$ as an element in $GL(m,\R)$
which transforms the coordinate basis $dx^{\alpha}$ of the cotangent space
$T_{x}^{*}(M)$, into an orthonormal basis of the same space, namely
\begin{equation}
  e^{a} = e^{a}_{\,\,\alpha}dx^{\alpha}.
\end{equation}
Similarly we identify $E_{a}^{\,\,\alpha}$ as a $GL(m,\R)$ matrix transforming the
basis $\partial_{\alpha}$ of tangent space $T_{x}(M)$ into an orthonormal basis:
\begin{equation}
  E_{a} = E_{a}^{\,\,\alpha}\partial_{\alpha},
\end{equation}
where one should note that even though $dx^{\alpha}$ always is an exact
form, $e^{a}$ need
not be exact and likewise the elements in the basis $E_{a}$ need not commute
in contrast to those in the basis $\partial_{\alpha}$.

The introduction of the so-called spin-connection one-form $\omega^{a}_{\,\,b}$ enables
one  to define the curvature and torsion two-forms $(R^{a}_{\,\,b},T^{a})$ with
purely internal indices. These are defined by the structure equations due to  Cartan
\begin{eqnarray}
  R^{a}_{\,\,b}& \equiv & d \omega^{a}_{\,\,b} + \omega^{a}_{\,\,c}\wedge\omega^{c}_{\,\,b}
                \equiv \frac{1}{2} R^{a}_{\,\,bcd} e^{c}\wedge e^{d}, \\
  T^{a} & \equiv & d e^{a} + \omega^{a}_{\,\, b} \wedge e^{b} \equiv \frac{1}{2} T^{a}_{\,\,bc}
  e^{b} \wedge e^{c}.
\end{eqnarray}
The Bianchi identities are isolated by taking the derivative of both of the
structure equations
\begin{equation}
  d R^{a}_{\,\,b} + \omega^{a}_{\,\,c}\wedge R^{c}_{\,\, b} - R^{a}_{\,\,c}\wedge
  \omega^{c}_{\,\,b} = 0. \label{Bianchi-2}
\end{equation}
\begin{equation}
  d T^{a} + \omega^{a}_{\,\,b}\wedge T^{b} = R^{a}_{\,\,b}\wedge e^{b},
\end{equation}
We now define a covariant derivative from the spin-connection with the action on
matrix-valued $p$-forms $V^{a}_{\,\,\,b}$:
\begin{equation}
  DV^{a}_{\,\,\,b} = d V^{a}_{\,\, b} + \omega^{a}_{\,\, c}\wedge
  V^{c}_{\,\,\, b} -
  (-1)^{p}     V^{a}_{\,\,\, c} \wedge \omega^{c}_{\,\, b}.
\end{equation}
This definition enables us to rewrite equation~(\ref{Bianchi-2}) as
\begin{equation}
  D R^{a}_{\,\, b} = 0.
\end{equation}

When expressing the metric through $m$-beins, we have more degrees of
freedom to describe the same geometry, so we have introduced a redundancy.
The metric tensor $g_{\alpha\beta}$ is symmetric and therefore it has
$m(m+1)/2$ independent components, while the $m$-bein has $m^2$ components.
That is, many different $m$-beins describe the same metric and these are related
to each other by local orthogonal transformations, i.e. local gauge
transformations:
\begin{equation}
e^{a} \mapsto \tilde{e}^{a}(x) = \Lambda^{a}_{\,\, b}(x) e^{b}(x);\,\, \forall x \in M.
\end{equation}
Since the space-time metric is rotation invariant the matrices $\Lambda^{a}_{\,\, b}(x)$
must satisfy
\begin{equation}
  \eta_{ab} \Lambda^{a}_{\,\, c}\Lambda^{b}_{\,\, d} = \eta_{cd}.
\end{equation}
This implies that\footnote{Or $SO(1,m-1)$ if the signature of $\eta^{\rm m }_{ab}$ is
  $(+,-,-,\dots,-)$ instead of $(-,+,+,\dots,+)$.}
\begin{equation}
  \Lambda^{a}_{\,\, b} \in
  \left\{  \begin{array}{ll} SO(m) & \mbox{if $(M,g)$ is Euclidean,} \\
  SO(m-1,1) & \mbox{if $(M,g)$ is Lorentzian.} \end{array} \right. 
\end{equation}
Under the local gauge transformations the space-time indices are left invariant
while the internal indices are rotated. This fits with the picture of the $m$-bein
(or tetrad) as a local orthonormal frame (i.e. coordinate system) over each
space-time point, whose basis vectors can be rotated freely.
The dimension of the
special orthogonal groups are
\begin{equation}
{\rm dim}[SO(m-1,1)] = {\rm dim}[SO(m)] = \frac{m(m-1)}{2} = m^{2} - \frac{m(m+1)}{2},
\end{equation}
which exactly was the difference in the number of independent components for the
$m$-bein and the metric.
Since $\Lambda \Lambda^{-1} = {\bf 1}$ it follows that
\begin{equation}
d\Lambda\cdot \Lambda^{-1} + \Lambda \cdot d(\Lambda^{-1}) = 0.
\label{trick-1}
\end{equation}
The action of the gauge transformation on the torsion reads
\begin{equation}
  \tilde{T}^{a} = d\tilde{e}^{a} + \tilde{\omega}^{a}_{\,\,b}\wedge \tilde{e}^{b},
\end{equation}
with $\tilde{T}^{a} = \Lambda^{a}_{\,\,b}T^{b}$ and using
equation~(\ref{trick-1}) one can isolate the transformed spin-connection
\begin{equation}
  \tilde{\omega}^{a}_{\,\,b} = \Lambda^{a}_{\,\,c}\omega^{c}_{\,\,d}
                                (\Lambda^{-1})^{d}_{\,\,b} +
                                \Lambda^{a}_{\,\, c}
                                (d\Lambda^{-1})^{c}_{\,\,b},
\end{equation}
in correspondence with equation~(\ref{nonabelian-gt}).
The the covariant derivative and the
curvature two-form transform covariantly
\begin{eqnarray}
  \tilde{R}^{a}_{\,\,b} &=& d \tilde{\omega}^{a}_{\,\,b} + \tilde{\omega}^{a}_{\,\,c}
  \wedge \tilde{\omega}^{c}_{\,\,b} = \Lambda^{a}_{\,\,c}R^{c}_{\,\,d}
  (\Lambda^{-1})^{d}_{\,\,b}, \\
  (\widetilde{DV})^{a}_{\,\,b} &=& \Lambda^{a}_{\,\,c}(DV)^{c}_{\,\,d}
  (\Lambda^{-1})^{d}_{\,\,b}.
\end{eqnarray}
The correspondence between the curvature and torsion two-forms in Latin and Greek
indices is given using the $m$-bein and its inverse, to obtain first the mixed
versions $(R^{a}_{\,\,b\alpha\beta},T^{a}_{\,\,\alpha\beta})$:
\begin{eqnarray}
  R^{a}_{\,\,b} &=& \frac{1}{2} R^{a}_{\,\,bcd}e^{c}\wedge e^{d} = \frac{1}{2}
  R^{a}_{\,\, b\alpha\beta} dx^{\alpha}\wedge dx^{\beta}, \\
  T^{a} &=& \frac{1}{2} T^{a}_{\,\,bc} e^{b}\wedge e^{c} = \frac{1}{2}
  T^{a}_{\,\,\alpha\beta} dx^{\alpha}\wedge dx^{\beta},
\end{eqnarray}
and finally the pure space-time versions:
\begin{eqnarray}
  R^{\alpha}_{\,\,\beta\gamma\delta} &=& E_{a}^{\,\,\alpha}e^{b}_{\beta}
                                         R^{a}_{\,\,b\gamma\delta}, \\
  T^{\alpha}_{\,\,\beta\gamma} &=& E_{a}^{\,\,\alpha} T^{a}_{\,\,\beta\gamma}.
\end{eqnarray}
The compatibility of the covariant derivative with respect to the metric
expressed by $D_{\alpha}g_{\beta\gamma}=0$, and the no-torsion requirement
\begin{equation}
  T^{\alpha}_{\,\,\beta\gamma} \equiv \Gamma^{\alpha}_{\beta\gamma} -
                                      \Gamma^{\alpha}_{\gamma\beta} = 0,
\end{equation}
fixes the connection (or Christoffel symbol)
$\Gamma^{\alpha}_{\beta\gamma}$ uniquely from the metric
\begin{equation}
  \Gamma^{\alpha}_{\beta\gamma} = \frac{1}{2} g^{\alpha\rho}\left(\partial_{\beta}g_{\rho\gamma} + \partial_{\gamma}g_{\beta\rho} - \partial_{\rho}g_{\beta\gamma} \right).
\end{equation}
These two requirements are translated to the following requirements on the spin-connection
and the $m$-bein
\begin{eqnarray}
  \mbox{Metric compatibility:} && \omega_{ab} = - \omega_{ba}, \\
  \mbox{No-torsion requirement:} && d e^{a} + \omega^{a}_{\,\,b}\wedge e^{b}
  = D e^{a} = 0.  \label{no-torsion-1}
\end{eqnarray}
For a spin-connection satisfying these requirements one can express
$\omega^{a}_{\,\,b\alpha}$ in terms of the $m$-bein and its inverse,
where $\omega^{a}_{\,\, b} = \omega^{a}_{\,\,b\alpha}dx^{\alpha}$.
Since $\omega_{ab}=-\omega_{ba}$ one can instead write
$\omega\epsilon^{a}_{\,\,b} = \omega^{a}_{\,\,b}$.
For use in General Relativity it is fruitful
to explicitly include the determinant
$e\equiv {\rm det}(e^{a}_{\,\,\alpha})$ in the expression for the
inverse $m$-bein 
\begin{equation}
E_{a}^{\,\,\alpha} = e^{-1} \eta_{ab}g^{\alpha\beta}e^{b}_{\,\,\beta}.
\end{equation}
From equation~(\ref{no-torsion-1}) we isolate $\omega^{a}_{\,\,b\alpha}$ as 
\begin{equation}
 \omega^{a}_{\,\,b\alpha} = - (e)^{-1}\epsilon^{\gamma\beta}
 \partial_{\gamma}e^{a}_{\,\,\beta}\cdot E_{b\alpha}.
\end{equation}
This clearly shows that the spin-connection is
uniquely determined from the no-torsion requirement if and only if
the $m$-bein is invertible. This result plays a central role in the formulation
of quantum theories of gravity and the relation to topological field theories.
We shall return to this fact several times in the following chapters.

\section{BF Gravity}
How to formulate a gauge invariant action for two dimensional gravity 
is a problem which has appealed to many authors. To our knowledge the specific gauge
invariant 
action which we study in this section, was first explored by
Fukuyama and Kamimura in two 
papers~\cite{Fukuyama-1, Fukuyama-2}. The gauge group considered was 
$O(2,1)$. Later, independently and with no reference to the work of Fukuyama 
and Kamimura, Isler and Trugenberger~\cite{Isler}, Chamseddine and 
Wyler~\cite{CW-1}, and Blau and Thompson~\cite{Blau-Thompson:ATF}, introduced the same action for
various gauge groups. In the latter approaches the gauge
groups $SO(2,1),\, SO(1,2), SO(3), ISO(1,1), ISO(2), PSL(2,\R)$ are discussed.

\noindent
In this section the common basis of these papers will be discussed and the
role of the different gauge groups are explained. 
Generally a Lie algebra valued connection one-form is written as
\begin{equation}
  A = A_{\alpha} dx^{\alpha} = A_{\alpha}^{i}T_{i}dx^{\alpha},
\end{equation}
where $T_{i}$ is a generator of the Lie algebra. When indices are suppressed
we always think of the situation defined above.

The common idea for all the approaches to two-dimensional gauge gravity is
to form a gauge connection one-form, with
values in the two-dimensional Poincar\'{e} algebra $ISO(1,1)$.
This connection is taken as a linear combination of the zwei-bein and 
spin-connection one-forms
\begin{equation}
A = e^a P_a + \omega J,
\end{equation}
where $a=\{1,2\}$. $P_a$ is the generator of translations 
and $J$ is the generator of Lorentz rotations.

These satisfy the relations:
\begin{equation}
[J, P_a] = \epsilon_{a}^{\, b}P_b \,\,;\,\, [P_a, P_b]=0.
\end{equation}
The indices are lowered and raised by the flat metric $\eta_{ab}$ -- here
taken to be of Euclidean signature such that $\eta_{ab}$=$\delta_{ab}$.
For Lorentzian signature we would take $\eta^{\rm m}_{ab}$.
In the case where there is a non-vanishing cosmological
constant $\Lambda \neq 0 $, (here taken as positive)
this algebra can be extended to the de Sitter
algebra $SO(2,1)$ (or $SO(1,2)$)
\begin{equation}
[J, P_a] = \epsilon_{a}^{\,\, b}P_b\,\,;\,\, [P_a, P_b]= - J
\epsilon_{ab} \Lambda. 
\end{equation}
This is more suited for a gauge theory description since there exists
an invariant trace on this algebra, in contrast to the situation for $ISO(1,1)$.
The Killing metric $g_{ij}$ equips this algebra with an
non-degenerated, invariant bilinear form $\langle \cdot,\cdot \rangle$
\begin{equation}
g_{ij} = \left[ \begin{array}{cc}  -1& 0\\  0 &\Lambda\, \eta_{ab}
\end{array}
\right], \label{killing-1}
\end{equation}
such that
\begin{equation}
  \langle P_{a}, P_{b} \rangle = J \delta_{ab} \,\,;
  \,\,  \langle J , J \rangle = 0.
\end{equation}
We introduce the indices $i,j = \{0,1,2\}$ and define the generators of the de Sitter 
algebra as
\begin{equation}
T_i \equiv \left\{T_0, T_1,T_2\right\} = \left\{J,P_{1},P_{2}\right\},
\end{equation}
the Lie Algebra can be expressed as
\begin{equation}
[T_i, T_j] = f_{ij}^{\,\,k} T_k = \epsilon_{ijk}g^{kl}T_l,
\end{equation}
with
\begin{equation}
g_{ij} = \frac{1}{2} f_{ik}^{\,\, l}f_{jl}^{\,\,k}, \,\, \epsilon_{012}=1.
\end{equation}
Note that the Killing metric clearly is degenerate for $\Lambda=0$,
illustrating that there is no invariant trace on $ISO(1,1)$.

Consider the following gauge invariant action
\begin{equation}
S_{BF}[A,\phi] = \int tr(\phi F) = \int d^{2}x\, \phi^{i}(x)
F_{\alpha\beta}^{j}(x)g_{ij}
\epsilon^{\alpha\beta}
, \label{phi-F}
\end{equation}
which defines a BF theory.
The trace on the Lie algebra is given by the Killing metric. The 
Lagrange multiplier $\phi$ is a zero-form with values in the Lie algebra.
The equations of motion are
\begin{eqnarray}
F^i [A] &=& 0, \label{flat-A} \\  D \phi^{i} &=& 0 \label{covariant-constant-phi}.
\end{eqnarray}
First the well-known statement that the classical solutions are the
flat connections.
Secondly the multiplier is a covariant constant with $F=0$ as consistency 
requirement, i.e. with respect to the covariant 
derivative $D=d +[A,\cdot]$, which is related to the flat connection $A$. 
We expand the action in the $i=(0,a)$-components
\begin{equation}
  S_{BF}[A,\phi] = \int d^{2}x\, tr\left( \phi^{0}F^{0} +  \phi^{a}F^{b}  
  \right) =  \int d^{2}x\, \left(  -
             \phi^{0}F^{0}_{\alpha\beta}\epsilon^{\alpha\beta} +
             \Lambda \delta_{ab}\phi^{a}F^{b}_{\alpha\beta}\epsilon^{\alpha\beta}\right).
           \label{bf-grav-1}
\end{equation}
The curvature components in terms of $(e^a,\omega)$ read
\begin{eqnarray}
  F^{0} &=& dA^{0} +  f_{jk}^{\,\,\, 0} A^{j}\wedge A^{k}
  \nonumber  \\ &=&
            d \omega - \frac{\Lambda}{2} e^{a}\wedge e^{b} \epsilon_{ab},
\end{eqnarray}
and
\begin{eqnarray}
  F^{a} &=& dA^{a} +  f_{jk}^{\,\,\, a} A^{j}\wedge A^{k}
  \nonumber \\  &= &
          d e^{a} + \omega^{a}_{\,\,b} \wedge e^{b} .
\end{eqnarray}
We do not always write the $(0)$ index, since it is only one component.
The action in equation~(\ref{bf-grav-1}) now reads
\begin{equation}
S_{BF}[e,\omega,\phi] = \int d^{2}x \,\left[
- \phi^{0} \left(d\omega - \frac{\Lambda}{2} e^{a}\wedge e^{b} \epsilon_{ab}
  \right) +  \Lambda \delta_{ab} \phi^{a}
\left( d e^{b} + \omega^{b}_{\,\, c} \wedge e^{c} \right)
  \right].
\end{equation}
The equations of motion read
\begin{eqnarray}
  d  \omega &=& \frac{\Lambda}{2} \epsilon_{ab}e^{a}\wedge e^{b},
  \label{d-omega} \\
  d e^{a} &=& - \omega^{a}_{\,\, b}\wedge e^{b}, \label{no-torsion-2} \\
  D\phi^{i} &=& 0.\label{d-phi} 
\end{eqnarray}
We recognise the form of these equations from the general
treatment on the first order formalism in the previous section.
Equation~(\ref{no-torsion-2}) is just the no-torsion requirement from
equation~(\ref{no-torsion-1}). When the zwei-bein is invertible we can
determine the spin-connection uniquely from $e^{a}$ (and its
derivative). Equation~(\ref{d-omega}) therefore states that
the scalar curvature
\begin{equation}
  R[\omega] = \frac{2 d\omega}{{\rm det}(e^{a})}= \Lambda,
\label{liouville-omega}
\end{equation}
with
\begin{equation}
  e = {\rm det}(e^{a}_{\,\,\alpha}) = \epsilon_{ab} e^{a}\wedge e^{b} =
  \epsilon^{\alpha\beta} \epsilon_{ab} e^{a}_{\,\,\alpha}e^{b}_{\,\,\beta}.
\end{equation}
Hence we find the Liouville equation from the Jackiw-Teitelboim model as
a result derived from the equations of motion in BF gravity.
One should also show that equation~(\ref{d-phi}) corresponds to
equation~(\ref{N-equation}). This was done by Fukuyama and
Kamimura~\cite{Fukuyama-1}. The trick is to identify the component
$\phi^{0}=N$ and identifying $\phi^{a}=\phi^{a}[\phi^{0}]$ from the
$i=0$ component of equation~(\ref{d-phi}). Then the $i=a$ component
of this equation translates into equation~(\ref{N-equation}), after one
transforms from first-order formalism back to metric
variables.

An action for a theory of gravity should be diffeomorphism invariant, so
we investigate the symmetries of the action~(\ref{phi-F}). Under an
infinitesimal gauge transformation the $A_{\alpha}$ transforms as
\begin{equation}
  \delta A_{\alpha} = D_{\alpha} \lambda\,\,;\,\,\lambda =
  \lambda^{i}T_{i}.
\end{equation}
The scalar field transforms in the adjoint representation
\begin{equation}
  \delta \phi^{i} = [\lambda, \phi]^{i} = f^{i}_{\,\,jk}\lambda^{j}\phi^{k}.
  \label{phi-gauge-transform}
\end{equation}
Under an infinitesimal diffeomorphism $x^{\alpha} \mapsto x^{\alpha} +
\epsilon^{\alpha}(x)$ such that $\delta x^{\alpha} = \epsilon^{\alpha}(x)$,
the gauge field transforms under the action of the Lie derivative along
$\epsilon^{\alpha}$~\cite{Jackiw-2}
\begin{equation}
  {\cal L}_{\epsilon^{\alpha}}A = \epsilon^{\beta}F_{\beta\alpha} + D_{\alpha}
  (\epsilon^{\beta}A_{\beta}) = D_{\alpha}\lambda,
\end{equation}
which is of the same form as a gauge transformation if
\begin{equation}
  \lambda \equiv \epsilon^{\beta}A_{\beta},
\end{equation}
and the equations of motion $F[A]=0$ are applied.
A similar result holds for the scalar field.
This is a very important result for most gauge formulations of 
gravity\footnote{This is not the case for the Ashtekar formalism
  (see e.g.~\cite{Ashtekar-book}).}, namely
that local gauge transformations equal diffeomorphisms on shell. We
will return to this point several times.

At this stage we discuss the various choices of gauge group, which
have been discussed in the literature. Except for the $O(2,1)$ study
in~\cite{Fukuyama-1, Fukuyama-2}, the remaining discussions are all
directly connected. At the level
of the Lie Algebra $PSL(2,\R)$, $SL(2,\R)$, $SO(2,1)$ and 
 $SO(1,2)$ are considered equal. Only when
considering global issues, will one
encounter a difference. Following the discussion of the
M\"{o}bius transformations in
chapter 1, $SL(2,\R)$ can be viewed as the double covering group of
$PSL(2,\R)$ according to equation~(\ref{psl2z-def}).
$PSL(2,\R)$ can also be identified with the component of
$SO(2,1)$ connected to the identity
\begin{equation}
  PSL(2,\R) \sim SO_{0}(2,1).
\end{equation}
For Euclidean signature on the Killing
metric we recover the groups $SO(3)$ from $SO(2,1)$ and $SO(1,2)$
while we get $ISO(2)$ from $ISO(1,1)$. 
The construction of a gauge invariant action for the latter needs special
care, since we noted that the Killing form became degenerate
when $\Lambda \rightarrow 0$, in equation~(\ref{killing-1}).
There exist methods to overcome this problem~\cite{Birmingham-review}
which we discuss here. If we take the limit $\Lambda \rightarrow 0$ in
equation~(\ref{phi-gauge-transform}) we find
\begin{eqnarray}
  \delta \phi^{a} &=& [\phi, \lambda]^{a} = -[\lambda,\phi]^{a}\nonumber \\
                  &=& -\lambda^{0}\phi{^b}[J,P_{b}]^{a} -
                       \lambda^{b}\phi^{0}[J, P_{b}]^{a} 
\nonumber \\      &=&  \epsilon^{a}_{\,\,b}(\lambda^{0}\phi^{b}-
                                             \lambda^{b}\phi^{0}),
\end{eqnarray}
is unchanged since it is independent of $\Lambda$. But
\begin{equation}
  \delta \phi^{0} = \phi^{a}\lambda^{b}[P_{a},P_{b}]^{0} = \Lambda \epsilon_{ab}
                    \lambda^{a}\phi^{b},
\end{equation}
will vanish for $\Lambda\rightarrow 0$.
This implies that the only gauge invariant $ISO(1,1)$ action is
\begin{equation}
  S = - \int d^{2}x\, \phi^{0}d\omega.
\end{equation}
This is invariant under $\delta \omega = d\lambda^{0}$ and
$\delta\phi^{0}=0$. 
By performing a rescaling of the component $\phi^{a}\mapsto\phi^{a}/\Lambda$
one obtains \label{ISO(2)}
\begin{eqnarray}
  \delta \phi^{0} &=& \epsilon_{ab}\lambda^{a}\phi^{b}, \label{iso-2-1-phi} \\
  \delta \phi^{a} &=& \epsilon^{a}_{\,\,b}\lambda^{0}\phi^{b}
  \label{iso-2-2-pphi} .
\end{eqnarray}
This is not the usual $ISO(1,1)$ transformation for $\phi^{i}$, but
it now fits with the $\Lambda\rightarrow 0$ limit of e.g. $SO(2,1)$.
The action, which is invariant under the new transformation reads
\begin{equation}
  S = \int d^{2}x\, \left[ -\phi^{0}d\omega + \delta_{ab}\phi^{a}
    \left( de^{b} - \omega^{b}_{\,\,c}\wedge e^{c}\right)\right].
\end{equation}
For this action the equations of motion are the $\Lambda\rightarrow 0$
limit of equations~(\ref{d-omega}, \ref{no-torsion-2}, \ref{d-phi}), which read
\begin{eqnarray}
 d  \omega &=& 0,  \label{R=0-first_order} \\
  d e^{a} &=& \omega^{a}_{\,\,b}\wedge e^{b},
  \label{notorsion-first_order}  \\
  D\phi^{i} &=& 0. 
\end{eqnarray}  
We see that the curvature $R[\omega]=0$ forces the genus of the
underlying space-time manifold to be one. Now the following picture
emerges: The choice of gauge group determines the genus of the surface,
on which we formulate a theory of gravity. By rescaling $P_{a}$
by $\Lambda$, when $\Lambda\neq0$, the value of the cosmological constant
can be set to $\pm 1$. Hence the sign of $\Lambda$ determines the signature
of the de Sitter algebra $SO(2,1)$ or $SO(1,2)$ and  $\Lambda=0$ implies
$ISO(1,1)$. Using our knowledge on the uniformization of Riemann
surfaces in chapter 1, together with the fact that the cosmological
constant also determines the curvature of the surfaces through the
Liouville equation, enables us to draw the following picture:
\begin{equation}
\begin{array}{|l|c|l|c|}\hline {\rm Genus}  & {\rm Curvature} & {\rm  Group}
  & \Lambda  \\ \hline
0 & 1 & SO(1,2) & -1 \\
1 & 0 & ISO(1,1) & \,\,\,\,0 \\
\geq 2 & -1 & SO(2,1) &\,\,\,\, 1 \\
\hline
\end{array}
\end{equation}
This is a central feature for this approach to gravity. One must specify
the gauge group accordingly to the genus of the surface one wishes to study.
This is problematic if one wants to make a sum over genus, and in the
next chapter we shall see how this can be
changed by allowing curvature singularities on the surfaces. Also here we note
that this feature is not shared with the Ashtekar gauge formalism of gravity,
where the gauge group is always fixed for each choice of space-time
dimension.

Since the action for this approach to gravity is of the BF type introduced
in chapter 2, all the general results regarding quantization of
the BF theory are valid for this theory of gravity. Clearly this is a
topological theory of gravity and one can translate all results on quantum
BF theory to the gravity variables $(e^{a},\omega)$ by expanding the
connection $A^{i}$ in these components. Therefore we do not need to
go through the quantization procedure again, but refer to chapter 2
for the discussion there. Instead we show how identifying the gauge theory 
with gravity singles out a connected component of the moduli space of flat
connections. 


\section{Quantum BF Gravity}

As always, it is possible to follow either the canonical or the covariant approach to
quantization of BF gravity. Both approaches offers different insights in to the
quantum nature of this theory of gravity and we discuss both in some detail.

\subsection{Canonical Quantization}
\begin{figure}[h]
\begin{center}
\mbox{
\epsfysize5cm
\epsffile{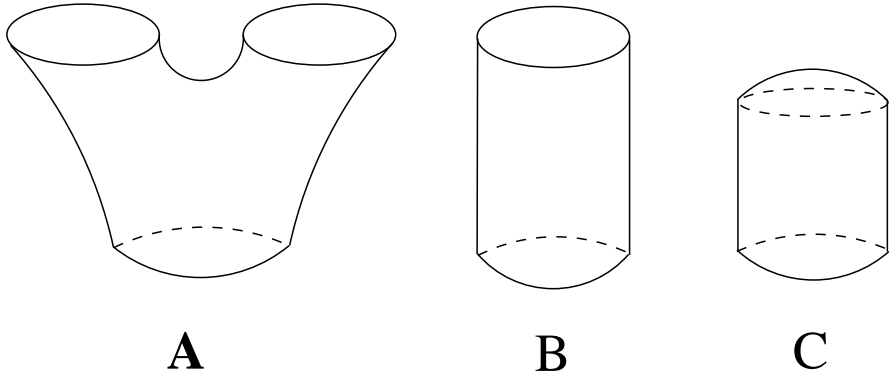}
}
\end{center}
\caption{ \label{Witten-fig-hh} A three-, two- and one-holed sphere}
\end{figure}

For the canonical approach it is natural to consider a foliation of
space-time such that the topology is of the form $\R\times S^{1}$. A
closed curve $C$ (with $S^{1}$ topology) represent an initial spatial
slice of space-time and the action reads
\begin{equation}
  S = \int dt\,\int_{C} dx\, \phi^{i}\left(2  F_{01}^{i}\right), \label{2dBF}
\end{equation}
where $F_{01}=F_{tx}$.
The Poisson brackets are given by
\begin{equation}
  \{ A_{1}^{i}(x,t), \phi^{j}(x',t) \} =
  \delta^{ij}\delta\left(x-x'\right),
\end{equation}
while the non-dynamical $A_{0}^{i}$ component can be viewed as a Lagrange multiplier
for the (Gauss-law) constraint
\begin{equation}
   D_{1}\phi^{i}=0, \label{constrain-1}
\end{equation}
which generates the gauge transformations. 
Considering $A_{1}^{i}$ as coordinate and $\phi^{i}$ as momentum, the Poisson
bracket is changed into a commutator of operators
\begin{equation}
  [ \hat{A}_{1}^{i}(x,t),\hat{\phi}^{j}(x',t)] = -i
  \delta^{ij}\delta\left(x-x'\right).
\end{equation}
It is possible to study the wave functions in the position $(A_{1})$ representation or
in the momentum $(\phi^{i})$ representation. In the position frame, the wave functions
will be of the form $\Psi(A_{1})$ and the conjugated operators $(A_{1}^{i},\phi^{i})$
act on the wave-functions in the usual way as
\begin{eqnarray}
  \hat{A}_{1}^{i}\cdot \Psi(A_{1}) &=& A_{1}^{i} 
  \Psi(A_{1}), \label{A-Psi} \\
  \hat{\phi}^{i}\cdot\Psi(A_{1}) &=& \frac{\delta}{\delta A_{1\,i}} 
  \Psi(A_{1}).  \label{phi-Psi}
\end{eqnarray} 
The physical states are isolated as those consisting the kernel of the
Gauss law constraint in equation~(\ref{constrain-1}). We can solve
this equation by choosing to represent $A_{1}^{i}$ as~\cite{Isler,Rajeev}
\begin{equation}
A_{1}(x) = S(x)\partial_{x} S^{-1}(x), \label{A-rep-1}
\end{equation}
where $S(x)$ is the Lie algebra valued function given by the holonomy around $C$
\begin{equation}
  S(x) \equiv {\cal P}\exp \left( - \int_{0}^{x} dx' A_{1}^{i}T_{i}(x')\right) .
\end{equation}
Differentiating with respect to $x$ pulls down the term
$\partial_{x}\int dx' A_{1}=A_{1}$ and the
exponentials cancel, which shows that the representation in
equation~(\ref{A-rep-1}) is consistent with the given choice of $S(x)$. As usual
${\cal P}$ denotes path ordering. To find a solution to equation~(\ref{constrain-1})
we should choose a gauge invariant wave function. The Wilson loop
$W(A_{1}) = tr S(L)$, where $L$ is the length of the closed loop
$C = \{x_{t}\in[0,L]\vert x_{0}=x_{L}\}$, is the obvious candidate since
\begin{equation}
  D_{1}\hat{\phi}^{i} \cdot W(A_{1}) = \left( \partial_{1}
    \frac{\delta}{\delta A_{1\,i}}  +
    f_{jk}^{\,\,\,i}A_{1}^{j}\frac{\delta}{\delta A_{1\,k}} \right)
  W(A_{1}) =  0.
\end{equation}
Hence the quantum states are taken as functions of the Wilson loops $\Psi(W(A_{1}))$.
The Wilson loops are gauge invariant such that $\Psi(A_{1}) = \Psi(\tilde{A}_{1})$,
where $\tilde{A}_{1}$ is the result of a gauge transformation. This implies that
the physical states are class functions of $W$, invariant under conjugation and
that they must be expressible as an expansion in characters of the gauge group
$G$~\cite{algebra,Witten:2DGTR}
\begin{equation}
\Psi(A_{1}) = \sum_{\alpha} c_{\alpha} \chi_{\alpha}(S(L)),\,\,\, c_{\alpha} \in \C,
\label{psi-sum-rep}
\end{equation}
where the sum runs over all isomorphism classes of irreducible representations
$\alpha$ of $G$. The function $\chi_{\alpha}(S(L))=tr_{\alpha} S(L)$, is the
Wilson loop taken in the $\alpha$ representation.
The Hilbert space of physical states ${\cal H}$ is thus of the form
\begin{equation}
 {\cal H} \simeq L^{2}\left( G/{\bf Ad}G \right).
\end{equation}
In the case of Minkowski space-time, the gauge groups of BF gravity are noncompact
like e.g. $SO(2,1)$ and one should include infinite dimensional representations
of the group in the sum in equation~(\ref{psi-sum-rep}). This approach to canonical
quantization has been used by Rajeev to quantize two-dimensional Yang-Mills theory
in the same space-time topology as considered here, and by Witten in his study
of two-dimensional gauge theories in~\cite{Witten:2DGTR}. It is interesting
to note that from this basic description of the wave functions of $2D$ quantum BF
theory, Witten derived the wave functions representing the geometric situations
in figure~(\ref{Witten-fig-hh})
which are representing A: a three-holed sphere, B: a two-holed sphere and C:
a one-holed sphere. The two-holed sphere, with two boundaries
$C_{1},C_{2}$ and two holonomies $S_{1}(L_{1}),S_{2}(L_{2})$ can for example be
represented by the amplitude
\begin{equation}
  \Psi_{\{2\}} = \sum_{\alpha_{1},\alpha_{2}} c_{\alpha_{1},\alpha_{2}}
  \chi_{\alpha_{1}}\left(S_{1}(L_{1})\right)\chi_{\alpha_{2}}\left(S_{2}(L_{2})\right)
. \label{two-holled-sphere-1}
\end{equation}
The complex number $c_{\alpha_{1},\alpha_{2}}$ can be related to the two-point function
\begin{equation}
 c_{\alpha_{1},\alpha_{2}} = 
 \langle \chi_{\alpha_{2}}\vert e^{-Ht} \vert
 \chi_{\alpha_{1}} \rangle = \delta_{\alpha_{1},\alpha_{2}}, \label{2-loop-c}
\end{equation}
since the Hamiltonian is vanishing for the BF action and the resulting
amplitude for the two-holed sphere is of the form
\begin{equation}
 \Psi_{\{2\}} = \sum_{\alpha} \chi_{\alpha}\left(S_{1}(L_{1})\right)\chi_{\alpha}\left(S_{2}(L_{2})\right)
. \label{two-holled-sphere-3}
\end{equation}
The Hamiltonian is vanishing (weakly) since it is proportional to the
Gauss law constraint
\begin{equation}
  H = - \int_{0}^{L} dx\, tr\left( A_{0}D_{1}\phi\right)\approx 0,
\end{equation}
which follows from the action in equation~(\ref{2dBF}).

A general genus $g$ Riemann surface can be constructed by gluing $2g-2$ three-holed
spheres together at the $3g-3$ boundaries, where the representations $\alpha$ at the
``gluing points'' are identified. Based on his studies, Witten~\cite{Witten:2DGTR}
derived the partition function for two-dimensional BF theory over a genus
$g$ surface $\Sigma_{g}$ to be
\begin{equation}
  Z(\Sigma_{g}) = \left( \frac{{\rm Vol}(G)}{(2\pi)^{{\rm dim}(G)}}\right)^{2g-2}
  \sum_{\alpha} \frac{1}{({\rm dim}\, \alpha)^{2g-2}}.
\end{equation}
The result above
depends directly on the gauge group and since we have to change the gauge group
of BF gravity for the different values $g=0,1,\geq 2$ of the genus, a genus
expansion of $Z(\Sigma_{g})$ like the one in matrix models (see chapter 5),
is not so attractive from this point of view.
The geometries in figure~(\ref{Witten-fig-hh}) are anyhow identical to those 
studied in string field theory (see
e.g.~\cite{Ishibashi-Kawai}) and in the
discussion of Hartle-Hawking wave-functions in matrix models (see e.g.~\cite{KKMW}).
In both approaches one considers a specified time direction, just like in the
canonical quantization of BF theory. In 2+1 dimensional Chern-Simons gravity, a
detailed analysis of topology changing amplitudes has been made using a representation of the wave
functions similar to those in the above discussion~\cite{Witten-2+1:2,Carlip}, but
to the author's knowledge no similar investigation has been made for 1+1 dimensional
BF gravity.
Note that if the length of the loops $C$ are taken to zero, they may be viewed as fixed
marked points (punctures) on the surface. The role of punctures in topological
gravity is important for the discussions in the next chapters.

As mentioned is it also possible to study the canonical quantization in the momentum
formalism, where the wave functions are functions $\Psi(\phi^{i})$ of the Lagrange
multiplier. This approach has been studied in~\cite{Amati}, and it offers no drastic new
insight compared to the position representation.

\noindent
It should be stressed that the consideration made by Witten in~\cite{Witten:2DGTR},
was not related to BF gravity, but instead it probed topological (Witten-type)
Yang-Mills theory from BF theory, through the relation between the two theories discussed
in section~\ref{section:YM-BF}. We conclude that the canonical
approach to two-dimensional quantum BF theory is completely determined by the
structure of the gauge group and its representation.

\subsection{Covariant Quantization}
In the covariant approach, we apply the results from chapter 2  directly, and
simply translate the components of $A_{\alpha}^{i}$ into the gravity variables
$(\omega_{\alpha},e_{\alpha}^{a})$. In chapter 2 the use of the Nicolai map 
reduced the partition function to an integral
over the moduli space of flat connections ${\cal M}_{F}$. The Ray-Singer torsion is
trivial in even dimensions, and the relevant ghost determinants therefore
cancel in this situation. The reduced phase space ${\cal N}$ in
equation~(\ref{reduced-phase-space-1}) is locally the product of the moduli
space of flat connections and the space ${\cal B}$ of gauge equivalence
classes of covariantly
constant $\phi^{i}$'s as seen from the equations of motion in
equations~(\ref{flat-A},\ref{covariant-constant-phi}). 
Recall the representation of the moduli space of flat
connections in equation~(\ref{moduli-flat-A-1}) as:
\begin{equation}
{\cal M}_{F} =  {\rm Hom}\biggl( \pi_{1}(M), G\biggr) / G,\label{moduli-flat-A-2}
\end{equation}
where the quotient of $G$ was taken as the adjoint action of $G$. Due to the
difference in gauge group depending on the value of the genus, we discuss the situation
for each of the values $g=0,1,\ge 2$.

\noindent
The fundamental group $\pi_{1}(\Sigma_{0})$ is trivial and the dimension of 
${\cal M}_{F}$ is zero. There are no obstructions to gauging the connection
$A_{\alpha}^{i}$ to zero and the reduced phase space is of the form (locally)
\begin{equation}
{\cal N} = {\rm pt. }  \times {\cal B},
\end{equation}
which is non-compact.
In genus one, $\pi_{1}(\Sigma_{g})$ is abelian and the dimension of ${\cal M}_{F}$
equals the dimension of $G$ according to equation~(\ref{dim-M_F}). It can be shown
that all connections are reducible in genus one~\cite{Birmingham-review}.
The reduced phase space is non-compact and can be given an explicit
representation~\cite{CW-2}.

\noindent
The most interesting situation is when $g\ge 2$ and the gauge group is taken to
be $SO(2,1)$. Since we identify $PSL(2,R)$ and $SO(2,1)$ at the  Lie algebra level,
we are able to use the description of the moduli space of flat $PSL(2,R)$ connections
over a Riemann surface $\Sigma_{g}$ developed by
Goldman~\cite{Goldman-Adv.math,Goldman-Springer,Goldman-Invent} and
Hitchin~\cite{Hitchin} in the mid-eighties. Goldman's approach was to study these
spaces as spaces of representations of the fundamental group over surfaces,
while Hitchin studied self-duality equations on Riemann surfaces and related their
solutions to the moduli space of flat connections. The topology of
${\rm Hom}(\pi_{1}(\Sigma_{g}), G)$
and ${\rm Hom}(\pi_{1}(\Sigma_{g}), G)/G$ depends on the choice of gauge group. For $G=SU(2)$ or
$G=SL(2,\C)$, ${\rm Hom}(\pi_{1}(\Sigma_{g}), G)$ is connected, while it has two connected
components for $G=SO(3)$ or $G=PSL(2,\C)$. For the group $PSL(2,\R)$ there are several connected
components which we consider in the following discussion.

\noindent
The group $G$ acts by conjugation and as we
discussed in chapter 2, the $G$-orbits parametrize equivalence classes of flat principal
$G$-bundles. The characteristic classes of the $G$-bundles determine invariants of representations
$\pi \mapsto G$, which for connected Lie groups $G$ are elements in
\begin{equation}
  H^{2}\biggl(\Sigma_{g},\pi_{1}(G)\biggr)\simeq \pi_{1}(G).
\end{equation}
The so-called obstruction map 
\begin{equation}
  \sigma_{2}: {\rm Hom}(\pi_{1}(\Sigma_{g}), G) \mapsto
  H^{2}\left(\Sigma_{g},\pi_{1}(G)\right)\simeq \pi_{1}(G), \label{obstruction-1}
\end{equation}
defines an invariant and if $G$ is a connected Lie group and $\pi_{1}(G)$ is finite, the map
$\pi_{0}(\sigma_{2}):\pi_{0}({\rm Hom}(\pi_{1}(\Sigma_{g}), G)) \mapsto \pi_{1}(G)$ is a bijection.
For $G=PSL(2,\R)$ one can form a flat $\R{\bf P}^{1}$ bundle associated to $\sigma$, with
an Euler class $e$. The connected components of ${\rm Hom}(\pi_{1}(\Sigma_{g}), G)$ are the
pre-images
$e^{-1}(k)$, where $k$ is an integer satisfying $\vert k \vert \leq 2g-2$. In this situation 
equation~(\ref{obstruction-1}) reads~\cite{Goldman-Invent}
\begin{equation}
e: {\rm Hom}\left( \pi_{1}(\Sigma_{g}),PSL(2,\R)\right) \mapsto  H^{2}\left(\Sigma_{g},\pi_{1}(G)\right)
\simeq \Z, \label{obstruction-2}
\end{equation}
and it has been proven that
\begin{equation}
  \vert e(h) \vert \leq \vert \chi(\Sigma_{g}) \vert, \,\,\,
  h\in {\rm Hom}\biggl( \pi_{1}(\Sigma_{g}),PSL(2,\R)\biggr),
\end{equation}
where $ \chi(\Sigma_{g})$ is the Euler-Poincar\'{e} characteristic of $\Sigma_{g}$.
The number $e(h)$ is known as the relative Euler number and we thus find $4g-3$
connected components ${\cal M}^{k}_{F}$ of the moduli space of flat $PSL(2,\R)$ connections.
The difference between the components with the same relative Euler number, but with
different signs, is a change of orientation. From this one can say that there
are (up to orientation) $2g-2$ different components of ${\cal M}_{F}$ for $G=PSL(2,\R)$.
The components ${\cal M}_{F}^{k}$ with $k\ne 0$ are smooth manifolds of dimension $6g-6$,
diffeomorphic to a complex vector bundle of rank $(g-1+k)$ over the symmetric product
$S^{2g-2-k}\Sigma_{g}$~\cite{Karoubi}, while the component $k=0$ corresponds to reducible connections~\cite{Hitchin}.

Goldman~\cite{Goldman-Adv.math} has shown that the moduli space of flat connections over a Riemann surface is a symplectic manifold, which corresponds perfectly with the fact that ${\cal M}_{F}$ is the phase space
for 2+1 dimensional Chern-Simons gravity.

We now return to 1+1 dimensional BF gravity and let us assume in the following that the genus of
the underlying Riemann surface is greater than or equal to two.
Let $\phi\in {\rm Hom}(\pi_{1}(\Sigma_{g}),PSL(2,\R))$ embed
$\pi_{1}(\Sigma_{g})$ as a subgroup $\Gamma$ in $PSL(2,\R)$. If $\Gamma$ is
a discrete subgroup, the quotient of the upper complex half-plane \Hu with
$\Gamma$:
\begin{equation}
\Hu/\Gamma \simeq \Sigma_{g}^{\phi},
\end{equation}
will be isomorphic to a Riemann surface $\Sigma_{g}^{\phi}$ with a
complex structure determined by $\phi$. This follows from the uniformization theorem
discussed in chapter 1. Recall equation~(\ref{moduli-teich-mpg}) to see that the collection
of all homomorphisms $\phi$ which embed $\pi_{1}(\Sigma_{g})$ into $PSL(2,\R)$ as
a discrete subgroup, will be isomorphic to Teichm\"{u}ller space \tich.

Two interesting results hold. First the restriction on the homomorphisms $\phi$
to embed the fundamental group as an discrete subgroup in $PSL(2,\R)$ singles out
exactly the homomorphisms in the component ${\cal M}_{F}^{k=2g-2}$ of the moduli
space of flat connections, while the remaining components $|k|< 2g-2$ correspond
to embeddings of $\pi_{1}(\Sigma_{g})$ as a continuous subgroup. For these components
the corresponding Riemann surfaces, formed accordingly by $\Hu/\Gamma$, will not
be smooth, but singular Riemann
surfaces~\cite{Witten-2+1:1}. These surfaces will contain conical
singularities~\cite{Hitchin-pc}. Witten offers in~\cite{Witten-2+1:1} an interesting visual
description of these singular
surfaces: Let $\phi\in {\cal M}_{F}^{k}$ for e.g. $k= 2g-2-2r$, the Riemann
surface $\Hu/\Gamma$ will then correspond to a singular Riemann surface with $r$ collapsed handles.
This description resembles that of pinched surfaces given in chapter one and we return
to a discussion of the similarities and differences in chapter 6.

Second, the presence of
conical singularities makes it possible to use a physical argument to exclude the
components ${\cal M}_{F}^{k}$ for $|k|<2g-2$. When identifying a gauge theory with
a gravitational theory, it is natural to require the defining principles of the gravitational theory
to be satisfied by the gauge theory. We previously discussed the requirement of
diffeomorphism invariance of the gravity action, and saw how it is satisfied on-shell
by the gauge invariance of the BF action. When proving the connection between Chern-Simons
theory and 2+1 dimensional general relativity, Witten~\cite{Witten-2+1:1} excluded the
singular surfaces by requiring the space-time metric to be positive definite. Since he
worked in a canonical formalism, the spatial slice is a Riemann surface, whose
induced metric should also be positive definite. Hence by requiring the $m$-bein to
be invertible, the presence of conical singularities are forbidden and the component
of ${\rm Hom}(\pi_{1}(\Sigma_{g}),PSL(2,\R))/PSL(2,\R)$ with relative Euler number $2g-2$ is 
singled out. By imposing this physical requirement, the only allowed component in the
moduli space of the gauge theory describing general relativity, becomes the component
isomorphic to Teichm\"{u}ller space. This offers a nice relation between the two theories.

In two-dimensional BF theory, the same arguments have been copied and while there are
no guidelines from general relativity to require a positive definite metric, it is
still enforced to single out the non-singular surfaces. By this line of arguments
Chamseddine and Wyler~\cite{CW-2} generalized Witten's arguments from 2+1 dimensions
to obtain a characterization of two-dimensional BF theory.  For genus $g\geq 2$ the
moduli space of flat connections ${\cal M}_{F}$ is identified with Teichm\"{u}ller
space and the reduced phase space ${\cal N}$ is obviously noncompact. Herman
Verlinde has found the general solutions to equations~(\ref{d-omega},
\ref{no-torsion-2}) for $G=SO(2,1)$ 
and they read~\cite{HV}
\begin{eqnarray}
  e^{a} &=& \exp (\pm \alpha) \frac{ df^{a} }{ f^{1}-f^{2} },  \label{e-result} \\
  \omega &=& d\alpha + \frac{ df^{1} + df^{2} }{ f^{1}-f^{2} }, \label{w-result}
\end{eqnarray}
where $\alpha,f^{a}$ are arbitrary functions. The sign in the exponential is plus for $a=1$
and minus for $a=2$. If one forms the connection $A^{i}$ from $(\omega,e^{a})$, these solutions
express the gauge field in the usual flat form
\begin{equation}
  A^{i}T_{i} = g^{-1}d g, \label{flat-A-g}
\end{equation}
where~\cite{HV}
\begin{equation}
  g = \frac{1}{\sqrt{f^{1}-f^{2}}}\left( \begin{array}{cc} \exp(\alpha)f^{1} & \exp(-\alpha)f^{2} \\
        \exp(\alpha)& \exp(-\alpha)\end{array} \right).
\end{equation}
The flat connection in equation~(\ref{flat-A-g}) is single-valued and real on $\Sigma_{g}$, whereas
the gauge parameters $g$ are multi-valued and possibly complex. The gauge parameters change
with constant transition function around the $2g$ homology cycles of $\Sigma_{g}$, with a
$PSL(2,\R)$ transformation
\begin{equation}
  g \mapsto M \cdot g, \hspace{1cm} M=\left(\begin{array}{cc} a & b \\ c& d \end{array} \right) \in PSL(2,\R).
\label{g-map-A}
\end{equation}
The transition functions around the cycles are precisely the holonomies of $A$.
The constant transition functions in equation~(\ref{g-map-A}) have an action on
the functions $(\alpha,f^{a})$ which reads
\begin{eqnarray}
  f^{a} &\mapsto& \frac{ a f^{a} + b}{c f^{a} + d}, \\
  \alpha &\mapsto & \alpha + \log \left( \frac{c f^{1} + d}{cf^{2} + d} \right) .
\end{eqnarray}
By interpreting $(f^{1},f^{2})$ as complex coordinates $(z,\overline{z})$ on $\Sigma_{g}$, the
space-time metric $g = e^{1}\otimes e^{2}$ will acquire the form
\begin{equation}
g = \frac{ df^{1} \otimes df^{2}}{(f^{1}-f^{2})^{2}},
\end{equation}
which is identical to the metric $(ds)^{2}_{-}$ on \Hu from equation~(\ref{uniform-metric}),
when defining $f^{1}=x+iy$ and $f^{2} = x-iy$, such that $f^{1}-f^{2}=2y = 2 {\rm Im}(f^{1})$.
The above identification of the metrics shows that the holonomies of $A$ generate a
discrete subgroup of $PSL(2,\R)$, identical to $\Gamma$, from
which $\Sigma_{g}\simeq \Hu/\Gamma$~\cite{HV}. Forcing the zwei-bein $e^{a}$ to be invertible,
prevents the differentials $df^{a}$ from vanishing at any point of $\Sigma_{g}$. 

From the results in equations~(\ref{e-result},\ref{w-result}) Chamseddine and Wyler~\cite{CW-2}
could derive a closed solution for equation~(\ref{d-phi}) in terms of $f^{a}$ and other new
arbitrary holomorphic functions. This makes it possible to describe the elements in
reduced phase space ${\cal N}$, but since both \m and ${\cal N}$ are noncompact, no topological
invariants are calculated on the basis of topological BF gravity. The partition function is only a
formal integral over \m or ${\cal N}$, as long as one does not take steps to compactify
these spaces. To the author's knowledge no attempts have been made to compactify the moduli space in order to investigate what topological invariants could be derived from two-dimensional
BF gravity. In contrast to this, as we shall see in the next chapters, the compactification
of \m is of vital importance for the Witten type topological gravity theories we study in the
next chapter. We return to a discussion of this and related questions in chapter 6.

At this stage we conclude that two-dimensional BF gravity offers a gauge invariant, and hence 
geometrical, formulation compatible with the Jackiw-Teitelboim model. The quantum theory is
rather rigid, being determined only by the representations of the gauge group $G$. The choice of
gauge group reflects the sign of the cosmological constant and ties the theory to a fixed genus
according to the uniformization theorem.

%% file: chap4.tex
\chapter{ $2D$ Topological Gravity}
\section{Introduction}
In this chapter we present the theory of Witten type 
topological gravity in two dimensions.
This subject is quite complicated to embrace at first hand, because
there exists a whole jungle of different formulations,
often tied to advanced mathematics, which makes it hard to
see what they have in common. One might also wonder what kind of physics
is related to these theories, when they can be written down
following so many, apparently different, first principles.

The main lesson is that all present formulations of topological
gravity in principle agree and that there exists a good explanation as to 
why so many different approaches are possible. The content of pure 
topological gravity is only mathematical, giving a method
to calculate topological invariants on the moduli space of Riemann surfaces.
But there exists a deep relation between topological gravity and ordinary
2D quantum gravity, when one enlarges the discussion to what we call
perturbed topological gravity. This will be exploited in the next chapter.

In chapter 2 we saw how Witten type theories
could be viewed as the result of a BRST-exact gauge fixing procedure
applied to a (trivial) classical action.
In the case of topological Yang-Mills this was
\begin{equation}
  S_{{\rm classical}} = \int_{M} tr(F\wedge F),
\end{equation}
which was trivial being a topological invariant. In gravity 
we consider the even more trivial action
\begin{equation}
  S[g_{\mu\nu}] = 0. \label{action_zero}
\end{equation}
This action was introduced in the paper by
Labastida, Pernici and Witten~\cite{LPW} which started the study of 2D
topological gravity.
Notice how one by writing the action as a functional of the metric has tried
to put information into the trivial action, reflecting that the phrasing ``zero action'' must not be taken too literally since we implicitly think of a given
situation, namely a Riemann surface with a given metric.

Due to the triviality of the action, it is invariant under a
much larger symmetry group than the usual actions of gravity. It is
invariant under all continuous transformations of the metric, in contrast to
the usual diffeomorphism invariance of gravitational theories. It is a
general feature of the Witten type theories that the symmetry group is
larger than in Schwarz type theories. It is common to phrase the
formulation so it reminds one of the formulation in ordinary gauge
theories and
one talks about local gauge
symmetries and topological shift symmetry at the same time.
This is the first reason for having so many different approaches to
topological gravity, because there is a freedom in the choice of local gauge
group. Since the topological symmetry includes the local gauge
symmetries, this split introduces a redundancy in the representation of the
symmetry group. This redundancy hinders quantization and one must lift it
by introducing ghost for ghosts as in topological Yang-Mills theory in
chapter 2.

In the following sections we discuss the two most important approaches
to topological gravity, namely the metric and the first order formulation.
Interestingly we shall see that it is only in the latter formulation that
one can write down explicit expressions for the observables in the theory. 
\section{Metric Formulation of Topological Gravity}
The most natural way to formulate a theory of gravity is in terms of the
space-time metric and this is also true for topological gravity,
even though the metric in some sense is irrelevant here.

The first approach to 2D topological gravity 
by Labastida, Pernici and Witten~\cite{LPW} used the
metric formulation and we follow this construction in detail.
We will clearly see how the theory describes the moduli
space of Riemann surfaces. But to do so, one must first choose a way of
representing the symmetries of the trivial action. The most transparent
way to present this situation is to follow the discussion by Henneaux and
Teitelboim in~\cite{Henneaux-1}.
It follows from the general theory of constrained
systems~\cite{Henneaux-Teitelboim-Book,Dirac},
that we must isolate the first class, primary constraints of the
theory, since they generate the gauge symmetry.
It is easy to find the conjugate momenta for the trivial action
in equation~(\ref{action_zero})
\begin{equation}
  \pi^{\alpha\beta}(x) = \frac{ \partial {\cal L}}{\partial
    \dot{g}_{\alpha\beta}(x) } = \frac{ \partial \, 0}{\partial
    \dot{g}_{\alpha\beta(x)}(x)} = 0.
\end{equation}
These form a set of first class constraints since their Poisson bracket
algebra closes
\begin{equation}
 \{ \pi^{\alpha\beta}(x), \pi^{\mu\nu}(y) \} = 0,
\end{equation}
which is obvious since the $\pi^{\alpha\beta}$'s are independent of the
coordinates $g_{\alpha\beta}$. To see whether the
$\pi^{\alpha\beta}$'s are primary we must study the Hamiltonian. The
Hamiltonian is rather trivial when $L=0$:
\begin{equation}
  H = \pi^{\alpha\beta} \dot{g}_{\alpha\beta},
\end{equation}
and the so-called total Hamiltonian is then
\begin{equation}
  H_{{\rm T}} = H + u_{\alpha\beta}\pi^{\alpha\beta},
\end{equation}
where one adds the sum of all constraints to the Hamiltonian with arbitrary
coefficients $u_{\alpha\beta}$. It follows from the general theory of
constrained systems, that we should analyse the results of
implementing the consistency condition $\dot{\pi}^{\alpha\beta} = 0$,
i.e.\ study the solution to
the equation
\begin{equation}
\{ \pi^{\alpha\beta}, H_{{\rm T}} \} =  \{ \pi^{\alpha\beta},
\pi^{\delta\gamma} \dot{g}_{ \delta\gamma} \} +
u_{\delta\gamma} \{  \pi^{\alpha\beta},
   \pi^{\delta\gamma} \} = 0.
\end{equation}
This leads to an equation of the type $0=0$ and we see that we find no
new constraints (these would have been secondary) and
no restrictions on the arbitrary coefficients $u_{\alpha\beta}$. Hence we can
conclude that the $ \pi^{\alpha\beta}$ are primary, first class constraints
for the Lagrangian $L=0$. It is easy to find the action of a gauge
transformation on the metric
\begin{equation}
  \delta g_{\alpha\beta} = \{ g_{\alpha\beta}, H_{{\rm T}}\} =
  \{ g_{\alpha\beta}, u_{\delta\gamma}\pi^{\delta\gamma} \} = u_{\alpha\beta},
\end{equation}
which mean that all metrics are in the same gauge equivalence class.

There exists no formulations of topological gravity,
based on this representation of the gauge symmetries. It is not clear whether
one could formulate such a version of topological gravity. 
All approaches to topological gravity build on the introduction of an
over-complete set of gauge generators, such that the gauge transformations
are no longer independent. This is done for several reasons, first to
introduce more fields in the theory in the hope of finding interesting relations
and secondly to make the theory look like other known theories, string theory for example.

\noindent
Let us consider the two different representations of the gauge
transformation of the metric in~\cite{LPW}. They both consist of the
combination of a local gauge group and a so-called topological shift
symmetry. The local gauge part is 
either the diffeomorphisms alone or the combined diffeomorphisms and
Weyl rescalings. In both cases the topological shift
invariance is the same, namely 
the infinitesimal transformations of the metric parametrized by arbitrary
$2\times2$ matrices $r^{\, \beta}_{\alpha}$
\begin{equation}
  \delta_{{\rm shift} }\,  g_{\alpha\beta} = (r^{\,\gamma}_{\alpha}
  g_{\gamma\beta} + r^{\,\gamma}_{\beta} g_{\gamma\alpha}).
\label{shift}
\end{equation}
Since the metric is symmetric it only has three degrees of freedom while the
$r^{\, \beta}_{\alpha}$ is arbitrary and has four degrees of freedom.
So already in this choice of representation, an over-completeness has been
introduced into the gauge algebra. The two different choices for the
local gauge group can be written as
\begin{eqnarray}
  \delta_{ {\rm diff} } \, g_{\alpha\beta} &=& D_{\alpha} v_{\beta} +
                                         D_{\beta} v_{\alpha},
                                         \label{diff} \\  
 \delta_{ {\rm diff} \times {\rm weyl} } \, g_{\alpha\beta} &=&
 D_{\alpha} v_{\beta} + D_{\beta} v_{\alpha} - \frac{1}{2}
 g_{\alpha\beta} D_{\gamma} v^{\gamma} + g_{\alpha\beta} \Phi,
 \label{diff_weyl}
\end{eqnarray}
where $v^{\alpha}$ is the vector field along which we perform the
diffeomorphism, $\Phi$ is the Weyl factor (or Liouville field) and
$D_{\alpha}$ is the covariant derivative with respect to $g_{\alpha\beta}$.
More details on these transformations in general are given in chapter 1.
\subsection{BRST Algebra for $G=\diff$}
We write the BRST algebra in the BRST-representation introduced in
chapter 2 and denote the BRST variation by the operator \br.
The full BRST multiplet reads~\cite{LPW}
\begin{eqnarray}
  \br \, g_{\alpha\beta} &=& D_{\alpha}c_{\beta}+
  D_{\beta}c_{\alpha} + \psi_{\alpha\beta}  , \label{LPW-g}\\
  \br \, \psi_{\alpha\beta} &=&
  c^{\gamma}D_{\gamma}\psi_{\alpha\beta} +
  D_{\alpha}c^{\gamma}\cdot\psi_{\gamma\beta} +
  D_{\beta}c^{\gamma}\cdot \psi_{\gamma\alpha} -
  D_{\alpha}\phi_{\beta} +D_{\beta}\phi_{\alpha} , \label{LPW-psi} \\
  \br\, c^{\alpha} &=& c^{\gamma}\partial_{\gamma}
  c^{\alpha} +  \phi^{\alpha} , \label{LPW-c} \\
  \br\, \phi^{\alpha} &=& c^{\gamma}D_{\gamma}
  \phi^{\alpha} -D_{\gamma}c^{\alpha}\cdot\phi^{\gamma}. \label{LPW-phi}
\end{eqnarray}
$c^{\alpha}$ is the Faddeev-Popov ghost for the vector field $v^{\alpha}$
in equation~(\ref{diff}) and it is
supplemented by the topological ghost $\psi_{\alpha\beta}$ which is
defined  as the symmetric combination
\begin{equation}
  \psi_{\alpha\beta} = - g_{\alpha\gamma}w^{\gamma}_{\beta} - g_{\beta\gamma}
  w^{\gamma}_{\alpha},
\end{equation}
where $w_{\alpha}^{\, \beta}$ is the ghost for the matrix
field $r^{\,\beta}_{\alpha}$ in equation~(\ref{shift}).
The fields $g_{\alpha\beta},\psi_{\alpha\beta}$ span the
superspace $\widehat{\met}$, representing forms on the space of
metrics \met. The ghosts generate the Weil algebra on the
basis of the Lie algebra of \diff, according to the
definition in equation~(\ref{Weil-algebra})
\begin{equation}
  \begin{array}{cccc}
  c^{\alpha}&\in& \Lambda(\underline{\diff}^{*}); & \#_{{\rm ghost}} =1,
  \\[.5cm]
  \phi^{\alpha} &\in& S(\underline{\diff}^{*});& \#_{{\rm ghost}} =2.
  \end{array}
\end{equation}
The next step is to construct the action, for which we need to write down
a gauge fermion $\Psi_{{\rm total}}$. Since the classical action is zero, 
there are no classical solutions spanning a moduli space to which we can
localize. Hence there is no localization gauge fermion. The total gauge
fermion
\begin{equation}
  \Psi_{{\rm total}} = \int_{\Sigma} \sqrt{g} \left( b^{\alpha\beta}(g_{\alpha\beta} -
    g_{\alpha\beta}^{0}) + B^{\alpha}D^{\beta}\psi_{\alpha\beta} \right),
\label{LPW-gauge-fermion-diff}
\end{equation}
is a sum of the gauge-fixing term and the projection term. The first term
is a background gauge fixing of the metric by setting it to
a given background value $g_{\alpha\beta}^{0} $. The second term is
the  usual projection term, ensuring that $\psi$ is horizontal.  
Two anti-ghost multiplets are introduced in the
definition of $\Psi_{{\rm total}}$, first an
anti-commuting symmetric anti-ghost $b^{\alpha\beta}$ and secondly a
commuting vector ghost $B^{\alpha}$. The full multiplet reads
\begin{equation}
  \begin{array}{cccc}
  \br b^{\alpha\beta} &=&  d^{\alpha\beta}; & 
  \#_{{\rm ghost}}(b^{\alpha\beta})=-1, \\[.5cm] 
  \br B^{\alpha} &=&  d^{\alpha}; 
  & \#_{{\rm ghost}}(B^{\alpha})=-2.
  \end{array}
\end{equation}
The form of the quantum action is found by $S_{{\rm q}}=\{Q,\Psi_{{\rm total}}\}$
\begin{eqnarray}
  S_{{\rm q}} &=& \int_{\Sigma} \sqrt{g}\left[ d^{\alpha\beta}(g_{\alpha\beta} -
    g_{\alpha\beta}^{0}) - b^{\alpha\beta}(D_{( \alpha}c_{\beta ) }
    + \psi_{\alpha\beta}) + d^{\alpha}D^{\beta}\psi_{\alpha\beta} \right.
  \nonumber \\  && \left.- B^{\alpha}D^{\beta}(D_{ (\alpha }\phi_{\beta )})
    + B^{\alpha} \left( -\phi^{\beta\gamma}D_{\gamma}\psi_{\alpha\beta}
      \nonumber \right. \right. \\ && \left.\left.
    +\frac{1}{2}\psi_{\beta\gamma}D_{\alpha}\psi^{\beta\gamma} -
    \frac{1}{2}\psi_{\alpha\beta}D^{\beta}\psi^{\gamma}_{\gamma}\,\,  \right)
\right] . \label{L-q-1}
\end{eqnarray}
One might consider whether the BRST variation should act on
volume element $\sqrt{g}$ in the gauge fermion in
equation~(\ref{LPW-gauge-fermion-diff}). In~\cite{LPW} it is argued that the
variation of the volume element can be absorbed in a shift of
$d^{\alpha\beta}$. While
this sounds reasonable, we choose to illustrate it by another method.

Consider an arbitrary function of the fields and
ghost $f(\Phi)$, and let $(X,Y)$ be an anti-ghost multiplet. We 
calculate
\begin{eqnarray}
 \lefteqn{ \br \left( \int_{\Sigma} \sqrt{g}\, X\cdot f(\Phi)\right) }
\nonumber \\  &=&\int_{\Sigma} (\br \sqrt{g})X\cdot f(\Phi) +
  \int_{\Sigma} \sqrt{g} \, \br (X) \cdot f(\Phi) +
  \int_{\Sigma} \sqrt{g}\, X\cdot \br (f(\Phi))  \nonumber \\&=&
  \int_{\Sigma} \sqrt{g}\,h(\Phi) A \cdot f(\Phi) +
  \int_{\Sigma} \sqrt{g}\, Y\cdot f(\Phi) +
  \int_{\Sigma} \sqrt{g}\, X \cdot \br (f(\Phi)), \label{metric-argument}
\end{eqnarray}
where we have used that $\br(\sqrt{g}) = \sqrt{g}h(\Phi)$, $h(\Phi)$
being some new function of the fields and
ghosts\footnote{See equation~(\ref{var-A-1}) for proof.}. Notice that the multiplier
$Y$ only enters in one term and integration over it produces a delta function
$\delta(f(\Phi))$ . With this knowledge we can drop the variation of
$\sqrt{g}$ since it will always be proportional to something which vanishes.

The partition function is defined as
\begin{equation}
  Z = \int {\cal D} \left[ g_{\alpha\beta},\psi_{\alpha\beta},b^{\alpha\beta},
  B^{\alpha}, d^{\alpha\beta},d^{\alpha}, c_{\alpha},\phi_{\alpha} \right]
\cdot \exp \left( -S_{{\rm q}} \right),
\end{equation}
where the ${\cal D} \left[ g_{\alpha\beta},\dots\right]$ denotes the
(formal) functional integration measure for all fields mentioned.
Integration over the field $b^{\alpha\beta}$ imposes a delta function
constraint
\begin{equation}
  D_{\alpha}c_{\beta} + D_{\beta}c_{\alpha} = - \psi_{\alpha\beta},
\end{equation}
while integration over $d^{\alpha\beta}$ ensures that the metric takes
its background value $g_{\alpha\beta}^{0}$.
After this the partition function is of the form
\begin{eqnarray}
  Z &=& \int {\cal D} \left[ g_{\alpha\beta},\psi_{\alpha\beta},
    B^{\alpha}, d^{\alpha}, c_{\alpha},\phi_{\alpha} \right]
    \cdot \, \delta \left(g_{\alpha\beta} -  g_{\alpha\beta}^{0}\right)
    \, \delta \left( \psi_{\alpha\beta}+D_{(\alpha}c_{\beta)} \right)
    \exp \left( - S_{{\rm q}} \right) \nonumber \\
    &=& \int {\cal D} \left[ B^{\alpha}, d^{\alpha}, c_{\alpha},\phi_{\alpha}
    \right] \exp \Biggl[ \int_{\Sigma}
 \sqrt{g^{0}} \biggl(  -d^{\alpha}\hat{D}^{\beta}\hat{D}_{(\alpha}c_{\beta)}
     - B^{\alpha}\hat{D}^{\beta}(\hat{D}_{ (\alpha }\phi_{\beta )})
  +   \nonumber \\
  &&\qquad  B^{\alpha} \bigl( - \hat{D}^{(\beta}c^{\gamma)}
     \hat{D}_{\gamma} \hat{D}_{(\alpha}  c_{\beta)}
 + \frac{1}{2} \hat{D}_{(\beta}c_{\gamma)}\hat{D}_{\alpha}\hat{D}^{(\beta}
 c^{\gamma)} - \hat{D}_{(\alpha}c_{\beta)}
 \hat{D}^{\beta}\hat{D}^{\gamma}c_{\gamma}
 \bigr) \biggr) \Biggr] . \nonumber \\ && \label{partition-LPW-1} 
\end{eqnarray}
where $\hat{D}_{\alpha}$ is the covariant derivative compatible with the
background metric.
By raising and lowering indices using $g_{\alpha\beta}^{0}$,
equation~(\ref{partition-LPW-1}) is seen to agree with
the result given in~\cite{LPW}. Note that no integration
over moduli space is needed to evaluate the partition function.
From the above form
of the partition function it is not easy to interpret this theory,
since it just contains kinetic terms in the ghosts $(c_{\alpha},\phi_{\alpha})$ and  higher
derivative
terms in $c_{\alpha}$. It is noted in~\cite{LPW} that the action above is
not invariant under rescalings of the background metric and thus not
conformally invariant, in addition to the clearly broken diffeomorphism
invariance.
The background for this statement is that the authors want to relate
topological gravity to the moduli space of Riemann surfaces \m.
A priori we have not chosen to model this moduli space,
since we have chosen only the diffeomorphisms as our gauge group, i.e. we
could only be modelling $\met/\diff$, but it turns
out that the formulation above also describes \m.
This can be clarified~\cite{CMR} by writing a model  
of \m only using $G=\diff$ as above. This is done
by localization to the subset of all metrics, which has constant
curvature. We have shown in chapter 1 that this requirement
fixes the Weyl invariance. 
The Weyl localization gauge fermion reads
\begin{equation}
  \Psi_{{\rm localization - Weyl}} = \int_{\Sigma} \sqrt{g} \kappa ( R - k),
\label{Weyl-loc-1}
\end{equation}
where $k$ is a real constant and we have introduced an anti-ghost
multiplet $(\kappa,\tau)$ with
$\#_{{\rm ghost}}(\kappa)=-1$.
One must make the BRST variation of the metric through the Ricci scalar and
we find
\begin{equation}
  S_{{\rm localization-Weyl}} = \int_{\Sigma} \sqrt{g} \left( \kappa (R-k) -
  \tau G^{\alpha \beta}
    \br(g_{\alpha\beta})\right),
\end{equation}
where $G^{\alpha\beta}$ is the following operator
\begin{equation}
  G^{\alpha\beta} = D^{\alpha}D^{\beta} - \frac{1}{2} g^{\alpha\beta}
  D^{\gamma}
    D_{\gamma} - \frac{1}{2}k g^{\alpha\beta}.
\end{equation}
This follows the discussion in chapter 1.
By explicitly breaking the Weyl symmetry one uses the identity
\begin{equation}
  \m \simeq \{ \met \mid R=k \} / \diff \label{moduli-weyl-fix}
\end{equation}
and formulates a theory of topological gravity for $G=\diff$.
Since both this formulation and the one in~\cite{LPW} effectively
describe the same moduli space they can be considered equivalent, but
the explicit breaking of Weyl invariance signals that one is really
interested in defining a theory which models \m and not just the
space of equivalence classes of metrics under diffeomorphisms.
In both approaches  the fact that the theory models \m is not
clear at all. This only  becomes transparent when we change the
gauge group to include the Weyl transformation, which is studied in the
next section.
\subsection{BRST Algebra for $G=\diff \times \w$}

To obtain the symmetry group $G=\diff \times \w$, we must extend the
BRST algebra to correspond to the bigger symmetry which we are trying to
gauge-fix. We replace the anti-ghost $B^{\alpha}$ with an anti-ghost
multiplet consisting of a symmetric
tensor $B^{\alpha\beta}$ with  $\delta_{B}B^{\alpha\beta}=f^{\alpha\beta} $
and introduce an extra ghost multiplet
$(\rho,\tau)$
\begin{equation}
  \begin{array}{llll}
  \br \, \rho & = & c^{\gamma}\partial_{\gamma}\rho +
                         \epsilon \tau; & \#_{{\rm ghost}}(\rho)=1,\\
  \br \, \tau & = & c^{\gamma} \partial_{\gamma} \tau
  - \phi^{\gamma}\partial_{\gamma}\rho; & \#_{{\rm ghost}}(\tau)=2.
  \end{array}
\end{equation}
This may be reformulated by redefining $\tau \mapsto \tau' = \tau +
 c^{\gamma}\partial_{\gamma}\rho$, to the standard form $\br \rho =
\tau'$ and $\br \tau' =0$. The first relation
is trivial to show by inspection. The second term is a bit more involved
but still straightforward
\begin{eqnarray}
  \br \, \tau' &=& \br \tau + \br(c^{\gamma}\partial_{\gamma}\rho) \nonumber \\ &=&
        c^{\gamma}\partial_{\gamma}\tau - \phi^{\gamma}\partial_{\gamma}\rho +
        (c^{\gamma}\partial_{\gamma}c^{\alpha} + \phi^{\alpha})\partial_{\alpha}\rho
        - c^{\gamma}\partial_{\gamma}(c^{\delta}\partial_{\delta}\rho + \tau) \nonumber \\
                                 &=&  0.
\end{eqnarray}
It is also convenient to redefine
$\psi_{\alpha\beta} \mapsto \psi_{\alpha\beta} - \rho g_{\alpha\beta}$.
Then the total BRST algebra reads
\begin{eqnarray}
  \br \, g_{\alpha\beta} &=& D_{\alpha}c_{\beta}+
  D_{\beta}c_{\alpha} + \psi_{\alpha\beta} - \rho g_{\alpha\beta}
   , \\
  \br \, \psi_{\alpha\beta} &=&
  c^{\gamma}D_{\gamma}\psi_{\alpha\beta} +
  D_{\alpha}c^{\gamma}\cdot\psi_{\gamma\beta} +
  D_{\beta}c^{\gamma}\cdot \psi_{\gamma\alpha}  \nonumber \\ && - 
  D_{\alpha}\phi_{\beta} +D_{\beta}\phi_{\alpha} -\tau
  g_{\alpha\beta} , \\
 \br \, \rho &=& c^{\gamma}\partial_{\gamma}\rho +
                         \tau,  \\
  \br \, \tau &=& c^{\gamma} \partial_{\gamma} \tau
  - \phi^{\gamma} \partial_{\gamma} \rho , \\ 
\br \, c^{\alpha} &=& c^{\gamma}\partial_{\gamma}
  c^{\alpha} +  \phi^{\alpha} , \\
  \br\, \phi^{\alpha} &=& c^{\gamma}D_{\gamma}
  \phi^{\alpha} - D_{\gamma}c^{\alpha}\cdot
  \phi^{\gamma}. \label{eq:brst-diff-weyl}
\end{eqnarray}
The new ghosts are in the Weil algebra for the Weyl transformations
\begin{equation}
  \begin{array}{llll}
    \rho & \in \Lambda(\underline{\w}^{*}); & \#_{{\rm ghost}}(\rho) &=1, \\
    \tau & \in S(\underline{\w}^{*}); & \#_{{\rm ghost}}(\tau) &=2. \\
  \end{array}
\end{equation}
We introduce the following gauge fermion
\begin{equation}
 \Psi = \int_{\Sigma} \sqrt{g} \left( b^{\alpha\beta}(g_{\alpha\beta} -
    g_{\alpha\beta}^{0}) + B^{\alpha\beta}\psi_{\alpha\beta} \right),
\label{LPW-gauge-fermion-conform}
\end{equation} 
where the first term is the gauge fixing part and the next term is the
projection part. There are no localization terms, due to the choice of
the symmetry group.
This relies on the representation of the moduli space of Riemann surfaces given in
equation~(\ref{moduli-1}). The
projection term is special for this gauge group. This is due to the fact that
the symmetry group is in conflict with the general framework of section~\ref{loc}.
The metric on \met given in equation~(\ref{met-metric}) 
as well as the metric on \diff in equation~(\ref{diff-metric}), are not invariant under
Weyl transformations. Hence this metric does not define a connection on the universal
bundle for $G=\diff\times\w$. In order to find the projection form $\Phi(P\mapsto M)$
which is fully vertical, we should use the projection operator to vertical forms. This
can be defined using the $P,P^{\dagger}$ operators from chapter 1. The projection
operator to horizontal forms is~\cite{CMR}
\begin{equation}
  \Pi_{\alpha\beta}^{\,\,\,\,\delta\gamma} = \left( 1 - P\frac{1}{P^{\dagger}P}P
    \right)_{\alpha\beta}^{\,\,\,\,\delta\gamma},
\end{equation}
and the projection gauge fermion should be of the form
\begin{equation}
\Psi_{{\rm projection}} = \int_{\Sigma} \sqrt{g} B^{\alpha\beta}
 \Pi_{\alpha\beta}^{\,\,\,\,\delta\gamma}
\psi_{\delta\gamma}.
\end{equation}
For the argument we wish to make, it is sufficient to assume that $\psi$ is
horizontal and that $\Pi\psi=\psi$. Then the gauge fermion in
equation~(\ref{LPW-gauge-fermion-conform}) imposes a delta function gauge for
$\psi$. If one does the calculation keeping the projection operator as in~\cite{CMR},
one should be careful due to problems regarding non-locality and genus
dependence of the kernel of $P^{\dagger}P$. Since we will not work with this gauge
group in the following, we do not wish to enter this discussion.

The gauge fermion in equation~(\ref{LPW-gauge-fermion-conform}) induces
the quantum action
\begin{eqnarray}
  S_{q} &=& \int_{\Sigma} \sqrt{g}\left( d^{\alpha\beta}(g_{\alpha\beta} -
    g_{\alpha\beta}^{0}) - b^{\alpha\beta}(D_{( \alpha}c_{\beta ) }
    + \psi_{\alpha\beta} - \rho g_{\alpha\beta})  \right.
  \nonumber \\ &&\left. + f^{\alpha\beta}\psi_{\alpha\beta} + 
   B^{\alpha\beta}(c^{\gamma}D_{\gamma}\psi_{\alpha\beta} +
   D_{\alpha}c^{\gamma} \cdot \psi_{\gamma\beta} +
   D_{\beta}c^{\gamma}\cdot \psi_{\gamma\alpha}
   -\rho \psi_{\alpha\beta}) \right. \nonumber \\ && \left. +
   B^{\alpha\beta}(\tau g_{\alpha\beta} - D_{(\alpha}\phi_{\beta )}) \right).
\end{eqnarray}
The partition function is  
\begin{equation}
  Z = \int {\cal D} \left[ g_{\alpha\beta},\psi_{\alpha\beta},b^{\alpha\beta},
  B^{\alpha\beta}, \rho, \tau, f^{\alpha\beta},d^{\alpha\beta}, c_{\alpha},
  \phi_{\alpha} \right]  \exp \left( - S_{{\rm q}} \right).
\end{equation}

\noindent
When we integrate over the multiplier field $f^{\alpha\beta}$ we get a delta
function $\delta(\psi_{\alpha\beta})$ while the
integration over $d^{\alpha\beta}$ forces the metric to equal the
background metric $g_{\alpha\beta}^{0}$.
This implies that 
\begin{eqnarray}
  Z &=& \int   {\cal D} \left[ g_{\alpha\beta},\psi_{\alpha\beta},
    b^{\alpha\beta},B^{\alpha\beta}, \rho, \tau, c_{\alpha},\phi_{\alpha} \right] 
 \delta \left(g_{\alpha\beta} -  g_{\alpha\beta}^{0}\right)\,
 \delta \left( \psi_{\alpha\beta} \right) \, \exp \left( -S_{{\rm q}} \right)
 \nonumber \\ &=& \int   {\cal D} \left[ b^{\alpha\beta},B^{\alpha\beta},
   \rho, \tau, c_{\alpha},\phi_{\alpha} \right] \exp \left(
   \int_{\Sigma} \sqrt{g^{0}} \left[ -b^{\alpha\beta}(
     \hat{D}_{(\alpha}c_{\beta)} - \rho g_{\alpha\beta}^{0}) \right.\right.
 \nonumber \\   && \left.\left.  +
     B^{\alpha\beta}(\tau g_{\alpha\beta}^{0} - D_{(\alpha}\phi_{\beta)})
   \right] \right) \nonumber \\ &=&
\int   {\cal D} \left[ \hat{b}^{\alpha\beta},\hat{B}^{\alpha\beta},
  c_{\alpha},\phi_{\alpha}\! \right]\! \exp\left(\!  
   \int_{\Sigma} \sqrt{g^{0}} \left[ -\hat{b}^{\alpha\beta}
     \hat{D}_{(\alpha}c_{\beta)} - 
     B^{\alpha\beta}D_{(\alpha}\phi_{\beta)}\right]
   \right),
\end{eqnarray}
where one in the final integration over $(\rho,\tau)$ forces
$(b^{\alpha\beta},B^{\alpha\beta})$ to be traceless
with respect to the background metric, indicated by the notation
$\hat{b}^{\alpha\beta},\hat{B}^{\alpha\beta}$ for the traceless components.
The resulting BRST exact action has the same form as the 
famous $(b,c,\beta(=B),\gamma(=\phi))$ type actions from string theory
\begin{equation}
S_{b-c,B-\phi} = \int_{\Sigma} \sqrt{g^{0}} \left(  \hat{b}^{\alpha\beta}
  \hat{D}_{\alpha} c_{\beta} -
  \hat{B}^{\alpha\beta}\hat{D}_{\alpha}\phi_{\beta} \right).
\label{LPW-bc-action}
\end{equation}
In order to understand this action we make a small detour into string theory
to see how the $b-c$ actions appear and how we should interpret them.
\subsection{Note on the $b-c$ ghost action}
We write a general bosonic string action, which is Poincar\'{e} and diffeomorphism
invariant~{\cite{GSW,SuperStringVol3}} as 
\begin{equation}
  S = -\frac{1}{2\pi}\int_{\Sigma} d^{2} \xi \sqrt{g}
  \left[ g^{\alpha\beta}\partial_{\alpha}X^{a} \partial_{\beta}X^{b}\eta_{ab}
    \right],
    \label{string_action}
\end{equation}
where the string is moving in a $d$-dimensional flat space $\tilde{M}$ with metric
$\eta_{ab}$. This is described by maps $X(\xi)$ from the two dimensional world-sheet with
coordinates $\xi$ and metric $g_{\alpha\beta}$ to $\tilde{M}$. These maps are given
by $d$ fields $X^{a}(\xi)$ for $a = 0,\dots,d-1$. The details of the string theory are
not so important for the arguments we want to emphasize. The partition function
is an integral over surfaces
\begin{equation}
  Z = \int {\cal D} [ g_{\alpha\beta},X^{a} ] e^{ -S(g,X) } .
\end{equation}
We must introduce a gauge fixing of the metric and it is common to use the
conformal gauge
\begin{equation}
  g_{\alpha\beta} = e^{\Phi} \hat{g}_{\alpha\beta},
\end{equation}
where we take the flat background metric $\hat{g}_{\alpha\beta}$
to be Euclidean.
Notice that this is exactly the isothermal coordinates described in chapter 1 and the gauge condition implies that
\begin{equation}
  g_{zz} = g_{\bar{z}\bar{z}} =0,
\end{equation}
where we have changed to holomorphic/anti-holomorphic coordinates introduced
in chapter 1.
The action of a diffeomorphism $f: \xi^{z(\bar{z})} \mapsto \xi^{z(\bar{z})} +
\sigma^{z(\bar{z}) }$
changes the metric according to equation~(\ref{diff}) which reduces to
\begin{eqnarray}
  \delta g_{zz} &=& 2 D_{z} \sigma_{z} , \\ 
  \delta g_{\bar{z}\bar{z}} &=& 2 D_{\bar{z}} \sigma_{\bar{z}} ,
\end{eqnarray}
by our choice of gauge. We use a standard trick in field theory and write
the following integral
\begin{equation}
  \int_{\diff} {\cal D}f(\xi) \, \delta \left(g_{zz}^{f} \right)
\delta \left(g_{\bar{z}\bar{z}}^{f} \right) \, {\rm det}
\left( \frac{g_{zz}^{f}}{\delta f} \right)
\left( \frac{g_{\bar{z}\bar{z}}^{f}}{\delta f} \right),
\end{equation}
which is independent of $f$. This can be inserted into the partition function
for free, since it just changes the constant in front of the path integral
which we always ignores. The resulting partition function is of the form
\begin{equation}
Z = \int {\cal D}[f(\xi),g(\xi),X] \, \delta \left(g_{zz}^{f} \right)
\delta \left(g_{\bar{z}\bar{z}}^{f} \right) \, {\rm det}
\left( \frac{g_{zz}^{f}}{\delta f} \right) {\rm det}
\left( \frac{g_{\bar{z}\bar{z}}^{f}}{\delta f} \right) \, e^{-S[g,X]}.
\end{equation}
Since the string action in equation~(\ref{string_action})
is diffeomorphism invariant we know that $S[g^{f},X] = S[g,X]$ which implies 
\begin{equation}
Z = \int {\cal D}[f(\xi),g(\xi),X] \,
\delta \left( g_{zz}^{f} \right)
\delta \left( g_{\bar{z}\bar{z}}^{f} \right) \, {\rm det}
\left( \frac{g_{zz}^{f}}{\delta f} \right) {\rm det}
\left( \frac{g_{\bar{z}\bar{z}}^{f}}{\delta f} \right) \, e^{-S[g^{f},X]}.
\end{equation}
Note that this integral only depends on the metric through the transformed
$g^{f}$ so we change integration from $g$ to $g^{f}$. This implies that the
${\cal D}f$ integration decouples to a multiplicative constant which we
can take out of the integral.
The part of the integral which is over the metric is of form
\begin{equation}
  \int {\cal D} g^{f} \delta \left(g_{zz}^{f} \right)
\delta \left(g_{\bar{z}\bar{z}}^{f} \right) \sim \int {\cal D}g_{z\bar{z}}^{f} =
\int {\cal D}\Phi,
\end{equation}
which reduces to an integral over $g_{z\bar{z}}$ identified with the 
Liouville field $\Phi$ in this gauge. To deal with the determinants we
introduce a Faddeev-Popov ghost/anti-ghost multiplet. The argument of
the determinant is
\begin{equation}
  \frac{g_{zz}^{f}}{\delta f} = \frac{g_{zz}^{f}}{ \delta \sigma_{z} } =
  \frac{2 D_{z}\sigma_{z}}{ \delta \sigma_{z}  } = 2 D_{z}\delta(\xi - \xi^{'}),
\end{equation}
and likewise for $z \mapsto \bar{z}$. The determinant of this argument is then
represented by the integral
\begin{equation}
  {\rm det} (D_{z}\delta(\xi - \xi^{'})) = \int {\cal D}[c^{\bar{z}}(\xi),b_{\bar{z}\bar{z}}(\xi)]
  \exp \left( -\frac{1}{\pi}
  \int_{\Sigma} d^{2}\xi \, \, c^{\bar{z}}D_{z}b_{\bar{z}\bar{z}} \right) ,
\label{det_1}
\end{equation}
and equivalently for the other determinant.
The partition function can now be written as the string action plus a
Faddeev-Popov gauge fixing action
\begin{equation}
Z = \int {\cal D}[\Phi(\xi), X(\xi),c(\xi),b(\xi)]  \exp \Biggl( -S[\phi,X]
  - S_{FP}[b,c] \Biggr),
\end{equation}
where we actually have some problems connected to the integration over the
Liouville field, due to an implicit metric dependence
of the norm of $\Phi(x)$. This will be ignored here
since it not does play a role for the arguments regarding the
$bc$-system, but it is of course central for the discussion of Liouville theory and
quantum gravity. For further information see e.g.~\cite{Hat,Ginsparg-Moore}. 
The Faddeev-Popov action can be written as~\cite{Fridan}
\begin{equation}
  S_{FP}[b,c] = \frac{i}{2\pi}\int_{\Sigma} d^{2}\xi \,\, \left(
    c^{z}\partial_{\bar{z}}b_{zz} + c^{\bar{z}}\partial_{z}b_{\bar{z}\bar{z}} \right),
\end{equation}
with equations of motions
\begin{eqnarray}
  \partial_{\bar{z}} c^{z} &=& \partial_{\bar{z}}b_{zz} = 0, \\
  \partial_{z} c^{\bar{z}} &=& \partial_{z}b_{\bar{z}\bar{z}} = 0.
\end{eqnarray}
The covariant derivative in equation~(\ref{det_1}) reduces to a
partial derivative in this gauge according to the discussion on
page~\pageref{isotherm-christoffel} in chapter 1.
From the discussions in chapter 1 we conclude that,
the equations for $b$ identify
it as a quadratic differential. We can also identify the $c$ field as being
a conformal killing vector, since its equation of motion resembles the
requirement $c^{z}\in {\rm ker}P_{1}$.
According to the Riemann-Roch 
theorem~\cite{SuperStringVol3}
\begin{equation}
  {\rm dim}\left({\rm Ker} P_{1}\right) -
  {\rm dim}\left({\rm Ker} P_{1}^{\dagger}\right) = \frac{3}{2}\chi(\Sigma)
  = 3 - 3g,
\end{equation}
and we see that the $bc$-system fits into this by the relation
\begin{equation}
  \# (\mbox{Quadratic Differentials}) -
  \#(\mbox{Conformal killing vectors}) = 3g -3.
\end{equation}
The $bc$-action models the moduli space of Riemann surfaces in the
sense that the Faddeev-Popov action gives the dimension of moduli space.
The quadratic differentials are exactly the Teichm\"{u}ller parameters, describing
the metric variations not arising from either Weyl rescalings or diffeomorphisms.
The $(B^{\alpha\beta},\phi_{\alpha})$'s are the commuting superpartners for $(b,c)$,
being ghosts for the super-diffeomorphisms.

\section{First Order Formulation of Topological Gravity}
As discussed in chapter 3 it is common to use the
first order formulation of general relativity when discussing quantum gravity.
The reasons for this is the more transparent connection to gauge
theories and the slightly enlarged number of degrees of freedom,
which one does not necessarily need to exclude. This opens the
possibility of discussing geometries where ${\rm det}g_{\alpha\beta} =0$,
which is not allowed in classical general relativity, but which may be
a natural situation in the quantum theory. Especially in the
path integral formulation it is natural to allow all values of the fields.

The construction of BF gravity from the first order variables
$(e^{a},\omega)$, can be extended to the case of so-called super BF theory.
In contrast to BF gravity this is a Witten type theory for two dimensional
topological gravity. This view has been advocated by E. and
H. Verlinde~\cite{VV} and also by Montano and Sonnenschein~\cite{MS}.
There are several important points in their presentation. First of all
the super BF action is of the form
\begin{equation}
  S[ B,A,\chi,\psi] = \int_{\Sigma_{g}} d^{2}x \, tr \left( B F[A] + \chi D
    \psi \right) , \label{super-BF-VV-1}
\end{equation}
where the supersymmetry transformation maps
\begin{equation}
  \delta_{{\rm S}} A = \psi \,\,\, ;\,\,\, \delta_{{\rm S}} \chi = B.
\end{equation}
Here  $\chi$ is a fermionic zero-form with
values in the Lie algebra and $\psi$
is a fermionic partner to the connection one-form $A$.
We denoted the Lagrange multiplier for $F[A]$ as $B$.
It is clearly a Witten type theory since the action is BRST exact
\begin{equation}
  S_{{\rm q}} =   S[ B,A,\chi,\psi] = \{Q,
  \int_{\Sigma_{g}} d^{2}x \,tr(\chi F[A])\} .
\end{equation}
In BF gravity we need to change the gauge group for each of the three situations
$g=0$, $g=1$ and $g\geq 2$. Montano and Sonnenschein~\cite{MS} gave a
description of topological gravity based on the super BF action in
equation~(\ref{super-BF-VV-1}).  
Their work was inspired by the developments in topological Yang-Mills and
tried to copy these to super BF theory. The
gauge field was then translated to gravity variables as in the previous
chapter with the same choice of gauge groups. They presented
formally analogues of the Donaldson polynomials
for genus $g\geq 2$,
based on the curvature $F[A]$, the ghost $\psi$, and the needed ghost for
ghost $\phi$. Even though this formally gives the form of the observables,
no expressions were derived in terms of local quantum fields.
Only general results based on the discussion of the universal bundle.
There are several other unanswered questions in relation to this
direct approach of super BF theory and topological gravity which we
discuss in chapter 6.

For genus one we had in the previous chapter the action of BF gravity 
\begin{equation}
  S = \int d^{2}x\, \left(  - B_{0} d\omega +  B_{a}De^{a}   \right),
\end{equation}
which corresponds to $G=ISO(1,1)$. The approach taken in~\cite{VV} was
to write a theory of topological gravity which models the moduli space
of Riemann surfaces for all genera, using an $ISO(2)$ version of super
BF theory. The change
to $ISO(2)$ from $ISO(1,1)$ changes the $\phi$ (here $B$) transformations given
in equations~(\ref{iso-2-1-phi},\ref{iso-2-2-pphi}) into
\begin{eqnarray}
 \delta B_{0} &=& -\epsilon_{ab}\lambda^{a}B^{b}, \\
  \delta B_{a} &=& \epsilon_{a}^{\,\,b}\lambda^{0}B_{b}.
\end{eqnarray}
This changes the sign on the $d\omega$ term of the action into a plus.
But the scalar curvature is zero, due to the
equations of motion and we seem tied to the torus. By
equation~(\ref{moduli-weyl-fix}) we identify 
\begin{equation}
\m \simeq
\frac{ \{ \met\vert R(g) = k \} }{ \{ \diff \} } \simeq
\frac{ \{ {\rm FRAMES}(e^{a})\vert d\omega(e)) = k \} }{ \{
  \diff\otimes{\rm LL} \}},
\end{equation}
where ${\rm FRAMES}$ denotes the set of zwei-beins and ${\rm LL}$ denotes the
local Lorentz transformations $SO(2)\simeq U(1)$. The Weyl symmetry is 
fixed by the constraint $d\omega=k$, but E. and H. Verlinde suggest that one
instead could set the scalar curvature to zero in all, but a fixed number
of points, where one inserts delta function singularities. 
Following string theory methods~\cite{GSW} one inserts 
vertex operators $\exp(-q_{i}B_{0}(x_{i}))$ at the position of
the singularities to ``screen'' the curvature. 
When these operators are inserted 
the bosonic part of the action reads 
\begin{equation}
S =  \int d^{2}x\, \left( B_{0}d\omega + B_{a}De^{a}  -
\sum_{i=1}^{s}q_{i}B_{0}(x_{i})\right) ,
\end{equation}
which changes the equation of motion for the spin-connection
\begin{equation}
  d \omega(x)  = \sum_{i=1}^{s}q_{i}\delta^{(2)}\left( x-x_{i}\right).
\label{changed-omega-constraint}
\end{equation}
The delta functions are considered as two-forms, such that the
integrated curvature equals the Euler number of the surface 
\begin{equation}
 \frac{1}{4\pi}\int \sqrt{g}R(g) =  2-2g = \chi(\Sigma_{g}) .
\end{equation}
Later we study how this
introduction of curvature singularities plays a vital role in the
geometry of topological gravity.
The curvature singularities ruin the gauge invariance of the equations of
motion at the points $(x_{i})$, where the curvature is situated. But
the equations of motion are invariant under the local Lorentz part of
$ISO(2)$, i.e. under $U(1)\subset ISO(2)$. It is only the inhomogeneous
part of $ISO(2)$ representing the diffeomorphisms, which breaks down
at these points. This is more or less obvious, since a diffeomorphism
could transform the position of a curvature singularity, into
a point where there is no singularity. One point of view is to say that the
super BF action is invariant under those gauge transformations which
vanish at the points $(x_{i})$,  such that $\delta A(x_{i}) = 0$. This
will on-shell correspond to those diffeomorphisms which leave the points
$(x_{i})$ inert. {\em This is the same as formulating the gauge theory on
a punctured Riemann surface}. Punctures play a vital role in
topological gravity and they are discussed at length later in this chapter.

We will continue the discussion of the first order formulation in the
following way: First we write down the BRST algebra corresponding to the
symmetries in this formulation. Next we show that the action of
this first order formulation, is directly related to the metric formulation.
Thereafter we enter the important topic of observables in
topological gravity. 

The BRST algebra is constructed as a
product of the local gauge transformations and the topological shift.
The replacement of gauge parameters to ghosts
reads $\lambda^{a} \mapsto c^{a}$ and $\lambda^{0}\mapsto c_{0}$.
The topological ghost $\psi^{i}$
can be viewed as the super partner for $A^{i}$, but independent of the point of
view one takes, it is expanded as $(\psi_{0},\psi^{a})$ which are then
the partners/topological ghosts for $(\omega,e^{a})$. The ghost for
ghost field $\phi^{i}$ is finally expanded as $(\phi_{0},\phi^{a})$.
The $U(1)$ part corresponds to the BRST transformations
\begin{eqnarray}
  \br \omega & = &  \psi_{0} + d c_{0} \label{d-omega-1},\\
  \br c_{0} &=&  \phi_{0}, \label{d-c-LL} \\
  \br \psi_{0} &=&  d \phi_{0}, \label{d-psi-LL} \\
  \br \phi_{0} &=& 0 \label{d-phi-LL}.
\end{eqnarray}
The inhomogeneous part reads
\begin{eqnarray}
  \br e^{a} &=& 
    \psi^{a} - dc^{a} +\epsilon^{a}_{\,\,b}\omega c^{b} +
  \epsilon^{a}_{\,\,b}e^{b}c_{0} ,\label{d-e-diff} \\
  \br \psi^{a} &=&   -\epsilon^{a}_{\,\,b}\psi^{b}c_{0} -
  d(\psi^{a}_{\alpha}c^{\alpha}) + \epsilon^{a}_{\,\,b}\omega
  \psi^{b}_{\alpha}c^{\alpha} + \epsilon^{a}_{\,\,b}\psi_{0}c^{b} 
\\ && + \,  \epsilon^{a}_{\,\,b}
  e^{b}\phi_{0} + d\phi^{a} -\epsilon^{a}_{\,\,b}\omega\phi^{b},
\label{d-psi-diff}\\
\br c^{\alpha} &=& \phi^{\alpha} +c^{\beta}\partial_{\beta}c^{\alpha}, \label{d-c-diff}\\
\br \phi^{\alpha}& =& c^{\beta}\partial_{\beta}\phi^{\alpha}-
\phi^{\beta}\partial_{\beta}c^{\alpha} \label{d-phi-diff} .
\end{eqnarray}
The ghost $(c^{\alpha})$ is defined as $c^{a}\equiv
c^{\alpha}e_{\alpha}^{\,\,a}$.

The fields $(e,\omega)$ are not independent and if the torsion constraint
is satisfied, one can express $\omega$ as a function of $e$, as shown in
chapter 3. Both fields are thus elements in the space of frames, and
together with $(\psi_{0},\psi^{a})$ (satisfying the super
torsion constraint) they span the superspace $\widehat{{\rm FRAMES}}$,
representing forms on ${\rm FRAMES}$. The ghosts generate the Weil
algebra of \diff and ${\rm LL}$
\begin{equation}
 \begin{array}{llll}
    c_{0} & \in \Lambda(\underline{{\rm LL}}^{*}); & \#_{{\rm ghost}}(c_{0})
    &=1, \\
    \phi_{0} & \in S(\underline{{\rm LL}}^{*}); & \#_{{\rm ghost}}(\phi_{0})
    &=2, \\
    c^{a} & \in \Lambda(\underline{\diff}^{*}); & \#_{{\rm ghost}}(c^{a})
    &=1, \\
    \phi^{a} & \in S(\underline{\diff}^{*}); & \#_{{\rm ghost}}(\phi^{a})
    &=2.
  \end{array}
\end{equation}
In contrast to the metric formalism, we now have an action so that we are
not gauge fixing ``zero''. We present one of
the many different discussions on gauge fixing in this framework, following
the original work in~\cite{VV}. E. and H. Verlinde did actually not include
the curvature singularities in the action as indicated above, but in the
definition of the 
observables, which we give in the next section and we exclude this question 
for the moment. By choosing a conformal gauge the super BF action can be
written as a conformal field theory, just as in the $\diff \otimes \w$
representation, but now also with a (super) Liouville sector.

We switch to complex coordinates $(z,\bar{z})$ and use the
isothermal coordinates and the relation between the metric and the zwei-beins,
to write (locally) 
\begin{equation}
e^{+}(z) = \exp ( \Phi_{+}(z) ) dz \,\, ; \,\, e^{-}(\overline{z}) =
\exp ( \Phi_{-}(\overline{z}) )
d\bar{z},
\end{equation}
where $\Phi_{+} = \overline{\Phi_{-}}$. So $e_{z}^{+}=e^{\Phi_{+}}$,
$e_{\overline{z}}^{-}=e^{\Phi_{-}}$ and $e_{z}^{-}=e_{\overline{z}}^{+}=0$
in this gauge.
The classical action reads
\begin{equation}
  S = \int d^{2} z\, \Bigl[ B_{0}d\omega + B_{+}De^{+} + B_{-}De^{-} +
                     \chi_{0} d\psi_{0} + \chi_{+}D\psi^{+} +
                     \chi_{-}D\psi^{-}\Bigr]. \label{VV-action-1}
\end{equation}
The conformal gauge parametrizes the
equivalence classes of zwei-beins under diffeomorphisms, with the
Liouville mode
and the moduli parameters given in the $bc$-action. The $c^{\alpha}(x)$ and
its anti-ghost $b^{\alpha}(x)$ are tied to the diffeomorphisms and the
$c_{0}(x)$ and its anti-ghost $b_{0}(x)$ to the local Lorentz
transformations.
From the form of the zwei-beins we see that the Lorentz
transformations are fixed by imposing the gauge condition
\begin{equation}
  \Phi_{+} = \Phi_{-},
\end{equation}
setting the Liouville mode to be real. Note that the modes are numerically
equal but that they depend on $z$ and $\overline{z}$ respectively.
When this invariance is fixed, the corresponding ghosts are non-dynamical
and by applying the BRST transformation of $e^{a}$ and identifying 
$\br (e^{+})=\br (e^{-})$, we can derive an equation for $c_{0}$.
At the same time we have the equivalent relations
for the partner $\psi$ so
\begin{equation}
  \psi^{+} = \exp(\Phi_{+})\psi_{+} dz \,\, ; \,\, d\psi^{-} = \exp (\Phi^{-})
  \psi_{-} d\bar{z},
\end{equation}
and the super Lorentz invariance is fixed through the requirement
\begin{equation}
  \psi_{+} = \psi_{-}.
\end{equation}
Calculate $\br (e^{+})=\br (e^{-})$ to obtain
\begin{equation}
  \psi_{+}e^{\Phi_{+}} + c_{0}e^{\Phi_{+}} + \partial_{z} c^{z}
  \cdot e^{\Phi_{+}}
  + c^{z}\partial_{z}\Phi_{+} =
  \psi_{-}e^{\Phi_{-}} - c_{0}e^{\Phi_{-}} + \partial_{\bar{z}}
  c^{\bar{z}} \cdot e^{\Phi_{-}} + c^{\bar{z}}\partial_{\bar{z}}\Phi_{-},
\end{equation}
and isolate $c_{0}$ to obtain
\begin{equation}
  c_{0} = \frac{1}{2} \left( \partial_{z} c^{z} + c^{z}\partial_{z}\Phi
    -  {\rm c.c.} \right) ,
\end{equation}
where ${\rm c.c.}$ abbreviates complex conjugation.
Since this Lorentz ghost is non-dynamical its super partner can be found
by calculating $\phi_{0}=\br c_{0}$:
\begin{equation}
  \phi_{0} = \left( \partial_{z}\phi^{z} + \phi^{z}\partial_{z}\Phi +
    c^{z}\partial_{z}\psi - {\rm c.c.}\right) .
\label{phi-relization}
\end{equation}
These gauge choices transform the action~(\ref{VV-action-1}) into a
quantum action of the form
\begin{equation}
  S_{{\rm q}} = \int_{\Sigma_{g}} B^{0}\partial_{z}\partial_{\overline{z}}
  \Phi + (b^{z\overline{z}}\partial_{\overline{z}}c_{z} + {\rm c.c.}) +
  \chi^{0}\partial_{z}\partial_{\overline{z}} \psi +
  (B^{z\overline{z}}\partial_{\overline{z}}\phi_{z} + {\rm c.c.}),
\end{equation}
which defines a conformal field theory. This action consists of a (super)
Liouville sector and a $(b,c,B,\phi)$\footnote{Warning! Here $B$ is an
  anti-ghost for $\phi^{\alpha}$, not the Lagrangian multiplier from the
  BF action.} ghost sector for the (super)
diffeomorphisms. In this way it models the moduli space of Riemann surfaces
just as in the metric formalism but the use of first-order variables
includes the dynamical (super) Liouville field.

\section{Observables in Topological Gravity}
All the way back to the first papers on topological gravity
it was expected that the elements of the BRST cohomology classes would be related to
topological invariants of \m. Both in the metric formulation~\cite{LPW}, and in the
traditional super BF approach by Montano and Sonnenschein~\cite{MS}, it was stated
that the observables should be identified with the so-called Mumford-Morita-Miller
classes. But it was first proved topologically by Witten in~\cite{Witten:340}
and the observables were first given explicitly,
in terms of the fields, in the theory by E. and H. Verlinde in~\cite{VV}

This section is organised as follows: First the definition of the observables
is given, then follows the topological definitions
given by Witten and how this fits with the general geometry of Witten type
TFT's and the topological invariants due to Mumford, Morita and Miller. Next follows a
detailed discussion on the relations between the observables, which are of
paramount importance, when we study the relation to two-dimensional
quantum gravity. Finally we give some hints on how to calculate
correlation functions of the observables and discuss other related topics.

\subsection{Definition of Observables}
Consider the BRST algebra in equations~(\ref{d-omega-1},\dots,\ref{d-phi-diff})
and note that the homogeneous part forms a closed subalgebra, on which
$\br^{2}=0$. This is easy to see when one remembers that \br and $d$ anti-commute
\begin{eqnarray}
  \br (\br \omega) &=& \br (\psi_{0} + dc_{0}) = d\phi_{0} -d(\br c_{0}) =
  d\phi_{0} - d\phi_{0} = 0, \\
  \br (\br c_{0} ) &=& \br \phi_{0} = 0 ,\\
  \br (\br \psi_{0} ) &=& \br d \phi_{0} = - d\br(\phi_{0}) = 0 ,\\
  \br(\br \phi_{0}) &=& 0.
\end{eqnarray}
Note that the BRST closed expressions are also BRST exact, which
a priori leads to trivial cohomology. This is expected since we work in the
BRST representation. To get non trivial cohomology
we should constrain ourselves to the basic subcomplex of the Weil algebra,
or work directly in the Cartan representation. From the point of
view of equivariant cohomology
there is no problem with triviality of observables in Witten type TFT's. Going to the Cartan
representation implies setting the Faddeev-Popov ghosts $c_{0}=c^{\alpha}=0$.
This clearly breaks the nilpotence of $\br$, just as expected.
We find explicitly that
\begin{equation}
  \br(\br \omega)\vert_{{\rm Cartan}} = \bc ^{2} = d\phi_{0}.
\end{equation}
The observables should be independent of $c_{0}$, and in order for 
the Cartan differential to be nilpotent the observables should
be gauge invariant, i.e.
BRST invariant. This led E. and H. Verlinde to define~\cite{VV}
\begin{equation}
  \sigma^{(0)} \equiv \phi_{0} \,\,\mbox{ and }\,\,\sn^{(0)}
  \equiv (\phi_{0})^{n}, \label{sigma-n-def}
\end{equation}
which of course are the gravity versions of the observables of topological
Yang-Mills theory. In contrast to topological Yang-Mills
we do not need to write the trace in equation~(\ref{sigma-n-def}),
since the local Lorentz part of $ISO(2)$ is $U(1)$.
Note also that there does not exist a trace on $ISO(2)$, which hinders
writing down gauge invariant observables for the full symmetry group.
The superscript $(0)$ marks the observables as zero-forms.
Using the realization of $\phi_{0}$ from equation~(\ref{phi-relization})
we can determine the observable from the field content of the action.

As in the discussion of topological Yang-Mills, one can construct higher
differential forms, related to the scalar observable via descent equations.
These are for topological gravity
\begin{eqnarray}
  \bc \sigma^{(2)} &=& d \sigma^{(1)}, \label{gravity-descent-1} \\
  \bc \sigma^{(1)} &=& d \sigma^{(0)}, \\
  \bc \sigma^{(0)} &=& 0.  \label{gravity-descent-3}
\end{eqnarray}
They also relate the different forms $\sn^{(0)},\sn^{(1)},\sn^{(2)}$,
and the relations were given by Becchi et al.
in~\cite{Becchi-Collina-Imbimbo-1}
\begin{eqnarray}
  \sn^{(1)}\! &=& \!n (\sn^{(0)})^{n-1} \,\wedge \, \sn^{(1)}, \\
  \sn^{(2)}\! &=&  \! n (\sn^{(0)})^{n-1}\! \wedge \sn^{(2)} + \frac{1}{2}n(n-1)
  (\sn^{(0)})^{n-2}\! \wedge \sn^{(1)} \!\wedge \sn^{(1)} .
\end{eqnarray}
For the BRST algebra given above one finds that
\begin{eqnarray}
  \sn^{(1)} &=& n \, \phi_{0}^{n-1}  \,\wedge \,\, \psi_{0}, \\
  \sn^{(2)} &=& n \, d\omega \,\wedge \, \phi_{0}^{n-1} + \frac{1}{2}n(n-1)\,
  \phi_{0}^{n-2}\wedge 
  \psi_{0}\wedge\psi_{0},
\end{eqnarray}
which can be verified by direct calculation:
\begin{equation}
  d\sn = \, d \phi_{0}^{n}\,  = \, n\,  \phi_{0}^{n-1}\wedge d\phi_{0},
\end{equation}
which is the same as
\begin{equation}
  \bc (n \, \psi_{0}\, \wedge \phi^{n-1}_{0}) = \,n\, d\phi_{0}\wedge\phi_{0}^{n-1}.
\end{equation}
Also
\begin{equation}
  d (n \,\psi_{0}\, \wedge\phi^{n-1}_{0}) = n\, d\psi_{0} \wedge \phi_{0}^{n-1}
  +  n \psi_{0} \wedge \left((n-1)\phi_{0}^{n-2}\wedge d\phi_{0}\right),
\end{equation}
which equals
\begin{eqnarray}
  \bc \bigl( n \, d\omega \, \phi_{0}^{n-1} + \frac{1}{2}n(n-1)\phi_{0}^{n-2}\wedge \psi_{0}
\wedge \psi_{0}\bigr) &=& \nonumber \\
 n d\left(\bc(\omega)\right)\wedge \phi_{0}^{n-1} +
 \frac{1}{2}n(n-1)\phi_{0}^{n-2} \left(
 \bc \psi_{0} \wedge \psi_{0} + \psi_{0}\wedge \bc \psi_{0}\right) &=&
\nonumber \\
 n \, d \psi_{0}\wedge \phi_{0}^{n-1} + n (n-1)\phi_{0}^{n-2}\wedge
 d\phi_{0}\wedge\psi_{0}. && 
\end{eqnarray}
Note that this would not be true if we had used \br instead of \bc, since
we would obtain a term $dc_{0}$ when acting on $\omega$. Note also that
\bc commutes with $d$, since we are restricted to the symmetric subalgebra
of the Weil algebra $W(\underline{g})$. 
It is obvious that special care should be taken when considering
$n=0,1,2$ for $\sn^{(1)}$ and $\sn^{(2)}$, since we get
negative powers of $\sigma^{(0)}$.
We return to this issue later, which is related to the notation of
punctures and the so-called ``picture
changing'' formalism from string theory~\cite{Fridan}\label{picture}.

For the BRST algebra derived in~\cite{LPW}, the general form of
$\sigma^{(i)}$ with $i=\{0,1,2 \}$ was given in~\cite{Becchi-Collina-Imbimbo-1}
and we list it below:
\begin{eqnarray}
\sigma^{(2)} &=& \frac{1}{2} \sqrt{g} R(g) \epsilon_{\alpha\beta}dx^{\alpha}
\wedge dx^{\beta}, \label{sigma-2-def}\\
\sigma^{(1)} &=& \sqrt{g}\epsilon_{\alpha\beta}\left( c^{\beta}R(g) +
  D_{\gamma}(\psi^{\beta\gamma}-g^{\beta\gamma}\psi_{\rho}^{\,\,\rho}) \right)
  dx^{\alpha}, \\
\sigma^{(0)} &=& \!\sqrt{g}\epsilon_{\alpha\beta}\!\left( c^{\alpha}c^{\beta}R(g)
  + c^{\alpha}D_{\gamma}( \psi^{\beta\gamma} - \!g^{\beta\gamma}\psi_{\rho}^{\,\,\rho})
  + D^{\alpha}\phi^{\beta} - \frac{1}{4}
  \psi_{\rho}^{\alpha}\psi^{\beta\rho}\right) .
\end{eqnarray}
We will not use this representation, but the formulation in~\cite{VV}
is in correspondence with~\cite{Becchi-Collina-Imbimbo-1}. Note that the
two-form version is defined as the Euler class. Also note that parallel with the
discussion in topological Yang-Mills theory, only the zero-form $\sigma^{(0)}$
is BRST invariant and hence a closed form. For the one-form $\sigma^{(1)}$, the integral
around a closed loop $\gamma$ (a one-cycle) is BRST invariant
\begin{equation}
  \{ Q, \oint_{\gamma} \sigma^{(1)} \} = 0, \label{sigma-etform-inv}
\end{equation}
in addition to the surface integral of the two-form $\sigma^{(2)}$
\begin{equation}
  \{ Q, \int_{\Sigma_{g}} \sigma^{(2)} \} = 0. \label{sigma-toform-inv}
\end{equation}
Equations~(\ref{sigma-etform-inv},\ref{sigma-toform-inv}) generalize to
$\sigma_{n}^{(1)}$ and $\sigma_{n}^{(2)}$.
The integrated one-form is BRST invariant, but fails to be invariant under diffeomorphisms
unless the closed loop is a boundary of the Riemann surface $\Sigma_{g}$~\cite{Huges-Montano,BS:TG}.
We have assumed our surfaces to be without boundaries and we may only use the zero-forms or the
integrated two-forms as observables of topological gravity. It is possible to
consider a restricted part of the diffeomorphism group known as the Torelli group,
which has a trivial action on the first cohomology group~\cite{jiggle}.
In the next chapter we return to the possibility of boundaries and
one-form observables.

We now present the mathematical definition of certain topological invariants
on the \m, due to Mumford, Morita and Miller. Mumford's work~\cite{Mumford}
relies heavily on algebraic geometry and has not been studied by the author. But
the invariants also have a geometric definition in terms of the universal
bundle for the action of the orientation preserving diffeomorphisms
$\diff^{+}$ on a genus $g$ Riemann surface $\Sigma_{g}$. This is due to
Miller~\cite{Miller} who gave the definition and to Morita~\cite{Morita} who showed that
the invariants are non-trivial.

The Mumford-Morita-Miller invariants are characteristic classes of a
certain surface bundle.
In general a surface bundle is a differentiable fibre bundle 
\begin{equation}
  \pi: E \mapsto X,
\end{equation}
with fibres being a closed orientable genus $g$ surface
$\Sigma_{g}$. It is common to consider $g\geq 2$ for simplicity.
Let $T\Sigma_{g}\subset TE$ be the subset of the tangent bundle of $E$, tangent
to the fibres. This space is assumed to be oriented and the bundle $E\mapsto X$
is thus also oriented. The Euler class can be defined on this bundle and it
is an element $e=e(T\Sigma_{g})\in H^{2}(E,\Z)$. By integration over the fibre,
the Euler class can be viewed as an element on $X$
\begin{equation}
  e_{i} = \pi_{*}(e^{i+1})\in H^{2m}(X,\Z),
\end{equation}
where $e^{i+1}$ is given by the $(i+1)$'th cup product~\cite{Karoubi}
\begin{equation}
  e^{i+1} \equiv \overbrace{  e \wedge\dots \wedge e}^{\mbox{$(i+1)$ times}}.
\end{equation}
By fixing a Riemannian fibre metric on $T\Sigma_{g}$, one induces a metric on
each fibre $\pi^{-1}(x)$ for $x\in X$. Since each fibre is an oriented two-dimensional
manifold, we know from chapter 1 that the metric induces a complex structure and
we can equivalently view the
surface bundle as a complex line bundle $\eta(\Sigma_{g})$ with the fibres
being one dimensional complex manifolds, namely Riemann surfaces $\Sigma_{g}$.
We define the $i$-th Chern Class $c_{i}$ of this complex line bundle
$\eta(\Sigma_{g})$ as an element in the cohomology class
\begin{equation}
  c_{i}=c_{i}(\eta(\Sigma_{g})) \in H^{2i}(X,\Z).
\end{equation}
For the definition of the Mumford-Morita-Miller invariants the relevant surface bundle is
the universal
bundle for the action of the orientation preserving diffeomorphisms $\diff^{+}$ on
$\Sigma_{g}$
\begin{eqnarray}
   & E\diff^{+}& \nonumber \\ &\downarrow& \\ &B\diff^{+}&
   \nonumber .
\end{eqnarray}
This bundle is known as the universal $\Sigma_{g}$-bundle. Recall that the
Mapping Class Group $\Gamma_{g}$ is the quotient of $\diff /{\rm Diff}_{0}(\Sigma_{g})$, i.e.
$\Gamma_{g}=\pi_{0}(\diff^{+})$.
It is known~\cite{Miller} that there exists an isomorphism between the 
cohomology classes
\begin{equation}
  H^{*}\left(B\diff^{+},\Z\right) \simeq H^{*}\left( B\Gamma_{g},\Z\right).
\end{equation}
Since $\m = \tich/\Gamma_{g}$ is the quotient of
Teichm\"{u}ller space with the discrete group $\Gamma_{g}$,
which identifies \m as an orbifold, it is possible to establish the following 
isomorphism between rational cohomology groups
\begin{equation}
  H^{*}\left(\m, {\bf Q}\right) \simeq H^{*}\left( B\Gamma_{g},{\bf Q}\right).
  \label{mumford-isomorphism}
\end{equation}
Let $c_{1}\in H^{2}(E\diff^{+})$ be the first Chern class of the complex line
bundle $\eta(\Sigma_{g})$ and define the Mumford-Morita-Miller invariant
as the cohomology class $y_{2n}$ obtained by integration over fibre in the
universal $\Sigma_{g}$-bundle
\begin{equation}
  y_{2n}\equiv \pi_{*}(c_{1}^{n+1})\in H^{2n} \label{Mumford-1}
  \left( B\Gamma_{g},{\bf Q}\right).
\end{equation}
By the isomorphism in equation~(\ref{mumford-isomorphism}) this defines analogous classes
\begin{equation}
  y_{2n}\in H^{2n}\left(\m,{\bf Q}\right). \label{Mumford-2}
\end{equation}
Just as the observables of topological Yang-Mills theory in equations~(\ref{witten-inv-1},
\dots, \ref{witten-inv-5})
gave rise to the intersection numbers of Donaldson invariants, the observables of
topological gravity are related to the intersection numbers of the
Mumford-Morita-Miller invariants.
Baulieu and Singer~\cite{BS:TG} have presented an alternative action of topological gravity,
in addition to those given by (LPW) and E. and H. Verlinde. We will not discuss this action
since it just adds to the confusion, but we present Baulieu and Singer's explanation
on how the observables of topological gravity can be made to fit with the picture
derived in chapter 2. This approach builds on both the metric and first order formalism and
is therefore of general interest.
Let $B$ be the fibre bundle of oriented frames on $\Sigma_{g}$ and let $\tilde{Q}\subset
B\times\met$ be the submanifold of $B\times \met$ consisting of the oriented frames
which are compatible to the metric $g\in \met$. This forms a principal $SO(2)$ bundle
\begin{eqnarray}
  &\tilde{Q}& \nonumber \\ &\downarrow& \\ & \Sigma_{g}\times \met. \nonumber
\end{eqnarray}
The diffeomorphisms act on $\tilde{Q}$ and 
\begin{eqnarray}
 &\tilde{Q}/\diff& \nonumber \\ &\downarrow& \\ & \left(\Sigma_{g}\times \met\right)/\diff&,
 \nonumber
\end{eqnarray}
is also a principal $SO(2)$ bundle. The base space
${\cal L}\equiv\left(\Sigma_{g}\times \met\right)/\diff$ is almost a surface bundle
\begin{eqnarray}
 & {\cal L}& \nonumber \\ &\downarrow& \\
 &\met/\diff&,
 \nonumber
\end{eqnarray}
with fibre $\Sigma_{g}$.
This is only partially true, since the diffeomorphisms act non-locally, i.e. they 
act both on $\Sigma_{g}$ and on the space of metrics \met. This construction
is actually a so-called orbifold bundle and the fibres are the quotient of
$\Sigma_{g}$ with the automorphism group of the metric $g_{\alpha\beta}$ on $\Sigma_{g}$.
If we restrict ourselves to the situation of surfaces with genus $g\geq 2$, we apply the
knowledge from chapter 1 where we learned that such surfaces can be mapped into 
surfaces with constant negative curvature $R=-1$, by conformal transformations.
We form the restricted total space of the orbifold bundle
\begin{equation}
  {\cal N} = \Sigma_{g} \times \met_{R=-1}/\diff \subset {\cal L},
\end{equation}
from which the orbifold bundle reduces to an orbifold bundle over the finite dimensional
moduli space
\begin{eqnarray}
  &{\cal N}& \nonumber \\ &\downarrow& \\ &\m&,  
\label{N-bundle}
\end{eqnarray}
with fibre $\Sigma_{g}/{\rm Aut}(g_{\alpha\beta})$.
The full picture can be seen from the following bundle map
\begin{equation}
  \begin{array}{clcclc}
    \tilde{Q} & & && \\[.4cm] \downarrow &\hspace{-1.5cm}\mbox{{\scriptsize $SO(2)$}} & \searrow  &
    \!\!\!\mbox{{\scriptsize \diff}} & \\[.4cm]
    \Sigma_{g}\times \met &&& &\hspace{-1cm} \tilde{Q}/\diff \\[.4cm]
    \downarrow &\hspace{-1.5cm} \!\!\! \mbox{{\scriptsize \diff}} & \swarrow&
    \!\!\! \mbox{ {\scriptsize $SO(2)$}} & \\[.4cm]
    {\cal L} & & && \\[.4cm]
    \downarrow & \hspace{-1.5cm} \!\!\! \mbox{{\scriptsize $\Sigma_{g}/{\rm Aut}(g_{\alpha\beta})$}} && &
    \\[.4 cm] \met/\diff. &&& &
   \end{array}
\end{equation}
As in topological Yang-Mills, there are two different gauge groups,
namely ${\rm Diff}(\Sigma_{g})$ and $SO(2)\simeq U(1)$. One can find a $U(1)$ connection on ${\cal N}$
with curvature ${\tilde F}$, which has three components $F_{(2,0)}$, $F_{(1,1)}$ and
$F_{(0,2)}$ according to the form degree on $\Sigma_{g}$ and $\met_{R=-1}$ respectively.
Baulieu and Singer~\cite{BS:TG} also identified the individual components of the
curvature, from the action they used. In that sense topological gravity is not different
from topological Yang-Mills and the ``map'' between the theories
is $\tilde{Q} \leftrightarrow P \times {\cal A}$, ${\cal L} \leftrightarrow M\times
{\cal A}/{\cal G}$. In topological gravity the
topological ghost is again a horizontal tangent vector to the moduli space and the ghost for
ghost is related to the curvature on moduli space. From the curvature $\tilde{F}$ we define
the first Chern class
\begin{equation}
  c_{1} = \tilde{F} \in H^{2}({\cal N}).
\end{equation}
Apply the cup product to form the classes $c_{1}^{i+1}$ and via integration over
fibres we find
\begin{equation}
  \tilde{n}^{i+1} \equiv \int_{\Sigma_{g}} c_{1}^{i+1}\in H^{2i}(\met/\diff),
\end{equation}
where we ignore the ${\rm Aut}(g_{\alpha\beta})$ correction. By restricting ourselves to
metrics of constant negative curvature $\m \hookrightarrow \met/\diff$ we
obtain the cohomology classes
\begin{equation}
  n^{i+1} = \tilde{n}^{i+1}\vert_{\met_{R=-1}} \in H^{2i}(\m).
\end{equation}
This construction presents one way to build a topological field theory, describing the
Mumford-Morita-Miller invariants.

The forms $n^{i+1}$ will be represented by
the observables $\sigma^{(2)}_{i}$.
The topological gravity version of equation~(\ref{vev-I-gamma}) in chapter 2, is
\begin{equation}
   \lan \int_{\Sigma_{g}} \sigma^{(2)}_{i}  \ra 
    = \int\limits_{\m} n^{i+1}.
\end{equation}
The closed form $\sigma_{i}^{(2)}$ is a $2i$-form on $M$ which we evaluate over the
fundamental homology cycle $[\Sigma_{g}]$, corresponding to an integration over fibres
in ${\cal N}_{g}$. If we extend  
the situation to punctured Riemann surfaces, we can also evaluate the zero-form observable
$\sigma_{i}^{(0)}$ against zero cycles (i.e. the marked points) to obtain
\begin{equation}
 \lan \sigma_{i}^{\,} ( x_{i} ) \ra = \int\limits_{{\cal M}_{g,1}} n^{i}.
\end{equation}

By ${\cal M}_{g,1}$ we mean the moduli space of Riemann surfaces with one puncture. The last
formula is not quite correct, since we should alter the definition of the invariants on moduli
space to the situation with punctures. But before we do that, we stress that one can construct
zero- and two-form observables whose expectation values give the intersection number of
closed forms on moduli space. In addition, the above integral is taken over a non-compact space
and it is necessary to compactify the moduli space in order for the expectation value of the observable to
be finite.  One result of the compactification is that the expectation values will be rational numbers
instead of integers.

\subsection{The Moduli Space of Punctured Riemann Surfaces}
We extend \m and \mb to surfaces with punctures. The moduli space of a Riemann surface
with genus $g$ and $s$ punctures is written as \ms and it is the moduli space of
configurations of $s$-marked points on $\Sigma_{g}$. We always assume that the
marked points are ordered. The Deligne-Mumford-Knudsen compactification 
\mbs is the moduli space of stable curves with $s$ marked points.
There is a rich mathematical structure on \mbs, which has played an important role
for identifying topological gravity and quantum gravity in two dimensions.

An element in \mbs is of form $\left\{\Sigma_{g}, x_{1},x_{2},\dots,x_{s}\right\}$ and
each marked point $x_{i}$ has a cotangent space $T^{*}_{x_{i}}\Sigma_{g}$. If the position
of $x_{i}$ is varied in \mbs, the cotangent spaces will vary holomorphically and hence there
exist $s$ holomorphic line bundles over \mbs
\pagebreak
\begin{eqnarray}
 &{\cal L}_{i}& \nonumber \\ &\downarrow& \\ &\mbs&, \label{linebundle-1}
\end{eqnarray}
with fibres $T_{x_{i}}^{*}\Sigma_{g}$. We also introduce the universal
curve~\cite{Nelson, Witten:SDG} ${\cal C}\mbs \mapsto \mbs$, which is a line
bundle over \mbs, with each fibre being a stable curve. The marked points on $\Sigma_{g}$ are
thus also points in ${\cal C}\mbs$, over $\Delta\subset \mbs$.
This universal curve is smooth (if orbifold points are excluded), since the singularities of
the noded surfaces lies in the fibres and not in ${\cal C}\mbs$ itself. The line bundle
${\cal N}\mapsto \m$ in equation~(\ref{N-bundle}) is an example of such a universal curve, but
where one has not yet compactified moduli space.

The complex dimension of \mbs is $3g -g + s$ and let $(d_{1},\dots,d_{s})$ be a set of
non-zero integers, which sums to the dimension of \mbs
\begin{equation}
  \sum_{i=1}^{s} d_{i} = 3g -3 + s. \label{sum-requirement-1}  
\end{equation}
The observables $\sigma_{n}$ of topological gravity are by construction related to
the cohomology pairing, i.e. intersection number, of the cup product of the first Chern classes
of ${\cal L}_{i}$
\begin{equation}
\lan \sigma_{d_{1}}\dots \sigma_{d_{s}} \ra \sim
\int\limits_{\mbs} \bigwedge_{i=1}^{s} c_{1}({\cal L}_{i})^{d_{i}},
\end{equation}
on \mbs. We denote the intersection number above as $\langle \tau_{d_{1}}\dots \tau_{d_{s}}
\rangle$, where $\tau_{n}$  relates to the definition of $\sigma_{n}$ in~\cite{Witten:340} as
\begin{equation}
  \sigma_{n} \equiv \tau_{n} n!.
\end{equation}
The difference is only important when we compare topological gravity with matrix models, and
we will shift between the $\tau$ and $\sigma$ notation. The requirement in
equation~(\ref{sum-requirement-1}) must be fulfilled for the intersection number to be
non-zero, i.e. when the cup product gives a top form on \mbs. The intersection number will later
be related to the expectation values of such products of local $\sigma_{n}$ observables
in topological gravity.

The Mumford-Morita-Miller classes $y_{2n}$ were defined on \mb, and to relate these to the
intersection numbers on \mbs consider the projection
\begin{equation}
  \tilde{\pi}: \overline{{\cal M}}_{g,1} \mapsto  \overline{{\cal M}}_{g,0},
\end{equation}
known as the forgetful map. This map erases a puncture, which here results in the
ordinary moduli space \mb. The Mumford-Morita-Miller class on
$\overline{{\cal M}}_{g,0}$ is
\begin{equation}
  \kappa_{n} \equiv y_{2n} = \tilde{\pi}_{*}\left( c_{1}({\cal L})^{n+1}\right),
\label{kappa-def}
\end{equation}
where ${\cal L}$ is the line bundle over $\overline{{\cal M}}_{g,1}$, with the fibre being the
cotangent space over the single marked point. The intersection number of these classes is 
\begin{equation}
  \lan \kappa_{n_{1}}\dots \kappa_{n_{s}} \ra \sim
\int\limits_{\mb} \bigwedge_{i=1}^{s} \kappa_{n_{i}}.
\end{equation}
In the simplest case we have
\begin{equation}
  \lan \tau_{n} \ra = \int_{ \overline{{\cal M}}_{g,1}} c_{1}({\cal L})^{n} =
  \int_{\overline{{\cal M}}_{g,0}} \kappa_{n-1} = \lan \kappa_{n-1} \ra, \,\, n\geq 1.
\end{equation}
This indicates that the intersection theory of the two types of classes agree and this can also
be shown~\cite{Witten:SDG}, but for more than one marked point the arguments are quite complicated.
We conclude that the $\tau_{n}$ classes and hence the $\sigma_{n}$ observables are related to the
Mumford-Morita-Miller classes. We return to the relation between the $\tau$ and $\kappa$ clases several
times in this chapter.

\subsection{Recursion Relation between Observables}
\begin{figure}[h]
\begin{center}
\mbox{
\epsfysize4cm
\epsffile{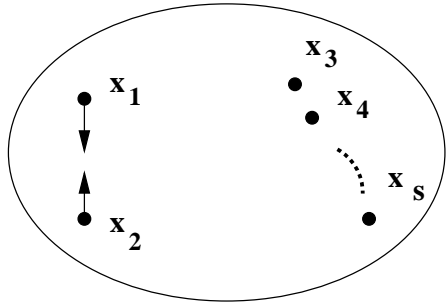}
}
\end{center}
\caption{ \label{approach} Stable degeneration of a punctured surface. Step 1.}
\end{figure}

\begin{figure}[h]
\begin{center}
\mbox{
\epsfysize2.5cm
\epsffile{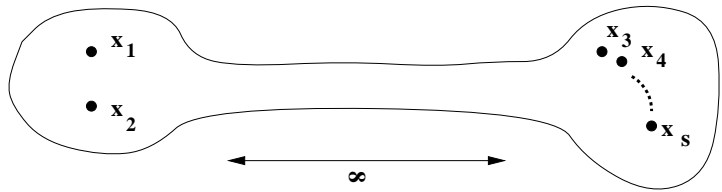}
}
\end{center}
\caption{ \label{approach-1} Stable degeneration of a punctured surface. Step 2.}
\end{figure}

\begin{figure}[h]
\begin{center}
\mbox{
\epsfysize3cm
\epsffile{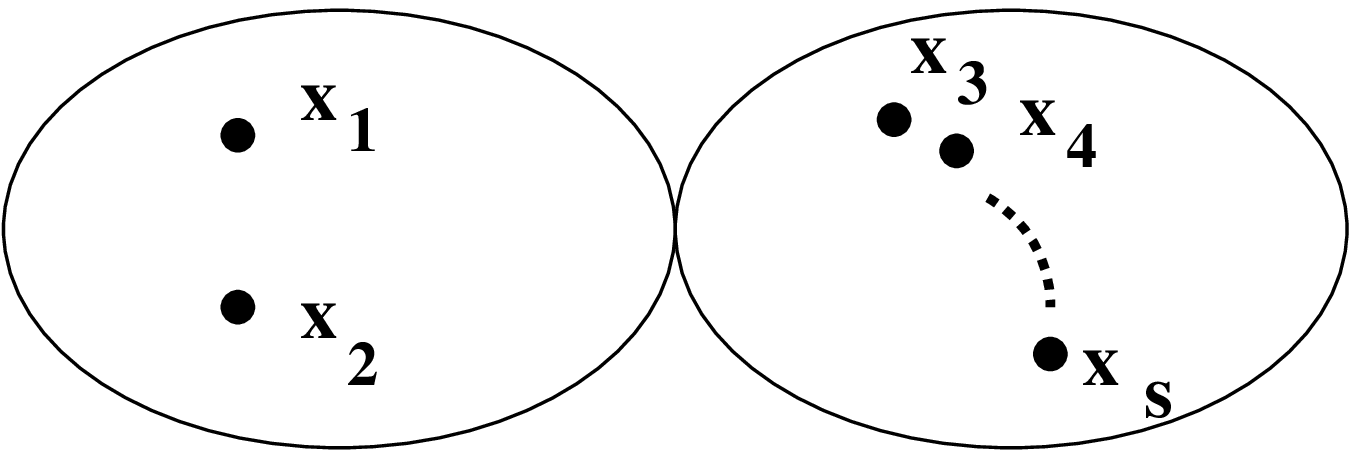}
}
\end{center}
\caption{ \label{approach-2} Stable degeneration of a punctured surface. Step 3.}
\end{figure}

\begin{figure}[h]
\begin{center}
\mbox{
\epsfysize2.5cm
\epsffile{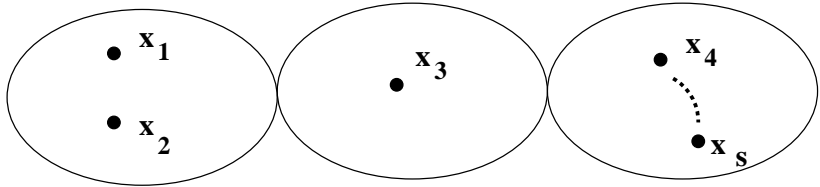}
}
\end{center}
\caption{ \label{approach-3} Stable degeneration of a punctured surface. Step 4.}
\end{figure}

First we study intersection numbers of $\sigma_{n}$ operators in genus zero.
It turns out that these obey a recursion relation, which reduces any correlators to
expressions involving one single correlation function. The geometrical reason for this
lies in the role of the punctures. When studying $\sigma_{n}\equiv \sigma_{n}^{(0)}$ we
needed to evaluate the zero-form at a point. But the compactification of \ms to \mbs,
constrains the choice of position for the marked points. Consider as an example a
genus zero surface with $s$ marked points.
Let two marked points
$x_{1}$ and $x_{2}$ approach each other, while the remaining
marked points $x_{3},x_{4},\dots,x_{s}$ are held at fixed positions, as illustrated
in figure~(\ref{approach}).
If one applies a Mobi\"{u}s transformation, this is identical to the situation in
figure~(\ref{approach-1}) where the distance between
$x_{1}$ and $x_{2}$ is kept fixed, while their distance to the remaining fixed points
goes to infinity. 
By use of the result from chapter 1 that an infinite cylinder is conformally equivalent
to the neighbourhood around a node, we obtain the situation in figure~(\ref{approach-2}).
This is then repeated for the double point and $x_{3}$, to produce a new ``blob''
as illustrated in figure~(\ref{approach-3}). These blobs are identified as
spheres with three marked points, if one remembers that the node is a double point.
We repeat this procedure until we only have spheres with three marked points, which can be
fixed under the $PSL(2,\C)$ automorphism group of the Riemann sphere. These blobs are
said to be stable (or conformally rigid) and they are allowed elements in $\overline{{\cal M}}_{0,s}$.
Note that the nodes are double points and they are counted as a marked point on each branch.
They are the only allowed singularities on the Riemann surfaces.
The Deligne-Mumford-Knudsen compactification tells us how the punctured Riemann
surface degenerates to noded surfaces, if any of the marked points coincide.
In general can one express ${\cal M}_{0,s}$ as the quotient of the space of
punctures $Z_{s} =  \left\{ (x_{1},\dots,x_{s})\in (\CP^{1})^{s-3} \vert x_{i}\ne 0,1,\infty,\,\, x_{i}\neq x_{j}\, \forall i\neq j \right\}$ where $s>3$, and the component-wise action
of the projective group such that~\cite{Takhtajan} 
\begin{equation}
  {\cal M}_{0,s} = Z_{s}/ PSL(2,\C),
\end{equation}
and the first examples are
\begin{equation}
   {\cal M}_{0,s} = \left\{ \begin{array}{cc} \emptyset & s=0, \\ {\rm pt} & s=1,2,3, \\
       \CP^{1}/\{0,1,\infty\} & s=4, \\ \vdots & s=5,6,\dots \end{array} \right. 
\end{equation}
and we see that ${\cal M}_{0,s}$ has complex dimension $s-3$.

Witten derived a very important result in~\cite{Witten:340}, explaining how 
general correlators of observables $\langle\sigma_{d_{1}},\dots,\sigma_{d_{s}}\rangle$ will
reduce to a sum of products of simpler correlators. First the result  is derived in
genus zero and later it is extended to higher genera. The main input to prove these
relations is the knowledge on how the surfaces degenerate if the marked points
coincide. The zero-form observable $\sigma_{d_{i}}(x_{i})$ is evaluated at the
marked point $x_{i}$ and we know that
\begin{equation}
  \lan \prod_{i=1}^{s}\sigma_{d_{i}} \ra = \int\limits_{\mbs} \bigwedge_{i=1}^{s}
  \left(c_{1}({\cal L}_{(i)})^{d_{i}}\cdot d_{i}!\right). \label{intersect-1}
\end{equation}
The first Chern class can be replaced by the Euler class, if we use a real surface bundle
instead of the complex line bundle in the definition of the observables.

The intersection number can also be given in an algebraic geometric formulation, which is
the one Witten used to prove his result. The Chern class $ c_{1}({\cal L}_{(i)})$
can be represented as a subvariety $W_{(i)}$ of \mbs. Let $w$ be a meromorphic section of
the line bundle $ ({\cal L}_{(i)})\mapsto \mbs$ over $x_{i}$, and let $[w=0]$ and $[w=\infty]$
be the divisors of zeros and poles\footnote{
Recall that a (principal) divisor of a meromorphic function $f$ is the assignment of
an integer to every point ${\bf a}\in\Sigma_{g}$: $[f]=\sum_{{\bf a}
  \in\Sigma_{g}}{\rm ord}_{{\bf a}}(f)\, {\bf a}$,
namely the order ${\rm ord}_{{\bf a}}(f)$ which is zero if $f({\bf a})\neq 0$
and $f$ is analytic in ${\bf a}$,
$k$ if $f({\bf a})=0$ with multiplicity $k$, $-k$ for a pole with
multiplicity $k$ and
$\infty$ if $f$ is identically zero. For a simple zero/pole the multiplicity
$k=1$.}. Then the subvariety is given~\cite{Witten:340,HG} as
\begin{equation}
  W_{(i)} \equiv [w=0]-[w=\infty].
\end{equation}

\begin{figure}[h]
\begin{center}
\mbox{
\epsfysize4cm
\epsffile{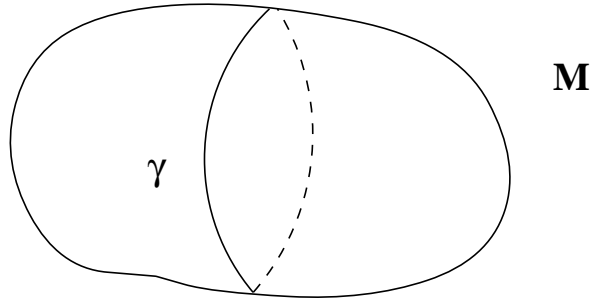}
}
\end{center}
\caption{ \label{jiggle-1} A manifold $M$ with a submanifold $\gamma$.}
\end{figure}

\begin{figure}[h]
\begin{center}
\mbox{
\epsfysize8cm
\epsffile{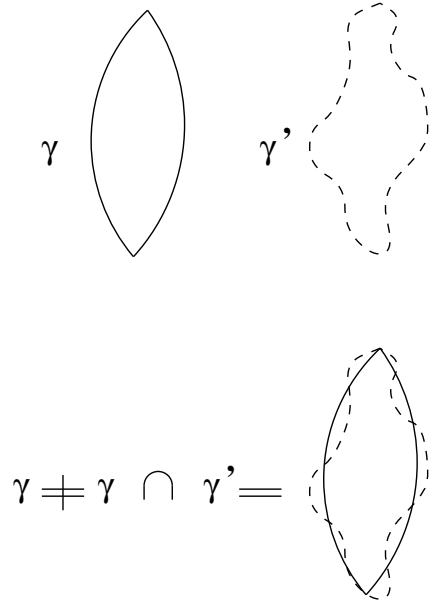}
}
\end{center}
\caption{ \label{jiggle-2} A nontrivial intersection between $\gamma$ and $\gamma'$.}
\end{figure}

\noindent
The cup product $c_{1}({\cal L}_{(i)})^{d_{i}}$ is represented by another subvariety
\begin{equation}
  H_{(i)} \equiv  W_{(i)}^{1} \cap  W_{(i)}^{2} \cap \cdots \cap   W_{(i)}^{d_{i}},
\end{equation}
where each individual term $W_{(i)}^{j}$ for $j\leq d_{i}$ is obtained by taking
$j$ copies of $W_{(i)}$ and perturbing them independently. The reason for
this can be illustrated by the following example. Consider a manifold $M$ and a
submanifold $\gamma$ as in figure~(\ref{jiggle-1}).
The intersection of two copies
of $\gamma$ is trivially just $\gamma$
\begin{equation}
  \gamma\cap\gamma=\gamma.
\end{equation}
If one slightly perturbes a copy of $\gamma$ to obtain a new submanifold $\gamma'$, 
the intersection between $\gamma$ and $\gamma'$ will be nontrivial as illustrated in figure~(\ref{jiggle-2}), even though
$\gamma$ and $\gamma'$ are taken to homologous.
This method covers what Witten calls to 
``jiggle'' the copies of $W_{(i)}$, in order to obtain a nontrivial intersection $W_{(i)}^{j}$.
This method can be made more precise by introducing vector fields on the normal bundle
to the submanifold~\cite{jiggle}.

The intersection number in equation~(\ref{intersect-1}) is then expressed as
\begin{equation}
   \lan \prod_{i=1}^{s}\sigma_{d_{i}} \ra = \#_{\rm intersection}
   \left( H_{(1)} \cap \cdots \cap H_{(s)} \right) \cdot \prod_{i=1}^{s} d_{i}!,
\end{equation}
which makes it possible to find a recursion formula for
$ \langle \sigma_{d_{1}}\cdots\sigma_{d_{i}}\rangle $. This is done by choosing e.g. 
\begin{equation}
  H_{(1)} = W_{(1)}^{1}\cap \left( W_{(2)}^{2}\cap\cdots \cap W_{(i)}^{d_{i}}\right)
  \equiv  W_{(1)}^{1}\cap H_{(1)}^{'}, \label{d-1}
\end{equation}
where the last equality sign defines $H_{(1)}^{'}$ as the intersection of the remaining
$d_{1}-1$ terms. Since the intersection product is associative, the intersection number
can be written in the final form
\begin{equation}
 \lan \prod_{i=1}^{s}\sigma_{d_{i}} \ra = \#_{\rm intersection}
   \left( W_{(1)} \cap ( H_{(1)}^{'}\cap H_{(2)}\cap \cdots \cap H_{(s)}) \right)
   \cdot \prod_{i=1}^{s} d_{i}!.
\label{intersect-2}
\end{equation}
The intersection number can thus be calculated by first restricting to $W_{(1)}$ and then 
calculating the intersection between $( H_{(1)}^{'}\cap H_{(2)}\cap \cdots \cap H_{(s)})$ and
$W_{(1)}$. If $W_{(1)}$ is chosen in the correct way, the intersection number reduces to
calculating a sum of similar intersection numbers where $d_{1}$ is replaced with $d_{1}-1$,
according to the definition given in equation~(\ref{d-1}).

\noindent
The result in equation~(\ref{intersect-2}) was used by Witten to prove the following genus
zero recursion relation:
\begin{equation}
  \lan \prod_{i=1}^{s}\sigma_{d_{i}} \ra = d_{1} \sum_{S=X\cup Y}
  \lan \sigma_{d_{1}-1} \prod_{j\in X}\sigma_{d_{j}} \sigma_{0} \ra 
 \lan \sigma_{0} \prod_{k\in Y}\sigma_{d_{k}} \sigma_{d_{s-1}}\sigma_{d_{s}} \ra.
\label{genus-zero-recursion}
\end{equation}
We now explain the notation and outline the proof. In the discussion on page~\pageref{picture}
we gave a warning regarding taking the first powers $(n=0,1,2)$ of the two-form versions
$\sigma_{n}^{(2)}$. The simplest zero-form operator
\begin{equation}
  \sigma_{0}(x) = (\phi(x))^{0} \equiv P \neq 1,
\end{equation}
is not trivial but instead defined to be the puncture creating operator, which inserts a
marked point at $x\in\Sigma_{g}$. We use $\sigma_{0}$ and $P$ interchangeably
for the puncture operator. When we consider a Riemann sphere with $s$ marked points,
we know that we can choose three points, say $x_{1},x_{s-1},x_{s}$, to be fixed under the
automorphism group. To be more precise, Witten considered an explicit section ${\cal Q}$ of the
line bundle ${\cal L}_{(1)}$
\begin{equation}
  {\cal Q} = dx_{1}\left( \frac{1}{x_{1}-x_{s-1}} -  \frac{1}{x_{1}-x_{s}}\right).
\end{equation}
This section has no poles or zeros on the dense open subset ${\cal M}_{0,s}$ of
$\overline{{\cal M}}_{0,s}$. But there might be poles or zeros on the
stable curves $\Delta = \overline{{\cal M}}_{0,s} \setminus {\cal M}_{0,s}$. Let
\begin{equation}
  w = dx\left( \frac{1}{x-x_{s-1}} -  \frac{1}{x-x_{s}}\right),
\end{equation}
be a meromorphic one-form on $\Sigma_{0}$.
The form $w$ will only have poles at $x_{s-1}$
and at  $x_{s}$ with residues plus and minus one, and no zeros on a smooth Riemann surface
since the points $x_{s-1}$ and $x_{s}$ are distinct.
A meromorphic one-form always
has total residue zero on a compact Riemann surface~\cite{FK}. Hence there
will be no zeros or poles for $w$ if $x \neq \{x_{s-1}, x_{s}\}$ on $\CP^{1}$.
If one considers a holomorphic form on a
Riemann surface, which then degenerates into two branches, the form will pick up a pole at
the node and the residues will be equal but of opposite sign on the two branches. This
form of poles are not counted in $[w=\infty ]$.

On a stable curve $\Sigma_{0,s}$ with two branches $\Sigma_{X},\Sigma_{Y}$, the one form $w$ is
a pair of one-forms $w_{1},w_{2}$, one on each branch, which each has a simple pole
with equal and opposite residues. This definition ensures that there exists a unique one-form
$w$ with poles at $x_{s-1}$ and $x_{s}$ and no zeros on the branches containing these two
points. If on the other hand $x_{1}$ is on one branch $\Sigma_{X}$ and $x_{s-1},x_{s}$ on
the branch $\Sigma_{Y}$, then ${\cal Q}$ will be identically zero at $\Sigma_{X}$ since
the only pole is at the double point, and this is in conflict with the total residue being zero
on each branch. We can obtain ${\cal Q}$ by evaluating $w$ at $x=x_{1}$ on this degeneration,
which results in ${\cal Q}$ having no poles in $\overline{{\cal M}}_{0,s}$. But ${\cal Q}$
vanishes
on this degeneration, which we denote $\Delta_{X,Y}$.
Denote the remaining points $S=\{ x_{2},\dots,x_{s-2}\}$. The divisor
of zeros $[{\cal Q} =0]$ is
a sum of components $\Delta_{X,Y}$, where $S=X\cup Y$ is a decomposition of $S$ into
disjoint subsets.
On the $X$ branch we have $r+2$ marked points namely $x_{1}$, the node and $r$ points from the
set $S$. On the $Y$ branch we have $s-r$ points namely $x_{s-1}$, $x_{s}$, the node and
$s-3 - r$ marked points from the set $S$ and we find 
\begin{equation}
  [{\cal Q}=0] = \sum_{S=X\cup Y} \Delta_{X,Y}.
\end{equation}
From the relation between holomorphic line bundles and divisors in algebraic geometry, one
can express the first Chern class of ${\cal L}_{(1)}$ as the divisor of zeros for
$W_{(1)}$\cite{Witten:340,HG}
\begin{equation}
  c_{1} \sim W_{(1)} = [{\cal Q}=0].
\end{equation}
The stable curve is isomorphic to the Cartesian product~\cite{Witten:340}
\begin{equation}
  \Delta_{X,Y} \simeq {\cal M}_{0,r+2}\times {\cal M}_{0,s-r},
\end{equation}
and the intersection number $\langle \sigma_{d_{1}}\cdots \sigma_{d_{s}} \rangle$ on
$\Sigma_{0,s}$ may be calculated by restricting to $W_{(1)}$ and then counting the
intersection points in $H^{'}_{(1)}\cap H_{(2)}\cap\dots\cap H_{(s)}$. The subvariety $W_{(1)}$
has the form
\begin{equation}
  W_{(1)} = \bigcup_{S=X\cup Y} \Delta_{X,Y},
\end{equation}
so the result will be a sum of terms, associated to each of the degenerated surfaces $\Delta_{X,Y}$
and have the form of the product
\begin{equation}
   \lan \sigma_{d_{1}-1} \prod_{j\in X}\sigma_{d_{j}} P(x^{*}) \ra 
 \lan P(x^{*}) \prod_{k\in Y}\sigma_{d_{k}} \sigma_{d_{s-1}}\sigma_{d_{s}} \ra,
\end{equation}
where the node is the fixed marked point $x^{*}$ created by the puncture operators on
each branch.
This outlines the idea behind the proof Witten gives, but the details are rather technical and
some issues, like whether this results holds after compactification, are not treated in this presentation. Since the operator $\sigma_{d_{1}}$ is replaced by a sum of terms including $\sigma_{d_{1}-1}$,
the recursion relation can be used to reduce any correlators
$\langle \sigma_{d_{1}}\cdots \sigma_{d_{s}} \rangle$ to a sum of correlators
involving only puncture operators. The dimensional requirement in
equation~(\ref{sum-requirement-1}), shows that the only non-zero possibility on a genus zero
surface is $\langle PPP\rangle$. The recursion relation implies
\begin{eqnarray}
  \lan \sigma_{d_{1}} \sigma_{d_{2}} \sigma_{d_{3}}\ra &=&
  \sum_{S=X\cup Y} d_{1}\lan \sigma_{d_{1}-1}P\ra\lan P \sigma_{d_{2}}\sigma_{d_{3}}\ra
  \nonumber \\
  &=&  d_{1}\lan \sigma_{d_{1}-1}P\ra\lan P \sigma_{d_{2}}\sigma_{d_{3}}\ra.
\end{eqnarray} 
There is only one decomposition since $S$ is empty. Applying the recursion relation
repeatly until $d_{i}-n\leq 0$ for $i=1,2,3$ and $n\in\Nat$, we end with only one type of
intersection number namely $\langle PPP \rangle$, since we define $\sn=0$ for $n<0$.

It might seem that there is not much to topological gravity if every genus zero correlator
reduces to products of $\langle PPP \rangle$. When we extend to perturbed topological gravity,
other genus zero correlators can exist, and we discuss these in the next chapter.
The fact that there exist three conformal killing vectors allowing us to
fix three points on $\CP^{1}$, but no moduli,
is illustrated by the fact that the correlator $\langle PPP \rangle$ is the only non-vanishing
contribution from a general
$\langle \sigma_{d_{1}}\cdots\sigma_{d_{s}}\rangle$ correlator.

The situation is slightly more complicated for genus one. Here again a recursion relation
can be derived and all correlators can be reduced to one of two different ``building blocks''.
We only sketch the proof for the following relation given by Witten in~\cite{Witten:340}
\begin{equation}
   \lan \prod_{i=1}^{s}\sigma_{d_{i}} \ra_{1} = \frac{1}{24}d_{1}
    \lan \sigma_{d_{1}-1}  \prod_{i=2}^{s}\sigma_{d_{i}} PP\ra_{0} +
    d_{1}\!\!\!\sum_{S=X\cup Y} \lan \sigma_{d_{1}-1} \prod_{j\in X}\sigma_{d_{j}}P\ra_{0}
     \lan P  \prod_{k\in Y}\sigma_{d_{k}}\ra_{1}.
\label{genus-one-recursion}
\end{equation}
We indicate by $\langle \dots \rangle_{1}$ and $\langle \dots \rangle_{0}$ correlators in
genus one and zero respectively. It is again the degeneration of punctured surfaces in the
Deligne-Mumford-Knudsen compactification scheme that determines how a general correlator
splits into the form above. A punctured genus one surface can either degenerate as a
pinched torus resulting in a genus zero surface with a node, which is represented by the
first term of equation~(\ref{genus-one-recursion}), or it can pinch off a sphere if any of
  the punctures approaches each other.
The first term represents an element in $\Delta_{0}$, while the last term, describing the 
process where a sphere is pinched off, represents an
element in $\Delta_{1}$. Witten used algebraic geometry to prove
equation~(\ref{genus-one-recursion}) by choosing a meromorphic section for the line bundle
${\cal L}_{(1)}$ over $\overline{{\cal M}}_{1,s}$ and calculated the first Chern class
via the same techniques as used in genus zero. The numerical factor of $1/24$ is
explained via the details of the proof and we do not enter this discussion. With the
topological arguments Witten used to prove these relations, he could not give results
for arbitrary genus, but E. and H. Verlinde~\cite{VV} derived such a result
from their approach to topological gravity. We first discuss the so-called puncture
equation which plays a special role in the following.

\begin{figure}[h]
\begin{center}
\mbox{
\epsfysize2.9cm
\epsffile{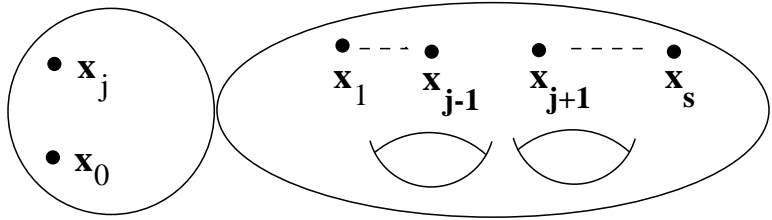}
}
\end{center}
\caption{ \label{stable} An element in $[D_{j}]$ which becomes unstable if $x_{0}$ is
  forgotten.}
\end{figure}

Consider the projection via the forgetful map $\tilde{\pi}:\overline{{\cal M}}_{g,s+1} \mapsto
\overline{{\cal M}}_{g,s}$ and define the following two line bundles
\begin{eqnarray}
 & {\cal L}_{(j)}& \nonumber \\ &\downarrow& \\  &\overline{{\cal M}}_{g,s+1},&\nonumber 
\end{eqnarray}
and
\begin{eqnarray} 
  &{\cal L}^{'}_{(j)}&\nonumber \\ &\downarrow& \\ 
  &\overline{{\cal M}}_{g,s}.& \nonumber 
\end{eqnarray}
The relation between these two bundles is significant in understanding why
$\langle \tau_{n} \rangle = \langle \kappa_{n-1} \rangle$. An important result is
that the first Chern class on ${\cal L}_{(i)}$ is not just the pull back of
${\cal L}^{'}_{(i)}$ via $\tilde{\pi}$. The correct result is~\cite{Witten:SDG,DW}
\begin{equation}
  c_{1}\left(  {\cal L}_{(j)} \right) = \tilde{\pi}^{*} \left( c_{1}\left(  {\cal L}^{'}_{(j)} \right)
    \right) + (D_{j}). \label{chern-divisor-1}
\end{equation}
Here $(D_{j})$ is the cohomology class which is dual to the divisor $[D_{j}]$. This divisor
represents the stable curves of the type illustrated in figure~(\ref{stable}) where there
is a noded surface with a genus zero branch containing the points $x_{j}$, $x_{0}$ and the node.
If the marked point $x_{0}$ is forgotten under the projection
$\tilde{\pi}:\overline{{\cal M}}_{g,s+1} \mapsto \overline{{\cal M}}_{g,s}$, the genus
zero branch
becomes unstable since there are only two marked points left. This is not an allowed element
in $ \overline{{\cal M}}_{g,s}$ and the genus zero branch is collapsed into a point.
The forgetful map $\tilde{\pi}$ induces a map between universal curves $\tilde{\pi}_{{\cal C}}:
{\cal C}\overline{ {\cal M}}_{g,s+1} \mapsto {\cal C}\overline{ {\cal M}}_{g,s}$. This map will
not just forget $x_{0}$, but must encode that resulting unstable curves are collapsed and due 
to this possibility $\tilde{\pi}_{{\cal C}}$ is not a fibration. The Chern class in
equation~(\ref{chern-divisor-1}) will receive contributions from the surfaces $(D_{j})$
of the type in figure~(\ref{stable}).
If the second term in equation~(\ref{chern-divisor-1}) was not present, the intersection
number
\begin{equation}
 \lan \sigma_{0}\prod_{j=1}^{s}\sigma_{d_{j}}\ra = \int\limits_{\overline{{\cal M}}_{g,s+1}}
 1 \cdot \bigwedge_{j=1}^{s} 
c_{1}\left(  {\cal L}_{(j)} \right)^{d_{j}} d_{j}! =
 \int\limits_{\overline{{\cal M}}_{g,s}} \tilde{\pi}^{*}\left(  \bigwedge_{j=1}^{s}  c_{1}\left(
     {\cal L}^{'}_{(j)} \right)^{d_{j}} d_{j}!\right)=0,
\end{equation}
would vanish since the pull back from $ \overline{{\cal M}}_{g,s}$ can never be a top form
on $ \overline{{\cal M}}_{g,s+1}$. The factor $1$ in the first line is due to the
term $c_{1}({\cal L}_{0})^{0}$. Only when the degenerated surfaces $(D_{j})$ are considered
will this intersection number be non-vanishing. When one evaluates the intersection number
$ \langle \sigma_{0}\prod_{j=1}^{s}\sigma_{d_{j}}\rangle$ using both terms in
equation~(\ref{chern-divisor-1}), one integrates over the positions of the marked points
including that of $x_{0}$. There are
$s$ possible contributions and we consider a single term below. The $n$'th cup product
of the Chern class can be expressed as
\begin{equation}
  c_{1}\left( {\cal L}_{(j)}\right)^{n} = \left[ \tilde{\pi}^{*}\left( c_{1}\left(
     {\cal L}^{'}_{(j)} \right)\right)   + (D_{j}) \right] \wedge  
c_{1}\left( {\cal L}_{(j)} \right)^{n-1}. \label{chern-divisor-2}
\end{equation}
The term
\begin{equation}
(D_{j})\wedge  c_{1}\left( {\cal L}_{(j)}\right)^{n-1} = 0,
\end{equation}
vanishes since  $(D_{j})$ corresponds to surfaces where $x_{j}$ is a marked point on a
genus zero branch with three marked points. This branch is conformally stable and the fibre
$T_{x_{j}}^{*}\Sigma_{0,3}$ over $D_{j}$ in the universal curve is a fixed vector space.
There are no moduli to vary and the restriction of ${\cal L}_{j}$ to $[D_{j}]$ is a
trivial vector bundle with vanishing first
Chern Class (recall that the characteristic classes in general are the obstructions
hindering a fibre bundle from being trivial). As long as $n-1>0$, $(D_{j})$ will not contribute to
$c_{1}({\cal L}_{(j)})^{n}$, but for $n=1$ the intersection number 
$ \langle \sigma_{0}\prod_{j=1}^{s}\sigma_{d_{j}}\rangle$ is evaluated
using $(D_{j})$ where one forgets the position of $x_{0}$ which means that
$\tilde{\pi}^{*}( c_{1} ({\cal L}^{'}_{(j)}))$ does not contribute and we replace
$d_{j}$ with $d_{j}-1$
according to equation~(\ref{chern-divisor-2}).
The intersection number reduces to
\begin{equation}
\lan \sigma_{0} \prod_{i=1}^{s} \sigma_{d_{i}} \ra = \sum_{j=1}^{s}\,\,
\int\limits_{\overline{ {\cal M}}_{g,s+1}}
(D_{j})\wedge \bigwedge_{i=1}^{s} c_{1}( {\cal L}_{i})^{d_{i}-\delta_{ij} },
\end{equation} 
and integration over the fibre of $\tilde{\pi}: \overline{{\cal M}}_{g,s+1}\mapsto
\overline{{\cal M}}_{g,s}$ gives the result
\begin{equation}
 \lan \sigma_{0}\prod_{j=1}^{s}\sigma_{d_{j}}\ra
 = \sum_{i=1}^{s} d_{i} \lan \prod_{j=1}^{s}\sigma_{d_{j}-\delta_{ij}} \ra .
\label{puncture-eq}
\end{equation}
This is known as the puncture equation and was first proved Dijkgraaf and Witten in~\cite{DW}
by arguments similar to those given here, which are due to Deligne.
Not all details are given here and further information can be found in~\cite{Witten:SDG,DW}.
Note that every correlator involving $\sn$'s with $n=0$ can be reduced to a sum of correlators with
one observable less. This is a very important result in topological gravity.

It is now possible to show how the $\tau_{n}$ and $\kappa_{n}$ classes are related, based on
the arguments from the previous discussion. We show how Witten could derive 
\begin{equation}
  \lan \tau_{d_{1}}\tau_{d_{2}} \ra = \lan \kappa_{d_{1}-1}\kappa_{d_{2}-1}\ra +
                                     \lan \kappa_{d_{1}+d_{2}-2} \ra.
\label{tau-kappa-2}
\end{equation}
Consider the following example of the forgetful map
\begin{equation}
  \tilde{\pi}_{i}: \overline{{\cal M}}_{g,2}\mapsto \overline{{\cal M}}_{g,1}^{(i)},
\end{equation}   
such that $\tilde{\pi}_{1}$ forgets $x_{1}$ and $\tilde{\pi}_{2}$ forgets $x_{2}$. The moduli
space $\overline{{\cal M}}_{g,1}^{(i)}$ is the one where $x_{i}$ is deleted. We need to consider
the following line bundles:
\begin{eqnarray}
  &{\cal L}_{(i)}& \nonumber \\ &\downarrow& \,\, \mbox{ with fibres: } T_{x_{i}}^{*}\Sigma_{g,2}
  \\ &\overline{{\cal M}}_{g,2},& \nonumber
\end{eqnarray}
and (due to the action of $\tilde{\pi}_{i}$)
\begin{equation}
  \begin{array}{ccc} {\cal L}^{'}_{(1)}&\hspace{2cm} & {\cal L}^{'}_{(2)} \\ \downarrow &
    \hspace{2cm} & \downarrow \\
    \overline{{\cal M}}_{g,1}^{(1)} &\hspace{2cm} &  \overline{{\cal M}}_{g,1}^{(2)},
  \end{array}
\end{equation}
with fibres $T_{x_{i}}^{*}\Sigma_{g,1}$ for $i=1,2$ respectively. According to
equation~(\ref{chern-divisor-1}) we have
\begin{equation}
  c_{1}( {\cal L}_{(2)} ) = \pi_{2}^{*}\left( c_{1}({\cal L}_{(2)}^{'})\right) + (D_{2}),
\end{equation}
where $(D_{2})$ corresponds to  the divisor $[D_{2}]$ in $\overline{{\cal M}}_{g,2}$
with a genus zero branch with two marked points $x_{1},x_{2}$ and a node separating the
branch from the rest of the genus $g$ surface. As long as $d_{i}>0$, ${\cal L}_{(i)}$
will be trivial when restricted to $[D_{i}]$ and one finds
\begin{equation}
  \lan \tau_{d_{1}}\tau_{d_{2}} \ra = \int\limits_{\overline{{\cal M}}_{g,2}}
  c_{1}({\cal L}_{(1)})^{d_{1}}\wedge \tilde{\pi}_{2}^{*}\left( c_{1}({\cal L}_{(2)}^{'})^{d_{2}}\right).
  \label{tau-tau-1}
\end{equation}
Similar to equations~(\ref{chern-divisor-1},\ref{chern-divisor-2}), we express the
$d_{1}$'th cupproduct of the first Chern class as
\begin{equation}
  c_{1}({\cal L}_{(1)})^{d_{1}} = \left( \tilde{\pi}^{*}_{1}\left( c_{1}({\cal L}_{(1)}^{'})\right)^{d_{1}}
  + (D_{1}) \right) \wedge \tilde{\pi}_{1}^{*}\left( c_{1}({\cal L}_{(1)}^{'})\right)^{d_{1}-1},
\end{equation}
which implies that equation~(\ref{tau-tau-1}) changes to
\begin{eqnarray}
  \lan \tau_{d_{1}}\tau_{d_{2}} \ra &=& \int\limits_{\overline{{\cal M}}_{g,2}}
 \tilde{\pi}_{1}^{*}\left( c_{1}({\cal L}^{'}_{(1)})\right)^{d_{1}}\wedge \tilde{\pi}_{2}^{*}
 \left( c_{1}({\cal L}^{'}_{(2)})\right)^{d_{2}}\nonumber \\ &&+ 
 \int\limits_{[D]\simeq \overline{{\cal M}}_{g,1}} \tilde{\pi}_{1}^{*}\left(
   c_{1}({\cal L}^{'}_{(1)})\right)^{d_{1}-1}\wedge \tilde{\pi}_{2}^{*}
 \left( c_{1}({\cal L}^{'}_{(2)})\right)^{d_{2}}.
\end{eqnarray}
According to the definition of the $\kappa$ classes in equation~(\ref{kappa-def}), the first term
gives a contribution $\langle \kappa_{d_{1}-1}\kappa_{d_{2}-1}\rangle$. The second term is a bit more
tricky, since when we restrict ourselves to $(D_{j})$, the pull back of the Chern classes over $x_{1}$ and
$x_{2}$ are equal, since the line bundles ${\cal L}_{1}^{'}$ and  ${\cal L}_{2}^{'}$ are isomorphic
when restricted to $(D)$ and we obtain~\cite{Witten:SDG}
\begin{equation}
   \int\limits_{[D]\simeq \overline{{\cal M}}_{g,1}}  \tilde{\pi}_{1}^{*}\left(
   c_{1}({\cal L}^{'}_{(1)})\right)^{d_{1}+d_{2}-1} = \lan \kappa_{d_{1}+d_{2}-2} \ra.
\end{equation}
This proves equation~(\ref{tau-kappa-2}). Similar, but more complicated results hold for
intersection numbers for three or more individual $\tau$ classes. The difference between the
$\tau$ and $\kappa$ classes is that the individual punctures not are allowed to intersect when
calculating products of $\tau$ classes due to the nature of \mbs, while they may collide for
products of $\kappa$ classes, when
we integrate the position of the puncture over the surface. This gives rise to so-called contact terms
which we discuss from the topological field theory point of view in the following discussion.

\noindent
There is a slight problem with the quantum field theory representation of the observables when we
work with punctured surfaces. The BRST algebra in equations~(\ref{d-omega-1},\dots,\ref{d-phi-diff})
does not respect the fact that the (super) diffeomorphisms do not act at the marked points.
There are two possibilities, either to change the BRST algebra or to change the definition
of the observables, where the latter is the one used in the literature.
The zero-form versions of the observables are redefined as
\begin{equation}
  \sn(x_{i}) \mapsto \sn(x_{i}) c^{\alpha}(x_{i})\delta\left(\phi^{\alpha}(x_{i})\right)
             \equiv \sigma_{n}(x_{i})\cdot P(x_{i}).
\label{sigma-redefine}
\end{equation}
The expectation value
\begin{equation}
  \lan \prod_{i=1}^{s}\sigma_{d_{i}}(x_{i})\ra = \int {\cal D}[B^{0},\Phi,
  \chi^{0},\psi,b^{\alpha\beta},c^{\alpha},B^{\alpha\beta},\phi^{\alpha}]
  e^{-S_{{\rm q}}} \prod_{i=1}^{s}\sigma_{d_{i}}(x_{i})
  c^{\alpha}(x_{i})\delta(\phi^{\alpha}(x_{i})),
\end{equation}
is evaluated by expanding the part of the action $S_{{\rm q}}(c)$ which has terms
depending on $c$, and the rules of grassmann integration
\begin{equation}
  \int \prod_{x\in \Sigma_{g}}\!\!dc^{\alpha}(x) \,\, c(x_{i}) \left( 1 + 
      S_{{\rm q}}(c)\right) =  \int \prod_{x\in \Sigma_{g}\setminus x_{i}}
   \!\! dc^{\alpha}(x) \,e^{-S_{{\rm q}}},
\end{equation}
ensure that the diffeomorphisms do not act at $x_{i}$. The delta functions
$\delta(\phi^{\alpha})$ ensure that the super diffeomorphisms do not act at
points where the diffeomorphisms do not act.
The redefinition in equation~(\ref{sigma-redefine}) also offers an explanation of
why $\sigma_{0}\neq 1$, since
\begin{equation}
  \sigma_{0}(x_{i}) = (\phi_{0}(x_{i}))^{0}\cdot c^{\alpha}(x_{i})
  \delta\left(\phi^{\alpha}(x_{i})\right)
  = 1\cdot P(x_{i}), \label{P-def}
\end{equation}
such that $(\phi_{0}(x_{i}))^{0}=1$ as expected. This expression for the puncture
operator relates to the string theory concept of ``picture changing''. In string theory
vertex operators can be viewed in two different pictures, either as operators which are integrated
over the world-sheet or as operators which are considered at a fixed position. These so-called
zero and minus one pictures also apply for this representation of the observables in topological
gravity. 
If one wants to calculate intersection numbers $\langle \sigma_{d_{1}}\cdots
\sigma_{d_{n}}\rangle$ in the topological field theory, instead of using
topological arguments like Witten, the approach by E. and H. Verlinde has an
advantage. Since the action is conformally invariant, the powerful methods of conformal
field theory can be applied.
If the topological ghost $\psi^{i}$ is viewed as a superpartner to $A^{i}$ and
not just as another ghost due to the enlarged gauge symmetry, methods from 
superstring theory can be applied.
By this approach E. and H. Verlinde define a general correlator
\begin{equation}
\lan \sigma_{n_{1}}\cdots\sigma_{n_{s}}\ra
\equiv \int\limits_{\m}\int\limits_{\widetilde{\m}} e^{-S}\prod_{i=1}^{s} e^{q_{i}B_{0}(x_{i})} \int_{\Sigma_{g}}
\sigma_{n_{i}}^{(2)}, \label{VV-def-1}
\end{equation}
as the integral over commuting moduli \m and anti-commuting $\widetilde{\m}$ moduli
together with integrated two-form versions of the observables. The presence of anti-commuting
moduli relies on the existence of superpartners to the zwei-beins, which corresponds to having
a supermetric. Instead of integrating over anti-commuting moduli one can insert $3g-3$ fermionic
operators $G_{i}$, defined as the convolution between the Beltrami differentials $\mu_{i}$ and
the supercurrents for the superconformal transformations, generated by the superpartner
of the stress-energy tensor. Both the stress-energy tensor and its partner consist of a Liouville-
and a ghost sector. The Beltrami differentials $\mu_{i}$ are dual to the
moduli $m_{i}$, for $i=1,\dots,3g-3$. We do not wish to enter into the details of all these
technicalities from superstring theory, but present the overall ideas.
The detailed form of equation~(\ref{VV-def-1}) written as equation (4.13) in~\cite{VV}, should
include several terms not written, which are only mentioned in the text. The missing
terms are four delta functions restricting the $b,\overline{b},B,\overline{B}$ integrations
to exclude zero modes and ensure the correct overall ghost number. The central point is
that when the puncture operator is inserted at a point, two $b$ and two $B$ zero 
modes are created due to the need of balancing the ghost number budget. These zero modes
relate to two real commuting moduli together with two real anti-commuting moduli.
The first two arise when fixing the point under diffeomorphisms and the last two for fixing the
superdiffeomorphisms. The complex dimension of \ms is $3g-3+s$ where $s$ is the number of
punctures, but the counting of anti-commuting supermoduli parameters are
in excess compared to \ms, since \ms is the space of commuting moduli.
Hence these anti-commuting moduli should be integrated away when performing the path integral
and at the same time the $B$ zero modes must be projected out by insertion of a delta
function. This is what is known as the operation of picture changing. If we were to write
equation~(\ref{VV-def-1}) in the minus one picture, we would include products of
zero-form observables such that 
\begin{equation}
\int\limits_{\mb} \prod_{i=1}^{3g-3}d^{2}m_{i}\dots \prod_{j=1}^{s}\int d^{2}z \, \sigma_{n_{j}}^{(2)}
  \mapsto \int\limits_{\mbs} \prod_{i=1}^{3g-3}d^{2}m_{i}\dots \prod_{j=1}^{s}
  \int d^{2}z \, d\theta d\overline{\theta} \, \sigma^{(0)}_{n_{j}}(x_{j}),
\end{equation}
where $m_{i}$ are commuting moduli parameters.
In the zero picture $z,\overline{z}$ are taken as the integration measure for the
integral over $\Sigma_{g}$, but in the minus one picture
$z,\overline{z}$ are the real moduli for the coordinates of the puncture and
$\theta,\overline{\theta}$ are the supermoduli (or super coordinates of the puncture).
The supermoduli are removed by insertion of a delta function in the so-called picture
changing operator $\Pi(z,\overline{z})$ and the real moduli $z,\overline{z}$ are
included in the integration measure on \mbs
\begin{equation}
 \int\limits_{\mbs} \prod_{i=1}^{3g-3}d^{2}m_{i}\dots
 \prod_{i=1}^{s} \int d^{2}z \sigma_{n_{i}}^{(0)}(x_{i}) =
 \int\limits_{\mbs} \prod_{i=1}^{3g-3}d^{2}m_{i} \prod_{j=1}^{s} d^{2}z_{j}\dots \prod_{k=1}^{s}
 \sigma_{n_{k}}^{(0)}(x_{k}),
\end{equation} 
representing the commuting moduli of the puncture.

The picture changing operator maps effectively $\sigma_{n}^{(0)}$ into $\sigma_{n}^{(2)}$,
by projecting out the $2s$ zero modes in $B$ and $\overline{B}$, leaving us with the
$2s$ commuting moduli which we can interpret as part of the measure on \mbs or as
the measure for the world-sheet integration.
It should be stressed though, that this transformation between $\sigma_{n}$ and $\sigma_{n}^{(2)}$
ignores some complications which may arise due to compactification of moduli space.

Related to the question of picture changing, is the choice of where to insert
the vertex operators $e^{-q_{i}B^{0}(x_{i})}$. If one considers the minus one
picture and expresses the expectation value in terms of the zero-form observables,
then one takes the vertex operators to be at 
the same position as $\sigma_{n}(x_{i})$.
This choice led E. and H. Verlinde to redefine the observable\footnote{E. and H. Verlinde~\cite{VV}
  used a different sign convention for the vertex operators, compared to our definition, such that
  the sign of the total curvature is changed.}
\begin{equation}
  \tilde{\sigma}_{n}(x_{i}) \equiv \sigma_{n}(x_{i})e^{\frac{2}{3}(n-1)\tilde{B}^{0}}(x_{i}).
\label{tilde-sigma-def}
\end{equation}
The $\tilde{B}$ signals that the original vertex operator is replaced by a BRST
invariant operator, whose explicit form can be found in~\cite{VV}.
The expectation value $\langle \sigma_{n_{1}}\cdots\sigma_{n_{s}}\rangle$
is only non-vanishing if the observables form a top form on moduli space, which happens if
\begin{equation}
  \sum_{i}^{s}(n_{i}-1) = 3g-3.
\end{equation}
According to the definition in equation~(\ref{sigma-n-def}) \sn $\,$ should have ghost number
$2n$, but the puncture operator has ghost number $-2$, from the anti-commuting moduli, and
we get a total ghost number $2n-2$. There has been some confusion in the literature regarding
the ghost number of the zero-form observable $\sigma_{n}^{(0)}$ and the two-form observable
$\sigma_{n}^{(2)}$. Witten's approach in~\cite{Witten:340} is that $\sigma_{n}^{(2)}$ has
ghost number $2n-2$ and $\sigma_{n}^{(0)}$ has $2n$, as we also would expect from
the definition of $\sigma_{n}^{(0)}$ and the descent
equations~(\ref{gravity-descent-1},\dots,\ref{gravity-descent-3}). But since the
zero-form is considered at a marked point according to the redefinition in
equation~(\ref{sigma-redefine}), we should subtract $2$ from its
ghost number. Hence the redefined zero-form operator $\sigma_{n}\mapsto \sigma_{n}\cdot P$
has the same ghost number as the two form $\sigma_{n}^{(2)}$.
This confusion is related to the question of whether one should integrate over the position
of the observable or not!

In the author's opinion this situation is unsatisfactory. Many times
one encounters calculations where zero-form and two-form observables are interchanged
during the same calculation and it is difficult to know exactly when one is considering the
one or the other. From the topological point of view advocated by Witten, this
relates to the difference between the $\tau_{n}$ classes and the $\kappa_{n}$ classes.
In that framework one always knows whether one is considering the one or the other type of
cohomology classes, simply from the difference between \m and \ms, but in the
topological field theory representation of these observables the situation is not so clear.

We note that the definition of the two-form observables signals that they may be
identified with the $\kappa$ classes,
just as the $\tau$ classes are related to the zero-form observable. The definition of the
$\kappa$ classes in equation~(\ref{kappa-def}) as the integrated first Chern class, is
equivalent to the integral over the Euler class, which defines $\sigma^{(2)}$ in
equation~(\ref{sigma-2-def}). The only difference lies in whether to view $\Sigma_{g}$ as
a complex or real manifold. The $\kappa_{n}$ and $\sigma_{n}^{(2)}$ are the cup product of
these integrated closed forms. In relation to punctures both the $\kappa_{n}$ and the
$\sigma_{n}^{(2)}$ classes are defined when the punctures are integrated away, which allows
for the appearance of contact terms. In this sense the roles played by these closed forms
are equivalent, but
even though the intersection theory of the $\tau$ and $\kappa$ classes are equal, there are 
relations as in equation~(\ref{tau-kappa-2}) between their intersection numbers.
Therefore when we include the possibility of curves becoming unstable, due to the integration over
the positions of the punctures, we can not have a direct $1-1$ map between $\sigma_{n}$
and $\sigma_{n}^{(2)}$ in the picture changing formalism as advocated in~\cite{VV}.
When we take products of $\sigma_{n}$ or $\tau$ observables the
punctures are not allowed to coincide while they may coincide for products of $\sigma_{n}^{(2)}$ or
$\kappa$ observables due to the integration over their position. We now show how the contact terms
give rise to a relation between the redefined $\sigma_{n}$ observables.

In order to give the recursion relation for general genera, derived in~\cite{VV} one
can consider the ghost number of $\langle \sigma_{n} \rangle$ as $2n-2$, either following the definition in
equation~(\ref{VV-def-1}) or by subtracting $2$ due to keeping the position fixed (i.e.
by being in the minus one picture). Since the
total (integrated) curvature must be $2g-2$, E. and H. Verlinde distributed the curvature
in $\frac{2}{3}(n_{i}-1)$ units at each curvature singularity, in the
exponent of the vertex operator. This explains the definition in
equation~(\ref{tilde-sigma-def}). When one integrates over the position of a
$\sigma_{n_{1}}^{(0)}(x_{1})$ operator there will be contributions when 
$\int dx_{1}$ hits the position of another operator, say
$\sigma_{n_{2}}^{(0)}(x_{2})$. This is what E. and H. Verlinde denote as ``contact term''
contributions. The compactification of Moduli space forbids the points to coincide and the
result of the integration of the $x_{1}$ position in a general correlator is
of the form~\cite{VV}
\begin{equation}
  \lan \tilde{\sigma}_{n_{1}}\prod_{i=2}^{s}\tilde{\sigma}_{n_{i}}\ra_{U_{j}} = \sum_{j=2}^{s}
  \frac{1}{3}(2n_{j}+1)\lan \tilde{\sigma}_{n_{1}+n_{j}-1}\prod_{i\neq j}\tilde{\sigma}_{n_{i}} \ra,
\label{contact-1}
\end{equation}
where $U_{j}$ are local neighbourhoods around the positions of the $x_{j}$ of the
remaining $s-1$ observables. The form of these contributions is very much like that of the
puncture equation~(\ref{puncture-eq}) and the changed numerical pre-factor is due to the
redefinitions of the observables. The result in equation~(\ref{contact-1}) holds for
arbitrary genus. There are also terms which results from integrating over a node, representing
both the degenerations in $\Delta_{0}$ and $\Delta_{i}$. The main result is the recursion
relation for a genus $g$ correlator~\cite{VV}
\begin{eqnarray}
  \lan \tilde{\sigma}_{n+1}\prod_{i=2}^{s}\tilde{\sigma}_{n_{i}}\ra &=& \sum_{j=2}^{s}
  \frac{1}{3}(2n_{j}+1)\lan \tilde{\sigma}_{n + n_{j}}\prod_{i\neq j}\tilde{\sigma}_{n_{i}}\ra
   + \alpha \sum_{k=1}^{n}\Biggl( \beta \lan \tilde{\sigma}_{k-1}\tilde{\sigma}_{n-k}
    \prod_{i=1}^{s}\tilde{\sigma}_{n_{i}}\ra \nonumber \\ &&
    + \sum_{S=X\cup Y}\lan \tilde{\sigma}_{k-1}\prod_{i\in X}\tilde{\sigma}_{n_{i}}
    \ra \lan\tilde{\sigma}_{n-k}\prod_{j\in Y}\tilde{\sigma}_{n_{j}}\ra\Biggr)  .
\label{general-sigma-recursion:TG}
\end{eqnarray}
Several comments should be made at this point. The above correlators are assumed to
have a genus expansion
\begin{equation}
  \langle \cdots \rangle = \sum_{g} \lambda^{2g-2}\langle \cdots \rangle_{g},
\end{equation}
where $\lambda$ is the so-called string-coupling constant. The first term in 
equation~(\ref{general-sigma-recursion:TG}) is the contribution from the contact terms
while the last terms represent the contributions from the $\Delta_{0}$ and $\Delta_{i}$
degenerations. Clearly the last term represents the degeneration of the surface into
contributions from two blobs separated by a node. There will be one of each type
of the three terms per genus. The parameters $\alpha,\beta$ are constants, which include
the value of $\lambda$. The proof of this recursion relation is based on the so-called
contact term algebra, which was derived in~\cite{VV}. From general arguments the
expectation value $\langle \sigma_{n+1}\prod_{i=2}^{s}\sigma_{n_{i}}\rangle$ should not
depend on the orders in which the integrations over the positions of the observables are
done, and this requirement can by fulfilled by letting the contact terms obey a certain algebra,
which actually is a subalgebra of the Virasoro algebra. The proof is not based on actual
computation of the individual terms, but on a mixture of topological arguments and string
theory techniques. It is clear from the form of equation~(\ref{general-sigma-recursion:TG})
that the compactification scheme is central to the derivation of this result. The
notation $\sigma_{n+1}\prod_{i=1}^{s}\sigma_{n_{i}}$ is chosen due to the
connection between this recursion relation and a similar relation for matrix model
observables, which we discuss in the next chapter. 

According to the genus $g$ recursion relation in equation~(\ref{general-sigma-recursion:TG}),
a general
expectation value will reduce to sums of products of two different types of building block,
namely $\langle PPP \rangle_{g=0}$ and $\langle \sigma_{1}\rangle_{g=1}$, where the
genus zero contribution is normalized such that
\begin{equation}
  \lan PPP \ra_{g=0} = 1,
\end{equation}
and the genus one contribution is normalized/calculated to be
\begin{equation}
  \lan \sigma_{1} \ra_{g=1} = \frac{1}{12}. \label{genus-1}
\end{equation}
The genus one contribution signals that there is one modulus together with one
conformal killing vector
in genus one and that one thus can fix the position of one operator, namely
$\sigma_{1}$. This operator is known as the dilaton operator, since it according to the definition
of the expectation value in equation~(\ref{VV-def-1}), is the integral of the Euler class over the
surface. Hence one expects the result
\begin{equation}
  \lan \sigma_{1}\prod_{i=1}^{s}\sigma_{d_{i}} \ra = \left(2g-2+s\right)
  \lan \prod_{i=1}^{s} \sigma_{d_{i}} \ra
  \label{dilaton-eq}
,
\end{equation}
known as the dilaton equation~\cite{Witten:SDG}. This operator measures the (minus)Euler number of the
punctured surface. But for genus one, the special result in equation~(\ref{genus-1}) applies.

If we consider $\sigma_{n}^{(2)}$ to be related to the $\kappa$ classes, we should expect 
relations as in equation~(\ref{chern-divisor-1}) between $\sigma_{n}$ and $\sigma_{n}^{(2)}$
correlators. But it is not clear how to prove such relations, even though
equation~(\ref{general-sigma-recursion:TG}) in principle gives such a result. The reason is
that the definition of the expectation values used in
equations~(\ref{contact-1},\ref{general-sigma-recursion:TG}) is not the same as given in
equation~(\ref{VV-def-1}), but instead of the form
\begin{equation}
  \lan \tilde{\sigma}_{d_{1}}\cdots \tilde{\sigma}_{d_{s}} \ra = \int\limits_{\mb}\int_{\Sigma}d^{2}x_{1}\cdots
  \int_{\Sigma}d^{2}x_{s} \lan \tilde{\sigma}_{n}(x_{1})\cdots \tilde{\sigma}_{n}(x_{s})\ra.
\end{equation}
In this definition the position of the zero-form observable $\tilde{\sigma}_{n}$ is integrated, such that
the measure $d^{2}x_{i}$ is viewed as a world-sheet measure instead of a part of the measure on
moduli space. If the recursion relation instead had been proven using the definition of the
correlators in equation~(\ref{VV-def-1}) we could test the relation in equation~(\ref{tau-kappa-2}).
Since the arguments leading to the recursion relation in equation~(\ref{general-sigma-recursion:TG}) are
rather complicated, it is not obvious how to rewrite it using the definition in equation~(\ref{VV-def-1}).

We conclude this discussion by stressing the similarity between the topological definition of the
Mumford-Morita-Miller classes $\kappa_{n}$ and the integrated two-form version of the observables in
topological gravity. We have not been able to test the relation derived by Witten between $\tau$ and
$\kappa$ classes, in terms of the local observables due to the use of different definitions at crucial
stages in~\cite{VV}. If one could use the two-form versions to calculate the intersection
numbers in the topological field theory, one should be careful to
control the contributions from integration over regions close to the nodes of the
surfaces to avoid infinities. The
emergence of contact terms has also been discussed from a slightly different point of
view in a series of papers by Becchi, Collina and Imbimbo~\cite{BCI,BI-1,BI-2}. In these
papers the appearance of contact terms are related to the Gribov problem. It is argued that the 
closed forms defining the observables fail to be globally defined forms on moduli space, and that
one instead should relate the forms in different local patches of moduli space to define a
global form. This relates also to the question of whether the observables depend on the gauge fixing
we perform to calculate the path integral. By applying so-called \v{C}hec-De Rham cohomology, a
globally defined form replaces the locally defined versions of the observables and the Ward
identities, which relate to the differences between local gauge choices, determine the contact terms between
the local observables. We do not enter these discussions, but instead change to the more
physically interesting theory of perturbed topological gravity and its relation to
two-dimensional quantum gravity in the next chapter.

%% file: chap5.tex
\chapter{ $2D$ Topological Quantum Gravity}
This chapter is devoted to discussing the surprising fact, that
two-dimensional topological gravity, in the form of a perturbation of the
theory presented in the last chapter, and two-dimensional quantum gravity,
in the form of the matrix model representation of non-critical string theory,
are identical in the continuum limit. We discuss this result from various points of view
and present the main results based on the work by Ambj\o rn, Harris and the author~\cite{AHW}.
The chapter is organized as follows. First
the perturbed
theory of topological gravity is introduced and the multi-critical
behaviour of correlators is discussed. The string equation and aspects of
KdV-theory are introduced together with the work of Kontsevich. Next
follows a brief introduction to discrete and continuum aspects of the
matrix model approach to quantum gravity. This includes the Ward identities
and Virasoro constraints, leading to matrix model versions of the recursion
relation derived at the end of the previous chapter. Finally we discuss 
the most important results in~\cite{AHW}.
\section{Perturbed Topological Gravity}
It is possible to perturb or deform pure topological gravity by introducing
a set of parameters $(t_{0},t_{1},\dots)$
representing coupling constants. We can chose any version of the quantum
action for the following discussion and will just denote it $S_{{\rm q}}$.
The only important thing is that the action should be BRST exact. Since the
various observables are BRST closed, it is possible to add one or more 
observables to the action as below
\begin{equation}
  \int e^{-S_{{\rm q}} } \mapsto \int e^{-S_{{\rm q}}+\sum_{i} t_{i}
    \int_{\Sigma} \sigma_{n_{i}}^{(2)}}.
\end{equation}
Since the observables are BRST closed the total action will still be
BRST invariant, but it fails to be BRST exact and therefore does not
satisfy the requirement for defining a Witten type TFT. But the damage is
not so big, since we only add BRST invariant terms. Here we have 
chosen to add the two-form observables, but the zero-form observables are
just as good. Also, at this point there has been some uncertainty regarding
the transformations between the different type of invariants, but this
discussion offers no new insight at this stage. Later we discuss some
new developments in that direction. 

To be in agreement with the conventions in~\cite{AHW} we will use the
following definitions
\begin{equation}
  \lan \prod_{i}\sigma_{d_{i}} \ra \equiv \int {\cal D}[X] e^{-S_{{\rm q}}}
\prod_{i}\sigma_{d_{i}},
\label{TG-QG-def-1}
\end{equation}
and
\begin{equation}
  \lan \prod_{i}\sigma_{d_{i}} \ra_{V} \equiv \int {\cal D}[X]
  e^{-S_{{\rm q}} + V(\{ \sigma\})}
\prod_{i}\sigma_{d_{i}}.
\label{TG-QG-def-2}
\end{equation}
The $V(\{ \sigma\}) $ symbolise the notation of a potential added to the action, with
\begin{equation}
  V(\{ \sigma\})  = \sum_{n=0}^{\infty} t_{n}\sigma_{n},
\end{equation}
being the most general choice. We use this general potential and
specify special situations by setting several of the coupling constants
$t_{i}$ to vanish. Especially one can study critical behaviour, by choosing
a point $t_{i}$ in the infinite dimensional space of coupling constants,
such that as a function of $t_{i}$ (for all other $t_{j}$'s fixed) the
expectation values are all powers of $t_{i}$. Such critical points are
labelled by an integer $k$.

The partition function of the perturbed theory can be expressed in terms
of correlators in pure topological gravity, by expanding the action in
the coupling constants to obtain
\begin{eqnarray}
\label{eq:part}
\partv & \equiv & \llang \exp \left[ \sum_{j=0}^{\infty} t_{j} \sigma_{j}
\right] \rrang_{V=0} \nonumber \\
&=& \sum_{n_0=0}^{\infty} \sum_{n_1=0}^{\infty} \cdots \left( 
\prod_{j=0}^\infty
 \frac{t_j^{n_j}}{n_j!} \right) \llang \overbrace{ \sigma_{0}\cdots\sigma_{0} }^{n_{0}-{\rm times}}
\overbrace{ \sigma_{1}\cdots\sigma_{1} }^{ n_{1}-{\rm times} }\cdots  \rrang_{V=0} \nonumber \\  
&=& \sum_{n_0=0}^{\infty} \sum_{n_1=0}^{\infty} \cdots \left(
\prod_{j=0}^\infty
 \frac{t_{j}^{n_j}}{n_j!} \right) \llang \sigma_{0}^{n_{0}} \sigma_{1}^{n_{1}} \cdots \ra_{V=0},
\end{eqnarray}
where we in the last step introduce a shorthand notation $\sigma_{i}^{n_{i}}$ for
$n_{i}$ copies of $\sigma_{i}^{(0)}$.
Most of the terms in equation~(\ref{eq:part}) will be zero, since the correlator should correspond
to a top form on moduli space. That is, the total ghost number should add
up to the dimension of moduli space. 

\noindent
The simplest choice $t_{0}\neq 0$ and $t_{i}=0$ for $i = 1,2,\dots$ is
known as the $k=1$ model and it has played a special role in the
development of the theory. The potential is of the form $V=t_{0}P(x_{0})$ and
a general correlator will be of the
form
\begin{equation}
  \lan \prod_{i=1}^{n} \sigma_{d_{i}}\ra_{V} = \sum_{r=0}^{\infty}
  \frac{t_{0}^{r}}{r!} \lan \left( \prod_{i=1}^{n}\sigma_{d_{i}}\right)
  \overbrace{PP\cdots PP}^{{\rm r-times}}\ra_{V=0}. \label{k=1-1}
\end{equation}
Even though we face an infinite sum, there will only be one contribution
to the sum from every fixed genus of the underlying surface $\Sigma_{g}$.
The total ghost number for the $\sigma_{d_{i}}$ operators is
$2\sum_{i} (d_{i}-1)$  and $-2r$  from the puncture operators. Since
the total number of marked points is $s=n+r$, there can only be contributions
to the sum when 
\begin{equation}
  r = \sum_{i=1}^{n}(d_{i}-1) - (3g-3),  \label{ghost-requirement-1}
\end{equation}
which makes
the total ghost number add up to the dimension $3g-3+s$ of \mbs.
A genus expansion of a
fixed correlator of the form in 
equation~(\ref{k=1-1}) receives only one contribution per genus, namely
the one where $r$ satisfies equation~(\ref{ghost-requirement-1}).
According to the genus zero recursion relation in
equation~(\ref{genus-zero-recursion}), a general correlator can be
reduced to calculating expectation values of products of puncture operators.
In pure topological gravity the only genus zero contribution is 
$\langle PPP \rangle$, which is usually normalized to the value one. In the $k=1$
model, equation~(\ref{k=1-1}) shows that the non-vanishing
genus zero correlators involving only puncture operators are as follows:
\begin{eqnarray}
  \lan 1 \ra_{V} &=& \frac{t_{0}^{3}}{3!}\lan PPP \ra_{V=0}=
  \frac{t_{0}^{3}}{6} , \label{<1>} \\
\lan P \ra_{V} &=& \frac{t_{0}^{2}}{2!}\lan P PP \ra_{V=0}=
\frac{t_{0}^{2}}{2}, \\
\lan PP \ra_{V} &=& \frac{t_{0}^{1}}{1!}\lan PP P \ra_{V=0}=
t_{0} , \\
\lan PPP \ra_{V} &=& \frac{t_{0}^{0}}{0!}\lan PPP \ra_{V=0}=1, \label{<4>}
\end{eqnarray}
while all genus zero correlators with four or more puncture operators
will vanish
\begin{equation}
\lan \overbrace{PP\cdots PP}^{(r\geq 4){\rm - times}} \ra_{V}=0.
\end{equation}

\noindent
In the following discussion we will restrict ourselves to genus zero
unless otherwise specified. Dijkgraaf has given the following useful
result~\cite{Dijkgraaf-1}
\begin{equation}
  \lan \prod_{i=1}^{s} \sigma_{d_{i}} \ra = \left( \sum_{i=1}^{s} d_{i} \right)
  !\,\, \delta \left( 3 + \sum_{i=1}^{s}(d_{i}-1)\right),
\label{dijkgraaf-1}
\end{equation}
by repeated use of the puncture and dilaton equations. We have changed the
normalization of his result, such that it corresponds to that of chapter 4.
Note that the correlator depends only on the total ghost number
$\sum_{i}d_{i}$ of all
the observables together with the total number of punctures $s$.

\noindent
From this we see that the partition function in
equation~(\ref{eq:part}) can be written as
\begin{equation}
\lan 1 \ra_{V} = 
\sum_{n_0=0}^{\infty} \sum_{n_1=0}^{\infty} \cdots \left(
\prod_{j=0}^\infty  \frac{t_j^{n_j}}{n_j!} \right)
\left( \sum_{k=0}^{\infty} k n_k \right) ! \ \ \delta \left( 3+
\sum_{k=0}^{\infty} (k-1) n_k \right). \label{partition-2}
\end{equation}
This is the starting point for one of the important results in~\cite{AHW}.
We now show how the partition function in equation~(\ref{partition-2}),
and hence any correlator, derived by differentiating the partition function with respect to
the couplings $t_{i}$, can be
expressed as a contour integral of a type similar to expressions known in
matrix models. This leads to an identification between the theories,
discussed later in this chapter. 

\noindent
We define the potential $V$ as
\begin{equation}
  V(z) = \sum_{i=0}^{\infty} t_{i}z^{i},
\end{equation}
where $z$ is a complex variable, with $z^{n}$ playing the role of
$\sigma_{n}$. The pre-factor to the delta function in
equation~(\ref{partition-2}) can be written in an economical manner
\begin{eqnarray}
  e^{V(\lambda z)/z} &=& \sum_{n_0=0}^{\infty} \frac{(t_{0})^{n_{0}}}{
    n_{0}! z^{n_{0}}} \sum_{n_1=0}^{\infty} \frac{(t_{1})^{n_{1}}}{
    n_{1}! }\lambda^{n_{1}} \sum_{n_2=0}^{\infty} \frac{(t_{2})^{n_{2}}}{
    n_{2}!}  z^{n_{2}} \lambda^{2n_{2}} \sum_{n_3=0}^{\infty} \frac{(t_{3})^{n_{3}}}{
    n_{3}!}  z^{2n_{3}} \lambda^{3n_{3}} \cdots \nonumber \\ &=& 
     \sum_{n_0=0}^{\infty} \sum_{n_1=0}^{\infty}\cdots \left[
       \frac{(t_{0})^{n_{0}}}{n_{0}!}\frac{(t_{1})^{n_{1}}}{n_{1}!}\cdots \right]
     z^{(-n_{0}+n_{2} +2n_{3}+\cdots)}\lambda^{\sum_{j=0}^{\infty}jn_{j}}.
\label{exponential-prefactor}
\end{eqnarray}
The exponent of $\lambda$ is clearly related to $(\sum_{k}kn_{k})!$ in
equation~(\ref{partition-2}), and it can be brought to this form by
differentiating repeatly with respect to $\lambda$ and afterwards setting $\lambda$
to zero
\begin{equation}
  \left[ 1 + \frac{\partial}{\partial \lambda} + \frac{\partial^{2}}{\partial \lambda^{2}}
    + \cdots \right] \lambda^{ \sum_{j=0}^{\infty} jn_{j} } \Bigg\vert_{\lambda=0}  =
  \left( 1- \frac{\partial}{\partial \lambda}\right)^{-1}
  \lambda^{\sum_{j=0}^{\infty}jn_{j}}\Bigg\vert_{\lambda=0} = \left(
      \sum_{j=0}^{\infty}jn_{j}\right)!,  \label{geometric-series-1}
\end{equation}
using that the sum over the $\lambda$-derivatives forms a formal geometric series.
With the pre-factor to the
delta function contained in
\begin{equation}
  \left( 1- \frac{\partial}{\partial \lambda}\right)^{-1} e^{V(\lambda z)/z}
  \Bigg\vert_{\lambda=0}
   ,
\end{equation}
we can find a way to isolate the term $z^{-3}$, such that the delta function constraint
is satisfied. This is done by expressing the partition function as the following
contour integral around the origin
\begin{equation}
  \lan 1 \ra_{V} = \Biggl[ \left( 1- \frac{\partial}{\partial \lambda}\right)^{-1} 
  \frac{1}{2\pi i} \oint_{{\rm origin}} dz\, z^{2} e^{\frac{V(\lambda z)}{z}} \Biggr]_{\lambda=0}.
\label{partition-3}
\end{equation}
Next we perform a series of manipulations to find a more convenient form for the partition
function. First perform the substitution $z'=\lambda z$ and then relabel the dummy integration
variable
$z'$ back to $z$ to obtain
\begin{equation}
  \lan 1 \ra_{V} = \Biggl[ \left( 1- \frac{\partial}{\partial \lambda}\right)^{-1}
  \frac{1}{2\pi i} \oint_{{\rm origin}} \frac{dz}{\lambda^{3}}\, z^{2} e^{\frac{\lambda V(z)}{z}}
  \Biggr]_{\lambda=0}.
\label{partition-4}
\end{equation}
Next we expand the exponential and pull out the sum $\sum_{r=0}^{\infty}\lambda^{r-3}$ in front
of the integral
\begin{eqnarray}
  \lan 1 \ra_{V} &=& \frac{1}{2\pi i} \Biggl[ \left( 1- \frac{\partial}{\partial \lambda}\right)^{-1}
  \sum_{r=0}^{\infty}\lambda^{r-3}
   \oint_{{\rm origin}} dz\, z^{2} \left(\frac{\lambda V(z)}{z}\right)^{r}\frac{1}{r!} \Biggr]_{\lambda=0}.
\nonumber \\
&=&  \frac{1}{2\pi i}  \sum_{r=0}^{\infty}(r-3)!
  \oint_{{\rm origin}} dz\, z^{2} \left(\frac{\lambda V(z)}{z}\right)^{r}\frac{1}{r!},
  \label{partition-5}
\end{eqnarray}
where we have evaluated the $\lambda$-derivatives in the last step.
The terms for which $r<3$ will only produce positive powers of $z$ and the contour
integral will vanish. Hence we start the sum at $r=3$ and cancel out the factorials to obtain
\begin{equation}
   \lan 1 \ra_{V} = \frac{1}{2\pi i}  \sum_{r=3}^{\infty}
  \oint_{\rm origin} dz\, z^{2} \left(\frac{ V(z)}{z}\right)^{r}\frac{1}{r(r-1)(r-2)}
  \label{partition-6} .
\end{equation}
At this stage it is convenient to make the substitution $V \mapsto \beta V$, where $\beta$
is a constant parameter introduced such that differentiating three times with respect to $\beta$
cancels the denominator $r(r-1)(r-2)$. Next we relabel $r'=r-3$ and find
the relation
\begin{eqnarray}
  \frac{\partial^{3}}{\partial \beta^{3}}
   \lan 1 \ra_{\beta V}& =& \frac{1}{2\pi i}  \sum_{r'=0}^{\infty}
  \oint_{\rm origin} dz\, z^{2} \left(\frac{ V(z)}{z}\right)^{r'+3}\beta^{r'}
\nonumber \\ &=& \frac{1}{2\pi i}\oint_{C} dz\, 
  V(z)^{3} \frac{1}{\left(z - \beta V(z)\right) },
  \label{partition-7} 
\end{eqnarray}
since the $r'$ sum constitutes a geometric series. We require that
\begin{equation}
  \bigg\vert \frac{\beta V(z)}{z} \bigg\vert < 1,
\end{equation}
ensuring that the denominator does not vanish. The contour $C$ should be deformed such that
this requirement holds and we assume this can be done. 
We want to eliminate the auxiliary parameter $\beta$ and do so by integrating three times
with respect to $\beta$ and then setting $\beta=1$, before we evaluate the contour integral.  
After the three-fold integration and dropping analytic terms in the integrand, we obtain the final result for
the partition function with a general potential
\begin{equation}
\lan 1 \ra_{V} =  -\frac{1}{2}\frac{1}{2\pi i}\oint_{C} dz\,  
  \biggl( z-V(z)\biggr)^{2}\, \ln \left(1-\frac{V(z)}{z}\right)  .
  \label{partition-8} 
\end{equation}
This expression was first presented in~\cite{AHW} and as we discuss later in this chapter, it
offers a new possibility to investigate the relation between perturbed topological gravity and
matrix models. 

\noindent
We use equation~(\ref{partition-8}) to express the partition function at certain critical
points. For a suitable choice of the coupling constants $(t_{0},t_{1},\dots)$ the
contour integral has a cut on the real axis, from the origin at $z=0$ to a point $z=u$. The
point $u$ is defined by the equation $u=V(u)$:
\begin{equation}
u = t_{0} + \sum_{k=1}^{\infty} t_{k}u^{k}, \label{stringeq-1}
\end{equation}
which allows one to express $u$ as a function of $t_{0}$ for fixed values of the
remaining $t_{k}$ $(k>1)$ parameters. Equation~(\ref{stringeq-1}) is known as the genus zero
string equation in two-dimensional quantum gravity, were it plays a central role in expressing the
genus zero structure.

Consider the $k=1$ model, which corresponds to the potential
\begin{equation}
  V(z) = t_{0} + t_{1}z, \label{k=1:1}
\end{equation}
where we have included a $t_{1}$ contribution, which at first hand disagrees with the previous description of
the $k=1$ model. It reflects that not only the puncture operator, but also the dilaton
operator have been added to the BRST exact action. However the only role of the dilaton operator
is to measure the Euler number of the Riemann surface, through its action in the dilaton
equation~(\ref{dilaton-eq}), and one can add this operator to the potential for free.

\noindent
The partition function based on the potential in equation~(\ref{k=1:1}) reads
\begin{equation}
\lan 1 \ra_{V} =  -\frac{1}{2}\frac{1}{2\pi i}\oint_{C} dz\,  
  \biggl( (1-t_{1})z-t_{0}\biggr)^{2}\,
  \ln \left(\frac{(1-t_{1})z-t_{0}}{z}\right),
  \label{partition-9} 
\end{equation}
which has a cut on the real axis from zero to the point $t_{0}/(1-t_{1})$ due
to the logarithm. By letting the contour enclose the cut in a anticlockwise manner,
the logarithm contributes a factor $(-2\pi i)$ which cancels part of the pre-factor to the
integral, and we find
\begin{eqnarray}
 \lan 1 \ra_{V} &=&  \frac{1}{2} \int_{0}^{t_{0}/(1-t_{1})} dx
 \biggl[ (1-t_{1})x - t_{0} \biggr]^{2} \nonumber \\ &=& \frac{1}{6}
 \frac{t_{0}^{3}}{1-t_{1}}.
\label{k=1:2}
\end{eqnarray}
Note that the $t_{1}$ contribution just gives a geometric series, reflecting that
the dilaton operator leaves the remaining operators unchanged. The factor $t_{0}^{3}/6$
is in direct correspondence with the result in equation~(\ref{<1>}), illustrating the
presence of the three conformal killing vectors on \Ci.

\noindent
The $k=2$ model is especially interesting, since it is known to correspond to
pure gravity in the matrix models. The simplest choice of potential is
\begin{equation}
  V(z) = t_{0} + z - z^{2},
\end{equation}
which is the first example of the choice of coupling constants made in the literature,
for the $k$'th multi-critical point
\begin{equation}
  t_{0}\neq 0,\,\,\,\,,t_{1}=1, \,\,\,\,t_{k}=-1. \label{convention-1}
\end{equation}
For the $k=1$ model there is a problem since $t_{1}$ should be both plus and minus one. This 
reflects the fact that the $k=1$ model plays a special role and we note also that the
result in equation~(\ref{k=1:2}) is singular for $t_{1}=1$.
In some aspects this model is too trivial to be included in the general
setting of multi-critical models since it has a trivial scaling behaviour.
We will return to this fact several times in this chapter.

\noindent
The $k=2$ partition function reads
\begin{equation}
\lan 1 \ra_{V} =  -\frac{1}{2}\frac{1}{2\pi i}\oint_{C} dz\,  
  \biggl( z^{2}-t_{0}\biggr)^{2}\,
  \ln \left(\frac{ z^{2}-t_{0}}{z}\right) .
  \label{partition-10} 
\end{equation}
We must study the cut structure for the logarithm: 
\begin{equation}
 \ln \left(\frac{ z^{2}-t_{0}}{z}\right) = \ln(z+\sqrt{t_{0}}) +  \ln(z-\sqrt{t_{0}})
 - \ln z,
\end{equation}
while enforcing $|V(z)/z|<1$, and we find two cuts on the real line, namely
from minus infinity to $-\sqrt{t_{0}}$ together with a cut from zero to
plus $\sqrt{t_{0}}$. We enclose the contour $C$ around the latter in an anticlockwise
direction to find
\begin{eqnarray}
\lan 1 \ra_{V} &=&  -\frac{1}{2}\frac{1}{2\pi i} \int_{0}^{ \sqrt{t_{0}}} dx
  (x^{2}-t_{0})^{2} (-2\pi i) \nonumber \\ &=&
  \frac{4}{15}t_{0}^{5/2}.
\label{k=2:1}
\end{eqnarray}
The exponent $5/2$ is in agreement with the partition function of pure two-dimensional
quantum gravity~\cite{Ginsparg-Moore}, since this is known to scale
like $t_{0}^{2-\gamma_{{\rm string}}}$, with $\gamma_{{\rm string}}=-1/2$ known as the string
susceptibility for pure gravity and $t_{0}$ being identified with the cosmological
constant $\Lambda$.
We return to this identification later
in this chapter.

\noindent
The general $k$'th multi-critical model is described by the potential
\begin{equation}
  V(z) = t_{0} + z - z^{k},
\end{equation}
in agreement with equation~(\ref{convention-1}). The partition function reads
\begin{equation}
\lan 1 \ra_{V} =  -\frac{1}{2} \frac{1}{2\pi i} \oint_{C} dz\,  
  \biggl( z^{k}-t_{0}\biggr)^{2}\,
  \ln \left(\frac{ z^{k}-t_{0}}{z}\right) ,
  \label{partition-11} 
\end{equation}
where the cut structure of course will depend on the specific choice of $k$.
There will be several cuts, not all on the real line, from plus/minus infinity and
into a certain point in the complex plane. But there will always be a cut
on the real line from zero to $t_{0}^{1/k}$ which the contour is chosen to enclose.
This is due to the singularities at $z=t_{0}^{1/k}w$, where $w$ is the $k$'th root of
unity. The final result is
\begin{eqnarray}
\lan 1 \ra_{V} &=&  \frac{1}{2} \int_{0}^{ t_{0}^{1/k}} dx
  (x^{k}-t_{0})^{2}  \nonumber \\ &=&
  \frac{t_{0}^{2+1/k}}{(2+1/k)(1+1/k)}.
\label{k=k:1}
\end{eqnarray}
Note that the partition function scales with the exponent $\gamma_{{\rm string}}= -1/k$.
This also signals that $k=1$ is a special choice, since $\gamma_{{\rm string}}=-1$ in
this model and the partition function does not have any critical behaviour.

Next, we derive closed expressions for various correlators in perturbed topological gravity
using the contour integral representation and discuss their critical behaviour.
The one-point $\langle \sigma_{l} \rangle$ function is derived by differentiating the
partition function with respect to the coupling constant $t_{l}$ and we find
\begin{eqnarray}
  \lan \sigma_{l} \ra_{V} &=& \frac{\partial}{\partial t_{l}}\Biggl[
   -\frac{1}{2} \frac{1}{2\pi i} \oint_{C} dz\,  
  \biggl( z - V(z)\biggr)^{2}\,
  \ln \left(\frac{ z-V(z)}{z}\right) \Biggr] \nonumber \\ &=&
  -\frac{1}{2} \frac{1}{2\pi i} \oint_{C} dz\, \Biggl[ 2 \biggl( z - V(z)\biggr) \left(-
    \frac{\partial V(z)}{\partial t_{l}}\right) \ln \left(\frac{ z-V(z)}{z}\right)
\nonumber \\ && +
     \biggl( z - V(z)\biggr)^{2} \frac{\left(-
    \frac{\partial V(z)}{\partial t_{l}}\right)}{\left(z-V(z)\right)}\Biggr] \nonumber \\ &=&
   \frac{1}{2\pi i} \oint_{C} dz\, z^{l} \biggl( z - V(z)\biggr)
   \ln \left(\frac{ z-V(z)}{z}\right), \label{sigma-l}
\end{eqnarray}
where the second term in the second line contributes only an analytical term to
the integrand, which can be
discarded. For the $k$'th multi-critical model the one-point function reads
\begin{equation}
 \lan \sigma_{l} \ra_{V} =   \frac{1}{2\pi i} \oint_{C} dz\, z^{l} \biggl( z^{k} - t_{0}\biggr)
   \ln \left(\frac{ z^{k}-t_{0}}{z}\right),
\end{equation}
where the potential is of the form $V(z)=t_{0}+z-z^{k}$. The cut structure is the
same as for the $k$'th multi-critical points for the partition function and we find
\begin{eqnarray}
   \lan \sigma_{l} \ra_{V} &=& - \int_{0}^{t_{0}^{1/k}}dx \,x^{l}\left(x^{k}-t_{0}\right)
   \nonumber \\ &=& \frac{t_{0}^{1+(l+1)/k}}{(l+1)(1+(l+1)/k)}.
\label{one-point-multicritical}
\end{eqnarray}
The two point function $\langle \sigma_{l_{1}}\sigma_{l_{2}}\rangle$ is found by taking the
derivative of  $\langle \sigma_{l_{1}}\rangle$ with respect to $t_{l_{2}}$.
\begin{equation}
   \lan  \sigma_{l_{1}}\sigma_{l_{2}} \ra_{V} = - \frac{1}{2\pi i} \oint_{C} dz\,  
 z^{l_{1}+l_{2}} \ln \left(\frac{z-V(z)}{z} \right)
= \int_{0}^{u} dx \, x^{l_{1}+l_{2}},
\label{two-point-function}
\end{equation}
For the $k$'th multi-critical model we set $u=t_{0}^{1/k}$ and find
\begin{equation}
  \lan  \sigma_{l_{1}}\sigma_{l_{2}} \ra_{V} =
  \frac{t_{0}^{(l_{1}+l_{2}+1)/k}}{l_{1}+l_{2}+1}. \label{two-point-multi}
\end{equation}
In the special case where $l_{1}=l_{2}=0$ equation~(\ref{two-point-multi}) reduces to
the important result
\begin{equation}
  u= \lan PP \ra_{V}.
\end{equation}

Obviously, the three-point function is the derivative of the two-point function with respect to
$t_{l_{3}}$
\begin{eqnarray}
\lan  \sigma_{l_{1}}\sigma_{l_{2}}\sigma_{l_{3}} \ra_{V} &=&  \frac{1}{2\pi i} \oint_{C} dz\,  
 \frac{ z^{l_{1}+l_{2}+l_{3}}}{z-V(z)} \\ &=&
\frac{\partial}{\partial t_{0}}\Biggl[ - \frac{1}{2\pi i} \oint_{C} dz\, z^{l_{1}+l_{2}+l_{3}}
\ln\left(\frac{ z-V(z)}{z}\right)\Biggr].
\end{eqnarray}
In the last step we use that $\partial V(z)/\partial t_{0}=1$ to see that
\begin{equation}
  \frac{\partial}{\partial t_{0}} \biggl[ \ln \left(\frac{z-V(z)}{z}\right) \biggr] =
  \frac{-1}{\left( z-V(z)\right)} .
\end{equation}
The higher $n$-point functions generalize the form of the three-point function. 
The $s$-point function reads
\begin{equation}
  \lan  \prod_{i=1}^{s}\sigma_{l_{i}} \ra_{V} =
  \left(\frac{\partial}{\partial t_{0}}\right)^{s-2} \Biggl[
\frac{-1}{2\pi i}\oint_{C} dz\,  
     z^{l_{1}+\cdots +l_{s}} \ln \left( \frac{z-V(z)}{z} \right) \Biggr] =
 \left(\frac{\partial}{\partial t_{0}}\right)^{s-2} \frac{u^{L+1}}{(L+1)}
     , \label{n-punkt}
\end{equation}
where
\begin{equation}
  L \equiv \sum_{i=1}^{s} l_{i}.
\end{equation}
The $k$'th multi-critical model is described by setting $u=t_{0}^{1/k}$, which leads
to the result
\begin{equation}
  \lan  \prod_{i=1}^{s}\sigma_{l_{i}} \ra_{V} =
  \left(\frac{\partial}{\partial t_{0}}\right)^{s-2} 
  \frac{t_{0}^{(L+1)/k}}{(L+1)}.
\end{equation}

The results for the partition function and the general $s$-point correlator
in the $k$'th multi-critical model reproduce those presented by Witten
in~\cite{Witten:340}, but in contrast to the derivation given there, we do not have
to guess any numerical constants.

It is possible to express the puncture and string equations as differential equations
for the partition function in the perturbed theory. The equation
\begin{equation}
  \frac{\partial}{\partial t_{0}} \lan 1\ra_{V} = \frac{t_{0}^{2}}{2} +
  \sum_{i=0}^{\infty}(i+1)t_{i+1}\frac{\partial}{\partial t_{i}} \lan 1\ra_{V},  \label{F-eq-1}
\end{equation}
corresponds to a recursion relation in term of $\sigma_{i}$ operators. By acting with
$\partial/\partial t_{0}$ on both sides of the equality sign of equation~(\ref{F-eq-1}) 
\begin{equation}
  \lan PP \ra_{V} = t_{0} +  \sum_{i=0}^{\infty}(i+1)t_{i+1} \lan P \sigma_{i} \ra_{V} =
  t_{0} + \sum_{j=1}^{\infty} j\,t_{j} \lan P \sigma_{j-1}\ra_{V} =
  t_{0} + \sum_{j=1}^{\infty} t_{j} u^{j},
\end{equation}
we obtain the string equation given in equation~(\ref{stringeq-1}). In the last step we have
applied equation~(\ref{two-point-function}). If we once again differentiate with respect to $t_{0}$
on both sides of equation~(\ref{F-eq-1}) and there after set
$t_{i}=0$ for $i>0$ we obtain the result~\cite{DW}
\begin{equation}
  \lan PPP \ra_{V} = 1,
\end{equation}
of equation~(\ref{<4>}). If we act with
\begin{equation}
  \prod_{i=1}^{s-1}\frac{\partial}{\partial t_{i}},
\end{equation}
on both sides of equation~(\ref{F-eq-1}) to pull down a product of $\sigma_{d_{i}}$'s from
the exponential, we obtain the puncture equation~(\ref{puncture-eq})
\begin{equation}
  \lan P \prod_{i=1}^{s-1}\sigma_{d_{i}}\ra_{V} = \sum_{j=1}^{s-1}d_{j} \lan \sigma_{d_{j}-1}
\prod_{i\neq j, \, i=1}^{s-1}\sigma_{d_{i}} \ra_{V},
\end{equation}
combined with the intial condition $\langle PPP \rangle_{V}=1$. In this way, the genus zero
string equation is identified with the $t_{0}$-derivative of the puncture equation.

At this stage we turn to a presentation of some
results in two-dimensional quantum gravity, which we will need in order to
discuss the identification between the theories.
\section{$2D$-Quantum Gravity}

In this section we discuss some important results of two-dimensional quantum
gravity, relevant for the comparison with perturbed topological gravity.
The quantum theory of gravity in two dimensions has been studied in hundreds of
papers through more than a decade. In this limited amount of space, we only scratch
the surface of this vast material. There are several approaches to quantum
gravity including: Liouville gravity, matrix models and dynamical triangulations.
These are not independent approaches, in particular the matrix models and
dynamical triangulation models are closely related, as explained below.
All the results derived in Liouville gravity have so far been in correspondence with the
results of the matrix models, but Liouville theory does not have the same predictive
power as the other two approaches.

\subsection{The Hermitian Matrix Model}
\begin{figure}[h]
\begin{center}
\mbox{
\epsfysize4cm
\epsffile{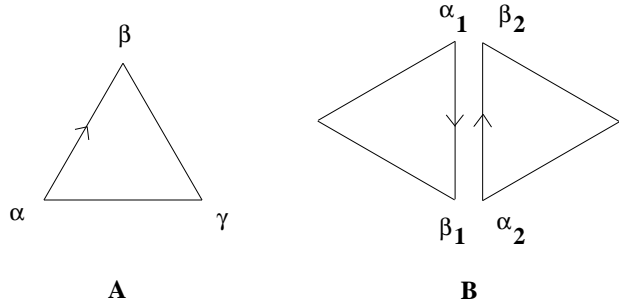}
}
\end{center}
\caption{ \label{hermitian} (A) Assigning indices to a triangle. (B) Gluing along a link.}
\end{figure}

In this section we primarily follow~\cite{Jan-lecture}.
The traditional approach to Euclidean quantum gravity is to write down the formal path integral
\begin{equation}
 Z = \int \frac{{\cal D}[g_{\alpha\beta}]}{{\rm Vol}(\diff)}
    \exp \Biggl( -\int \sqrt{g}\bigl(\frac{R[g]}{4\pi G} +\Lambda\bigr) \Biggr),
\end{equation}
where we have included the gravitational coupling $G$.
In order to calculate this integral, one possibility is to introduce a reparametrization
invariant cutoff~\cite{Jan-lecture}, to make the integral well-defined. This is the idea behind the
theory of dynamical triangulation. The cutoff is taken by building up the
surfaces using small equilateral
triangles with edge length $a$. Since a closed surface can be built in many different
ways by gluing triangles, the integration over inequivalent metrics turns out to correspond to
a sum over inequivalent (abstract) triangulations ${\cal T}$. In this approach, a triangulation
corresponds to a diffeomorphism equivalence class of metrics. The path integral is discretized
such that
\begin{equation}  
 Z[\Lambda,G] = \int \frac{{\cal D}[g_{\alpha\beta}]}{{\rm Vol}(\diff)}
    e^{-\int \sqrt{g}(\frac{R[g]}{4\pi G}+\Lambda)} \rightarrow
   Z[\Lambda,G]  =  \sum_{T\in {\cal T}} e^{-\Lambda N_{T} + \chi(T)/G},
\end{equation}
where we sum over all classes of abstract triangulations. $N_{T}$ is the number of triangles,
and $\chi$ is the Euler-Poincar\'{e} characteristic of the triangulated surface. In order to control
the contributions from different topologies we rewrite the partition function as a sum of
triangulations of fixed genus plus a sum over genus:
\begin{equation}
  Z[\Lambda,G] = \sum_{g=0}^{\infty} e^{(2-2g)/G}\sum_{K}e^{-\Lambda K} {\cal N}(K,g), \label{action-matrix}
\end{equation}
where $\chi=2-2g$ and ${\cal N}(K,g)$ is the number of triangulations of fixed genus $g$, built from
$K$ triangles. There is an exponential bound on ${\cal N}(K,g)$~\cite{Jan-lecture} and the
partition function is well defined for fixed topology
\begin{equation}
  Z_{g}[\Lambda,G] = e^{(2-2g)/G} \sum_{K}e^{-\Lambda K} {\cal N}(K,g), \label{partition-DT-1}
\end{equation}
where the sum over genus is not well-defined due to an exponential growth of the number of
triangles.
The critical cosmological constant $\Lambda^{c}$ is the lowest value of $\Lambda$, for
which the partition function in equation~(\ref{partition-DT-1}) is convergent. This value is
independent of the genus and by approaching the critical value from above, it is possible to
obtain a continuum limit~\cite{Jan-lecture}. The limit $a\rightarrow 0$, i.e. 
removing the cutoff, corresponds to approaching the continuum theory. In the matrix model
approach, we can automatically
calculate how many different ways we can combine a fixed number of triangles to create a closed
manifold. Consider a triangulation where we label the vertices of the $i$'th triangle with the
numbers $\alpha_{i},\beta_{i},\gamma_{i}$. To each edge we assign a hermitian
$N\times N$ matrix 
$\phi_{\alpha_{i}\beta_{i}}$, which makes it possible to assign a factor
\begin{equation}
  \phi_{\alpha_{i}\beta_{i}} \phi_{\beta_{i}\gamma{i}}
  \phi_{\gamma{i}\alpha_{i}} =
  tr \phi^{3},
\end{equation}
to every oriented triangle.
The process of gluing two triangles together along an 
edge, is represented as the Wick contraction of two hermitian matrices of the form
$\phi_{\alpha_{1}\beta_{1}} \phi_{\alpha_{2}\beta_{2}}$ as illustrated in
figure~(\ref{hermitian}).
The gaussian matrix integral
\begin{equation}
  \lan \phi_{\alpha_{i}\beta_{i}}\phi_{\alpha_{j}\beta_{j}} \ra
  \equiv \int {\cal D}\phi e^{-\frac{1}{2}\vert \phi_{\alpha\beta}\vert^{2}}
\phi_{\alpha_{i}\beta_{i}} \phi_{\alpha_{j}\beta_{j}} =
  \delta_{ \alpha_{i}\beta_{j} } \delta_{ \beta_{i}\alpha_{j} } , \label{Wick}
\end{equation}
represents the Wick contraction corresponding to the gluing of two triangles. The measure
consists of the product of the real and the imaginary part of the matrices
\begin{equation}
  {\cal D}\phi \equiv \prod_{\alpha\leq \beta} d \left({\rm Re}\phi_{\alpha\beta}\right)
\prod_{\alpha < \beta} d \left({\rm Im}\phi_{\alpha\beta}\right),
\end{equation}
since the diagonal elements are required to be real for a hermitian matrix.
In this sense the integral
\begin{equation}
  \int {\cal D}\phi  e^{-\frac{1}{2}\vert \phi_{\alpha\beta}\vert^{2}}\frac{1}{K!}
  \left( tr \phi^{3} \right)^{K}, \label{matrix-K}
\end{equation}
corresponds to all possible ways of gluing $K$ triangles to create surfaces, which may have
disconnected parts. By standard field theory methods, the connected diagrams, i.e. surfaces
are generated by the free energy
\begin{equation}
  Z(\Lambda,G) = \log\frac{Z(g,N)}{Z(0,N)} \label{Jan-result-1}
\end{equation}
where $Z(g,N)$ is the matrix model partition function:
\begin{equation}
  Z(g,N) = \int {\cal D}\phi \exp\left( -\frac{1}{2} tr\phi^{2} +
    \frac{g}{3\sqrt{N}}tr\phi^{3}\right), \label{phi-3-action} \label{Jan-2}
\end{equation}
with coupling constant $g$ in front of the intersection part $(\phi^{3})$ of the action.
In order for the result of equation~(\ref{Jan-result-1}) to hold, we identify the parameters
\begin{equation}
  N = \exp \left(\frac{1}{G}\right), \,\,\, \Lambda = - \log g.
\end{equation}
We have rescaled the matrices in equation~(\ref{Jan-2}) such that
\begin{equation}
  tr \phi^{3} \mapsto  \frac{g}{3\sqrt{N}}tr\phi^{3}, \label{rescale-matrix}
\end{equation}
where $N$ is the number of indices of the matrices. By expanding the interaction $(\phi^{3})$ part of the action, we generate a sum of terms like
that in equation~(\ref{matrix-K}) for $K=1,2,\dots$. The gaussian part of the action ensures
the Wick contraction of the indices such that we obtain a closed surface.

\begin{figure}[h]
\begin{center}
\mbox{
\epsfysize6cm
\epsffile{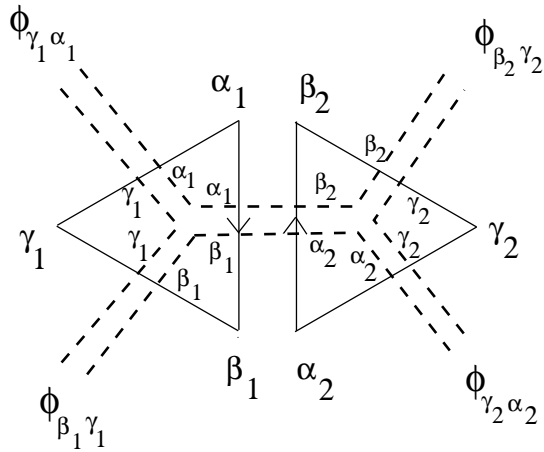}
}
\end{center}
\caption{ \label{fat} A dual fat $\phi^{3}$ graph.}
\end{figure}

\noindent
When we construct a closed surface by gluing triangles together, we will pick up a
factor from the Wick contractions that transform an external vertex into an internal vertex.
Since all vertices are internal for a closed surface we end up with a factor $N^{V}$,
$V$ being the total number of vertices in the triangulation.
By the rescaling in equation~(\ref{rescale-matrix}) the
total pre-factor becomes $N^{V-K/2}=N^{\chi}$, since the Euler-Poincar\'{e}
characteristic for a closed surface is
\begin{equation}
  \chi = V-E + K = V - \frac{K}{2},
\end{equation}
where $E$ is the number of edges. 
The redefinition of $\phi$ ensures that the weight in the action for the term $tr\phi^{3}$
only depends on the the genus of the surface. The factor $N$ is always removed at
the end of a calculation by taking the limit $N\rightarrow \infty$. This is done to
ensure that the individual vertices are labelled by independent indices~\cite{Jan-lecture}.
The identified edges which are glued together
constitute the links of a three-valent graph, forming a ``fat'' $\phi^{3}$ graph
as illustrated in figure~(\ref{fat}). The summation over triangulations correspond to
the summation over all closed trivalent graphs, where an individual graph corresponds to
a surface. The graphs are dual to the triangulation as indicated in figure~(\ref{fat}).
The graphs are closed by Wick contracting the open ends of the graph. The graph is fat due to the
presence two indices along each link. The action in equation~(\ref{phi-3-action}) is of course
of the familiar $\phi^{3}$ form from ordinary field theory and is thus known to generate
three-valent Feynman diagrams. 

Instead of gluing together triangles all types of polygons can be used and a general potential
can be introduced with a set of arbitrary coupling constants $\{ g_{i} \}$
\begin{equation}
V(\{g_{n}\}) = \sum_{n=1}^{\infty} \frac{g_{n}}{n}\phi^{n}. \label{potential-1}
\end{equation}
The factor $1/n$ in is a symmetry factor for the $n$ ways to label
the legs of an $n$-valent vertex.
It is common to identify $g_{0}$ as the cosmological constant $\Lambda$ such that the
term $N g_{0}tr\phi^{0} = N\Lambda$ corresponds to the cosmological term in
the action in equation~(\ref{action-matrix}). We interchange between
the $g_{0}$ and $\Lambda$ in the following discussions. The expectation value $\langle
tr\phi^{n}\rangle$ corresponds to differentiation of $Z$ with respect to $g_{n}$. One can
interpret $\langle tr\phi^{n}\rangle$ as the result of summing over all distinct surfaces, which
have an $n$-polygon of length $l= an$ as boundary. In this sense, expressions like  e.g. $\langle
tr\phi^{n}\rangle$ and $\langle tr\phi^{n}tr\phi^{m}\rangle$, correspond to different geometries,
namely with one or two boundaries. 

\subsection{Matrix Models and the KdV Equation}
After these general comments, we focus on some aspects of the continuum version of
the hermitian matrix model. Following Dijkgraaf, E. and H. Verlinde~\cite{DVV} consider the
general hermitian matrix model, where we integrate over $N\times N$ hermitian
matrices $\phi$ to obtain the partition function:
\begin{equation}
  Z = \int {\cal D}\phi\, e^{-N tr V(\phi)},
\label{general-hermitian}
\end{equation}
for a general potential
\begin{equation}
  V(\phi) = \sum_{n=2}^{\infty} g_{n} \phi^{n}, \label{matrix-1}
\end{equation}
with coupling constants $g_{n}$. Compared to the previous discussion the symmetry factor $1/n$
and the $g_{1}$ term have been removed, which renormalizes the remaining coupling constants.
In the so-called double scaling limit ( see e.g.~\cite{DSL}) two
simultaneous limiting procedures are used. The limit $N\rightarrow \infty$ is taken
while at the same time tuning the values of the coupling constants to their critical values
$g_{n}\rightarrow g_{n}^{c}$, transforms the theory into its continuum formulation.
The free energy of the continiuum theory
\begin{equation}
  F = - \log Z,
\end{equation}
has a genus expansion through the  string
coupling constant $\lambda\sim 1/N$
\begin{equation}
  F = \sum_{g=0}^{\infty} \lambda^{2g-2} F_{g},
\end{equation}
where $F_{g}$ is the genus $g$ contribution to the free energy. The coupling constants in the
continuum are denoted $(t_{0},t_{1},\cdots)$ and they couple to continuum versions of
$tr\phi^{n}$. The discrete matrix model observables $tr \phi^{n}$ correspond to
microscopic loops around a vertex in the triangulation. In the continuum they are
replaced by local scaling operators $\sigma_{n}$, coupled to $t_{n}$ for $n\geq 0$.

\noindent
The following differential equation~\cite{DVV}
\begin{equation}
  u(t_{0},t_{1},\dots) = - \lambda^{2}\frac{\partial^{2}}{\partial t_{0}^{2}} F[t_{0}
  ,t_{1},\dots],
\end{equation}
defines $u=\langle PP \rangle_{V}$ as the specific heat of the theory. One of the
important results in matrix models is that the specific heat
\begin{equation}
  u(t_{0},t_{1},\dots) = \lan PP\ra_{V} = \sum_{g=0}^{\infty}
  \lambda^{2g-2}\lan PP \ra_{V,g},
\end{equation}
satisfies the equations of the Korteweg-de Vries (KdV) hierarchy~\cite{BDSS,GM}.
The insertion of a $\sigma_{n}$ operator in $\langle PP\rangle_{V}$ by differentiating
with respect to $t_{n}$, is identified with the so-called $n$'th KdV flow of
$u$~\cite{DW,DVV}
\begin{equation}
  \lan \sigma_{n} PP \ra = \frac{\partial u(t_{0})}{\partial t_{n}} =
  \frac{\partial}{\partial t_{0}} R_{n+1}[u]. \label{KdV-eq}
\end{equation}
The $R_{n}[u]$'s are known as Geldfand-Dikii polynomials, which are
polynomials of $u$ and its derivatives with respect to $t_{0}$, determined from the 
recursion relation
\begin{equation}
  \frac{\partial}{\partial t_{0}} R_{n+1}[u] = \frac{n}{2n+1}\left( \frac{1}{2}\lambda^{2}
  \frac{\partial^{3}}{\partial t_{0}^{3}} + 2 u(t_{0}) \frac{\partial}{\partial t_{0}}
+ \frac{\partial}{\partial t_{0}}u(t_{0}) \right) R_{n}[u]. \label{KdV-1}
\end{equation}
The first polynomials are $R_{0}=1$ and $R_{1}=u$. The string equation
\begin{equation}
  u = t_{0} + \sum_{n=1}^{\infty} n t_{n} R_{n}[u], \label{stringeq-2}
\end{equation}
determines $u(t_{0})$ for a fixed set of values of the remaining couplings $\{t_{i}\}$, $i>0$.
By tuning the values of the parameters $\{t_{n}\}$
it is possible to interpolate between different multi-critical points in the space of
coupling constants. The form of the string
equation in terms of the Gelfand-Dikii polynomials, generalises the genus zero form in
equation~(\ref{stringeq-1}). This follows from the genus expansion of the $R_{n}$
polynomials~\cite{DW}
\begin{equation}
  R_{n} = \sum_{g=0}^{\infty} \lambda^{2g}R_{n}^{(g)},
\end{equation}
which has the following genus zero contribution
\begin{equation}
  R_{n}^{(g=0)}[u] = \frac{1}{n}u^{n}.  
\end{equation}
One of the major results based on this formulation is that the partition function is
the so-called tau function for the KdV hierarchy, which means that for $Z=\tau$,
\begin{equation}
  u = \frac{\partial^{2}}{\partial t_{0}^{2}}\log \tau[t_{0},t_{1},\dots]
\end{equation}
is a solution to the (generalized) KdV equation~(\ref{KdV-eq}). There has been much
confusion in the literature, regarding whether it is the partition function or the
square root of the partition function, which is the tau function for the KdV hierarchy.
Based on detailed matrix model calculations~\cite{Jan-Charlotte} it has been shown to
be the former which is correct.  The controversy is related to
how the double scaling limiting procedure is performed. For more information
on the various points of view regarding this problem see e.g.~\cite{Witten:SDG,Charlotte}

The so-called loop equation~\cite{Jan-bog} offers another powerful approach to matrix
models, from which many important results have been derived. The loop equation is expressed in
terms of the loop functional $w(l)$, which can be defined in both the discrete and the
continuum regime. The approach taken in~\cite{DVV} is to prove that the string
equation~(\ref{stringeq-2}) together with the KdV equation~(\ref{KdV-eq}) can
be reformulated to give the loop equation. The loop functional $w(l)$ is then viewed
as representing a macroscopic loop, derived as the result of the double scaling limiting
procedure on a microscopic loop $tr \phi^{n}$ with length $l=na$.
When taking the continuum limit the power $n\rightarrow \infty$ simultaneously with
the edge length $a\rightarrow 0$, such that $l$ is kept at a fixed value.
The result is a macroscopic loop, expressed through the loop functional~\cite{DVV}
\begin{equation}
  w(l) \equiv \sum_{n=0}^{\infty} \frac{l^{n+\frac{1}{2}}\sigma_{n}}{\Gamma(n+\frac{3}{2})}.
\label{et-loop-w-cont.}
\end{equation}
Since $w(l)$ is proportional to a sum of $\sigma_{n}$ operators, it is not surprising that
relations between expectation values of $\sigma_{n}$ operators, can be translated into
relations between expectation values of $w(l)$'s. The loop equation is exactly such a relation.
We do not need the specific form of the loop equation for the following discussion.
At this stage it should be noted, that
it is more natural from the discrete approach to matrix models, to begin from the
discrete loop equation and derive the
string equation and the KdV equation as consequences from this~\cite{Jan-bog}.

An important result in~\cite{DVV} is that the loop equation translates into a set of
linear differential equations when expressed in term of the $\sigma_{n}$ operators
through equation~(\ref{et-loop-w-cont.}), which thereafter are translated into constraints 
on $\tau$, when identifying $\sigma_{n}$ correlators with $t_{n}$ derivatives of $Z=\tau$.
The final constraints are of the form
\begin{equation}
  L_{n}\tau = 0,\,\,\, n =  -1,0,1,2,\dots,
\end{equation}
which for example takes the form for $n=-1$:
\begin{equation}
  L_{-1} = \sum_{m=0}^{\infty} (m+\frac{1}{2})t_{m}\frac{\partial}{\partial t_{m-1}}
  + \frac{1}{8}\lambda^{-2}t_{0}^{2}.
\label{virasoro-ex}
\end{equation}
The set of these $L_{n}$ operators forms  a subalgebra of the Virasoro algebra
\begin{equation}
  [ L_{n},L_{m}] = (n-m)L_{n+m}, \,\,\, n\geq -1.
\end{equation}
These constraints have a physical interpretation in the discrete matrix model approach, which
we give below.

Consider the following approach to hermitian matrix model by Ambj\o rn, Jurkiewicz and
Makeenko~\cite{JJM}, where the theory is defined with a
general potential as in equation~(\ref{potential-1}).
The partition function 
\begin{equation}
  Z = \int {\cal D}\phi^{-NV(\phi)},
\end{equation}
is obviously invariant under redefinitions of the integration variable $\phi$.
This is reflected in the Dyson-Schwinger equation~\cite{JJM}
\begin{equation}
  \lan \sum_{k=0}^{n} tr \phi^{k} tr \phi^{n-k} \ra =
  N \lan \sum_{j=1}^{\infty}g_{j}tr \phi^{n+j} \ra, \label{DS-1}
\end{equation}
which holds for $n=-1,0,1,\dots$, when restricting negative powers of $tr \phi^{n}$
to vanish. 
It has been shown by Fukuma, Kawai and Nakayama~\cite{FKN} that equation~(\ref{DS-1})
can be written as a set of differential operator constraints
\begin{equation}
  L_{n}Z=0,\,\,\,n=-1,0,1,2,\dots
\end{equation}
on the partition function.
The operators $L_{n}$ are expressed in term of the coupling constants:
\begin{eqnarray}
  L_{n}&=&\sum_{k=1}^{\infty} \frac{k(n-k)}{N^{2}}
    \frac{\partial^{2}}{\partial g_{k}\partial g_{n-k}} - 2n \frac{\partial}{\partial g_{n}}
      \nonumber \\ && + \sum_{k=0}^{\infty} (k+n)g_{k}\frac{\partial}{\partial g_{n+k}}
      - \delta_{-1,n}Ng_{1} + \delta_{0,n}N^{2}.
\label{discrete-virasoro}
\end{eqnarray}
All derivatives with respect to $g_{-n}$, for $n>0$ are omitted since they correspond
to negative powers of $\phi$. The discrete $L_{n}$ operators also form a subalgebra of
the Virasoro algebra. It was shown in~\cite{FKN} how to take the
continuum limit of these constraints
such that the resulting equations annihilate the continuum partition function.
The choice of normalization in equations~(\ref{virasoro-ex},\ref{discrete-virasoro})
is not identical.

Dijkgraaf, E. and H. Verlinde derived in~\cite{DVV} a
recursion relation for the $\sigma_{n}$ operators in the $k$'th multi-critical model,
based on the continuum Virasoro constraints. This was done by translating the 
differentiations with respect to $t_{n}$, to insertions of $\sigma_{n}$ operators. 
The recursion relation holds for general genus $g$, by expanding the Virasoro
constraints in the string coupling constant $\lambda$
and it reads:
\begin{eqnarray}
  \lan \sigma_{n+k}\prod_{i=1}^{s} \sigma_{n_{i}}\ra_{g} &=&
  t_{0}\lan \sigma_{n}\prod_{i=1}^{s} \sigma_{n_{i}}\ra_{g} +
  \sum_{j=1}^{s}j\lan \sigma_{j+n}\prod_{j\neq i\,\, i=1}^{s}\sigma_{n_{i}}\ra_{g}
  \nonumber \\ && +
  \sum_{j=1}^{n}\Biggr[ \lan\sigma_{j-1}\sigma_{n-j}\prod_{i=1}^{s} \sigma_{n_{i}}\ra_{g}
  \nonumber \\ && + \frac{1}{2}\!\!\! \sum_{\begin{array}{l} {\scriptstyle S=X\cup Y} \\[-.2cm]
      {\scriptstyle g=g_{1}+g_{2}}\end{array} }
  \lan \sigma_{j-1}\prod_{i\in X}\sigma_{n_{i}}\ra_{g_{1}}\!\!\!
  \lan \sigma_{n-j}\prod_{ i\in Y}\sigma_{n_{i}}\ra_{g_{2}}\Biggr],
\label{DVV-recursion}
\end{eqnarray}
where $S=\{1,2,\dots,s\}$ and $X,Y$ represent the degeneration of a genus $g$ surface
into a genus $g_{1}$ branch $X$ and a genus $g_{2}$ branch $Y$. The terminology is
similar to the one used in chapter 4 and this recursion relation is one of the
results which has been used to identify topological and quantum gravity. Note that
the operators $\sigma_{n}$ for $n\geq k-1$ can be eliminated such that a general
correlator can be expressed in terms of a finite set of operators
$\sigma_{n}$, for which $n\leq k-2$. These operators should correspond
to the primary fields in the $(2,2k-1)$ minimal model, in conformal field theory~\cite{DVV}.
We explain this in some detail.

The higher multi-critical models $k>2$ are known to correspond to statistical systems
coupled to gravity. At their critical points, such systems are known to be described by 
conformal field theories. The coupling to gravity comes about by formulating the
statistical model on a random triangulation, instead of a cubic lattice. The coupling
to gravity dresses the scaling exponents, characterising the system.

The so-called minimal conformal field theories are labelled by a central charge $c$, which 
can be expressed by two integers $(p,q)$, with no common divisor.
The central charge is expressed as~\cite{Ginsparg-Moore}
\begin{equation}
  c = 1 - 6\frac{\left(p-q\right)^{2}}{pq}; \,\,\,\,\, 2\leq p< q,
\end{equation}
where the discrete unitary models are described by the subset of the parameters $(p,q)=(m,m+1)$
for $m=2,3,\dots$. A famous example is when $m=3$ leading to $c= 1/2$ describing the
two-dimensional Ising model, and the general result for $c$ in the unitary models reads
\begin{equation}
  c  = 1-\frac{6}{m\left(m+1\right)}.
\end{equation}
The string susceptibility $\gamma_{{\rm string}}$ is related to the central
charge~\cite{Ginsparg-Moore}
\begin{equation}
  \gamma_{{\rm string}} = \frac{1}{12} \Biggl(c-1-\sqrt{\left(c-25\right)\left(c-1 \right)}\Biggr),
\end{equation}
leading to $\gamma_{{\rm string}} = -1/(p+q-1)$ for the general $(p,q)$ models and
$\gamma_{{\rm string}} = -1/m $ for the unitary models $(m,m+1)$.
For the $k$'th multi-critical model, the hermitian matrix model is identified with the
$(2,2k-1)$ models, which are non-unitary for $k\neq 2$. For $k=2$ (pure gravity) the
matrix model corresponds to the unitary series $(k,k+1)\vert_{k=2}$ coupled to gravity, and
we see that $c=0$ since there are no matter fields in pure gravity.

\section{Identifying the Theories}

We first review various different arguments, on how perturbed topological gravity and
two-dimensional quantum gravity have been identified. Thereafter we review some of the
main points in the identification presented in~\cite{AHW}. We also discuss some aspects
related to the role of the metric, based on these discussions.

\subsection{Old Results}
Dijkgraaf, E. and H. Verlinde suggested in~\cite{DVV} that two-dimensional quantum gravity
and pure topological gravity can be identified, based on the similarity between the 
recursion relation in equation~(\ref{DVV-recursion}) and the recursion relation in
equation~(\ref{general-sigma-recursion:TG}). This is only true for $k=1$ and $\Lambda=0$.
In general, the cosmological constant $t_{0}=\Lambda$ is kept non-zero in quantum gravity,
such that the expected scaling behaviour is valid. If $\Lambda=0$, the cosmological term
vanishes and the action is zero. Even though the action used to derive
equation~(\ref{general-sigma-recursion:TG}) in some sense is identical to the
zero action used by~\cite{LPW}, there is an important difference. Topological
gravity is the result of gauge fixing the zero action, not just having a zero action.
The reason for choosing $k=1$ is that one needs to get the correct index in the second term
of equation~(\ref{DVV-recursion}), such that it corresponds to the part
of equation~(\ref{general-sigma-recursion:TG}), which encodes the puncture equation.
Being only able to identify the $k=1$
multi-critical model with topological gravity, is a meager result. As discussed before, the
$k=1$ model is a rather simplified model, compared to the higher multi-critical models.
The reason for setting $\Lambda=0$ is first and foremost to get rid of the term
$t_{0}\langle \sigma_{n}\prod_{i=1}^{s} \sigma_{n_{i}}\rangle_{g}$, since it does not
appear in equation~(\ref{general-sigma-recursion:TG}). Here one should be a bit careful, as
one could argue that the correlators in equation~(\ref{DVV-recursion}) should be taken as
$\langle \cdots \rangle_{V}$ correlators in perturbed topological gravity, which then by
equation~(\ref{k=1-1}) should be transformed into $\langle \cdots \rangle_{V=0}$ correlators.
In this way, one can actually get rid of the $t_{0}$ term, since it will appear with the same
power in every term (at least in genus zero) by insertion of the correct number of puncture
operators. But in the end, nothing is gained from this exercise, since one just ends up
showing that the puncture equation equals the puncture equation. Actually there is a
slight difference in the numerical constants, due to a mismatch between the normalizations
used in the various approaches. The deeper reason for why $t_{0}=0$ must be taken, is that the
super BF theory used in~\cite{VV} was based on the $G=ISO(2)$ gauge group, which we know
corresponds to vanishing cosmological constant through the Liouville
equation~(\ref{Liouville-2}).
With this choice of gauge group it will never be possible to derive the relation in
equation~(\ref{DVV-recursion}), and we conclude that the identification of recursion relations
offers only a minimal overlap $(k=1,t_{0}=\Lambda=0)$ between topological gravity and
quantum gravity. It is not clear that this could be helped by changing to $G=SO(2,1)$ in
the super BF theory, since we would encounter problems in defining a genus
expansion. In this sense, $ISO(2)$ play a special role in the super BF theory.
The curvature singularities that were inserted in the form of vertex operators, gave rise to
the transition to punctured Riemann surfaces, which is a central feature for the identification
with quantum gravity. We discuss the role of punctures in more detail later in this section.

A stronger identification was presented by Witten~\cite{Witten:340}, based on the scaling
behaviour for a general correlator $\langle \prod_{i=1}^{s} \sigma_{d_{i}}\rangle_{V}$
identical to that presented in equation~(\ref{n-punkt}) for the $k$'th multi-critical model.
As a symbolic identification, Witten wrote
\begin{equation}
  \lan \sigma_{n} \ra_{V} \sim \lan tr \phi^{2n}+ \dots \ra, \label{symbolic}
\end{equation}
where $\dots$ were not specified in~\cite{Witten:340}. This identification
is based on the results proved by Gross and Migdal~\cite{GM}, showing that
the continuum  limit of multi-critical matrix model correlators scales in
the same way as topological gravity in the multi-critical models.
From this correspondence between the continuum limit of the hermitian matrix model and
topological gravity, Witten conjectured that
the partition functions of the two theories are identical.
The details of this identification are more involved than just noting an identical scaling
behaviour, since one should define the continuum observables in the matrix model with
some care. The conjecture put forward by Witten, has been proved by Kontsevich in a
seminal paper~\cite{Kont}.
We first present the idea behind Kontsevich's proof and thereafter we discuss the new
perspectives on this identification, presented in~\cite{AHW}.

The main result in~\cite{Kont} is that the intersection numbers on \mbs, generated by 
the partition function of perturbed topological
gravity in equation~(\ref{eq:part}),
can be described by a matrix integral.

Kontsevich showed~\cite{Kont} that the partition function is given by 
the following matrix integral
\beqj{*8}
Z_N(M) = \frac{\int d\phi \; \exp\Bigl\{ -tr \Bigl( \frac{M \phi^2}{2}+
\frac{i \phi^3}{6} \Bigr)\Bigr\}}{\int d\phi \;
 \exp\Bigl\{ -tr \Bigl( \frac{M \phi^2}{2}\Bigr)\Bigr\}},
\eeqj
where the integration is over $N\times N$ hermitian matrices $\phi$.  
The $N\times N$ matrix $M$ is assumed to be symmetric and positive definite, 
and one can show that $Z_N(M)$ only depends on 
\beqj{*9}
t_k = - \ \frac{(2k-1)!!}{k!} \ tr M^{-(2k+1)}, \mgap k \geq 0,
\eeqj
where by definition $(-1)!!=1$.
For  $N \to \infty$ the tau-function of the KdV-equation~\cite{Dijkgraaf-1,Kont} is then
\beqj{*9a}
u(\{t_i\}) = \frac{\prt^2}{\prt t_0^2} \log Z(\{t_i\}).
\eeqj
One can write (for $N \to \infty$):
\beqj{*10}
\log Z (\{t_i\}) = \lan\exp \sum_{j=0}^\infty t_j \sg_j \ra
= \sum_{n_0=0}^\infty \sum_{n_1=0}^\infty \cdots 
\lan \sigma_{0}^{n_0} \sigma_{1}^{n_1} \cdots \ra \prod_{j=0}^\infty 
\left( \frac{t_j^{n_j}}{n_j !}\right) . \
\eeqj
This expression has a large $N$ expansion which at the same 
time is a genus expansion generating  the intersection indices for each genus
$ \langle \sigma_{0}^{n_0} \sigma_{0}^{n_1} \cdots \rangle_g$.

It seems a mysterious result, that the simple action in equation~(\ref{*8}),
with a gaussian and a cubic term, should generate the intersection numbers on \mbs.
This result builds on the existence of a cell decomposition, or triangulation of
\mbs, which was cleverly used by Kontsevich. We do not enter this discussion and
refer the reader to the literature~\cite{Dijkgraaf-1,Kont,alg-geom,IZ,Distler-Penner} for
details.

The Kontsevich matrix model in equation~(\ref{*8}) is very different from the form of
the usual hermitian matrix models, as presented in equation~(\ref{general-hermitian}) with a
general potential as the one in equation~(\ref{potential-1}). Anyhow it has been proven by
Ambj\o rn and Kristjansen~\cite{Jan-Charlotte}, that the double scaling limit of the
partition function of the hermitian matrix model coincides with the tau function for
the KdV hierarchy of the Kontsevich model. All loop correlators in the matrix model,
also corresponds with the correlators of the Kontsevich model, when the double scaling
limit is taken. In this way, it is possible to calculate the intersection numbers
on \mbs, by ordinary matrix model calculations, for all genera~\cite{matrix-beyond}.
This is a very strong result, since these numbers are hard to calculate using
algebraic geometry for genera higher than two.

Finally, let us mention the work Distler~\cite{Distler} who has found similar
results, derived from the quantum gravity perspective. He showed these by coupling Liouville
gravity to conformal matter fields, with central charge $c=-2$. From the
relation~\cite{GM}
\begin{equation}
  c = 1 - \frac{3(2k-3)^{2}}{2k-1},
\end{equation}
we find that this corresponds to the $k=1$ multi-critical model. In this model,
Distler showed that the total action (Liouville + matter) via bosonization recovers
the form of the Labastida, Pernici and Witten~\cite{LPW} action in
equation~(\ref{LPW-bc-action}). By adding operators to the action, which correspond
to gravitationally dressed primary matter fields, Distler could investigate the scaling
behaviour of general correlators in the $k=1$ model, where he found results in accordance
with those found by Witten. Also the general $k$ scaling behaviour was investigated
in~\cite{Distler}, but the relation between topological and gravitational continuum
observables was never explicitly defined. It seems that it has played a role for several of
these investigations, that the results from the matrix model were known before the derivations.

\subsection{New Perspective on the Identification}

We present the new identification between the theories, reported in~\cite{AHW}. The result has
only been proved for genus zero, but in contrast to all previous identifications this result
does not involve taking the double scaling limit of the matrix model. In this way, we have
found the first identification, at the discrete level, between topological gravity and matrix
models. An important step in this work was to realize, that the well-known expressions for
the free energy in the so-called reduced hermitian matrix model, which only contain even
powers of $\phi$ in the potential, could be written as a contour integral similar to that in
equation~(\ref{partition-8}).

The study of the reduced hermitian matrix model goes back to  Bessis, Itzykson and
Zuber~\cite{BIZ}. They used the following potential
\begin{equation}
  \ubar(\phi) = \half \phi^2 + \sum_{p=2}^{\infty} \frac{\overline{g_p}}{N^{p-1}}
\phi^{2p}, \label{eq:defpot}
\end{equation}
for the $N\times N$ hermitian matrices $\phi$, with the partition function defined
as the following matrix integral
\begin{equation}
\label{eq:defmat}
\znbar = \int {\cal D}\phi \ \exp \left[ -  tr \ubar \right].
\end{equation}
This definition is different from the one used previously in this section, due
to the over-all factor of $N$, which is missing from the action. The result needed
from~\cite{BIZ} is the explicit solution of
the genus zero free energy
\begin{equation}
\egbar = \lim_{N \to \infty} - \frac{1}{N^2} \log \left(\frac{\znbar}{
\overline{Z_N}(0)}\right),
\end{equation}
which was shown to be
\begin{equation}
  \label{eq:egbarw}
\egbar = \int_{0}^{a^2} dr \ \frac{dw(r)}{dr} \left[ 1 -w(r) \right] \log \left(
\frac{w(r)}{r} \right) .
\end{equation}
The function $w(r)$ is the given by
\begin{equation}
w(r) = r + \sum_{p=2}^{\infty} \overline{g_p} \frac{(2p)!}{p!(p-1)!} r^p 
\end{equation}
and the integration limit $a^{2}$ is defined as the solution to the
equation
\begin{equation}
  1=w(a^2), \mgap a^2>0 .
\end{equation}
The first manipulation is to integrate equation~(\ref{eq:egbarw}) by parts
to obtain 
\begin{equation}
\label{eq:ebarint}
\egbar = \half \int_{0}^{a^2} dr \ (1-w)^2 \frac{d}{dr} \left[
\log\left( \frac{w(r)}{r}\right)\right].
\end{equation}
At this stage we note that by relabelling the coupling constants, it is possible to write
$\egbar$ as a contour integral similar to that of topological gravity. The coupling
constants $\{\overline{g}_{n}\}$ are transformed in to a new set of parameters $\{ t_{n} \}$
according to the definition
\begin{equation}
\label{eq:relate}
\frac{\overline{g_p}(2p)!}{p!(p-1)!} \equiv - \frac{t_p
t_{0}^{p-1}}{(1-t_1)^p} \mgap {\rm for \ } p \ge 2.
\end{equation}
While there were $\{ \overline{g}_{2},\overline{g}_{3},\dots\}$ parameters in the old set
of coupling constants, there are $\{ t_{0},t_{1},t_{2},\dots\}$ in the new set, i.e.
two more parameters which we may vary. The reason for choosing the definition in
equation~(\ref{eq:relate}) becomes clear if one expresses $w(r)$ in term of the new coupling
constants, where we find
\begin{equation}
  w(r) = r - \sum_{p=2}^{\infty} \left(\frac{t_{p}t_{0}^{p-1}}{(1-t_{1})^{p}}
  \frac{p!(p-1)!}{(2p)!}\right)\frac{(2p)!}{p!(p-1)!}r^{p} =
r - \sum_{p=2}^{\infty}\frac{t_{p}t_{0}^{p-1}}{(1-t_{1})^{p}}r^{p}.
\end{equation}
Replacing $r$ with $z$ and introducing a new set of coordinates
\begin{equation}
  z' = \frac{t_{0}}{1-t_{1}} z,
\end{equation}
we see that
\begin{equation}
  w(z) = z - \frac{1}{t_{0}} \sum_{p=2}^{\infty} t_{p} (z')^{p}.
\label{w-intermid}
\end{equation}
By defining a potential $V(z)$ of the same form as in topological gravity
\begin{equation}
  V(z)\equiv \sum_{p=0}^{\infty}t_{p}z^{p}, \label{eq:pot}
\end{equation}
we notice the identification
\begin{equation}
  \Bigl( w(z) -1 \Bigr) = \frac{1}{t_{0}}\Bigl( z' - V(z')\Bigr), \label{wV}
\end{equation}
since the right hand side can be expanded as
\begin{eqnarray}
 \frac{1}{t_{0}}\Bigl( z' - V(z')\Bigr) &=& \frac{1}{t_{0}} \Bigl( z'
 - t_{0} - t_{1}z' - \sum_{p=2}^{\infty} t_{p} (z')^{p}\Bigr) \nonumber \\ &=&
 z - \frac{1}{t_{0}}\sum_{p=2}^{\infty} t_{p} (z')^{p} - 1 \nonumber \\ &=& w(z)-1,
\end{eqnarray}
where we have applied equation~(\ref{w-intermid}) in the last step.
Note that the integration limits $[0,a^{2}]$ in
equation~(\ref{eq:ebarint}), can be translated, in the new coordinates,
to $[z'=0,z'=u]$ where $u$ is defined as the solution $u=V(u)$. This makes the cut structure
for $\egbar$ identical to that of topological gravity. Hence we rewrite equation~(\ref{eq:ebarint})
as
\begin{equation}
\egbar =  - \half \ \iint \left[1-w(z) \right]^2
\left[\frac{d}{dz} \log \left(\frac{w(z)}{z}\right) \right] \log
\left( \frac{w(z)-1}{z} \right) .
\end{equation}
The contour is taken to encircle the cut, which appeared
when we inserted the term $\log[(w(z)-1)/z]$. This is done since we identified the cut structure
to fit with the above integral. Using equation~(\ref{wV}) we obtain the final expression
for the free energy of the reduced hermitian matrix model at genus zero
\begin{equation}
  \overline{E_0}= - \frac{1}{2 t_{0}^2} \pint
(z-V)^2 \left[ \frac{d}{dz} \log \left( \frac{z-V+t_{0}}{z} \right)
\right] \zvlog .
\end{equation}
The contour is taken to enclose the cut $[0,u]$ on the real axis in a anticlockwise direction.
Even though this integral resembles the partition function of perturbed topological gravity
in equation~(\ref{partition-8}), we rescale the reduced hermitian matrix model, in order
to find the simplest possible identification between the theories. This is done to repair the
lack of an overall $N$ factor in equation~(\ref{eq:defmat}). We make the following changes:
introduce a matrix $\Phi \equiv \phi(1-t_{1})^{-\frac{1}{2}}N^{-\frac{1}{2}}$ and redefine the
couplings
\begin{equation}
  g_{p} \mapsto \left\{ \begin{array}{ll} g_{1}= -\frac{1}{2}t_{1}, & \\
      g_{p} = \overline{g_{p}}\left( 1-t_{1}\right)^{p} & p\geq 2 .\end{array}\right. 
\end{equation}
By these redefinitions, the partition function of the reduced hermitian matrix model is of the
form:
\begin{equation}
\label{eq:zred}
Z_N(g) = \int {\cal D} \Phi \ \exp \left[ - N tr U(\Phi) \right],
\end{equation}
with the potential 
\begin{equation}
\label{eq:upot}
U(\Phi) = \half \Phi^{2} + \sum_{p=1}^{\infty} g_{p} \Phi^{2p},
\end{equation}
including an gaussian term and a interaction part. Note that there also is a gaussian
term in the intersection part, such that the total $\Phi^{2}$ dependence reads
\begin{equation}
 \Phi^{2}\sim  N\left[ \frac{1}{2} \Phi^{2} + g_{1} \Phi^{2} \right] =
 \frac{1}{2}\Phi^{2}\left( 1-t_{1}\right)N = \frac{1}{2}\phi^{2},
\end{equation}
which illustrates the result of the rescaling.

\noindent
The genus zero free energy is
\begin{equation}
\label{eq:freeenergy}
E_0(g) = \lim_{N \to \infty} - \frac{1}{N^2} \log \left(
\frac{Z_N(g)}{Z_N(0)} \right) .
\end{equation}
The contour integral representation of the free energy changes to
\begin{equation}
E_0 = \half \log(1-t_1) - \frac{1}{2 t_{0}^2} 
\pint (z-V)^2 \left[ \frac{d}{dz} \log \left( 
\frac{z-V+t_{0}}{z} \right) \right] \zvlog ,
\end{equation}
with $V(z)$ defined by (\ref{eq:pot}) as before. The relationship
between the two sets of parameters is 
\begin{equation}
\label{eq:newrelate}
\frac{g_p(2p)!}{p!(p-1)!} = - {t_p
t_{0}^{p-1}} \mgap {\rm for \ } p \ge 1.
\end{equation}
For a given $U(\Phi)$ we may freely choose $t_{0}$ (except $t_{0}
\neq 0$) and then all the other $t_p$ are determined.

The identification can be simplified further  by
differentiating $E_0$ with respect to $t_l$. For $l\ge 1$,
\begin{equation}
\frac{\partial E_0}{\partial t_l} = 
\frac{- \delta_{l,1}}{2(1-t_1)} - \frac{1}{2 t_{0}^2} 
\pint \left[ 2 z^{l-1} (z-V) \zvlog - \frac{z^{l-1}
(z-V)^2}{z-V+ t_{0}}  \right] , \label{free-energy-derivative}
\eeq
after taking the $t_{l}$-derivative under the integral sign,
integrating by parts and dropping analytic terms in the integrand.
The last part of the integral can only have a pole (i.e. $1/z$ dependence) if $l=1$ and hence the
only non-analytic term is of the form:
\begin{equation}
\frac{1}{2 t_{0}^2}\pint \frac{ t_{0}^{2}}{z- t_{0} - t_{1}z + t_{0}} =
\frac{1}{2}\frac{1}{(1-t_{1})} \pint \frac{1}{z} = \frac{1}{2}\frac{1}{(1-t_{1})},
\end{equation}
which cancels the first term in equation~(\ref{free-energy-derivative}) when we
insert a delta function $\delta_{l,1}$ in the integral
\begin{equation}
 \frac{1}{2 t_{0}^2} 
\pint \frac{z^{l-1}(z-V)^2}{z-V+t_{0}}.
\end{equation}
This simplifies equation~(\ref{free-energy-derivative}) considerably. Note that the remaining
integral is identical to equation~(\ref{sigma-l}), such that we can write
\begin{equation}
\label{eq:mainresult}
\frac{\partial E_0}{\partial t_l} = - \frac{1}{t_{0}^2}
\llang \sigma_{l-1} \rrang_{_V}
\end{equation}
and hence for $l \ge 1$,
\begin{equation}
\label{eq:oneptrelate}
\llang \sigma_{l-1} \rrang_{_V} = \lim_{N \to \infty}
\frac{t_{0}^{l+1}}{N} \ \frac{l!(l-1)!}{(2l)!} \llang tr
\Phi^{2l} \rrang .
\end{equation}
This equation relates, at genus zero, the one-point correlation
functions in topological gravity with correlation functions in the
corresponding reduced hermitian matrix model. Setting $l=1$ gives
\begin{equation}
\llang P \rrang_{_V} = \lim_{N \to \infty} \frac{1}{N}
\frac{t_{0}^2}{2} \llang tr \Phi^{2} \rrang ,
\end{equation}
which shows that the correlation function for genus zero surfaces with
a single puncture in topological gravity, is proportional to the
partition function for triangulated surfaces with a marked link, with
a suitable identification of the coupling constants in the two
theories, that is, equation (\ref{eq:newrelate}). Note that this identification
does not include taking the double scaling limit of the matrix model. Taking the
limit $N \to \infty$, just ensures that only genus zero terms contributes to the
matrix model expectation value.

Using the same methods, one can show by differentiating $E_0$ with
respect to $t_{0}$ that,
\begin{equation}
\label{eq:dedeps}
t_{0}^3 \frac{\partial E_0}{\partial t_{0}} = -
\sum_{l=1}^{\infty} l \ t_{l+1} \llang \sigma_l \rrang_{_V} .
\end{equation}
Equation (\ref{eq:oneptrelate}) can be generalized to multi-point
correlators. Define a new set of operators, $\{\Sigma_l\}$, in the matrix
model as follows,
\begin{equation}
\Sigma_l \equiv N t_{0}^{l-1} \ \frac{l! (l-1)!}{(2l)!} \ tr \! \! \left( 
\Phi^{2l} \right) \mgap {\rm for \ } l \ge 1 .
\end{equation}
Then we have, by repeatedly differentiating (\ref{eq:oneptrelate})
with respect to $t_{l_i}$,
\begin{equation}
\llang \sigma_{l_1-1} \prod_{i=2}^{s} \sigma_{l_i}
\rrang_{\! \! _V} = \lim_{N \to \infty} \left( \frac{t_{0}}{N}
\right)^2 \llang \prod_{i=1}^s \Sigma_{l_i} \rrang_{\rm
\! \! conn.} \mgap {\rm for \ } l_i \ge 1 , \label{sigma-Sigma}
\end{equation}
where $\llang \cdots \rrang_{\rm conn.}$ denotes the
connected part of the expectation value in the matrix model, for example 
\begin{equation}
\llang \Sigma_{l_1} \Sigma_{l_2} \rrang_{\rm conn.}
\equiv \llang \Sigma_{l_1} \Sigma_{l_2} \rrang -
\llang \Sigma_{l_1}  \rrang
\llang \Sigma_{l_2}  \rrang .
\end{equation}
Note the special role of the first operator $\sigma_{l-1}$ in the lhs correlator
in equation~(\ref{sigma-Sigma}). Since $l_{i}\geq 1$, the only way to produce, say
a $\langle PP\rangle_{V}$ correlator, is to set $l_{1}=1$ and $s=1$ and thereafter take
a $t_{0}$ derivative. The resulting term on the rhs will not be so simple, since we
have an $t_{0}$ factor in front of $\langle \Sigma_{1} \rangle_{{\rm conn.}}$. We
obtain two contributions, when taking the $t_{0}$ derivative, and hence we do not find a
nice relation, between a single correlator on both sides of the equality sign, as 
in equation~(\ref{sigma-Sigma}) when $l_{i}\geq 1$. 
The relation between $\sigma_{l}$ in perturbed topological gravity  and
$\Sigma_{l}$ in the reduced hermitian matrix model is a new result and it is
different from the previous suggestions made in the literature~\cite{Witten:340, DVV, Distler}.
Several papers confuse $\sigma_l$ with an
operator $O_l$, which is related to the $l$-th multi-critical model, which is part of the
the reason for Witten writing the unspecified $\dots$ in
equation~(\ref{symbolic}). The relation in equation~(\ref{sigma-Sigma}) is exact in
genus zero, but is unknown for higher genera.

The identification between the two theories, makes it possible to calculate the topological
gravity version of the known loop correlators from the matrix model. We define 
loop and multi-loop correlators in the matrix model as
\begin{equation}
W(p_1) = \frac{1}{N} \sum_{k_1=0}^\infty \frac{ \llang tr \Phi^{2k_1}
\rrang}{p_1^{2k_1 +1}} \label{w-et-punkt},
\end{equation}
\begin{equation}
W(p_1, \cdots , p_s) = N^{s-2} \sum_{k_1,\cdots,k_s =1}^\infty
\frac{\llang tr \Phi^{2k_1} \cdots tr \Phi^{2k_s} \rrang_{\rm
conn.}}{ p_1^{2k_1 +1} \cdots p_s^{2k_s +1}}\,; \,\,\,\,\,\, s\geq 2 .
\end{equation}
Since we only have the identification between the theories in genus zero, we let $N\rightarrow
\infty$ and denote the genus zero contribution $W_{0}(p_{1},\cdots, p_{s})$.
By performing a Laplace transformation of all the individual $p_{i}^{2}$, $(i=1,2,\dots,s)$ factors
we obtain:
\begin{equation}
w_0(l_1) = \frac{1}{N} \sum_{n_1=0}^\infty
\frac{l_1^{n_1-\half}}{\Gamma (n_1+\half )} 
\llang tr \Phi^{2 n_1} \rrang, \label{et-loop-w-discrete}
\end{equation}
and for $s \geq 2$,
\begin{equation}
w_0(l_1,\cdots,l_s)= N^{s-2} \sum_{n_1,\cdots,n_s=1}^\infty
\frac{l_1^{n_1-\half}}{\Gamma (n_1+\half )} \cdots
\frac{l_s^{n_s-\half}}{\Gamma (n_s+\half )} 
\llang tr \Phi^{2 n_1} \cdots tr \Phi^{2 n_s}\rrang_{\rm conn.} .
\end{equation}
These loop functionals are generating functionals for the connected
$tr \Phi$ correlators, and we note that equation~(\ref{et-loop-w-discrete}) is the discrete
version of the continuum loop functional in equation~(\ref{et-loop-w-cont.}). 

By applying equation~(\ref{eq:oneptrelate}) we can express the loop functionals as
contour integrals in topological gravity. For example we find~\cite{AHW} the one loop
functional in equation~(\ref{w-et-punkt}) to be of the form:
\begin{equation}
W_0(p_1) = - \left( \frac{1}{p_1 t_{0}} \right) \pint \left(1-
\frac{z}{(t_{0}p_{1}^{2})/4} \right)^{-\half} \left(1-\frac{dV}{dz} \right) \zvlog ,
\end{equation}
in correspondence with matrix model results in the literature~\cite{Jan-review}.

By considering the multi-critical reduced hermitian matrix models, we are able to
clarify some of the confusion in the literature, on the relation between
scaling operators in topological gravity and matrix models. As stated earlier,
the scaling behaviour for a general $\langle \sigma_{d_{1}}\cdots \sigma_{d_{s}}\rangle_{V}$
correlator, was found by Witten~\cite{Witten:340} to equal that of matrix model
scaling operators found by Gross and Migdal~\cite{GM}. The formulation of the
reduced hermitian matrix model in~\cite{GM} is based on the partition function
\begin{equation}
\label{eq:hermdef}
Z_N = \int {\cal D} \Phi \ \exp \left[ - \frac{N}{t_{0}}
tr U(\Phi) \right],
\end{equation}
where an overall factor $1/t_{0}$ is taken out in front of the potential
\begin{equation}
\label{U-def}
U(\Phi) = \half \Phi^2 - \sum_{n=1}^\infty \overline{\mu_n} U_n(\Phi),
\end{equation}
with a new set of coupling constants $\{ \overline{\mu_{n}}\}$.
This formulation is special in the way that the functions $U_n(\Phi)$ are
derived such that the model describes the $n$'th multi-critical model.
The explicit form of these potentials was derived by Gross and Migdal in~\cite{GM}
\begin{eqnarray}
U_1(\Phi) &=& \half \Phi^2 , \\
U_2(\Phi) &=&  \Phi^2 - \frac{1}{12} \Phi^4, \\
& \vdots & \\
U_n(\Phi) &=& \sum_{p=1}^n (-1)^{p-1} \Phi^{2p} \frac{n!
(p-1)!}{(n-p)!(2p)!} \ .
\end{eqnarray}
The effect of having the overall $1/t_{0}$ factor, changes the relation between
topological and matrix model couplings to
\begin{equation}
\label{eq:gptotp}
t_p = - \frac{g_p (2p)!}{p!(p-1)!} \mgap {\rm for \ } p \geq 1 .
\end{equation}
The potential for the topological model corresponding to the above matrix model is
\begin{equation}
\label{eq:potmultipert}
V(z) = t_{0} - \sum_{n=1}^\infty \overline{\mu_n} \left[ (1-z)^n -1 \right], 
\end{equation}
and the $k$-th multi-critical model is obtained by choosing $t_{0}
\neq 0$, $\overline{\mu_1}=1$ and $\overline{\mu_k}=-1$; that is,
\begin{equation}
\label{eq:vkth}
V(z)= t_{0} + z + (1-z)^k -1 \equiv V_k.
\end{equation}
The approach by Gross and Migdal was to study the
$k$'th multi-critical model, by adding perturbations around the $k$'th multi-critical
point from the other multi-critical points. In equation~(\ref{eq:potmultipert})
we expand around the $k=1$ point, which is illustrated by the presence of $t_{0}$
in equation~(\ref{eq:vkth}).
Note that this is different from the potential usually chosen in
topological gravity for the $k$'th multi-critical model, namely $V=
t_{0} +z - z^k$; this potential, which has $t_1=1$, would give
$g_1=-\half$ and hence $U(\Phi)=\sum_{k=2}^{\infty} g_k \Phi^{2k}$. Lacking a
$\Phi^2$ in the exponential 
term to ensure the Wick contractions as in equation~(\ref{Wick}), the potential
$U(\Phi)=\sum_{k=2}^{\infty} g_k \Phi^{2k}$ defines a matrix model, which 
does not correspond to random surfaces made from gluing together polygons.
This is the source of much of the confusion that
has arisen when comparing the multi-critical matrix and topological
models. The reason for this fact not being clarified a long time ago, is that
all prior identifications of the theories have been made in the continuum limit
of the matrix model, where the scaling behaviour erases any sign of the difference
in the potential of the $k$'th multi-critical model.

We consider this in some detail, by studying the scaling behaviour for the $k$'th
multi-critical topological gravity model, using the potential in equation~(\ref{eq:vkth}).
The end of the cut $u$, defined by $u=V_{k}(u)$ is of the form
\begin{equation}
  u = 1 - (1-t_{0})^{\frac{1}{k}} \equiv 1 - (\Delta t_{0})^{\frac{1}{k}}.
\label{GM-u}  
\end{equation}
Insert this expression for $u$ in equation~(\ref{partition-8}) to obtain
\begin{eqnarray}
  \lan 1 \ra_{V_{k}} &=&  -\frac{1}{2}\frac{1}{2\pi i}\oint_{C} dz\,  
  \biggl( z- \bigl( t_{0} + z +(1-z)^{k}-1\bigr)\Biggr)^{2}\,
  \ln \left(1-\frac{ t_{0} + z +(1-z)^{k}-1  }{z}\right) \nonumber \\ &=&
  \frac{1}{2} \int_{0}^{u} dx\,  \left( \Delta t_{0} - (1-x )^{k}\right)^{2} \nonumber \\ &=&
  -\frac{1}{2} \int_{1}^{(\Delta t_{0})^{1/k} } dy\,  \left( \Delta t_{0} - y^{k} \right)^{2}
   \nonumber \\ &=& -\frac{1}{2} \Biggl(
   \biggl[ (\Delta t_{0})^{2}y \biggr]_{1}^{(\Delta t_{0})^{1/k}}
   + \biggl[ \frac{1}{2k+1}y^{2k+1} \biggr]_{1}^{(\Delta t_{0})^{1/k} }
 -2 \biggl[  (\Delta t_{0}) \frac{1}{k+1} y^{k+1} \biggr]_{1}^{ (\Delta t_{0})^{1/k} }
 \Biggr) \nonumber \\ &=&  -\frac{1}{2} \biggl( (\Delta t_{0} )^{2+1/k} - (\Delta t_{0})^{2} 
    +   \frac{1}{2k+1}  (\Delta t_{0})^{2+1/k} - \nonumber\\&& \qquad \frac{1}{2k+1} -
  \frac{2}{k+1} (\Delta t_{0})^{2+1/k} - \frac{2}{k+1} (\Delta t_{0}) \biggr)
     \nonumber \\ 
     &\sim & -\frac{k^{2}}{(2k+1)(k+1)}(\Delta t_{0})^{2+1/k} .
\end{eqnarray}
In the first step, the contour is taken to enclose the cut $[0,u]$ in a anticlockwise
direction and in the second step we perform the substitution $y=1-x$. The individual
terms are integrated and evaluated and we drop the non-critical terms in the last step
to obtain the scaling behaviour of the topological partition function in the $k$'th
multi-critical model, based on the potentials used by Gross and Migdal~\cite{GM}.
We see that the scaling behaviour is the same as the one derived in
equation~(\ref{k=k:1}) for the simpler potential $t_{0}\neq 0, t_{1}=1, t_{k}=-1$.

It is possible to define a new set of scaling operators, which scales in the
correct way in the $V_{k}$ potential. 
These are denoted $\{O_l\}$:
\begin{equation}
\label{eq:defol}
O_l = - \sum_{n=0}^l (-1)^n \left( {l \atop n} \right) \sigma_n ,
\end{equation}
and were first introduced in~\cite{GM}.
Compared to $\langle 1 \rangle_{V_{k}}$, the operator $O_l$ produces a
factor of $-(1-z)^l$ in the
contour integral and by direct evaluation of equation~(\ref{sigma-l}) we find
\begin{equation}
\lan O_l \ra_{_{V_k}}
 = - \int_1^{(\Delta t_{0})^{1/k}} \back dy \ y^l (\Delta t_{0} -
y^k ) \sim   \frac{-k}{(l+k+1)(l+1)} (\Delta t_{0})^{1+(l+1)/k},
\end{equation}
where we have dropped the regular terms in the last step.
Using equation~(\ref{n-punkt}), the general case (for $s \geq 2$) is given
by
\begin{eqnarray}
\label{eq:genocorr}
\lan \prod_{i=1}^s O_{l_i} \ra_{V_k}
 &=& - \left( - \dde
\right)^{s-2} \int_1^{(\Delta t_{0})^{1/k}} dy \ y^L \\
& \sim & - \left(\frac{\partial}{\partial (\Delta t_{0})} \right)^{s-2}
\left[ \frac{1}{(L+1)} (\Delta t_{0})^{(L+1)/k} \right] ,
\end{eqnarray}
where $L=\sum\limits_{i=1}^s l_i$. We see that up to regular terms, 
the scaling with respect to
$\Delta t_{0}$ of correlation functions of
$O_l$ for the potential
\begin{equation}
  V_k= t_{0}  + z + (1-z)^k -1,
\end{equation}
is the same as the scaling with respect to $t_{0}$ of
correlation functions of $\sigma_l$
using the simpler potential
\begin{equation}
V= t_{0} + z - z^k.
\end{equation}
Part of the confusion in earlier papers on the identification between topological and
matrix model scaling operators, lies in the wrong assumption that $\sigma_{l}=O_{l}$.
But we have shown that the usual $\sigma_{l}$ operators does not correspond to the
matrix model, describing the $k$'th multi-critical point for random surfaces, and that
one should instead use the $O_{l}$ operators defined to fit the potentials used
by Gross and Migdal. We see that it is possible to make a redefinition of the
operators in topological gravity, such that the scaling behaviour corresponds to the
point in coupling constant space, where the matrix model corresponds to gluing polygons to
form random surfaces. Further relations involving $O_{l}$ and matrix model correlators
can be studied in~\cite{AHW}.

The identification between the reduced hermitian matrix model and perturbed topological
gravity, has only been proved at genus zero. There are several reasons for this result,
not being generalized to higher genera. First, for higher genus we lack the nice result
from equation~(\ref{dijkgraaf-1}), which enabled us to write $\langle 1 \rangle_{V}$ in
the simple form of equation~(\ref{partition-2}). This was very important for deriving the
contour integral representation of $\langle 1 \rangle_{V}$ in equation~(\ref{partition-8}).
Another problem is that the explicit expressions for the free energy of the matrix model
are more complicated at higher genus. For these reasons, it has not been possible to
extend the results derived in~\cite{AHW} to higher genera. If this had been possible, it
would have opened for a new interpretation of the intersection numbers on the moduli
space of Riemann surfaces, but at the present stage we do not have a geometric understanding
of why the new identification is true.

At this stage, we turn our attention to a question which arises when one tries to
interpret the connection between topological and quantum gravity. We have
seen that it is possible to form topological counter parts to the loop functional of quantum
gravity. But whereas these have a direct physical interpretation in quantum gravity, as
expectation values for surfaces with macroscopic boundaries of a certain length, the
topological versions seem to offer no such interpretation. In the topological theory,
one should not expect any information of metric quantities such as the geodesic
length of a boundary, to be available. This controversy has been raised by Dijkgraaf, E. and H.
Verlinde in~\cite{DVV,DV,DVV-1}, but without offering any explanation.

There are two possible ways to address this question. One is to enlarge the discussion
of topological gravity, to involve Riemann surfaces with boundaries, which allows for the
introduction of BRST invariant observables
\begin{equation}
\oint_{\partial \Sigma_{g}}  \sigma_{n}^{(1)}.
\end{equation}
In contrast to the general situation without boundaries, the integrated one-form observable
will be invariant under diffeomorphisms preserving the boundary, and is a possible
invariant. This quantity can be added to the action as a perturbation of $S_{{\rm q}}$, where
the coupling constant, related to the boundary operator, plays the role of a boundary cosmological constant. This approach is similar to that used in Liouville theory~\cite{Ginsparg-Moore}
and it has been advocated by Huges and Montano in~\cite{Huges-Montano}. An
alternative approach to boundary contributions has been studied by Myers~\cite{Myers}, but
none of the above methods have had any real
success. In~\cite{AHW}, we advocated a simpler approach which relates to the question
of whether there is any paradox at all, or whether it is natural to have global metric
information in a theory of topological gravity.

We showed in~\cite{AHW}, that it is possible to formulate, at least formally, a
Witten type topological field theory for a restricted set of metrics. The particular
model, is based on the metrics, which corresponds to a certain fixed value for the
area of the underlying Riemann surface. Or more directly, we want to localize the
elements in \met, to those in the subset
\begin{equation}
 \met \rightarrow \left\{g_{\alpha\beta} \in \met \left\vert
     \int_{\Sigma_{g}} \sqrt{g}=A^{0} \right.\right\}.
\end{equation}
Thus we can consider a localization function 
 \begin{equation}
{\cal F}[g] = \int_{\Sigma} \sqrt{g} \, -  A^{0} = 0, \label{area-loc}
\end{equation}
which clearly is gauge invariant, i.e.\ diffeomorphism invariant. We know from 
general principles~\cite{Birmingham-review} that such a gauge invariant statement
can be implemented in the action by the use of Lagrange multipliers.
For this we need to know the action of the BRST operator, using equations~(\ref{LPW-g},
\dots, \ref{LPW-phi}) on ${\cal F}[g]$,
\begin{eqnarray}
  \br \int_{\Sigma_{g}}\sqrt{g}  &=&  \int_{\Sigma_{g}} \br \left(
   \sqrt{det(g_{\alpha\beta})} \right) =
   \int_{\Sigma_{g}} \frac{1}{2}\frac{1}{\sqrt{g}}  \br 
  \left(e^{tr \, \ln \, g_{\alpha\beta}}\right) \nonumber \\ &=&
\int_{\Sigma_{g}} \frac{1}{2}\sqrt{g}\,
tr\, ( g^{\mu\gamma}\br (g_{\gamma\nu})) =
\int_{\Sigma_{g}} \frac{1}{2}\sqrt{g}\,
 (2 D^{\mu}c_{\mu} + \psi^{\mu}_{\mu})  \nonumber \\
 &=& \int_{\Sigma_{g}} \frac{1}{2}\sqrt{g}\,\psi^{\,\,\mu}_{\mu}.\label{var-A-1}
\end{eqnarray}
In the last step we used the fact that the area is diffeomorphism invariant,
reflected by the integral over the covariant derivative being a total
derivative. The integral of the trace of the topological
ghost is generally non-vanishing and hence the area is not BRST invariant.
This is a consequence of the non-invariance of the area under a
topological shift transformation in equation~(\ref{shift}), which is proportional to
\begin{equation}
  \delta_{{\rm shift}} \int_{\Sigma_{g}}\sqrt{g} = -\int_{\Sigma_{g}}\frac{1}{2}\sqrt{g}
  \, tr \, (r^{\gamma}_{\mu}g_{\gamma\nu}+
  r^{\gamma}_{\nu}g_{\gamma\mu}) = -\int_{\Sigma_{g}} \sqrt{g} \, r^{\mu}_{\mu} .
\end{equation}
We see that the action of the BRST operator on the area directly reflects
this fact. Note that compared to the discussion in chapter 4, we now need to
take the BRST derivative of the $\sqrt{g}$ factor, since it is not multiplied by
an arbitrary expression $f(\Phi)$ of the field in the theory, as in
equation~(\ref{metric-argument}).

In order to restrict the integration over metrics, we must insert the
constraint in equation~(\ref{area-loc}) together with its superpartner through
equation~(\ref{var-A-1}). Symbolically we have the situation
 \begin{equation}
Z[A^{0}]\equiv
\int {\cal D}[\Phi] \, e^{-S_{{\rm q}}} \delta \left(\int_{\Sigma} \sqrt{g} \, - A^{0}
\right) \left( \int_{\Sigma} \sqrt{g} \, \psi^{\mu}_{\mu} \right), \label{partition-loc}
\end{equation}
where $\Phi$ symbolises all relevant fields in the action and where we have inserted a
bosonic and a fermionic delta function. In order to incorporate the delta function
as part of the action, we introduce a set of Lagrange multipliers $(\chi,\mu)$.
The usual relation for a delta function
\begin{equation}
  \int_{-\infty}^{\infty} \frac{dy}{2\pi} e^{-iyx} = \delta(x),
\end{equation}
involves an imaginary unit $i$ in the exponential. We move this factor to the $\mu$
integration limit by an analytical continuation, and rewrite the partition function from
equation~(\ref{partition-loc}) as
\begin{equation}
  Z[A^{0}] = \int {\cal D}[\Phi]d\mu d\chi \, \exp \biggl( -S_{{\rm q}}
- \mu \left(\int_{\Sigma_{g}} \sqrt{g}- A^{0}\right) + \chi \int_{\Sigma_{g}}\frac{\sqrt{g}}{2}
\psi_{\mu}^{\,\,\mu}\biggr) . \label{partition-loc-2}
\end{equation}
As reported in~\cite{AHW}, we note that by letting $(\chi,\mu)$ form a BRST anti-ghost
multiplet
\begin{eqnarray}
  \br \chi &=& \mu, \\
  \br \mu &=& 0,
\end{eqnarray}
it is possible to write the exponential of equation~(\ref{partition-loc-2}) in a BRST
exact way
\begin{equation}
   Z[A^{0}] = \int {\cal D}[\Phi]d\mu d\chi \, \exp \biggl( -S_{{\rm q}} -
\left\{ Q, \chi\left( \int_{\Sigma_{g}}\sqrt{g} - A^{0} \right) \right\}\biggr).
\end{equation}
This action defines, at least formally, a Witten type theory of topological gravity which
directly involves global metric information about the underlying surface, namely the global
area. The Lagrange multipliers are not fields, but functions, since the expressions to
which they couple in the action, are global expressions and they constitute a
so-called constant anti-ghost multiplet~\cite{Baulieu-Bellon}.

As a consistency requirement, we must impose the constraint, that the chosen area must
be compatible with the background metric, which is included in $S_{{\rm q}}$, such that
\begin{equation}
  \int_{\Sigma_{g}} \sqrt{g^{0}} = A^{0}.
\end{equation}
Even though the action satisfies the requirement for a Witten type theory by having a
BRST exact action, it is not clear what kind of topological invariants it models. The
space of equivalence classes of metrics, with a fixed area, under diffeomorphisms is
still an infinite dimensional space, in contrast to \m. But, by imposing a curvature
localization term like equation~(\ref{Weyl-loc-1}), it is possible to model \m through
this method, if we integrate over all possible values of $A^{0}$ in the end.
This is a central point. Intuitively, it should be possible to include metric information
such as the area in the partition function, and then at the end integrate
over all possible choices for the area to recover the original theory. By performing a
Laplace transformation in $A^{0}$ we find
\begin{eqnarray}
  Z[\mu'] &=& \int_{0}^{\infty}e^{-\mu'A^{0}}Z[A^{0}] dA^{0} \nonumber \\
          &=& \int {\cal D}[\Phi] d\chi d\mu \delta\left(\mu-\mu^{'}\right)
            \exp\biggl( -S_{{\rm q}} - \mu\int_{\Sigma_{g}}\sqrt{g} + \chi\int_{\Sigma_{g}}
            \frac{\sqrt{g} }{2} \psi_{\mu}^{\,\,\mu} \biggr) \nonumber.
\end{eqnarray}
By integrating over $\mu$ we find
\begin{equation}
          Z[\mu] =\int  {\cal D}[\Phi] d\chi  \exp \biggl( -S_{{\rm q}} -
          \left\{ Q, \chi \int_{\Sigma_{g}}\sqrt{g} \right\} \biggr),
\end{equation}
where we have relabel the integration variable $\mu'\mapsto \mu$. Hence
by integrating over all areas, we find a topological action with a BRST
exact ``cosmological'' term. We put cosmological in quotation marks, because we
have two terms, namely the cosmological term and its superpartner. It is well-known
in quantum gravity that the area and the cosmological constant are ``dual'' under Laplace
transformation.

This calculation shows that there is no rule against defining a Witten type theory of
topological gravity, which involves metrical information. Recall that the metric, in contrast to
the Schwarz theories, is usually present in the action of a Witten type theory. The
theory is topological, through the BRST exact nature of the action, which ensures that
all physical quantities, only change inside the same cohomology class under an
infinitesimal transformation of the metric. Hence, from this point of view there is 
nothing wrong when we have an identification of quantum gravity with a topological theory.
Here it is important to stress that the metrical information present in quantum gravity,
refers to global metrical information, such as the area or the geodesic length of a boundary
or between two marked points. The observables of quantum gravity are precisely objects
which depend on such global parameters. For example the Hartle-Hawking wave-functionals
are functions of the geodesic boundary length of the surface. If we consider the continuum
one loop functional from equation~(\ref{et-loop-w-cont.})
\begin{equation}
  \lan w(l) \ra \sim \sum_{n=0}^{\infty} l^{n+1/2}\lan \sigma_{n} \ra,
\end{equation}
we note that the functional dependence factorizes between the metric parameter $l$ and
the topological part $\sigma_{n}$. One should of course make the proper translation from
the continuum matrix model operator $\sigma_{n}$ to that of topological gravity, but it
is clear from the previous discussions, that this part is related to the topological
information.
Hence we face the situation, where an observable of quantum gravity has an expansion in
a global metric parameter, where the individual coefficients of the expansion are
topological invariants. This is a very natural situation. If a function $F(m)$ only depends
on some global metrical quantity $m$, and we can make an expansion, say a power series
expansion, where the coefficients $a_{n}$ are independent of $m$
\begin{equation}
  F(m) \sim \sum_{n=0}^{\infty} a_{n}m^{n},
\end{equation}
then the coefficients $a_{n}$ can only depend on the underlying topology, since all metrical
dependence per definition lies in $m$.
This interpretation offers an explanation for the relation between
quantum gravity and topological gravity. Recall also that we had to perturb topological
gravity, in order to find the identification with quantum gravity, which destroys the
BRST exact nature of the action. In this sense, additional information is included in
the theory through the coupling constants. But we know that through a Laplace
transformation, the coupling constants of quantum gravity can be translated to global
metric parameters and hence we see a possible explanation for the appearance of global
metric information appearing in perturbed topological gravity.

%% file: chap6.tex
\chapter{Discussion}

In this chapter, we present discussions and conclusions related to the topics
presented in the previous chapters. As stated in the introduction, the most prominent
questions are the role of the Ashtekar formalism in two dimensions and the relation
between the BF and super BF approaches to $2D$ quantum gravity. The last section, focuses on
a problem in the super BF formulation of $2D$ quantum gravity which has been overlooked
in the literature.

\section{Ashtekar Gravity in Two Dimensions}
The first topic we discuss, is the role of Ashtekar gravity in two dimensions
and whether this formalism of gravity has any relation to $2D$ quantum gravity.

Let us try to characterize the fundamental nature of the Ashtekar
formulation of gravity. The first important thing to note is that
Ashtekar gravity is a reformulation and enlargement of general relativity,
which simplifies the form of the hamiltonian constraint, known
as the Wheeler-DeWitt equation. The formulation is
a gauge theory, built on the Lorentz group and the
Ashtekar connection $A_{\mu}^{i}$ takes values in the Lie algebra of the Lorentz
group. The $m$-bein $e_{\mu}^{\,\,i}$ is the conjugated momentum
to $A_{\mu}^{i}$, which signals that the Ashtekar formulation is constructed
to formulate a canonical theory. This is also obvious, since the hamiltonian
constraint naturally appears in the canonical formulation of general
relativity. But it is possible to work in a covariant formulation of
Ashtekar gravity and by doing so, it becomes clear that that the
two-dimensional situation is special. Consider the action of Ashtekar gravity
in four, three and two dimensions:
\begin{equation}
  \begin{array}{llll}
    S[A,e] &=& \int d^{4}x \,tr\left( e\wedge e\wedge F[A]\right); & \mbox{
      $4$ dimensions} ,\\
     S[A,e] &=& \int d^{3}x \,tr\left(e\wedge F[A]\right); & \mbox{
      $3$ dimensions} ,\\
     S[A] &\stackrel{?}{=}& \int d^{2}x \, tr\left( F[A]\right); & \mbox{
      $2$ dimensions},    
\end{array}
\end{equation}
where the two-dimensional action is a proposal made by extrapolation from
the four and three dimensional results. The problem is that general
relativity does not exist in two dimensions due to the vanishing
Einstein equations. When there are no Einstein equations, there is no
hamiltonian constraint and the Ashtekar formulation has 
no role in two dimensions from this point of view.

In three dimensions, the action of Ashtekar gravity is clearly of the BF
type~\cite{Pullin-Gonzales} and the theory offers a topological
field theory description of general relativity. This is closely related
to Witten's description of Chern-Simons gravity~\cite{Witten-2+1:1}, where
the only difference lies in the choice of gauge group. While Ashtekar gravity
is formulated using the Lorentz group, Witten uses the Poincar\'{e} group,
and a detailed map between the theories has been known for some
time~\cite{Romano}. The proposed two-dimensional action involves only the $A$ field
and offers no direct possibility to introduce Ashtekar
variables $(A,e)$ in two dimensions.

But, there are several possibilities to
study the question in an indirect manner. One possibility is to study the
role of BF gravity in two dimensions, since the three-dimensional
action of Ashtekar gravity is of the BF type.
Another approach is to say that the fundamental feature of the Ashtekar
formulation is the canonical nature of the variables, and in that way
generalize the question to investigate the role of canonical quantization
in two-dimensional gravity. Here one still needs to define what kind
of gravitational theory to study. We discuss two different choices, the
canonical approach to Jackiw-Teitelboim gravity and a canonical gauge
fixing of the metric in the path integral of $2D$ quantum gravity.

The canonical formulation of general relativity goes back to the
seminal work by Arnowitt, Deser and Misner~\cite{ADM}, who introduced
a split of space-time into space and time, i.e. a foliation of
the space-time manifold, such that a canonical formulation can be defined.
In this process, the two-dimensional space-time metric $g_{\mu\nu}$ is parametrized
by the spatial metric $h$ plus two additional fields, known as the
shift $N$ and lapse $\lambda$
\begin{equation}
  ds^{2} = g_{\mu\nu}dx^{\mu}dx^{\nu} = \left(N(x^{0},x^{1})dx^{0}\right)^{2} +
  h(x^{0},x^{1})\left( \lambda(x^{0},x^{1}) dx^{0} + dx^{1}\right)^{2}.
\end{equation}
The lapse represents the vector field generating diffeomorphisms in
the time $(x^{0})$ direction, while the shift generates diffeomorphisms in
the space direction $(x^{1})$. The metric tensor is thus of the form
\begin{equation}
  g_{\mu\nu} = \left[ \begin{array}{cc} N^{2} + h\lambda^{2} & h\lambda  \\
      h \lambda & h \end{array} \right].
\end{equation}
A direct approach to canonical quantization of the Jackiw-Teitelboim model has
been made by Henneaux~\cite{Henneaux-JT}, where he isolates the constraints
of the theory, which generate the symmetries. These are the vector constraint generating
the spatial diffeomorphisms and the hamiltonian constraint generating the time diffeomorphisms.
But the quantization is only made for open universes, which is somehow unrelated to our
previous discussions. A slightly more indirect approach was made by Banks and
Susskind~\cite{Banks-Susskind}, where they considered a canonical approach to the
Liouville action, descending from Polyakov's approach to $2D$ quantum gravity
in the conformal gauge. The Liouville
action, which has the same equation of motion as the Jackiw-Teitelboim model is studied in
the so-called synchronous gauge (in terms of the zwei-bein)
\begin{eqnarray}
  e_{0}^{\,\,0} &=& 1  ,\\
   e_{0}^{\,\,1} =  e_{1}^{\,\,0} &=& 0, \\
  e_{1}^{\,\,1} &\equiv& e.
\end{eqnarray}
This gauge choice corresponds to $g_{00}=1$, $g_{01}=g_{10}=0$, $g_{11}=h$, in terms of the
metric. For a space-time manifold $M$ of the form $\R \times S^{1}$, the Liouville
action
\begin{equation}
  S_{{\rm L}} = \rho\int_{M} \biggl[  \frac{\left(\partial_{\mu}\Phi\right)^{2}}{2} -
  \Lambda e^{2\Phi}\biggr];\,\,\,\,\,\, \rho = \frac{26}{24\pi},
\end{equation}
leads to the following lagrangian and hamiltonian functions in the synchronous gauge
\begin{equation}
  L =  \rho \int dx^{1}  \left( \frac{\dot{e}^{2}}{2e} - \Lambda e\right), \label{BS-action}
\end{equation}
and
\begin{equation}
  {\cal H}(x^{1}) =  \rho \left( \frac{\dot{e}^{2}}{2e} + \Lambda e\right) =
  \frac{1}{2\rho} e \Pi_{e}^{2} + \rho\Lambda e,
\end{equation}
where $\Pi_{e}= 2\rho \dot{e}/e$ is the conjugated momentum to $e$. The Wheeler-DeWitt
equation is formally solved, but the resulting quantum theory suffers from 
problems with non-normalizable wave-functions. It is claimed that $\Lambda<0$ leads to
a universe where the wave-functions represent superpositions of expanding and contracting
De Sitter space-times. Banks and Susskind~\cite{Banks-Susskind} concluded that there is no
possibility for encountering spatial quantum fluctuations.

\noindent
A different canonical approach to two-dimensional quantum gravity
has been pursued in relation to developments in string field theory, made
by Ishibashi and Kawai~\cite{Ishibashi-Kawai}. In this formulation, one has creation
and annihilation operators creating and removing closed loops, which can then propagate
in time, due to a hamiltonian time evolution. Interactions can occur between
strings, which may join or split to form a lower or higher number of propagating strings.
In the paper~\cite{Fukuma}, the loop-loop correlator $f(l,l',D)$, for an initial loop $C$ of
length $l$ and an exit loop $C'$ of length $l'$, separated by the geodesic distance $D$, was studied
in a path integral approach
\begin{eqnarray}
  f(l,l',D)&\equiv& \int \frac{{\cal D}[g_{\mu\nu}]}{{\rm Vol}(\diff)}
  \exp\left(-\Lambda\int d^{2}x \sqrt{g}\right)\delta\left(\int_{C}\sqrt{g_{\mu\nu}dx^{\mu}dx^{\nu}}
   - l\right)\nonumber \\ && \cdot\delta\left(\int_{C'}\sqrt{g_{\mu\nu}dx^{\mu}dx^{\nu}}
   - l'\right).
\end{eqnarray}
This is supplemented with boundary conditions ensuring the propagating loop to be homeomorphic
to the initial loop at all geodesic separations $d<D$ while the propagating loop should be
homeomorphic to $C'$ when the distance $d$ becomes identical to $D$. The relevance of this
approach in a discussion of canonical quantization, is the choice of an ADM gauge
in~\cite{Fukuma}, leading to the gauge fixed metric
\begin{equation}
  \overline{g}_{\mu\nu} =
\left[ \begin{array}{cc} 1 + l(x^{0})^{2}\overline{\lambda}(x^{0},x^{1})^{2} & l(x^{0})^{2}\\
 l(x^{0})^{2}\overline{\lambda}(x^{0},x^{1})^{2}  &  l(x^{0})^{2} \end{array} \right].
\end{equation}
Here the length $l(x^{0})^{2}\equiv h(x^{0},x^{1})$ due to the gauge choice
$\partial_{1}h(x^{0},x^{1})=0$. The space-time topology is also of the canonical
form $\R\times S^{1}$.
The path integral is evaluated by similar methods to
that used in the derivation of the Polyakov string~\cite{Hat} and the integration
over metrics gives rise to integration over $l$ and $\overline{\lambda}$ together with two fields
$v(x^{0},x^{1})$, $c(x^{0})$ parametrizing the diffeomorphism. The computations are
rather complicated and involve the introduction of an auxiliary positive constant $\beta$,
which is used to transform a certain differential operator into an elliptic operator, which
eases the calculations. The resulting path integral over $l$, after integrating over the remaining
fields, is then given in the limit $\beta\rightarrow 0$:
\begin{equation}
  f(l,l',D) = \lim_{\beta\rightarrow 0} \int\frac{dl}{l}e^{-S_{\beta}[l]} \delta\Biggl(
    \bigl[l(x^{0})=0\bigr] - l\Biggr)\delta\Biggl( \bigl[l(x^{0})=D\bigr] - l'\Biggr).
\end{equation}
The regularized action is of the form
\begin{equation}
  S_{\beta}[l] = \int dx^{0} \left( \frac{\dot{l}^{2}}{2\beta l} + \Lambda_{R}l
    \right), \label{Fukuma-action}
\end{equation}
where $\Lambda_{R}$ is the renormalized cosmological constant. After the $\beta\rightarrow 0$
limit is taken, the hamiltonian reduces to:
\begin{equation}
H = \Lambda_{R}l,
\end{equation}
which gives the loop-loop correlator the following simple form
\begin{equation}
 f(l,l',D) = \gamma \langle l' \vert e^{-DH}\vert l\rangle = \gamma e^{-\Lambda_{R}lD}\delta\left(
   l-l'\right), \label{cylinder}
\end{equation}
where $\gamma$ is some constant.
The nontrivial part of the theory lies in possibility for loops to join and split~\cite{Fukuma}.
Note that the length of $l\sim e_{1}^{\,\,1}(=e)$, since the zwei-bein formally is the square
root of the metric, which makes it appealing to relate the two actions in
equations~(\ref{BS-action},\ref{Fukuma-action}). But there are differences, first of all in the
sign of the cosmological term and in the fact that $\Lambda_{R}$ is the renormalized cosmological
constant. The resemblance of the actions is due to the role played by the Liouville action,
which is involved as a Jacobian in the $\overline{\lambda}(x^{0},x^{1})$ integration when a
Weyl rescaling is performed to ease the calculations~\cite{Fukuma}. This is the 
reason for the $\dot{l}^{2}/l$ term to appear in equation~(\ref{Fukuma-action}).
Related discussions using the ADM gauge for the metric can be found in~\cite{Nakayama,Ghoroku},
where it is shown that the loop-loop correlator discussed here (with slight changes in the action)
is in agreement with the results derived using matrix models. Note that the trivial cylinder
amplitude in equation~(\ref{cylinder}) resembles the result in equation~(\ref{2-loop-c})
from the canonical quantization of BF theory.

The conclusion one can draw from the above discussion, is that it is indeed possible to
use the canonical formalism to study quantum gravity in two dimensions, but that it is through
the use of the ADM gauge in the Euclidean path integral over metrics, that one finds results
which can be reproduced by other methods. The string field theory approach by Ishibashi and
Kawai~\cite{Ishibashi-Kawai} represents a second quantizaion of string theoy in a hamiltonian
formulation, and not the usual canonical quantization of the space-time metric.

Both approaches illustrate the fact, that two-dimensional quantum gravity reduces to the quantum
mechanics of closed loops of geodesic length $l(x^{0})$, but the quantum version of the
Jackiw-Teitelboim theory is too simple to catch the real contents of quantum gravity.
It is only in the path integral approach, where it is possible for a propagating loop to
split into two loops, or to dissapear, that topology changing amplitudes can be found.

Another way to broaden the question of the role of Ashtekar gravity in two
dimensions, is to study the role of BF gravity in relation to
quantum gravity. This is part of the discussion in the next section.

\section{BF Gravity vs. Super BF Gravity}
When one considers the classical actions for BF and super BF theory 
\begin{eqnarray}
  S_{{\rm BF}}[B,A] &=& \int_{\Sigma_{g}}  tr\left( BF\right), \\
 S_{{\rm S-BF}}[B,A,\chi,\psi] &=& \int_{\Sigma_{g}}  tr\left( B F + \chi D \psi\right),
\end{eqnarray}
it is natural to wonder whether the BF theory is in some way embedded in the more complicated
super BF theory. Especially one can ask whether the observables of BF gravity are possible
observables in super BF gravity? We have learned that topological gravity, which could be formulated
as a super BF theory, is closely related to $2D$ quantum gravity, but at the same time we know that quantum
gravity in three dimensions can be described as a BF theory. From this perspective, it is very
interesting to find a better understanding of the relation between BF and super BF theory, since this
might help understanding the difference of two and three dimensional quantum gravity.

First, we consider the difference between the symmetries of the two theories. The Witten type
theory has the largest possible symmetry group, namely the topological shift symmetry, in
contrast to BF theory
which only has the ordinary (Yang-Mills like) local gauge symmetry. The topological shift symmetry
includes the local gauge symmetry as a subset of the transformation, which leads to the
overcompletenes of the BRST algebra and the introduction of a 2nd generation ghost.
From this point of view, the two theories are
quite different, but they both share the property that the equivalence between the gauge symmetry and
the diffeomorphisms only holds on shell, i.e. when the curvature $F[A]$ is vanishing.

Next, compare the topological invariants that the theories enable us to calculate. In BF theory,
like other Schwarz type theories, one can calculate link invariants as expectation
values of products of Wilson loops, and their generalizations as Wilson surfaces.
These invariants
contain information on the intersection/linking of homology cycles embedded in the space-time
manifold $\Sigma_{g}$. In contrast, super BF theory is directly related to topological invariants
on the moduli space of flat connections. The observables are closed differential forms on moduli space,
which are expressed by local operators in the field theory. These operators carry a ghost number
since they are functionals of the 2nd order ghosts. 

Finally, the role of BRST invariance is different in the two theories. As a direct consequence of the
different symmetries, the Witten type BRST algebra is an enlargement of the one for the Schwarz type
 theory. If one sets the topological ghost $\psi$ and its 2nd order ghost $\phi$ to zero, the
 original Faddeev-Popov BRST algebra is recovered. While it is possible to perform a
 BRST quantization of two-dimensional BF theory, it is not essential due to the simplicity of the
 theory, in contrast to higher dimensional BF theory, which has a symmetry in the $B$ field as well
 as in the $A$ field. The Wilson loop is gauge invariant and therefore also BRST invariant under the
 Faddeev-Popov BRST transformations. The situation is very much
different for super BF theory, which, as a Witten type theory, has BRST invariance and exactness
as its defining properties. The topological invariance of the BRST invariant observables, is due to the BRST exact
nature of the action.
Note that the Wilson loops of BF theory, fail to be BRST invariant in the larger BRST algebra
of super BF theory, due to the topological shift symmetry. From this it is clear that observables of
the $2D$ BF gravity do not form a subset of the observables of super BF gravity.
This result goes both ways, since the observables of super BF gravity are
functionals of fields not present in the simpler BF theory.

One perspective on the difference between the theories, is related to the geometric interpretation
of the BRST operator. Recall the discussion in section~\ref{section:math-behind-phys}, where
the topological ghost $\psi$ extended the Faddeev-Popov BRST transformation of the gauge field
$A$ to the BRST transformation \br
\begin{equation}
  \delta_{{\rm BRST}}\, A^{i} = D c^{i} \rightarrow \br A^{i} = Dc^{i} + \psi^{i}.
\end{equation}
This projection requirement, which is imposed on $\psi$ such that it becomes a horizontal tangent vector to \A,
gave rise to the interpretation of
\br as having a vertical and a horizontal part in equation~(\ref{BRST-split}):
\begin{equation}
  \br = \br^{{\rm horizontal}} + \br^{{\rm vertical}} =
  \dw + \delta_{{\rm BRST}}.
\end{equation}
From this split, the original $\delta_{\rm BRST}$ is identified as the exterior derivative
in the \G direction of the universal bundle, while \dw is the covariant exterior derivative
on \AMG and accordingly on the moduli space of flat connections ${\cal M}_{{\rm F}}$.
We see that the fermionic symmetry $(\dw A = \psi)$ is responsible for isolating the
covariant exterior derivative on moduli space, i.e. for isolating the closed forms on moduli space.
From the ordinary Faddeev-Popov BRST algebra, this information would not be available. {\it Hence we
conclude that it is the fermionic sector of super BF theory, which makes it possible to study
topological invariants on moduli space, instead of topological invariants of space-time.}

It might seem that we can simply ignore the simpler BF theory when discussing $2D$ quantum gravity,
but there are several reasons for taking a harder look at some of the details in the previous
discussions. First we note that BF theory has not been developed 
to answer the same questions as topological gravity. In chapter 3, we initially derived the partition function
of BF gravity as an integral over \m and finally over the reduced phase space ${\cal N}$. These
spaces are non-compact and the integrals are divergent. In the original approach by Chamseddine and
Wyler~\cite{CW-2} the study of the quantum theory was terminated by concluding that the
partition function, in the presence of a nontrivial solution to equation~(\ref{covariant-constant-phi}),
could not be calculated. Here it is of course possible to perform the
compactification of the moduli space $\m \rightarrow \mb$, as in the study of topological
gravity, such that these integrals are finite. In addition, the curvature constraint in equation~(\ref{d-omega}),
can be changed into the form given in equation~(\ref{changed-omega-constraint}), by
introduction of curvature singularities represented by
vertex operators. However the resulting action fails to be gauge invariant due to the
missing action of the diffeomorphisms at the marked points. This was also the case in
topological gravity, but there one still has the topological shift symmetry to ensure
an overall gauge invariance of the action~\cite{Birmingham-review}. Finally, even if one
performed the compactification of \m, to allow for the computation of the partition function
of BF gravity, the formalism would not be able to isolate closed forms on moduli space since
these were the BRST invariant objects with respect to the \dw derivative.
It is clear that the missing role of punctures in BF gravity, is an important part of the failure to
reproduce $2D$ quantum gravity, since these played a vital role for the identifications presented in chapter 5.

We should perhaps mention that several authors have tried to relate BF gravity to $2D$ quantum gravity
through the Liouville equation, present in both theories. We refer the reader to the
papers~\cite{Mohammedi,Berger}, but do not feel that these discussions offer any new
insight. In addition, several authors have tried to map BF theory to a Witten type theory,
by extensions of the BRST algebra. The papers of Brooks et al.\ offer research
along those lines~\cite{Brooks-1,Brooks-2,Brooks-3,Brooks-4}, but we do
not have time to enter these discussions.

Even though it seems impossible to probe the topological invariants on \mbs from
BF theory, there are two results which hint that the Schwarz type theory has some
relation to the Witten type super BF theory anyhow. The first indication comes from the result,
noted by Blau and Thompson in~\cite{Blau-Thompson:ATF}, that the equations of motion for
the fields $(\omega,e)$ of BF gravity in equations~(\ref{d-omega}, \ref{no-torsion-2}):
\begin{eqnarray}
  d  \omega &=& \frac{\Lambda}{2} \epsilon_{ab}e^{a}\wedge e^{b},
  \label{d-omega-disc} \\
  d e^{a} &=& \omega^{a}_{\,\, b}\wedge e^{b}, \label{no-torsion-2-disc}
\end{eqnarray}
can be translated into a set of self-duality equations due to Hitchin~\cite{Hitchin}. A two-dimensional
version of the self-duality equation $F[A^{{\rm H}}]=*F[A^{{\rm H}}]$ for the Hitchin
connections $A^{{\rm H}}$ in a $SO(3)$ bundle over
$\R^{4}$ is isolated by dimensional reduction. Let $F[A^{{\rm H}}]$ be the curvature on a $SO(3)$ bundle
over a Riemann surface $\Sigma_{g}$ and consider a given complex structure on $\Sigma_{g}$.
For this fixed structure, let $D''=d'' + A_{\overline{z}}^{{\rm H}}d\overline{z}$ be the anti-holomorphic
part of the covariant exterior derivative $D$ on $\Sigma_{g}$. Introduce two complex forms,
a holomorphic one-form $\Phi\in \Omega^{(1,0)}$ with values in the adjoint bundle and a
corresponding anti-holomorphic one-form $\Phi^{*}\in\Omega^{(0,1)}$. These complex forms, represent the
$\mu=3,4$ components of $A_{\mu}^{i\, {\rm H}}$, which have been dimensionally reduced from four to two
dimensions. Hitchin's self-duality equations over a Riemann surface are of the form:
\begin{eqnarray}
F[A^{{\rm H}}] &=& -[\Phi,\Phi^{*}], \label{Hitchin-1}
\\
D''\Phi &=&0. \label{Hitchin-2}
\end{eqnarray}
These have been identified with the relevant equations of motion of the twice dimensional reduced action of
$4D$ topological Yang-Mills theory~\cite{KO}, since this action is build to model the solutions to the
same self-duality equations.

The dimensionally reduced connection can be reduced to a $U(1)$ connection, and the composed $PSL(2,\R)$
connection
\begin{equation}
\tilde{A} =  A^{{\rm H}} + \Phi + \Phi^{*},
\end{equation}
can be proved to be flat~\cite{Hitchin}. Donaldson~\cite{Donaldson} has proved the equivalence
between an $SO(2,1)$ connection fulfilling the self-duality equation and a flat $PSL(2,\R)$ connection.
By identifying the complex one-forms with the components of the zwei-bein, one finds that
\begin{equation}
\tilde{A} =  A^{{\rm H}} + P_{a}e^{a}_{z}dz + P_{a}e^{a}_{\overline{z}}d\overline{z}=
J\omega + P_{a}e^{a} = A^{{\rm BF}},
\end{equation}
if one identifies the $U(1)$ connection $A^{{\rm H}}=\omega J$, with the $SO(2)\simeq U(1)$
spin connection. By these identifications, the 
self-duality equations~(\ref{Hitchin-1},\ref{Hitchin-2}) transform to the equations of motion
for the BF theory in equations~(\ref{d-omega-disc},\ref{no-torsion-2-disc}). 
This implies that the two formulations are identical at the level of equations of motion, if we
restrict ourselves to study irreducible connections for which equation~(\ref{d-phi}) has no
nontrivial solutions. This result might indicate that BF gravity could be described as a
$U(1)$ gauge field coupled to a Higgs field $\Phi$, which should be identified with the zwei-bein
as seen above. In this interpretation, the appearance of the space-time metric would be a result of this
coupling, since it is composed from the zwei-beins\footnote{This is the same philosophy as in Ashtekar
  gravity where the $m$-bein is the fundamental object and the metric is an object which is derived from
  the theory~\cite{Ashtekar-book}.}.
It is interesting to note that the relevant sector of super BF gravity, was
encoded in the $U(1)$ sub-sector corresponding
to the homogeneous part of the BRST algebra. This could indicate that by mapping BF gravity into the
self-duality equations, it would be possible to search for a relation to the $U(1)$ sector of
topological gravity. 

Another surprising relation between BF theory and Witten type TFT's, has been proved by Witten in relation
to the study of topological Yang-Mills theory in two dimensions. Witten was able to prove
that the expectation values for certain observables of $2D$ topological Yang-Mills theory, would agree with
the same observables calculated in physical $2D$-Yang-Mills theory, up to exponentially small corrections.
These special observables are constructed from the BF action, plus the symplectic two-form on the moduli
space of flat connections, together with a gaussian term in the $B$ field. The relation between BF theory and
Yang-Mills theory described in section~\ref{section:YM-BF} plays an important role in the proof of
this result. We refer the reader to~\cite{Witten-2dGTR} for details and
just want to stress that the role of the BF action in two-dimensional TFT's is more subtle than would
be expected at first sight.

The relation between Schwarz vs. Witten type TFT's, is a complicated problem from the mathematical
point of view, but some research indicates that the Witten type invariants, known as Casson
invariants~\cite{Birmingham-review,AJ,Witten-2+1:1,Casson}, which are derived from $3D$ super BF theory
with gauge group $SU(2)$, in some complicated way should be included in the most general knot
invariants, which formally are derived from the Schwarz type theories~\cite{jiggle}. This could be
the first step towards a unified interpretation of topological field theories, but no results
have been proven at the present stage.

\section{Perspectives and Conclusions}

In this final section, we discuss two different topics. The first is related to a problem in the
discussion of topological gravity as a super BF theory, which has been overlooked in the literature.
The second is related to new developments in relating the $\tau_{n}$ and $\kappa_{n}$ observables
in topological gravity.
\subsection{A Problem for Super BF Gravity}
Recall the discussion of BF gravity in chapter 3, where we saw that the moduli space of flat
$PSL(2,\R)$ connections, had $4g-3$ connected components ${\cal M}^{k}_{F}$ labelled by the
integer $|k|\leq 2g-2$. The top component $k=2g-2$ corresponded to homomorphisms $\phi$ mapping the
fundamental group $\pi_{1}(\Sigma_{g})$ into $PSL(2,\R)$ as a discrete subgroup, leading to a smooth
Riemann surface. This led to the identification (for $g\geq 2$) between ${\cal M}^{2g-2}_{F}$
and Teichm\"{u}ller space \tich.
The remaining elements of ${\cal M}^{k}_{F}$, for $|k|< 2g-2$ corresponded to
Riemann surfaces with certain conical singularities, which were described by Witten as surfaces with
collapsed handles~\cite{Witten-2+1:1}. These connected
components of  ${\cal M}^{k}_{F}$ should thus correspond to
conformal equivalence classes of singular Riemann surfaces. In the discussion of BF gravity, the
singular contributions were excluded by requiring the zwei-bein to be invertible, in order for the
metric to be positive definite. This requirement was imposed because one makes a similar requirement
in three dimensions where there is an Einstein theory, which the gauge theory should be identified with.

But, when one studies the $2D$ super BF approach to topological gravity, the situation is rather
different, because we inserted curvature singularities on the Riemann surface in order to generalize
the $ISO(2)$ action to arbitrary genus. This introduction led to the notation of punctures, which we
later saw played a vital role in the map of topological gravity to $2D$ quantum gravity. It is clear that
the requirement
\begin{equation}
  {\rm det} \left(e_{\mu}^{\,\,a}(x)\right) \neq 0; \,\,\,\,\,
  \mbox{ for all points } x\in \Sigma_{g},\label{positive}
\end{equation}
will not hold at the nodes on the Riemann surfaces, which appear after the compactification of the
moduli space. Therefore it is no longer possible to exclude all components 
 ${\cal M}^{k}_{F}$ for $|k|<2g-2$, by imposing the constraint of equation~(\ref{positive}).
This problem has been completely overlooked in the literature on topological gravity. The situation can be
illustrated in the following way:
\begin{equation}
  \begin{array}{ccc}
    \mbox{ Gauge Theory }& & \mbox{ $2D$ Quantum Gravity}   \\
    {\cal M}^{k=(2g-2)}_{F}&   \rightarrow &\tich  \\
    {\cal M}^{k< (2g-2)}_{F}& \stackrel{?}{\rightarrow} & \Delta.
  \end{array} \label{missing-link}
\end{equation}
Here $\Delta$ is meant to illustrate the conformal equivalence classes of noded
Riemann surfaces. It is not necessarily a problem that one should include the components
of the moduli space of flat connections with $|k|<2g-2$, when one wishes to connect
super BF theory to topological gravity, since it offers a more natural setting for discussing the
singular geometries. Even though Witten's pictorial description of the singular surfaces, seems
to fit with the elements in $\Delta$, there exists no mathematical proof for such an identification
of the singularities. Actually there are reasons to believe that these singular components are not
related~\cite{Goldman-pc}.

It should be stressed that no concrete realization of the map,
indicated by $\stackrel{?}{\rightarrow}$ in equation~(\ref{missing-link}), ever has been presented.
As the identification between the theories is presented in the literature, the super BF theory is
identified with a gravitational theory for the $k=2g-2$ component of the moduli space of flat
connections, and the remaining components are ignored. After this, the restriction in
equation~(\ref{positive}) is then lifted and the moduli space \m is compactified with the
introduction of the degenerated Riemann surfaces $\Delta = \mbs\setminus \ms$. This
is in the author's opinion not a trustworthy approach and the role of the gauge formulation
of topological gravity must be taken into renewed consideration. As long as the map
$\stackrel{?}{\rightarrow}$ in equation~(\ref{missing-link}), proposed for the first time
in this thesis, is unknown, the identification of the theories is unsatisfactory. There is of course
a lot of truth in the identification, as is hopefully clear after the discussions in
chapter 4 and 5, but one should be careful with saying that the gauge theory is identical to
$2D$ quantum gravity. For this statement to be true, there should exist a well-defined
map between the relevant moduli spaces ${\cal M}_{{\rm F}}$ and \mbs, since they per
  definition represent the physical states of the theory.
 
The above discussion presents an unsolved problem in the gauge theoretical formulation
of quantum gravity in two-dimensions. This should be noted together with the fact that the
gauge theory offers much less computational power than the matrix models, where the
intersection numbers of Riemann surfaces can be calculated at any genus in theory.

\subsection{New Developments in Topological Gravity}

There have been some new developments at the mathematical front of topological
gravity~\cite{Kabanov}, which
relate to the role of the $\tau_{n}$ and $\kappa_{n}$ classes, which were discussed in
chapter 4. The new developments have been to introduce a generating functional for
combined $\langle \tau_{n}\kappa_{m}\rangle$ correlators. In chapter 5, we studied the expansion
of $\sigma_{n} = \tau_{n}n!$ in equation~(\ref{eq:part}), and using a compressed symbolic notation,
we rewrite equation~(\ref{eq:part}) as
\begin{equation}
  \lan 1 \ra_{V} \equiv F(t_{0},t_{1},\cdots) \sim \sum_{\{m_{i}\}}
  \lan\prod_{i=0}^{\infty} \sigma_{i}^{m_{i}} \frac{t^{m_{i}}}{m_{i}!}\ra_{V=0}.
\end{equation}
By the methods used in chapter 5, one can derive a similar expression for the $\kappa_{n}$ classes.
The composit correlator is defined by the integral
\begin{equation}
\lan \prod_{i=0}^{\infty}\sigma_{i}^{m_{i}} \prod_{j=0}^{\infty}\kappa_{j}^{p_{j}} \ra
\equiv \int\limits_{\mbs} c_{1}\left({\cal L}_{1}
\right)^{d_{1}}\wedge
\cdots  \wedge c_{1}\left({\cal L}_{s}\right)^{d_{s}}\wedge \kappa_{1}^{p_{1}}\wedge
  \kappa_{2}^{p_{2}} \wedge
  \cdots,
\end{equation}
where the numbers $\{d_{1},\dots,d_{s}\}$ contains $m_{0}$ zeros, $m_{1}$ ones, etc.
The generating functional for these invariants
depends on two sets of coupling constants $\{ t_{i} \}$ and $\{ s_{j} \}$, and it is formally
defined as:
\begin{equation}
  H(t;s) \equiv \sum_{\{m_{i}\},\{p_{j}\}} \lan
  \prod_{i=0}^{\infty}\sigma_{i}^{m_{i}} \prod_{j=0}^{\infty}\kappa_{j}^{p_{j}}
  \ra  \prod_{i=0}^{\infty}\frac{t^{m_{i}}}{m_{i}!}\prod_{j=0}^{\infty}
  \frac{s^{p_{j}}}{p_{j}!}.
\end{equation}
This expression has been introduced in~\cite{Kabanov}, where it is also shown that
there exist recursion relations relating 
these composite correlators with differential operators, which annihilate the function
$\exp(H(t;s))$. These operators led to generalizations of the Virasoro constraints
$L_{n}\tau=0$, which characterised the tau-function of the KdV hierarchy and they
offer a possible generalization of the whole framework, which ties topological gravity and
$2D$ quantum gravity together. It is still to early too say, what applications these
new results may have in the study of $2D$ quantum gravity.

\subsection{Final Remarks}

Several topics could have been included in this thesis, but have been left out in order to
save space-time. These include among other topics, Atiyah's axioms for topological field
theories~\cite{Atiyah}, and the role of twisted supersymmetry in topological gravity~\cite{thesis}.
This is in no way a measure of these topics being less interesting than those discussed in this
thesis.

The author hopes, that this presentation has illustrated the important and exciting
role topological field theories and
topological invariants have played for our understanding of lower-dimensional quantum gravity.

%% file: appendix1.tex
\appendix
\chapter{Conventions}
We list some useful definitions and conventions used in the thesis.
\begin{description}
  \item{{\bf Convention}:} The space-time signature is taken to be Euclidean unless  otherwise specified.
  \item{{\bf Convention}:} We use the notation of graded forms, which means that we do not
    explicitly distinguish between bosonic and fermionic variables. In general
    all forms $(X_{p},Y_{q})$ have a grading, i.e. a grassmann parity,
    of grade $(p,q)$, such that
    \begin{equation}
      X_{p}Y_{q} = (-1)^{pq} Y_{q}X_{p}.
    \end{equation}
    The graded commutator is defined such that
    \begin{equation}
      \biggl[ X_{p}, Y_{q}\biggr] = X_{p}Y_{q} + (-1)^{pq} Y_{q}X_{p},
    \end{equation}
    is the ordinary commutator if $pq$ are even, and the
    anti-commutator if $p$ and $q$ are odd. If the forms both have a form degree $p_{1},q_{1}$ and a
    grasmann parity $p_{2},q_{2}$, we take $p=p_{1}+p_{2}$ and $q=q_{1}+q_{2}$ above.
\end{description}
We list some basic definitions of differential forms on an
  $m$ dimensional manifold $M$:

 \begin{description}
\item{{\bf Definition}:}
A differential form of order $r$ is a totally antisymmetric covariant 
tensor of rank $r$.
\end{description}
\begin{description}
\item{{\bf Definition}:} 
The wedge product $\wedge$ is the totally antisymmetric tensor pro\-duct.
\begin{equation}
  dx^{\mu} \wedge dx^{\nu} = dx^{\mu} \otimes dx^{\nu} - dx^{\nu} \otimes dx^{\mu} .
\end{equation}
\end{description}
\begin{description}
\item{{\bf Definition:}} Denote the vector space of $r$-forms at $p\in M$ by 
$\Omega_{p}^{r}(M)$. This space will be spanned by a basis of one-form wedge products
\begin{equation}
dx^{\mu_{1}} \wedge dx^{\mu_{2}}\wedge \cdots \wedge dx^{\mu_{r}}  = 
\sum_{ \mbox{{\scriptsize permutations}}\, P} \mbox{sign}(P) dx^{\mu_{P(1)}}
\otimes dx^{\mu_{P(2)}} 
\otimes \cdots \otimes dx^{\mu_{P(r)}},
\end{equation}
\end{description}
\noindent
and a general element $\omega \in \Omega_{p}^{r}(M)$ is expanded as 
\begin{equation}
\omega = \frac{1}{r!} \omega_{\mu_{1} \mu_{2}\cdots \mu_{r}}dx^{\mu_{1}}\wedge \cdots 
\wedge dx^{\mu_{r}}. \label{def:r-form}
\end{equation}
\noindent
where $\omega_{\mu_{1} \mu_{2}\cdots \mu_{r}}$ is totally antisymmetric, just 
as the basis.

\noindent
The dimension of $\Omega_{p}^{r}(M)$ is 
\begin{equation}
dim(\Omega_p^r(M))= \left( \begin{array}{c} m \\ r \end{array} 
\right), 
\end{equation}
because we should choose the coordinates $(\mu_1,\mu_2,\cdots,\mu_r)$ out 
of the set $(1,2,\cdots,m)$, when creating the basis. By the relation for Binomial coefficients
\begin{equation}
\left( \begin{array}{c} m \\ r \end{array} 
\right) =  \left( \begin{array}{c} m \\ m-r \end{array} 
\right),
\end{equation}
we see that 
\begin{equation}
dim(\Omega_p^r(M)) = dim(\Omega_p^{m-r}(M)),
\end{equation}
and from the theory of vector spaces we know that the two spaces are isomorphic. 
By definition $\Omega_p^0(M) \equiv \R$.
\begin{description}
\item{{\bf Definition}:} The exterior product $\wedge$ of a $q-$form and an $r-$form 
\begin{equation}
\wedge: \Omega_{p}^{q}(M) \times  \Omega_{p}^{r}(M) \mapsto \Omega_{p}^{q+r}(M)
\end{equation}
enables us to define the algebra 
\begin{equation}
\Omega_{p}^{*}(M) \equiv \Omega_{p}^{0}(M)\oplus \Omega_{p}^{1}(M)\oplus \cdots 
\oplus \Omega_{p}^{m}(M).
\end{equation}
Thus $\Omega_{p}^{*}(M)$ is the space of all differential forms at $p$ and it 
is closed under the exterior product.
\end{description}
\begin{description}
\item{{\bf Definition:}} The exterior derivative is a map $d_r: \Omega_{p}^{r}(M)\mapsto 
\Omega_{p}^{r+1}(M)$ whose action on an $r-$form $\omega$ reads 
\begin{equation}
d_{r} \omega = \frac{1}{r!} \left( \frac{\partial}{\partial x^{\nu}}
\omega_{\mu_{1} \mu_{2}\cdots \mu_{r}}\right)dx^{\nu}\wedge dx^{\mu_{1}}\wedge \cdots 
\wedge dx^{\mu_{r}}.
\end{equation}
\end{description}
Usually one drops the subscript $r$ and writes $d$ as the exterior derivative. 
A very important result is that $d^2 = 0$, which is proven by acting twice with $d$ on  
$\omega$ above and we find
\begin{equation}
d(d \omega) = \frac{1}{r!} \left( \frac{\partial^2}{\partial x^{\rho}x^{\nu}}
\omega_{\mu_{1} \mu_{2}\cdots \mu_{r}}\right)dx^{\rho}\wedge dx^{\nu}\wedge dx^{\mu_{1}}
\wedge \cdots \wedge dx^{\mu_{r}},
\end{equation}
but since the partial derivative is symmetric in $\rho,\nu$ the contraction with the 
total antisymmetric basis vanishes. Thus $d^{2}=0$.
\begin{description}
\item{{\bf Definition:}} If $M$ is endowed with a metric $g_{\alpha\beta}$, 
one can define a 
natural isomorphism between $\Omega_p^r$ and $\Omega_p^{m-r}$ called the Hodge star 
$(*)$ operation. This is a map $* : \Omega_p^r \mapsto\Omega_p^{m-r}$ with the 
following action on a basis vector of $ \Omega_p^{r}$
\begin{equation}
*(dx^{\mu_{1}} \wedge \cdots \wedge dx^{\mu_{r}}) = 
\frac{\sqrt{\mid g \mid }}{(m-r)!} \epsilon^{\mu_1 \cdots \mu_r}_{\nu_{r+1}\cdots 
\nu_{m}} dx^{\nu_{r+1}}\wedge \cdots \wedge dx^{\nu_{m}}.
\end{equation}
Here $g$ is the determinant of the metric. 

Let $\omega$ be an $r$-form, defined as in equation~(\ref{def:r-form}), then
\begin{equation}
  * \omega = \frac{\sqrt{g}}{r!(m-r)!}\omega_{\mu_{1}\mu_{2}\dots\mu_{r}}
  \epsilon^{\mu_{1}\mu_{2}\dots\mu_{r}}_{\nu_{r+1}\dots\nu_{m}} dx^{\nu_{r+1}}\wedge
  \cdots \wedge dx^{\nu_{m}}.
\end{equation}
In Euclidean space one has the result:
\begin{equation}
  ** \omega = (-1)^{r(m-r)}\omega\,;\,\,\,\,\, \omega\in \Omega^{r}(M). \label{**}
\end{equation}
\end{description}
In order to perform integration on a manifold, we need to 
have a substitute for the usual Lebesgue measure. We introduce the notation of 
a volume element, which is an everywhere non-vanishing $m-$form, known to exist for 
all orientable $m$ dimensional manifolds $M$. This volume form $\omega$ plays 
the role of the measure. Consider the $m-$form 
\begin{equation}
\omega = h(p) dx^1 \wedge \cdots \wedge dx^m,
\end{equation}
where $h(p)$ is positive definite over all of $M$. Strictly speaking this is valid on 
a chart $(U,\phi)$ with coordinates $x=\phi(p)$, but $\omega$ can be extended to all of 
$M$ such that $h(p)$ is positive definite in any chart on $M$. The form $\omega$ is 
a volume element and it is unique up to the choice of orientation and normalisation.
\begin{description}
\item{{\bf Definition}:}
Integration of a function $f:M\mapsto \R$ is defined in a neighbourhood $U_i$ 
with coordinates $x$ as
\begin{equation}
\int_{U_i} f\omega = \int_{\phi(U_i)} f(\phi_{i}^{-1}(x))h(\phi_{i}^{-1}) 
dx^1\cdots dx^m.
\end{equation}
By partition of unity the integration is taken from a single $U_i$ to all of $M$. 
\end{description}
\begin{description}
\item{{\bf Definition:}} In the presence of a metric $g_{\alpha \beta}$ there exists a 
natural volume element, which is invariant under coordinate transformations
\begin{equation}
\Omega_M = \sqrt{\mid g\mid} dx^1\wedge\cdots\wedge dx^m.
\end{equation}
The volume form equals the action of the $*$ operation on $1$ and is 
\begin{equation}
 \sqrt{\mid g\mid} dx^{1} \wedge \cdots \wedge dx^{m} 
= \frac{\sqrt{\mid g\mid}}{m!}\epsilon_{\mu_1\cdots\mu_m}dx^{\mu_1}\wedge\cdots 
\wedge dx^{\mu_{m}} = * 1 .
\end{equation}
\end{description}
\begin{description}
\item{{\bf Definition:}}
Integration on $M$ with a metric structure is defined by use of $\Omega_M$
\begin{equation}
\int_M f \Omega_M = \int_M f \sqrt{\mid g\mid} dx^1dx^2\cdots dx^m,
\end{equation}
which is invariant under change of coordinates.
\end{description}
\begin{description}
\item{{\bf Definition:}} A differential form $\omega$ is said to be 
closed if $d\omega = 0$. An $r-$form $\omega$ is said to be exact 
if there exists an $r-1$ form $\rho$ such that $\omega = d \rho$.
\end{description}
Since $d^2=0$ we have that ${\rm Im}\, d_r \subset {\rm Ker}\, d_{r+1}$, for example
take $\omega \in \Omega_p^r(M)$. Because $d_r \omega \in {\rm Im}\, d_r$ and
$d_{r+1}(d_r\omega) =0$ implies $d_r \omega \in {\rm Ker}\, d_{r+1}$ we conclude
that ${\rm Im}\, d_r \subset {\rm Ker}\, d_{r+1}$.
\begin{description}
\item{{\bf Definition}:} The quotient space
  \begin{equation}
  H^r(M) = {\rm Ker} \, d_{r} / {\rm Im} \, d_{r-1},
  \end{equation}
  is called the $r$'th de Rahm cohomology group. 
\item{{\bf Definition:}} The adjoint exterior derivative $d^{\dagger}:\Omega^{r}(M)\mapsto \Omega^{r-1}(M)$,
is  defined in Euclidean signature as
  \begin{equation}
    d^{\dagger} \equiv (-1)^{mr+m+1} *d*.
  \end{equation}
  It follows from equation~(\ref{**}) that $(d^{\dagger})^{2} = 0$.
\item{{\bf Definition:}} The Laplacian $\Delta:\Omega^{r}(M)\mapsto\Omega^{r}(M)$ is defined by
  \begin{equation}
    \Delta \equiv \left( d+d^{\dagger}\right)^{2} = dd^{\dagger} + d^{\dagger}d.
   \end{equation}
\end{description}